\documentclass[iop, twocolumn, appendixfloats]{emulateapj}
\usepackage{apjfonts}
\usepackage{natbib}
\usepackage{amsfonts}
\usepackage{amsmath}
\usepackage[usenames]{color}
\usepackage{ulem}
\usepackage{verbatim}
\usepackage{longtable}
\usepackage{array}
\usepackage{epsfig}
\usepackage{scrextend}
\usepackage{mathrsfs}
\usepackage{rotating}
\newcommand\msun{{M_\odot}}
\newcommand\lsun{{L_\odot}}
\newcommand\rsun{{R_\odot}}
\newcommand\logage{\rm \log(Age)\;[yr]}
\newcommand\teff{T_{\rm eff}}
\newcommand\mbol{M_{\rm bol}}
\newcommand\ergs{\rm{erg~s^{-1}}}
\newcommand\amlt{\alpha_{\rm MLT}}
\newcommand\msunperyr{M_\odot~\rm{yr^{-1}}}

\shorttitle{}
\shortauthors{}

\begin{document}

\title{MESA Isochrones and Stellar Tracks (MIST). I: Solar-scaled models}
\author{Jieun Choi\altaffilmark{1}}
\author{Aaron Dotter\altaffilmark{2}}
\author{Charlie Conroy\altaffilmark{1}}
\author{Matteo Cantiello\altaffilmark{3}}
\author{Bill Paxton\altaffilmark{3}}
\author{Benjamin D. Johnson\altaffilmark{1}}

\altaffiltext{1}{Harvard-Smithsonian Center for Astrophysics, Cambridge, MA 02138, USA}
\altaffiltext{2}{Research School of Astronomy and Astrophysics, The Australian National University, Weston Creek, ACT 2611, Australia}
\altaffiltext{3}{Kavli Institute for Theoretical Physics, University of California, Santa Barbara, CA 93106, USA}

\begin{abstract}
This is the first of a series of papers presenting the Modules for Experiments in Stellar Astrophysics (MESA) Isochrones and Stellar Tracks (MIST) project, a new comprehensive set of stellar evolutionary tracks and isochrones computed using MESA, a state-of-the-art open-source 1D stellar evolution package. In this work, we present models with solar-scaled abundance ratios covering a wide range of ages ($5 \leq \logage \leq 10.3$), masses ($0.1 \leq M/\msun \leq 300$), and metallicities ($-2.0 \leq \rm [Z/H] \leq 0.5$). The models are self-consistently and continuously evolved from the pre-main sequence (PMS) to the end of hydrogen burning, the white dwarf cooling sequence, or the end of carbon burning, depending on the initial mass. We also provide a grid of models evolved from the PMS to the end of core helium burning for $-4.0 \leq \rm [Z/H] < -2.0$. We showcase extensive comparisons with observational constraints as well as with some of the most widely used existing models in the literature. The evolutionary tracks and isochrones can be downloaded from the project website at http://waps.cfa.harvard.edu/MIST/.
\end{abstract}

\maketitle

\tableofcontents
\clearpage

\section{Introduction}
Stars are ubiquitous objects. Individually, they are both hosts to rich exoplanet systems and progenitors of some of the most spectacular transients in the distant Universe. As an ensemble, they are both cosmic engines that transformed the state of the early Universe as well as fossils bearing clues of galaxy formation and evolution. The interpretation of these systems and phenomena hinges on our understanding of stellar physics and well-calibrated stellar evolution models across a wide range of masses, metallicities, and evolutionary states.

Decades since the golden era of stellar astrophysics \citep[e.g.,][]{Burbidge1957, BohmVitense1958, Schwarzschild1958, Henyey1964, Paczynski1970}, the field has enjoyed a renaissance in recent years, largely due to technological advances in both computing and observational astronomy. Improvements in computers and numerical algorithms have resulted in a tremendous speedup in solving the nonlinear, coupled differential equations of stellar structure and evolution. Another important factor was the availability of increasingly precise tabulated opacities, nuclear reaction rates, and equations of state. Accordingly, a large number of stellar evolution models have been published to tackle a wide variety of problems in astrophysics. Studies of old, low-mass stellar populations in globular clusters and quiescent galaxies have relied on models such as BaSTI \citep{Pietrinferni2004}, DSEP \citep[Dartmouth;][]{Dotter2008}, GARSTEC \citep{Weiss2008}, Lyon \citep{Baraffe1998, Baraffe2003, Baraffe2015}, Padova/PARSEC \citep{Girardi2002, Marigo2008, Bressan2012}, Y$^2$ \citep{Yi2001, Kim2002, Yi2003, Demarque2004}, Victoria-Regina \citep{VandenBerg2006}, and more. On the other hand, studies of young, massive stellar populations in clusters and star-forming environments have made use of e.g., Geneva \citep{Ekstrom2012, Georgy2013}, STARS \citep{Eggleton1971, Pols1995, Eldridge2004}, and STERN \citep{Brott2011, Koehler2015} stellar evolution models. 

On the observational front, large quantities of precise data from recent and ongoing space- and ground-based programs have initiated an explosive growth in fields such as asteroseismology (e.g., {\it CoRoT}; \citealt{Baglin2006}, {\it Kepler}; \citealt{Gilliland2010}), time-domain astronomy (e.g., {\it Palomar Transient Factory}; \citealt{Law2009}, {\it Pan-STARRS}; \citealt{Kaiser2010}), galactic archaeology (e.g., {\it APOGEE}; \citealt{Zasowski2013}, {\it GALAH}; \citealt{DeSilva2015}), and resolved stellar populations (e.g., the {\it Hubble Space Telescope (HST)} program {\it PHAT}; \citealt{Dalcanton2012}). Moreover, future surveys, e.g., {\it LSST} \citep{Ivezic2008}, and next-generation observatories such as {\it JWST} and {\it Gaia} will provide an unprecedented volume of high-quality data whose analysis demands uniform models covering all relevant phases of stellar evolution.

In order to fully exploit these new and upcoming datasets, we have set out to construct stellar evolution models within a single, self-consistent framework using Modules for Experiments in Stellar Astrophysics \citep[MESA;][]{Paxton2011, Paxton2013, Paxton2015}, a popular open-source 1D stellar evolution package.\footnote{There exists another database of isochrones computed also using MESA. See \cite{Zhang2013} for more details.} MESA is well-suited for this purpose due to its flexible architecture and its capability to self-consistently model the evolution of different types of stars to advanced evolutionary stages in a single continuous calculation. Furthermore, its thread-safe design enables parallel computing, which greatly reduces the computation time and makes the large-scale production of models feasible.

This is the first of a series of papers presenting MESA Isochrones and Stellar Tracks (MIST), a new set of stellar evolutionary tracks and isochrones. In this paper, we present models with solar-scaled abundances covering a wide range of ages, masses, phases, and metallicities computed within a single framework. We will subsequently present models with non-scaled-solar abundances, including e.g., alpha-enhanced, carbon-enhanced metal-poor, and CNONa-enhanced for modeling globular clusters, in Paper II.

The paper is organized as follows. Section~2 gives an overview of MESA, focusing on the high-level architecture of the code and its time step and spatial mesh controls. Section~3 reviews the treatment of the relevant physical processes as implemented in a 1D framework, all of which is summarized in Table~\ref{tab:param_summary}. Calibration of the model against the properties of the Sun is discussed in Section 4, and a short overview of the model outputs, the method for constructing isochrones, and the treatment of bolometric corrections is presented in Section~5. Section~6 presents an overview of the properties of the models, and Section~7 features comparisons with some of the most widely used existing databases. Sections~8 and 9 separately present comparisons with data for low-mass ($\lesssim 10~\msun$) and high-mass stars ($\gtrsim 10~\msun$). Finally, Section~10 concludes the paper with a discussion of caveats and future work. 

We define some conventions and assumptions adopted in the paper. We use $M_{\rm i}$ throughout to refer to the initial stellar mass of a model. Both $Z$ and $\rm [Z/H]$ are used to refer to metallicities, where $Z$ is the metal mass fraction. For the models presented in this paper, $\rm [Z/H]~=~[Fe/H]$ since we adopt solar-scaled abundances. The $\rm [Z/H]$ notation assumes the \cite{Asplund2009} protosolar birth cloud bulk metallicity, not the current photospheric metallicity, as the reference value (see \ref{section:abun} for more details). Magnitudes are quoted in the Vega system unless noted otherwise. Where necessary, we adopt a Kroupa initial mass function (IMF) \citep{Kroupa2001}. Lastly, in accordance with the XXIX$^{\rm th}$ IAU resolutions B2 and B3,\footnote{http://astronomy2015.org/sites/default/files/IAU\_XXIX\_GA\_Final\_Resolutions\_B1-B4.pdf} we adopt the following two conventions. First, we use the following nominal values to express stellar properties in solar units: $\msun = 1.988 \times 10^{33}$~g, $\lsun= 3.828 \times 10^{33}$~erg\;s$^{-1}$, $\rsun=6.957\times10^{10}$~cm, and $T_{\rm eff,\;\odot}=5772$~K. Formally, the IAU published these values in SI units, but we report them in cgs units here to be consistent with the convention adopted in this work. Second, the zero point of the absolute bolometric magnitude scale is set by enforcing that $\mbol=0$, which corresponds exactly to $L_{\circ}=3.0128\times10^{35}$~erg\;s$^{-1}$. This zero point was chosen such that the nominal solar luminosity $\lsun$ has an $M_{\rm bol,\;\odot}=4.74$~mag, a value commonly adopted in the literature. 

\section{The MESA Code}
\label{section:MESA}
MESA\footnote{http://mesa.sourceforge.net/} is an open-source stellar evolution package that is undergoing active development with a large user base worldwide \citep{Paxton2011, Paxton2013, Paxton2015}. Its 1D stellar evolution module, MESAstar, has been thoroughly tested against existing stellar evolution codes and databases, including BaSTI/FRANEC \citep{Pietrinferni2004}, DSEP \citep[Dartmouth;][]{Dotter2008}, EVOL \citep{Bloecker1995, Herwig2004a, Herwig2004b}, GARSTEC \citep{Weiss2008}, Lyon \citep{Baraffe1998, Baraffe2003, Baraffe2015}, KEPLER \citep{Heger2005}, and STERN \citep{Petrovic2005, Yoon2005, Brott2011}. Its highly modular and therefore flexible infrastructure combined with its robust numerical methods enable its application to a wide range of problems in computational stellar astrophysics, from asteroseismology to helium core flash in low-mass stars, as well as the evolution of giant planets, accreting white dwarfs (WD), and binary stars.

MESAstar simultaneously solves the fully coupled Lagrangian structure and composition equations using a Newton-Raphson solver. The required numerics (e.g., matrix algebra, interpolation) and input physics (e.g., opacity, mass loss) are organized into individual thread-safe ``modules,'' each of which is an independently functional unit that generates, tests, and exports a library to the general MESA libraries directory. This modular structure is unique among stellar evolution programs. One of its main advantages is that experimentation with different available physics or even implementation of new physics is easy and straightforward. The user input is given at runtime via the {\tt \small inlist} file, which contains the user's choice for parameters for input physics, time step, mesh, and output options. The {\tt \small run\_star\_extras.f} file, a Fortran module that is compiled at runtime, allows the user to introduce new routines that hook into the source codes in order to adapt MESA to the problem of interest. Examples include the introduction of new physics routines, modification of model outputs, and customization of time step adjustments.

Here we provide an overview of some of the features in MESAstar, namely its time step controls, adaptive mesh refinement, and parallelization. We refer the reader to \cite{Paxton2011, Paxton2013, Paxton2015} for more detailed information.

Choosing an appropriate time step throughout various stages of stellar evolution is critical to the accurate evolution of a model. It must be both sufficiently small to allow convergence and sufficiently large to carry out evolution calculations in a reasonable amount of time. In MESA, a new time step is first proposed using a scheme based on digital control theory \citep{Soederlind2006}. Next, the proposed time step undergoes a series of tests to check if the hypothetical changes to various properties of the model (e.g., nuclear burning rate, luminosity, central density) in a single time step are adequately small, as excessively large changes can lead to convergence or accuracy issues in subsequent evolution.

At the beginning of each evolution step, MESAstar checks whether or not a spatial mesh adjustment is necessary. Similar to time step controls, there is a trade-off between having sufficiently small cells to properly resolve physical processes locally while avoiding an unnecessarily fine grid that is computationally expensive. Remeshing involves the splitting and merging of cells, and each remesh plan aims to minimize the number of cells affected in order to reduce numerical diffusion and improve convergence. At the same time, the remesh plan must meet the criteria for the tolerated cell-to-cell changes in relevant variables, which are specifiable by the user in addition to the basic variables, e.g., mass, radius, and pressure. For instance, cells near a convective boundary might be split in order to better resolve its location, while cells well within the convective zone might be merged if they are sufficiently similar in, e.g., composition and temperature. We refer the reader to the appendix for temporal and spatial convergence tests.

MESA is optimized for parallelization and uses OpenMP\footnote{http://openmp.org/wp/} to carry out computations in parallel. Since \cite{Paxton2011}, MESA's performance has been greatly improved mainly due to the implementation of a new linear algebra solver---the single-most computationally expensive component---that is compatible with multicore processing. As demonstrated in Figure 48 of \cite{Paxton2013}, many key components of MESAstar, such as the linear algebra solver and the evaluation of the nuclear reaction network, closely obey the ideal scaling law.

For this work, we use MESA version v7503 compiled with GNU Fortran version 4.9.3 installed as part of the MESA SDK version 245.\footnote{http://www.astro.wisc.edu/~townsend/static.php?ref=mesasdk}

\section{Adopted Physics}
\label{section:adopted_physics}
In this section we review the relevant physics adopted in the models and their implementation in MESA. Readers who are interested in the most salient points can skip to Table~\ref{tab:param_summary}, which presents a summary of the adopted physics. For the effects of varying some key physics ingredients, see Section~\ref{section:nonstandard_models}.

\subsection{Abundances}
\label{section:abun}
The accurate determination of solar abundances has been an ongoing effort for decades. Within the last decade, there has been a systematic downward revision of the solar metallicity from $Z_{\odot}\sim0.02$ \citep[e.g.,][]{Anders1989} to $Z_{\odot}\lesssim0.015$ \citep[e.g.,][]{Asplund2009, Lodders2009, Caffau2011}. Recently, the abundances of C, N, O, and Ne have experienced dramatic revisions as a collective result of improved atomic and molecular linelists and the introduction of 3D non-local thermodynamic equilibrium (NLTE) hydrodynamical modeling techniques (see \citealt{Asplund2009} for a review). In this paper, we adopt the {\it protosolar} abundances recommended by \cite{Asplund2009} as the reference scale for all metallicities, unless noted otherwise. In other words, $\rm [Z/H]$ is computed with respect to $Z = Z_{\odot,\; \rm protosolar} = 0.0142$, not $Z = Z_{\odot,\; \rm photosphere} = 0.0134$. The difference between the two is a consequence of diffusion of heavy elements out of the photosphere over time. We emphasize that this is not an attempt to redefine $Z_{\odot,\; \rm photosphere}$---the current photospheric abundances are determined by spectroscopy---but rather to clarify what a ``solar metallicity model'' entails.

To compute $X$ and $Y$ for an arbitrary $Z$, we use the following formulae:
\begin{eqnarray}
Y_{\rm p} &=& 0.249 \\
Y &=& Y_{\rm p} + \left (\frac{Y_{\odot,\;\rm protosolar}-Y_{\rm p}}{Z_{\odot,\;\rm protosolar}} \right) Z \\ 
X &=& 1-Y-Z \;\;.
\end{eqnarray}
This approach adopts a primordial helium abundance $Y_{\rm p}=0.249$ \citep{Planck2015} determined by combining the {\it Planck} power spectra, {\it Planck} lensing, and a number of ``external data'' such as baryonic acoustic oscillations. In the above equations, we assume a linear enrichment law to the protosolar helium abundance, $Y_{\odot,\;\rm protosolar}=0.2703$ \citep{Asplund2009}, such that $\Delta Y/\Delta Z = 1.5$. Once $Y$ is computed for a desired value of $Z$, $X$ is trivial to compute. We assume the isotopic ratios $\rm ^2H/^1H=2\times10^{-5}$ and $\rm ^3He/^4He=1.66\times10^{-4}$ from \cite{Asplund2009}.

\begin{table*}
\centering
\begin{longtable}{l l l l} \\ 
\caption{Summary of the Adopted Physics.} \\
\hline
\hline \noalign{\smallskip}
Ingredient & Adopted Prescriptions and Parameters & Section & Reference \\
\hline \noalign{\smallskip}
Solar Abundance Scale & $X_{\odot} = 0.7154$, $Y_{\odot} = 0.2703$, $Z_{\odot} = 0.0142$ & \ref{section:abun} & \cite{Asplund2009} \\
Equation of State & OPAL+SCVH+MacDonald+HELM+PC & \ref{section:eos} & \cite{Rogers2002}; \\
\; & \; & \; & \cite{Saumon1995}; \\
\; & \; & \; & \cite{MacDonald2012} \\
Opacity & OPAL Type I for $\log T \gtrsim 4$; Ferguson for $\log T \lesssim 4$; & \ref{section:kappa} & \cite{Iglesias1993, Iglesias1996}; \\
\;& Type I $\rightarrow$ Type II at the end of H\;burning & \; & \cite{Ferguson2005} \\
Reaction Rates & JINA REACLIB & \ref{section:nuclear} & \cite{Cyburt2010} \\
Boundary Conditions & ATLAS12; $\tau=100$ tables + photosphere tables + gray atmosphere & \ref{section:boundary_conditions} & \cite{Kurucz1970}, \cite{Kurucz1993}\\
Diffusion & Track five ``classes'' of species; MS only & \ref{section:diffusion} & \cite{Thoul1994}; \\
\; & \; & \; & \cite{Paquette1986} \\
Radiation Turbulence & $D_{\rm RT} = 1$ & \ref{section:diffusion} & \cite{Morel2002} \\
Rotation & solid-body rotation at ZAMS with $v_{\rm ZAMS}/v_{\rm crit} = \Omega_{\rm ZAMS}/\Omega_{\rm crit} = 0.4$ & \ref{section:rotation} & \cite{Paxton2013} \\
Convection: Ledoux + MLT & $\alpha_{\rm MLT} = 1.82$, $\nu=1/3$, $y=8$ & \ref{section:convection} & \cite{Henyey1965} \\
Overshoot & time-dependent, diffusive, $f_{\rm ov,\;core} = 0.0160$, $f_{\rm ov,\;env} = f_{\rm ov,\;sh} = 0.0174$ & \ref{section:overshoot} & \cite{Herwig2000} \\
Semiconvection & $\alpha_{\rm sc} = 0.1$ & \ref{section:sc_th} & \cite{Langer1983} \\
Thermohaline & $\alpha_{\rm th} = 666$ & \ref{section:sc_th} & \cite{Ulrich1972}; \\
\; & \; & \; & \cite{Kippenhahn1980} \\
Rotational Mixing & Include SH, ES, GSF, SSI, and DSI with $f_\mu = 0.05$ and $f_c = 1/30$  & \ref{section:rot_mixing} & \cite{Heger2000} \\
Magnetic Effects & Not currently implemented & \ref{section:magnetic} & \\
Mass Loss: Low Mass Stars & $\eta_{\rm R} = 0.1$ for the RGB & \ref{section:mass_loss_lm} & \cite{Reimers1975} \\
\; & $\eta_{\rm B} = 0.2$ for the AGB & \; & \cite{Bloecker1995} \\
Mass Loss: High Mass Stars & $\eta_{\rm Dutch} = 1.0$ & \ref{section:mass_loss_hm} & \cite{Vink2000, Vink2001} \\
\; & a combination of wind prescriptions for hot and cool stars & \; & \cite{Nugis2000} \\
\; & and a separate WR wind prescription & \; & \cite{deJager1988} \\
Mass Loss: Rotational & $\xi = 0.43$, boost factor capped at $10^4$, $\dot{M}_{\rm max} = 10^{-3}~\msun~{\rm yr^{-1}}$ & \ref{section:rot_mass_loss} & \cite{Langer1998} \\
\noalign{\smallskip}
\hline \hline
   \label{tab:param_summary}
\end{longtable}
\end{table*}

\subsection{Microphysics}
\label{section:microphysics}
\subsubsection{Equation of State (EOS)}
\label{section:eos}
The EOS tables in MESA are based on the OPAL EOS tables \citep{Rogers2002}, which smoothly transition to the SCVH tables \citep{Saumon1995} at lower temperatures and densities. The extended MESA EOS tables cover $X=0.0, 0.2, 0.4, 0.6, 0.8$, and 1.0, and $Z = 0.0, 0.02$, and 0.04. At higher metallicities, MESA switches to the MacDonald EOS tables \citep{MacDonald2012} for $Z = 0.2$ and 1.0, which, unlike the HELM EOS tables \citep{Timmes2000} used in the earlier versions of MESA, allow for partially ionized species. The HELM and PC tables \citep{Potekhin2010} are used at temperatures and densities outside the range covered by the OPAL + SCVH + MacDonald tables, and assume full ionization. The EOS tables in MESA also cover the late stages of WD cooling, during which the ions in the core crystallize, although the current MIST models do not reach such conditions. The low-mass models are evolved until $\Gamma$, the central plasma interaction parameter or Coulomb coupling parameter, reaches 20 (see Section~\ref{section:ages_masses_phases_metallicities}), and crystallization occurs at $\Gamma \approx 175$ for pure oxygen.

\subsubsection{Opacities}
\label{section:kappa}
MESA divides the radiative opacity tables into two temperature regimes, high ($\log T \gtrsim 4$) and low ($\log T \lesssim 4$), and treats them separately. This system allows for the user to choose, for the low temperature opacities, between either \cite{Ferguson2005} or \cite{Freedman2008} with updates to ammonia opacity from \cite{Yurchenko2011} and the pressure-induced opacity for molecular hydrogen from \cite{Frommhold2010}. The high temperature opacity tables come from either OPAL \citep{Iglesias1993, Iglesias1996} or OP \citep{Seaton2005}. The OPAL tables are split into two types, Type I and Type II: Type I tables are used for $0.0\leq X \leq 1.0-Z$ and $0.0 \leq Z \leq 0.1$ for a fixed abundance pattern; Type II tables are optionally available which allow for enhanced carbon and oxygen abundances in addition to those already accounted for in $Z$, covering $0.0 \leq X \leq 0.7$ and $0.0 \leq Z \leq 0.1$. Type II opacities are particularly important for helium burning and beyond. The electron conduction opacity tables are originally based on \cite{Cassisi2007}, but they have been extended to cover temperatures up to $10^{10}$~K and densities up to $10^{11.5}~{\rm g\;cm^{-3}}$ \citep{Paxton2013}.

We use the \cite{Ferguson2005} low temperature tables and the OPAL Type I tables, then gradually switch to the OPAL Type II opacities starting at the end of hydrogen burning, smoothly interpolating between $X<10^{-3}$ and $X<10^{-6}$. Note that we use the \cite{Asplund2009} protosolar mixture where available to be consistent with our choice of the solar abundance scale, but the opacity tables implemented in MESA were computed for the \cite{Asplund2009} photospheric abundances.

\subsubsection{Nuclear Network}
\label{section:nuclear}
We import the nuclear reaction rates directly from the JINA REACLIB database,\footnote{10/2015; https://groups.nscl.msu.edu/jina/reaclib/db/} a compilation of the latest reaction rates in the literature \citep{Cyburt2010}. For example, the $^{15}$N(p,$\alpha$)$^{12}$C reaction rate comes from \cite{Angulo1999}, while the triple-$\alpha$ reaction rate comes from \cite{Fynbo2005}. We use the JINA reaction rates for p--p chains, cold and hot CNO cycles, triple-$\alpha$ process, $\alpha$-capture up to $^{32}$S, Ne--Na and Mg--Al cycles, and C/O burning. We adopt the {\tt \small{mesa\_49.net}} nuclear network in MESA.

The nuclear network tracks and solves for the abundances of the following 52 species: n,  $^{1}$H, $^{2}$H, $^{3}$He, $^{4}$He, $^{7}$Li, $^{7}$Be, $^{9}$Be, $^{10}$Be, $^{8}$B, $^{12}$C, $^{13}$C, $^{13}$N, $^{14}$N, $^{15}$N, $^{14}$O, $^{15}$O, $^{16}$O, $^{17}$O, $^{18}$O, $^{17}$F, $^{18}$F, $^{19}$F, $^{18}$Ne, $^{19}$Ne, $^{20}$Ne, $^{21}$Ne, $^{22}$Ne, $^{21}$Na, $^{22}$Na, $^{23}$Na, $^{24}$Na, $^{23}$Mg, $^{24}$Mg, $^{25}$Mg, $^{26}$Mg, $^{25}$Al, $^{26}$Al, $^{27}$Al, $^{27}$Si, $^{28}$Si, $^{29}$Si, $^{30}$Si, $^{30}$P, $^{31}$P, $^{31}$S, $^{32}$S, $^{33}$S, $^{34}$S, $^{40}$Ca, $^{48}$Ti, $^{56}$Fe. The three heaviest elements $^{40}$Ca, $^{48}$Ti, and $^{56}$Fe are our modifications to the default {\tt \small{mesa\_49.net}} network and are inert species---they do not participate in any nuclear reactions---that are thus only affected by mixing and diffusion processes. 

Electron screening is included for both the weak and strong regimes. We use the default option in MESA which computes the screening factors by extending the classic \cite{Graboske1973} method with that of \cite{Alastuey1978}, and adopting plasma parameters from \cite{Itoh1979} for strong screening.\footnote{http://cococubed.asu.edu/code\_pages/codes.shtml}

\subsection{Boundary Conditions}
\label{section:boundary_conditions}
The pressure and temperature in the outermost cell of a stellar model calculation must be specified as a set of boundary conditions in addition to the trivial boundary conditions at the center of the star. There is a multitude of options that ranges from simple analytic approximations to tables based on full atmospheric structure models.

The simplest choice, {\tt \small{simple\_photosphere}}, uses the Eddington $T(\tau)$ relation to obtain $T$:
\begin{equation}
\label{equation:eddington}
T^4(\tau) = \frac{3}{4}T^4_{\rm eff}\left(\tau + \frac{2}{3}\right)\;\;,
\end{equation}
where $T_{\rm eff}$ is calculated directly from the MESA interior model. Similarly, $P$ is computed as follows:
\begin{equation}
\label{equation:pressure}
P = \frac{\tau g}{\kappa}\left[ 1+ P_0 \frac{\kappa}{\tau}\frac{L}{M}\frac{1}{6 \pi c G} \right]\;\;.
\end{equation}
The second term in the square brackets accounts for the nonzero radiation pressure \citep[e.g.,][]{Cox1968} which can be significant in high-mass stars. $P_0$ is a dimensionless factor of order unity used to scale up the radiation pressure in order to help convergence in massive stars and post asymptotic giant branch (post-AGB) stars radiating close to or at super-Eddington luminosities. We adopt $P_0=2$ for this work.

In most cases, the {\tt \small{simple\_photosphere}} option is a poor choice as there is no guarantee that $\kappa$ and $P$ from the interior model are consistent according to Equation \ref{equation:pressure}, assumed to be the correct relation at the stellar surface. For this work, we adopt realistic model atmospheres as the outer boundary conditions for most locations in the Hertzsprung--Russell (HR) diagram. We have computed a new grid of 1D plane-parallel atmosphere models specifically for this project using the ATLAS12 code \citep{Kurucz1970, Kurucz1993}. The atmospheres are computed to a Rosseland optical depth of $10^3$ with $\amlt=1.25$ following the implementation of convection as outlined in \cite{Castelli1997}.\footnote{Note that this value of $\amlt$ adopted for the model atmosphere cannot be directly compared to $\amlt$ adopted for the stellar interior in Section~\ref{section:convection} due to differences in the details of the implementation of convection.} We employed the latest atomic and molecular line lists from R. Kurucz, including molecules important for cool stars such as TiO and H$_2$O. Individual models are calculated for $\log(g)=0$ to 5 and $T_{\rm eff}=2500$~K to $50,000$~K for $\rm {[Z/H]}=-7.0$ to $+0.5$ on the \cite{Asplund2009} abundance scale. Beyond these limits the tables have been smoothly extrapolated to encompass all possible locations of the stellar tracks. This is a satisfactory solution since the few phases that fall into these extrapolated regimes (e.g., post-AGB) are typically very short-lived.

With model atmosphere tables in hand, one is left to choose where (in terms of Rosseland depth) to use the tables as boundary conditions for the models. The standard convention, which we adopt for most stars, is the photosphere, i.e., where $T=T_{\rm eff}$. However, for cooler dwarfs a more sensible choice is to set the boundary condition deeper in the atmosphere, i.e., $\tau=100$. This option will result in more realistic models for cool low-mass stars whose atmospheres are heavily influenced by molecules that are not included in the MESA interior model calculations. This issue is less critical for the cool giants because the structure of these stars is overall less sensitive to the boundary condition (i.e., the pressure at the photosphere for giants is much closer to zero than for dwarfs). We refer the reader to Section~\ref{section:nonstandard_models} for additional discussion on this topic.

For our grid of models, the {\tt \small tau\_100\_tables} option is used for $0.1\textrm{--}0.3~\msun$, {\tt \small photosphere\_tables} is used for $0.6\textrm{--}10~\msun$, and {\tt \small simple\_photosphere} is used for $16\textrm{--}300~\msun$. To facilitate a smooth transition between different regimes, we run both {\tt \small tau\_100\_tables} and {\tt \small photosphere\_tables} for $0.3\textrm{--}0.6~\msun$ and {\tt \small photosphere\_tables} and {\tt \small simple\_photosphere} for $10\textrm{--}16~\msun$. The tracks in this transition region are then blended (see Section~\ref{section:isochrone_construction} for more details). At the highest masses, {\tt \small simple\_photosphere} is a sufficient approximation due to the flattening of opacity as a function of temperature for $\teff \gtrsim10^4$~K.

\subsection{Diffusion}
\label{section:diffusion}
Microscopic diffusion and gravitational settling of elements are essential ingredients in stellar evolution models of low-mass stars, leading to a modification to the surface abundances and main sequence (MS) lifetimes, as well as a shift in the evolutionary tracks toward lower luminosities and temperatures in the HR diagram \citep[e.g.,][]{Michaud1984, Morel1999, Salaris2000, Chaboyer2001, Bressan2012}. Calculations of diffusion and gravitational settling are implemented in MESA following \cite{Thoul1994}. All species are categorized into one of five ``classes'' according to their atomic masses, each of which has a representative member whose properties are used to estimate the diffusion velocities. MESA's default set of representative members for the five classes are $^{1}$H, $^{3}$He, $^{4}$He, $^{16}$O, and $^{56}$Fe. Atomic diffusion coefficients are calculated according to \cite{Paquette1986}: the representative ionic charge for each class is estimated as a function of $T$, $\rho$, and free electrons per nucleon, while the diffusion velocity of the representative member is adopted. The diffusion equation is then solved using the total mass fraction within each class.

However, the inclusion of microscopic diffusion alone cannot reproduce observations of surface abundances in the Hyades open cluster and OB associations including the Orion association \citep[e.g.,][]{Cunha1994, Varenne1999, Daflon2001}. Models with diffusion predict an over-depletion of helium and metals in the outer envelopes of stars with $M_{\rm i}>1.4~\msun$, a problem that appears to worsen with increasing mass due to a disappearing outer convection zone and a concomitant steepening of the temperature and pressure gradients \citep{Morel2002}. The solution to this problem requires additional forces to counteract gravity. Radiative levitation \citep{Vauclair1983, Hu2011} can help to reduce the gravitational settling of highly charged elements, such as iron, via radiation pressure. However, it is thought to be mostly important in hot, luminous massive MS stars or helium-burning stars \citep[e.g., hot subdwarf stars;][]{Fontaine2008}, and to have only a modest effect for solar-type stars \citep[e.g.,][]{Alecian1993, Turcotte1998}. We employ radiation turbulence \citep{Morel2002} to reduce the efficiency of diffusion in hot stars, though there exist other explanations for the observed surface abundances, including turbulent mixing due to differential rotation \citep[e.g.,][]{Richard1996}. We adopt the radiative diffusivity parameter $D_{\rm RT} = 1$, which relates the strength of radiative diffusivity (the deposition of photon momentum into the fluid, resulting in radiative mixing) to the kinematic radiative viscosity.
 
Since the effects of elemental diffusion are most significant in the absence of more efficient mixing processes, such as convection, diffusion is neglected for fully convective MS stars in the MIST models. Additionally, diffusion is expected to play a reduced role in massive stars and during some post-MS phases (which are also associated with large convective envelopes) for which the evolutionary timescales are comparable to or much shorter than the timescale for diffusion and gravitational settling \citep[e.g.,][]{Turcotte1998}. Thus the effects of microscopic diffusion are considered only for the MS evolution. However, it may have a notable impact on both the atmospheres and interiors of cooling WDs by modifying the surface abundances and lengthening cooling times through the release of gravitational energy. The \cite{Thoul1994} formalism, which assumes isolated interactions between two particles at a time, breaks down in the regime of strongly coupled plasmas. This is a particularly relevant issue for the interiors of cooling WDs, and there is ongoing effort in MESA to update the diffusion implementation to account for this. The inclusion of diffusion during the post-MS evolution, especially the WD cooling phase, is one of the priorities for Paper II.

To summarize, the implementation of diffusion is limited to MS stars above the fully convective limit for which it is most effective in terms of both the relevant timescales and the relative importance compared to other mixing processes.

\subsection{Rotation}
\label{section:rotation}
The effect of rotation on stellar models has been studied for decades \citep[e.g.,][]{Strittmatter1969, Fricke1972, Tassoul1978, Zahn1983, Pinsonneault1997, Heger2000, Maeder2000, Palacios2003, Talon2005, Denissenkov2007, Hunter2007, Chieffi2013, Cantiello2014}, but it remains as one of the most challenging and uncertain problems in stellar astrophysics. Rotation is particularly important for massive stars, as rotationally induced instabilities, combined with the non-negligible effects of radiation pressure, can significantly alter their evolution \citep[e.g.,][]{Heger2000, Meynet2000, Hirschi2004, Woosley2006, Yoon2006, deMink2010, Georgy2012, Langer2012, Koehler2015}. Although the overall importance of rotation in models---lifetimes, surface abundances, evolutionary fates, to name a few---has been explored, the details are not yet fully understood.

Rotation is inherently a 3D process, but the so-called ``shellular approximation'' allows stellar structure equations to be solved in 1D \citep{Kippenhahn1970, Endal1976, Meynet1997, Heger2000, Paxton2013}. This approximation is valid in the regime where strong anisotropic turbulence arises from differential rotation and smears out both chemical composition and velocity gradients along isobars \citep{Zahn1992, Meynet1997}. As a result, the standard stellar structure equations are simply modified by centrifugal acceleration terms in the presence of rotation. More details on the implementation of rotation in MESA can be found in \cite{Paxton2013}.

Our models are available in two varieties, with and without rotation. All rotating models are initialized with solid body rotation on the zero age main sequence (ZAMS), which is the standard choice in stellar evolution codes \citep{Pinsonneault1989, Heger2000, Eggenberger2008}. As discussed extensively in \cite{Heger2000}, pre-main sequence (PMS) stars achieve rigid rotation due to convection, and once they settle onto ZAMS, they establish close-to-rigid rotation mainly via Eddington--Sweet (ES) circulation and Goldreich--Schubert--Fricke (GSF) instability. However, it is worth noting that there are important exceptions. First, this rigid rotation approximation fails in the solar convection zone as inferred from helioseismology observations \citep[e.g.,][]{Brown1989}. Second, current detailed evolutionary models \citep[e.g.,][]{Bouvier2008, Gallet2013} suggest that low-mass stars ($\lesssim1.2~\msun$), particularly those with slow and moderate rotation rates, have strong differential rotation profiles at ZAMS.

Currently, surface magnetic fields are not included in MESA calculations, which can couple to mass loss and give rise to magnetic braking \citep[e.g.,][]{Weber1967, Mestel1968, udDoula2002, Meynet2011}, a mechanism for winding down surface rotation over time in stars with appreciable convective envelopes (\citealt{Kraft1967}; see also Section~\ref{section:magnetic}). 

Since magnetic braking is not presently modeled in MIST, we do not include rotation for stars with $M_{\rm i}\leq1.2~\msun$ in order to reproduce the slow rotation rate observed in the Sun and in other low-mass stars. Over the mass range $1.2\textrm{--}1.8~\msun$, the rotation rate is gradually ramped up from 0 to the maximum value of $v_{\rm ZAMS}/v_{\rm crit}=\Omega_{\rm ZAMS}/\Omega_{\rm crit}=0.4$, where $v_{\rm crit}$ and $\Omega_{\rm crit}$ are critical surface linear and angular velocities, respectively (See Equation~\ref{equation:omegacrit}). This rotation rate, also adopted in the Geneva models \citep{Ekstrom2012},\footnote{Note that $v_{\rm crit}$ and $\Omega_{\rm crit}$ are defined differently in MESA and in the Geneva models. In the Geneva models, the equatorial radius is 1.5 times larger than the polar radius when $\Omega = \Omega_{\rm crit}$ but this distinction is not made in MESA. See Section 2.1 in \cite{Georgy2013} for more details.} is motivated by both recent observations of young B stars \citep{Huang2010} and theoretical work on rotation rates in massive stars \citep{Rosen2012}. A comparison with observed rotation rates of both MS and evolved stars in the mass range $1.2\textrm{--}1.5~\msun$ \citep{Wolff1997, CantoMartins2011} reveals that our ramping scheme produces velocities that are reasonable ($\sim10\text{--}25~{\rm km\;s^{-1}}$ during the main sequence for $1.3\textrm{--}1.35~\msun$) even in the absence of magnetic braking.

Chemical mixing and angular momentum transport due to rotationally induced instabilities are discussed in Section~\ref{section:rot_mixing}, and rotationally enhanced mass loss is discussed in Section~\ref{section:rot_mass_loss}.

\subsection{Mixing Processes}
\subsubsection{Convection}
\label{section:convection}
Mixing Length Theory (MLT), whose modern implementation in stellar evolution codes was pioneered by \cite{BohmVitense1958}, describes the convective transport of energy in the stellar interior. There is a vexing yet crucial free parameter of order unity, $\amlt$, that determines how far a fluid parcel travels before it dissolves into the background, $l_{\rm MLT}$, in units of the local pressure scale height, $H_{\rm P}$ ($l_{\rm MLT} = \amlt H_{\rm P}$). In other words, it parametrizes how efficient convection is, because a large $\amlt$ means that the parcel travels a large distance before it deposits its energy into the ambient medium.

Convective mixing of elements is treated as a time-dependent diffusive process with a diffusion coefficient computed within the MLT formalism, which may later be modified by overshoot mixing across convection boundaries (see Section~\ref{section:overshoot}). Convective heat flux is computed by solving the MLT cubic equations to obtain the temperature gradients (e.g., Equation 42 in \citealt{Henyey1965}). We adopt the modified version of MLT from \citet{Henyey1965} instead of the standard MLT prescription \citep{Cox1968}, as the latter assumes no radiative losses from fluid elements and is therefore applicable only at high optical depth.\footnote{We note that neither prescription is adequate for treating the radiation dominated envelopes of very massive stars, for which 1D stellar evolution calculations must be considered uncertain. See \cite{Jiang2015} for more details.} In addition to $\amlt$, there are two free parameters, $\nu$ and $y$, which are multiplicative factors to the mixing length velocity and the temperature gradient in the convective element. The latter two parameters are set to 8 and 1/3, respectively \citep{Henyey1965}. This particular framework allows for convective efficiency to vary with the opacity of the convective element, an important effect to take into account in the layers near the stellar surface. The empirical calibration of $\amlt$ is discussed in Section~\ref{section:solar_calib}

Classically, the location of the convective region is determined using the Schwarzschild criterion, which implies that a region is convectively stable if
\begin{equation}
\nabla_T < \nabla_{\rm ad}\;\;,
\end{equation}
where $\nabla_T$ is the local background temperature gradient (in practice, it is set to the radiative temperature gradient $\nabla_{\rm rad}$) and $\nabla_{\rm ad}$ is the adiabatic temperature gradient. Alternatively, the Schwarzschild criterion can be replaced by the Ledoux criterion, which also takes into account the composition gradient, $\nabla_{\mu}$. In this case, a region is convectively stable if
\begin{eqnarray}
\nabla_T &<& \nabla_{\rm L} \\
\nabla_{\rm L} &\equiv& \nabla_{\rm ad} - \frac{\chi_{\mu}}{\chi_T}\nabla_{\mu} \\
\chi_{\mu} &\equiv& \left[ \frac{\partial \ln (P)}{\partial \ln (\mu)}\right]_{\rho, T} \\
\chi_{T} &\equiv& \left[ \frac{\partial \ln (P)}{\partial \ln (T)}\right]_{\rho, \mu}\;\;,
\end{eqnarray}
where the thermodynamic derivatives $\chi_{\mu}$ and ${\chi_T}$ are equal to $-1$ and 1 for an ideal gas, respectively. We adopt the Ledoux criterion for convection in our models to account for the composition effects in the stellar interiors.

\subsubsection{Overshoot Mixing}
\label{section:overshoot}
Rather unsurprisingly, the MLT framework, which relies on a 1D diffusive model in place of a full 3D hydrodynamical treatment, offers an incomplete description of convection. To model the mixing occurring at convective boundaries, also known as overshoot mixing, one must turn to yet another set of parameterizations. Typically, a convective region is extended beyond the fiducial boundary determined by either the Schwarzschild or Ledoux criterion in order to account for the nonzero momentum of the fluid element approaching the edge of the convective zone as well as its subsequent penetration into the non-convective region \citep[e.g.,][]{Bohm1963, Shaviv1973, Maeder1975, Roxburgh1978, Bressan1981}. This overshoot action leads to enhanced mixing and it can account for both the observed properties of asymptotic giant branch (AGB) and post-AGB stars \citep{Herwig2000}, the observed MS width \citep[e.g,][]{Schaller1992}, and the main sequence turn off (MSTO) morphology in clusters such as M67 \citep[e.g.,][]{Magic2010}.

There are two prescriptions for convective overshoot available in MESA. The first method, which we call step overshoot, is to simply extend the fiducial convective boundary by a fraction, usually $\sim 0.2$, of the local pressure scale height. This instantaneous treatment is often calibrated to fit the observed MSTO of stellar clusters and associations, and is a commonly adopted scheme in many stellar evolution codes \citep[e.g.,][]{Demarque2004, Pietrinferni2004, Dotter2008, Brott2011, Bressan2012, Ekstrom2012}.         

The second method, adopted in the present work, was motivated by the plume-like nature of convective elements seen in 2D and 3D radiation hydrodynamic simulations, where coherent downward and upward flows were observed rather than a hierarchy of blob-like eddies \citep[e.g.,][]{Freytag1996}. This led to a picture in which the turbulent velocity field decays exponentially away from the fiducial convective boundary and the convective element eventually disintegrates in the overshoot region through a diffusion process. Following the parametrization discussed in \cite{Herwig2000}, the resulting diffusion coefficient in the overshoot region is given by
\begin{equation}
D_{\rm OV} = D_0 \exp{\left(\frac{-2z}{H_{\rm v}}\right)}\;\; ; \;\; H_{\rm v} = f_{\rm ov} H_{\rm P}\;\;,
\end{equation}
where $H_{\rm v}$ is the velocity scale height of the overshooting convective elements at the convective boundary, $f_{\rm ov}$ is a free parameter that essentially determines the efficiency of overshoot mixing, $H_{\rm P}$ is the local pressure scale height, and $D_0$ is the diffusion coefficient in the unstable region ``near'' the convective boundary (more specifically at a depth of $f_{0,\;\rm ov} H_{\rm P}$ from the convective boundary). For simplicity, we adopt two sets of ($f_{\rm ov}$, $f_{0,\;\rm ov}$) values, one for the core and another for shell/envelope, irrespective of the type of burning taking place in the overshoot region. For further simplicity, $f_{0,\;\rm ov}$ is set to $0.5 f_{\rm ov}$. The temperature gradient in the overshoot region is assumed to be equal to the radiative gradient as in the step overshoot approach.

Other models in the literature make use of an additional parameterization to both avoid a physically unrealistic size of the overshoot region outside a small convective core as well as to account for the possibility that convective overshoot efficiency is smaller in lower mass stars \citep{Demarque2004, Pietrinferni2004, Dotter2008, Bressan2012}. In the step overshoot formalism where the size of the convective core is extended by a fraction $f_{\rm ov,\;step}$ of $H_{\rm P}$, one could end up with a physically unrealistic situation where the size of the overshoot region exceeds the size of the convective core itself. This can occur when the convective core is very small, e.g., for the critical mass around $1.1\text{--}1.2~\msun$ when the CNO cycle begins to dominate over the pp-chain and the hydrogen-burning core becomes convective instead of radiative. Since the convective core boundary is not far from the center and $H_{\rm P}$ formally diverges as $r\rightarrow0$, the size of the overshoot region, $f_{\rm ov,\;step}H_{\rm P}$, also diverges. Thus, to avoid the excessive growth of the convective core for low-mass MS stars, the common solution is to gradually ramp up the overshoot efficiency from $M_{1}\sim1~\msun$ to $M_{2}\sim1.7~\msun$, with no convective overshoot below $M_{1}$ and maximum efficiency above $M_{2}$. These boundary masses vary with metallicity due to opacity effects \citep{Demarque2004, Bressan2012}. In the exponential overshoot formalism adopted in this work, we bypass this issue, and thus do not require a secondary parameterization involving $M_{1}$ and $M_{2}$.

We adopt a modest overshoot efficiency $f_{\rm ov,\;core}=0.016$ for the core which is roughly equivalent to $f_{\rm ov,\;step}=0.2$ in the step overshoot scheme \citep{Magic2010}. This value is calibrated to reproduce the shape of the MSTO in the open cluster M67 (Section~\ref{section:star_clusters}). However, it is worth noting that the strength of core convective overshoot depends on numerous other factors as well. For instance, \cite{Stothers1991} explored the role of opacities in models with core overshoot and found that the overall increase in radiative opacities from the OPAL group \citep{Iglesias1991} compared to the older values from the Los Alamos groups \citep[e.g.,][]{Cox1965, Cox1970} reduced the overshoot efficiency required to reproduce observations of intermediate- and high-mass stars. \cite{Magic2010} studied how variations in the solar abundances (\citealt{Asplund2005} vs. \citealt{Grevesse1998}), element diffusion, overshooting, and nuclear reaction rates influence the MSTO morphology in M67. The authors concluded that the appearance of the characteristic MSTO hook (also known as the ``Henyey hook'') depends sensitively on the choice of the input solar abundances and that the effects of uncertain input physics on the Henyey hook morphology are degenerate.

We emphasize that the strength of convective overshoot is calibrated purely empirically: the overshoot efficiency in the core is constrained by matching the MSTO in M67, and the overshoot efficiency in the envelope, $f_{\rm ov,\;env}$, is chosen along with $\amlt$ during solar calibration (Section~\ref{section:solar_calib}). As noted in \cite{Bressan2012}, envelope overshoot has a negligible effect on the evolution, e.g., the MS lifetime, though it is believed to affect the surface abundances of light elements, the location of the red giant branch (RGB) bump, and the extension of the blue loops. We also remind the reader that the overshoot efficiency in shells, e.g., hydrogen-burning shells during the RGB, is set to $f_{\rm ov,\;env}$ for simplicity.

\subsubsection{Semiconvection and Thermohaline Mixing}
\label{section:sc_th}
As noted in Section~\ref{section:convection}, we adopt the Ledoux criterion for convection in our models. Due to the additional composition gradient term, a region that is formally convectively unstable to Schwarzschild criterion may be stable according to the Ledoux criterion (i.e., a thermally unstable medium with a stabilizing, positive composition gradient), which leads to a type of mixing called semiconvection. The importance of semiconvection on the evolution of massive stars has been studied for many decades, e.g., during the core helium burning phase (CHeB) \citep{Stothers1975, Langer1985, Grossman1996}. The fraction of a star's core helium burning lifetime spent on the Hayashi line relative to that in the blue part of the HR diagram, in other words the ratio of red supergiants to blue supergiants, is found to depend sensitively on the inclusion of semiconvection in the model. Additionally, the resulting core mass has significant implications for the supernova progenitors, from their ability to undergo a successful explosion \citep[e.g.,][]{Sukhbold2014} to the actual nucleosynthetic yields \citep[e.g.,][]{Langer1989, Heger2002, Rauscher2002}. Semiconvection also operates in lower-mass stars with convective cores on the MS and it can have an important effect on the actual size and appearance of the core \citep[e.g.,][]{Faulkner1973, Silva2011, Paxton2013}.

Alternately, a thermally stable medium may have a negative, destabilizing composition gradient, which triggers a different type of instability called thermohaline mixing. An inverted chemical composition gradient is rare in stars, since fusion usually occurs inside out and synthesizes lighter elements into heavier products. This phenomenon can occur due to mass-transfer in binaries \citep{Stothers1969, Stancliffe2007}, off-center ignition in semi-degenerate cores \citep{Siess2009}, or the $^{3}$He($^{3}$He,2p)$^{4}$He reaction taking place just beyond the hydrogen-burning shell during the RGB, horizontal branch (HB), and AGB \citep{Eggleton2006, Charbonnel2007, Cantiello2010, Charbonnel2010,  Stancliffe2010}. Thermohaline mixing is thought to be responsible for the modification of surface abundances of RGB stars near the luminosity bump that otherwise cannot be explained using standard models. However, more recent work suggests that thermohaline mixing alone cannot account for the observed surface abundance anomalies \citep{Denissenkov2010, Traxler2011, Wachlin2011, Wachlin2014}.

In MESA, semiconvection and thermohaline mixing are both implemented as time-dependent diffusive processes. The diffusion coefficient for semiconvection is computed following \cite{Langer1983}:
\begin{equation}
D_{\rm sc} = \alpha_{\rm sc}\left(\frac{K}{6C_P\rho}\right)\frac{\nabla_{T}-\nabla_{\rm ad}}{\nabla_{\rm L}-\nabla_T}\;\;,
\end{equation}
where $K$ is the radiative conductivity, $C_P$ is the specific heat at constant pressure, and $\alpha_{\rm sc}$ is a dimensionless free-parameter. Similarly, the diffusion coefficient for thermohaline mixing is computed following \cite{Ulrich1972} and \cite{Kippenhahn1980}:
\begin{equation}
D_{\rm th} = \alpha_{\rm th}\left(\frac{3K}{2C_P\rho}\right)\frac{-\frac{\chi_\mu}{\chi_T}\nabla_{\mu}}{\nabla_{T}-\nabla_{\rm ad}}\;\;,
\end{equation}
where $\alpha_{\rm th}$ is a dimensionless number that describes the aspect ratio of the mixing blobs or ``fingers'' (a large $\alpha_{\rm th}$ corresponds to slender fingers).

As summarized in \cite{Paxton2013}, the range of $\alpha_{\rm sc}$ and $\alpha_{\rm th}$ adopted by various authors spans several orders of magnitude, partially due to differences in the implementation in various codes. We adopt $\alpha_{\rm sc} = 0.1$ though values as small as 0.001 or as large as 1 can be found in the literature \citep{Langer1991, Yoon2006}. For thermohaline mixing we adopt $\alpha_{\rm th}=666$, as this value has been shown to reproduce the surface abundances anomalies in RGB stars past the luminosity bump \citep{Charbonnel2007, Cantiello2010}. Note however that in the literature $\alpha_{\rm th}$ spans the range $1\textrm{--}1000$ \citep{Kippenhahn1980, Charbonnel2007, Cantiello2010, Stancliffe2010, Wachlin2011}. There are ongoing theoretical efforts aimed at eliminating these free parameters with full 3D simulations \citep[e.g.,][]{Traxler2011, Brown2013, Spruit2013, Wood2013}.

\subsubsection{Rotationally Induced Instabilities}
\label{section:rot_mixing}
In MESA, the transport of both chemicals and angular momentum arising from rotationally induced instabilities are treated in a diffusion approximation \citep{Endal1978, Zahn1983, Pinsonneault1989, Heger2000, Yoon2005} in place of the alternative diffusion-advection approach \citep{Maeder2000, Meynet2000, Eggenberger2008, Potter2012a}. The five rotationally induced instabilities included in our models are: dynamical shear instability, secular shear instability, Solberg--H{\o}iland (SH) instability, ES circulation, and GSF instability \citep{Heger2000, Paxton2013}. Of these, ES circulation and shear instabilities have the largest impact on the evolution of a rotating star. We refer the reader to \cite{Heger2000} and \cite{Maeder2000} for excellent overviews of these phenomena.

Once the diffusion coefficients for these rotational mixing processes are computed, they are combined with the diffusion coefficients for convection, semiconvection, and thermohaline. This grand sum enters the angular momentum and abundance diffusion equations solved at each time step. There are two free parameters in this implementation, first introduced by \cite{Pinsonneault1989} to model the Sun: $f_c$, a number between 0 and 1 that represents the ratio of the diffusion coefficient to the turbulent viscosity, which scales the efficiency of composition mixing to that of angular momentum transport; and $f_\mu$, a factor that encodes the sensitivity of rotational mixing to the mean molecular weight gradient, $\nabla_\mu$. A small $f_c$ corresponds to a process that transports angular momentum more efficiently than it can mix material, and a small $f_\mu$ means that rotational mixing is efficient even in the presence of a stabilizing $\nabla_\mu$. We adopt $f_c=1/30$ and $f_\mu=0.05$ following \cite{Pinsonneault1989}, \cite{Chaboyer1992}, and \cite{Heger2000}. As we demonstrate in Section~\ref{section:n_abun}, these parameters produce surface nitrogen enhancements that are in reasonable agreement with the observations.

\subsubsection{Magnetic Effects}
\label{section:magnetic}
There is a growing body of evidence that our understanding of internal angular momentum transport in stars is not complete. For example, the observed spin rates of WDs and neutron stars \citep{Heger2005, Suijs2008} and the angular velocity profiles inferred from asteroseismic observations of red giants \citep{Eggenberger2012, Cantiello2014} cannot be reproduced with models that only include rotational mixing from hydrodynamic instabilities and circulations.

Spruit--Tayler (ST) dynamo is a mechanism for the amplification of seed magnetic fields in radiative stellar interiors in the presence of differential rotation \citep{Spruit2002}. Stellar models including torques from ST dynamo fields can reproduce the flat rotation profile in the solar interior \citep{Mestel1987, Charbonneau1993, Eggenberger2005} and the observed spin rates of WDs and neutron stars \citep{Heger2005, Suijs2008}, although they still cannot explain the slow rotation rates of cores in red giants \citep{Cantiello2014}. The chemical mixing and the transport of angular momentum due to internal magnetic fields are not included in our models, though this is implemented in MESA following KEPLER \citep{Heger2005} and STERN \citep{Petrovic2005}. We note that the very existence of the ST-dynamo loop is still under debate \citep{Braithwaite2006, Zahn2007}, and there are ongoing efforts to understand the role of angular momentum transport via magnetic fields in radiative stellar regions \citep[e.g.][]{Rudiger2015, Wheeler2015}.

Magnetic fields are also observed near the stellar surface, which are thought to be either of fossil origin \citep[e.g.,][]{Braithwaite2004} or generated through dynamo operating in convective zones in the outer layers of low-mass stars \citep[e.g.,][]{Brandenburg2005}. However, as discussed in Section~\ref{section:rotation}, magnetic braking due to the coupling between winds and surface magnetic fields is not yet included in MESA.

\subsection{Mass Loss}
\label{section:mass_loss}
The implementation of mass loss in stellar evolution calculations is based on a number of observationally and theoretically motivated prescriptions. It is frequently cited as one of the most uncertain ingredients in stellar evolution, and is thought to play a crucial role in the advanced stages of stellar evolution for low-mass stars and in all phases of evolution for massive stars (see \citealt{Smith2014} for a recent review on this topic). In this section we review our treatment of mass loss across the HR diagram. We note that the total mass loss rate is capped at $10^{-3}~\msunperyr$ in all models to prevent convergence problems.

\subsubsection{Low Mass Stars}
\label{section:mass_loss_lm}
Mass loss for stars with masses below $10~\msun$ is treated via a combination of the \cite{Reimers1975} prescription for the RGB and \cite{Bloecker1995} for the AGB. Both mass loss schemes are based on global stellar properties such as the bolometric luminosity, radius, and mass: 
\begin{eqnarray}
\dot{M}_{\rm R} &=& 4\times10^{-13}\eta_{\rm R} \frac{(L/\lsun)(R/\rsun)}{(M/\msun)}\;\;{\rm \msunperyr}\;\;, \\
\dot{M}_{\rm B} &=& 4.83\times10^{-9}\eta_{\rm B} \frac{(L/\lsun)^{2.7}}{(M/\msun)^{2.1}}\frac{\dot{M}_{\rm R}}{\eta_{\rm R}}\;\;{\rm \msunperyr}\;\;,
\end{eqnarray}
where $\eta_{\rm R}$ and $\eta_{\rm B}$ are factors of order unity. These free parameters have been tuned to match numerous observational constraints, including the initial--final mass relation (IFMR) (Section~\ref{section:ifmr}; see also \citealt{Kalirai2009}), AGB luminosity function (Section~\ref{section:agbcs_lf}; see also \citealt{Rosenfield2014}), and asteroseismic constraints from open cluster members in the {\it Kepler} fields \citep{Miglio2012}. The \cite{Bloecker1995} mass loss scheme was proposed as an alternative to the classic \cite{Reimers1975} prescription to account for the onset of the superwind phase found in dynamical simulations of atmospheres of Mira-like variables. However, we emphasize that these are still empirically motivated recipes and therefore remain agnostic on the subject of the actual mechanism driving the winds (see a review by \citealt{Willson2000} for a discussion on, e.g., dust-driven winds).

For simplicity, we turn on Reimers mass loss at the beginning of the evolution, but a negligible amount of mass loss occurs throughout the MS ($\sim 10^{-13}~\msunperyr$ for a solar metallicity $1~\msun$ star). Once core helium is depleted, the mass loss rate is chosen to be max[$\dot{M}_{\rm R}$, $\dot{M}_{\rm B}$] at any given time. We adopt $\eta_{\rm R} = 0.1$ and $\eta_{\rm B} = 0.2$ in order to reproduce the IFMR (Section~\ref{section:ifmr}) and the AGB luminosity functions in the Magellanic Clouds (Section~\ref{section:agbcs_lf}).

\subsubsection{High Mass Stars}
\label{section:mass_loss_hm}
For hot and luminous massive stars, mass loss is thought to arise from the absorption of ultraviolet photons by metal ions in the atmosphere, resulting in a preferentially outward momentum transfer from the absorbed photons to the plasma \citep[line-driven winds;][]{Lucy1970, Castor1975}. For our models, mass loss for stars above $10~\msun$ uses a combination of radiative wind prescriptions, collectively called the {\tt \small Dutch} mass loss scheme in MESA, inspired by \cite{Glebbeek2009}. There is an option for an overall scaling factor $\eta_{\rm Dutch}$ analogous to $\eta_{\rm R}$ and $\eta_{\rm B}$ for the low-mass stars, but we adopt $\eta_{\rm Dutch} = 1.0$. For prescriptions that include metallicity-scaling, we retain the reference $Z_{\odot}$ adopted by each author. We expect this difference to have a negligible effect relative to the large overall uncertainties in mass loss prescriptions.

We now describe our mass-loss scheme for high-mass stars in each region of the HR diagram:

\begin{enumerate}
\item For $\teff > 1.1\times10^4$~K and $X_{\rm surf}$ (surface hydrogen mass fraction) > 0.4, the mass loss prescription from \cite{Vink2000, Vink2001} is used, which is appropriate for the early phases of the evolution prior to the stripping of the hydrogen-rich envelope. The authors computed mass loss rates using a Monte Carlo radiative transfer code, taking into account multiple scatterings and assuming that the loss of photon energy is coupled to the momentum gain of the wind. The Vink mass loss rate is specified by five parameters; $M$, $L$, $\teff$, $v_{\infty}/v_{\rm esc}$, and $Z_{\rm surf}$.

For $2.75\times10^4 < \teff < 5\times10^4$~K,
\begin{equation}
\begin{split}
\dot{M}_{\textrm{V, hot}} =\; & 10^{-6.697} \times (L/10^5\lsun)^{2.194} \times \\
& (M/30~\msun)^{-1.313} \times \left( \frac{v_{\infty}/v_{\rm esc}}{2.0} \right)^{-1.226} \times \\
& (\teff/4\times10^4~{\rm K})^{0.933} \times \\
& 10^{-10.92[\log(\teff/4\times10^4~{\rm K})]^2} \times (Z_{\rm surf}/Z_{\odot})^{0.85} \;\;. \\
\end{split}
\end{equation}
The ratio of terminal flow velocity to the escape velocity increases with metallicity following $v_{\infty}/v_{\rm esc} = 2.6(Z_{\rm surf}/Z_{\odot})^{0.13}$.

For $1.1\times10^4 <\teff < 2.25 \times10^4$~K,\footnote{In \cite{Vink2001}, mass loss rates were computed for $\teff\geq1.25\times10^4$~K, but the prescription is extended down to $\teff = 1.1\times10^4$~K in MESA.}
\begin{equation}
\begin{split}
\dot{M}_{\textrm{V, cool}} =\; & 10^{-6.688} \times (L/10^5\lsun)^{2.210} \times \\
& (M/30~\msun)^{-1.339} \times \left( \frac{v_{\infty}/v_{\rm esc}}{2.0} \right)^{-1.601} \times \\
& (\teff/4\times10^4~{\rm K})^{1.07} \times (Z_{\rm surf}/Z_{\odot})^{0.85} \;\;, \\
\end{split}
\end{equation}
where $v_{\infty}/v_{\rm esc} =1.3(Z_{\rm surf}/Z_{\odot})^{0.13}$.

For $ 2.25 \times 10^4 \leq \teff \leq 2.75\times10^4$~K, either $\dot{M}_{\textrm{V, hot}}$ or $\dot{M}_{\textrm{V, cool}}$ is adopted depending on the exact position of the so-called bi-stability jump, a phenomenon in which $\dot{M}$ increases with decreasing $\teff$ due to the recombination of metal lines:

\begin{equation}
T_{\textrm{eff, jump}} = 61.2 + 2.59 \log\langle\rho\rangle\;\;,
\end{equation}
where $\langle\rho\rangle$ corresponds to the characteristic wind density at $50\%$ of the terminal velocity of the wind. The successful predictions of the mass loss rates and the bi-stability jump near $\teff\sim2.5\times10^4$~K for the Galactic and SMC O-type stars have made the Vink prescription a popular choice among massive star modelers (but see the discussion below).

\item Once the star reaches $\teff > 10^4$~K and $X_{\rm surf} < 0.4$, it is formally identified as a Wolf--Rayet (WR) star and we switch over to the \cite{Nugis2000} empirical mass loss prescription which depends strongly on luminosity and chemical composition:
\begin{equation}
\label{equation:nl}
\dot{M}_{\rm NL} = 10^{-11}(L/\lsun)^{1.29} X_{\rm surf}^{1.7} Z_{\rm surf}^{0.5}\;\;{\rm \msunperyr}\;\;.
\end{equation}
This formula has been shown to reproduce the properties of a large sample of Galactic WR stars (WN, WC, and WO subtypes) that have well-constrained stellar and wind parameters \citep{Nugis2000}. With Equation~\ref{equation:nl}, we are able to reproduce the observed ratio of WC to WN subtypes as a function of metallicity (see Section~\ref{section:highM_ratios}).

\item For all stars with $\teff < 10^4$~K,\footnote{In MESA, mass loss rate for $10^4<\teff<1.1\times10^4$ is computed by smoothly transitioning between the low temperature prescription (de Jager) and high temperature prescription (Vink or Nugis \& Lamers).} including stars in the red supergiant (RSG) phase, we utilize the \cite{deJager1988} empirically derived wind prescription:
\begin{equation}
\dot{M}_{\rm dJ} = 10^{-8.158}(L/\lsun)^{1.769}\teff^{-1.676}\;\;{\rm \msunperyr}\;\;.
\end{equation}
Although a quantitative model of mass loss in RSGs does not exist yet, it is believed that the main mechanism is dust-driven outflows. The low temperatures and pulsations in the outer layers lead to the condensation of dust at large radii, which is then driven out due to radiation pressure on grains \citep{Mauron2011}. In a recent work comparing different wind prescriptions with mass loss rates estimated from a sample of RSGs in the Galaxy and in the Magellanic Clouds, \cite{Mauron2011} found that the de Jager rates agree to within a factor of 4 with most estimates derived from the 60~$\mu \rm m$ flux. Furthermore, the authors concluded that the de Jager wind recipe performs better overall compared to more recent prescriptions, though they recommended an additional metallicity scaling $(Z_{\rm surf}/Z_{\odot})^{0.7}$. For simplicity, we adopt the original de Jager prescription available in MESA.
\end{enumerate}

A recent review by \cite{Smith2014} explores some of the shortcomings of these prescriptions, one of which is that they fail to account for the clumpiness and inhomogeneity in outflows.\footnote{An exception is the \cite{Nugis2000} prescription which does take clumping effects into account.} For instance, mass loss inferred from H$\alpha$ emission or free-free continuum excess that assumes a homogeneous wind \citep[e.g.,][]{deJager1988, Nieuwenhuijzen1990} is believed to overestimate the true rate by a factor of 2 to 3. However, the reduced mass loss rate corrected for clumpiness may be problematic for the formation of WR stars, which requires the removal of their hydrogen-rich envelopes. Eruptive episodic mass loss episodes and/or binaries are likely to play a role \citep[e.g.,][]{Smith2006, Yoon2010, Sana2012}, but neither phenomenon can be realistically captured in a simple recipe for implementation in a 1D stellar evolution code. Other forms of enhanced mass loss rates include super-Eddington winds when the stellar luminosity exceeds Eddington luminosity \citep{Grafener2011, Vink2011}. 

\subsubsection{Rotationally Enhanced Mass Loss}
\label{section:rot_mass_loss}
Observations of O and B type stars have long argued for rotationally enhanced mass loss rates \citep{Gathier1981, Vardya1985, Nieuwenhuijzen1988}. It is now a standard ingredient in modern stellar evolution models with rotation \citep[e.g.,][]{Heger2000, Brott2011, Potter2012a}. In MESA, mass loss rates are enhanced in models as a function of surface angular velocity $\Omega$ as follows:
\begin{equation}
\label{equation:rotboost}
\dot{M}(\Omega) = \dot{M}(0)\left(\frac{1}{1-\Omega/\Omega_{\rm crit}}\right)^\xi\;\;,
\end{equation}
where $\dot{M}(0)$ is the standard mass loss rate (Reimers, Bl{\"o}cker, or ``Dutch''), $\xi$ is assumed to be 0.43 \citep{Friend1986, Bjorkman1993, Langer1998}, and $\Omega_{\rm crit}$ is the critical angular velocity at the surface:
\begin{equation}
\label{equation:omegacrit}
\Omega_{\rm crit}^2 = \left(1-\frac{L}{L_{\rm Edd}}\right)\frac{GM}{R^3}\;\;.
\end{equation}
The Eddington luminosity, $L_{\rm Edd}$, is a mass-weighted average over the optical depth $\tau$ between 1 and 100. For stars close to the Eddington limit, $\Omega_{\rm crit}$ approaches 0 and therefore even a small $\Omega$ will result in a dramatic boost according to Equation~\ref{equation:rotboost}. To prevent the mass loss from becoming too catastrophic, we cap the rotational boost to $10^4$.

\section{Solar Model}
\subsection{Solar Calibration}
\label{section:solar_calib}
As mentioned in Section~\ref{section:convection}, it is customary to calibrate the mixing length parameter, $\amlt$, using helioseismic data and surface properties of the Sun. We utilize the MESA test suite {\tt \small{solar\_calibration}} which conducts an extensive parameter search using the simplex method to obtain a set of input parameters that reproduces the solar observations. We vary the initial composition of the Sun, $\amlt$, and convective overshoot in the envelope. For each iteration, a new set of parameters is drawn and the star is evolved from the PMS to 4.57~Gyr.\footnote{As noted by \cite{Bahcall2006}, there is some ambiguity in the exact definition of the ``age of the Sun'' as the PMS contraction is estimated to last approximately 0.04~Gyr. For simplicity, we adopt the commonly assumed age of 4.57~Gyr.} A global $\chi^2$ value is computed by summing over the $\log L$, $\log R$, surface composition, $R_{\rm cz}$ (the location of the base of the convection zone), and $\delta c_{\rm s}$ (model $-$ observed sound speed) terms with non-uniform user-defined weights. This process is repeated until $\chi^2$ ceases to change considerably and the tolerance parameters are met. For simplicity and consistency, the target solar values we adopt are the same nominal values recommended by the IAU.

\begin{figure}
\centering
\includegraphics[width=0.45\textwidth]{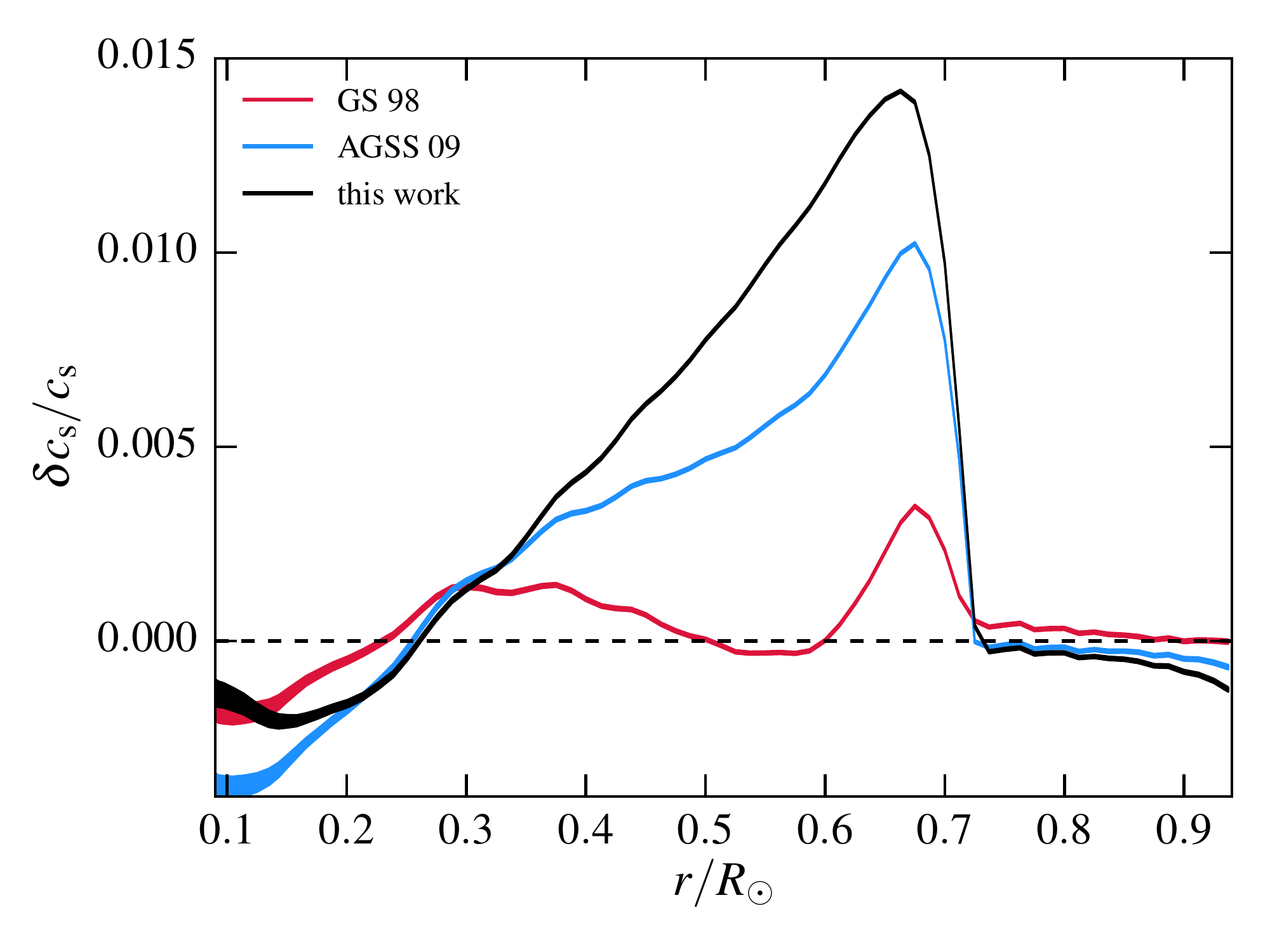} 
\caption{The fractional error in sound speed compared to helioseismology observations from \cite{Rhodes1997} for the present model (black) and two \cite{Serenelli2009} models each with \citealt{Grevesse1998} (GS98; red) and \citealt{Asplund2009} (AGSS09; blue) meteoritic abundances.}
\label{fig:csound_profile}
\end{figure}

Table~\ref{tab:solar_calib} summarizes the solar calibration results and Figure~\ref{fig:csound_profile} shows $\delta c_{\rm s}/c_{\rm s}$, the fractional error in sound speed compared to that from helioseismic observations \citep{Rhodes1997}, as a function of radius. The best-fit MIST solar-calibrated model is shown in black and two \cite{Serenelli2009} models adopting \citealt{Grevesse1998} (GS98) and \citealt{Asplund2009} (AGSS09) protosolar abundances are shown in red and blue, respectively.

\begin{table}
\centering
\begin{longtable}{l l l l} \\ 
\caption{Solar calibration results.} \\
\hline
\hline \noalign{\smallskip}
Parameter & Target & Model Value & Fractional Error (\%) \\ \noalign{\smallskip}
\hline
\noalign{\smallskip}
$\lsun (10^{33}~\ergs$) & 3.828\footnote{\label{IAU}XXIX$^{\rm th}$ IAU resolutions B2 and B3} & 3.828 & $4.1\times10^{-3}$ \\
$\rsun (10^{10}~\rm{cm})$ & 6.957\footref{IAU} & 6.957 & $1.2\times10^{-3}$ \\
$T_{\rm eff, \odot}$ (K) & 5772\footref{IAU} & 5772 & $2.4\times10^{-3}$ \\ 
 $X_{\rm surf}$ & 0.7381\footnote{\label{A09}\citet{Asplund2009}} & 0.7514 & $1.8$ \\ 
 $Y_{\rm surf}$ & $0.2485$\footnote{\label{B04}\citet{Basu2004}} & 0.2351 & $5.4$ \\
 $Z_{\rm surf}$ & 0.0134\footref{A09} & 0.0134 & $0.3$ \\ 
 $R_{\rm cz}$ & 0.7133\footnote{\citet{Basu2004}} & 0.7321 & $2.6$ \\ \noalign{\smallskip}
\hline
\noalign{\smallskip}
$\amlt$ & - & 1.82 & - \\
$f_{\rm ov,\;env}$ & - & 0.0174& - \\
X$_{\rm initial}$ & - & 0.7238 & - \\
Y$_{\rm initial}$ & - & 0.2612 & - \\
Z$_{\rm initial}$ & - & 0.0150 & - \\
\noalign{\smallskip}
\hline \hline
   \label{tab:solar_calib}
\end{longtable}
\end{table}

Although the model $\log L$ and $\log T_{\rm eff}$ are in excellent agreement with the observed values, there are noticeable discrepancies in the surface helium abundance, the location of the base of the convection zone, and the sound speed profile. Although many initial guesses and different weighting schemes were explored, we were unable to obtain a solar model that satisfies all available observational constraints. This is a well-known problem for solar models that adopt the AGSS09 abundances \citep[e.g.,][]{Asplund2009, Serenelli2009}.

The helium surface abundance at 4.57~Gyr in the best solar model is much lower compared to the helioseismologically inferred value of $0.2485\pm0.0034$ \citep{Basu2004}. While most elemental abundances are determined through 3D spectroscopic modeling, the helium abundance is inferred indirectly from helioseismology, relying on the change in the adiabatic index in the \ion{He}{2} ionization zone near the surface \citep{Asplund2009}. The tension between the helioseismic value and the inferred abundance from interior modeling was noted in \cite{Asplund2009}, and remains an unsolved problem to this day.

The sound speed profile comparison shows that the GS98 model is in good agreement while the models adopting the AGSS09 abundances show a large deviation at $\sim0.6~\rsun$. This is partly due to the discrepancy between the predicted and observed locations of the convective boundary. In particular, the lower oxygen and neon abundances in AGSS09 relative to the older models like GS98 (or an even newer model like \citealt{Caffau2011}) imply a smaller mean opacity below the convective zone, which shifts the inner convective boundary further out in radius. The abundance of oxygen, one of the most abundant and important elements, has undergone a striking overall downward revision over the past few decades. Still, there are likely lingering uncertainties in the surface abundance determinations due to the challenges associated with spectroscopic modeling, such as non-LTE effects, line blending, and uncertainties in the atomic and molecular data \citep{Asplund2009}.

The \cite{Serenelli2009} AGSS09 sound speed profile is in better agreement with the observed profile, likely because their model matches the location of the base of the convection zone more closely than the MIST model predicts. However, their model prefers an even lower surface helium abundance compared to the best-fit helium abundance in our model, and the present-day luminosity, radius, and effective temperature are not included as part of their fit.

\begin{figure}
\centering
\includegraphics[width=0.45\textwidth]{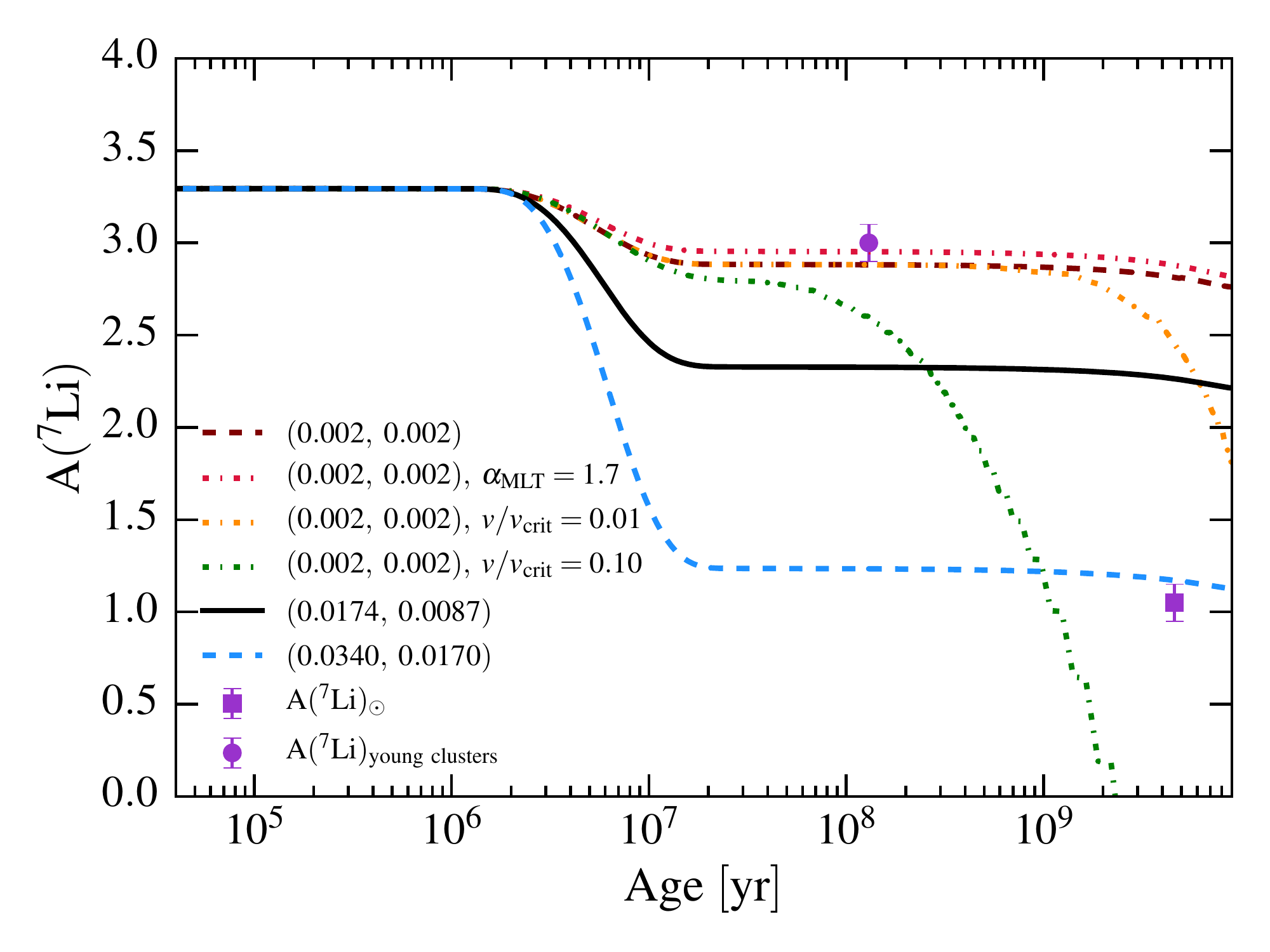}
\caption{The evolution of surface lithium abundance relative to hydrogen, ${\rm A(^7Li)}=\log(N_{\rm ^7Li}/N_{\rm ^1H})+12$, as a function of time for several $1~\msun$ models. Each pair of numbers in parentheses corresponds to the two envelope overshoot efficiency parameters $f_{\rm ov,\;env}$ and $f_{\rm ov,\;env,\;0}$, respectively. The solid black line represents the fiducial model that adopts the solar-calibrated envelope overshoot parameters (0.0174, 0.0087) and $\amlt=1.82$. The two dashed maroon and blue lines are models with less and more efficient overshoot, resulting in reduced and enhanced lithium depletion, respectively. The three dotted--dashed lines show additional variations in input physics: the red line correspond to $\amlt=1.7$, while the orange and green lines are models that include PMS rotation with $v/v_{\rm crit}=0.01$ and 0.10, respectively. The purple square and circle are surface lithium abundance for the present-day Sun \citep{Asplund2009} and the typical surface lithium abundance for nearby solar-metallicity, young clusters \citep[e.g.,][]{Jeffries2005, Sestito2005, Juarez2014}. The current models cannot simultaneously reproduce both observed lithium abundances.}
\label{fig:lithium}
\end{figure}

Several explanations have been offered to reconcile the mismatch between the standard AGSS09 solar model and helioseismology results. One resolution invokes increased opacity, an idea that has been quantitatively explored by several authors \citep[e.g.,][]{Bahcall2005a, ChristensenDalsgaard2009, Serenelli2009}. These authors concluded that a $\sim10\%$ increase is required to match the observations, but that current atomic physics calculations do not leave room for such substantial change in opacities. However, there was a recent upward revision of iron opacities, based on new experimental data, that might account for roughly half the increase in the total mean opacity required to resolve this problem \citep{Bailey2015}. Other possible resolutions include more efficient diffusion processes in the radiative zone \citep[e.g.,][]{Asplund2004}, increased neon abundance to compensate for decreased oxygen abundance \citep[e.g.,][]{Antia2005, Bahcall2005b, Drake2005}, the introduction of new physics currently missing from stellar evolution calculations \citep[e.g., convectively induced mixing in radiative zone;][]{Young2005}, and improved implementations of the current input physics \citep[e.g., a replacement for MLT;][]{Arnett2015}. 

Although the widely adopted practice is to fix the solar-calibrated $\amlt$ across all masses, evolutionary phases, and abundances, there has been a recent effort to map out $\amlt$ as a function of $\log g$ and $\teff$ \citep{Trampedach2007, Trampedach2014} as well as metallicity \citep{Magic2015} from full 3D radiative hydrodynamic calculations of convection in stellar atmospheres. Recently, \cite{Salaris2015} included variable $\amlt$ and $T(\tau)$ boundary condition from \cite{Trampedach2014} in their stellar evolution calculations and found that varying $\amlt$ has a small effect on the evolution and surface properties. We adopt a constant value of $\amlt = 1.82$ for the present work, but discuss the implications of this assumption in more detail in Section~\ref{section:star_clusters}

In summary, we adopt solar-calibrated $\alpha_{\rm MLT}$ and convective overshoot efficiency in the envelope $f_{\rm ov,\;env}$ ($f_{\rm 0,\;ov,\;env}$ is fixed to $0.5f_{\rm ov,\;env}$). As noted earlier in the Introduction, we adopt the \cite{Asplund2009} abundance mixture and ``solar metallicity'' in this paper refers to the initial bulk $Z = Z_{\odot,\; \rm protosolar} = 0.0142$. 

\subsection{Lithium Depletion}
Lithium is a very fragile element that is burned via proton capture at temperatures as low as $2.5\times10^{6}$~K. Mixing processes such as convection that transport lithium from the outer layers to the interior where the temperature is sufficiently high lead to the depletion of surface lithium on very short timescales. The time evolution of surface lithium abundance therefore depends very sensitively on the initial stellar mass (proxy for temperature) and mixing physics.

Standard stellar evolution models so far have not been able to successfully reproduce the solar surface lithium abundance, indicating the need to include extra physical mechanisms. One way to account for missing physics in the models is to vary $\amlt$ \cite[see e.g., Lyon models;][]{Baraffe1998, Baraffe2003, Baraffe2015}. Alternatively, models that incorporate the effects of rotation and internal gravity waves \citep[e.g.,][]{Charbonnel2005} are able to reproduce both the solar interior rotation profile and surface lithium abundances for the Sun and other galactic cluster stars. Relatedly, \cite{Somers2014} found that radius dispersion on the PMS (correlated with rotation and chromospheric activity) can explain the spread in lithium abundances in young clusters such as the Pleiades.

In Figure~\ref{fig:lithium} we show the evolution of surface lithium abundance relative to hydrogen, ${\rm A(^7Li)}=\log(N_{\rm ^7Li}/N_{\rm ^1H})+12$, as a function of time for several $1~\msun$ models. The purple square and circle are surface lithium abundance for the present-day Sun \citep{Asplund2009} and the typical surface lithium abundance for nearby solar-metallicity, young clusters \citep[e.g.,][]{Jeffries2005, Sestito2005, Juarez2014}. Each pair of numbers in parentheses corresponds to the two envelope overshoot efficiency parameters $f_{\rm ov,\;env}$ and $f_{\rm 0,\;ov,\;env}$, respectively. The surface lithium abundance decreases over time in all of the displayed models due to the inclusion of diffusion. The solid black line represents the fiducial model that adopts the solar-calibrated envelope overshoot parameters (0.0174, 0.0087) and $\amlt=1.82$. The fiducial model burns too much lithium early on and then does not deplete lithium efficiently on the MS. The two dashed maroon and blue lines are models with less and more efficient overshoot, resulting in reduced and enhanced lithium depletion, respectively. The three dotted--dashed lines show additional variations in input physics: the red line correspond to $\amlt=1.7$, while the orange and green lines are models that include PMS rotation with $v/v_{\rm crit}=0.01$ and 0.10, respectively. As expected, a lower $\amlt$ produces a puffier, cooler star, resulting in less lithium depletion. The inclusion of rotation during the PMS produces very different, potentially more promising behavior, though the current absence of magnetic braking (see Sections~\ref{section:magnetic} and \ref{section:rotation}) in MESA results in an unrealistically high rotation speed ($v/v_{\rm crit}$ between 0.1 and 1) on the MS. The models presented here fail to simultaneously match both the young and present-day solar surface lithium abundances. The inclusion of rotation on the PMS, the implementation of magnetic braking, as well as the exploration of a variable $\amlt$ to mimic the effects of non-standard physics are planned for follow-up investigations in the near future.

\section{Model Outputs and Bolometric Corrections}
In this section, we provide an overview of the two principal model outputs: evolutionary tracks and isochrones, which can be downloaded from http://waps.cfa.harvard.edu/MIST/. There are both theoretical and observational isochrones available. We offer both packaged models for download and a web interpolator that generates models with user-specified parameters.

\begin{table*}
\centering
  \begin{longtable}{ll} \\
  \caption{Primary EEPs and corresponding evolutionary phases.} \\
    \hline \hline \noalign{\smallskip}
    Primary EEP & Phase \\
    \noalign{\smallskip}
    \hline
    \noalign{\smallskip}
	 1& pre-main sequence (PMS) \\
	 2& zero age main sequence (ZAMS) \\
	 3& intermediate age main sequence (IAMS ) \\
	 4& terminal age main sequence (TAMS) \\
	 5& tip of the red giant branch (RGBTip) \\
	 6& zero age core helium burning (ZACHeB)\footnote{i.e., zero age horizontal branch; ZAHB for low-mass stars.} \\
	 7& terminal age core helium burning (TACHeB)\footnote{terminal age horizontal branch; TAHB.} \\

    \noalign{\smallskip}
    \hline
    \noalign{\smallskip}
    	Low Mass Type & \; \\ 
    \noalign{\smallskip}
    \hline
    \noalign{\smallskip}
	 8& thermally pulsating asymptotic giant branch (TPAGB) \\
	 9& post asymptotic giant branch (post-AGB) \\
	 10& white dwarf cooling sequence (WDCS) \\
    \noalign{\smallskip}
    \hline
    \noalign{\smallskip}
	High Mass Type & \; \\ 
    \noalign{\smallskip}
    \hline
    \noalign{\smallskip}
	 8& carbon burning (C-burn) \\
    \noalign{\smallskip}
    \hline \hline
\label{tab:EEPs}
  \end{longtable}
\end{table*}

\begin{figure*}
\centering
\includegraphics[width=0.92\textwidth]{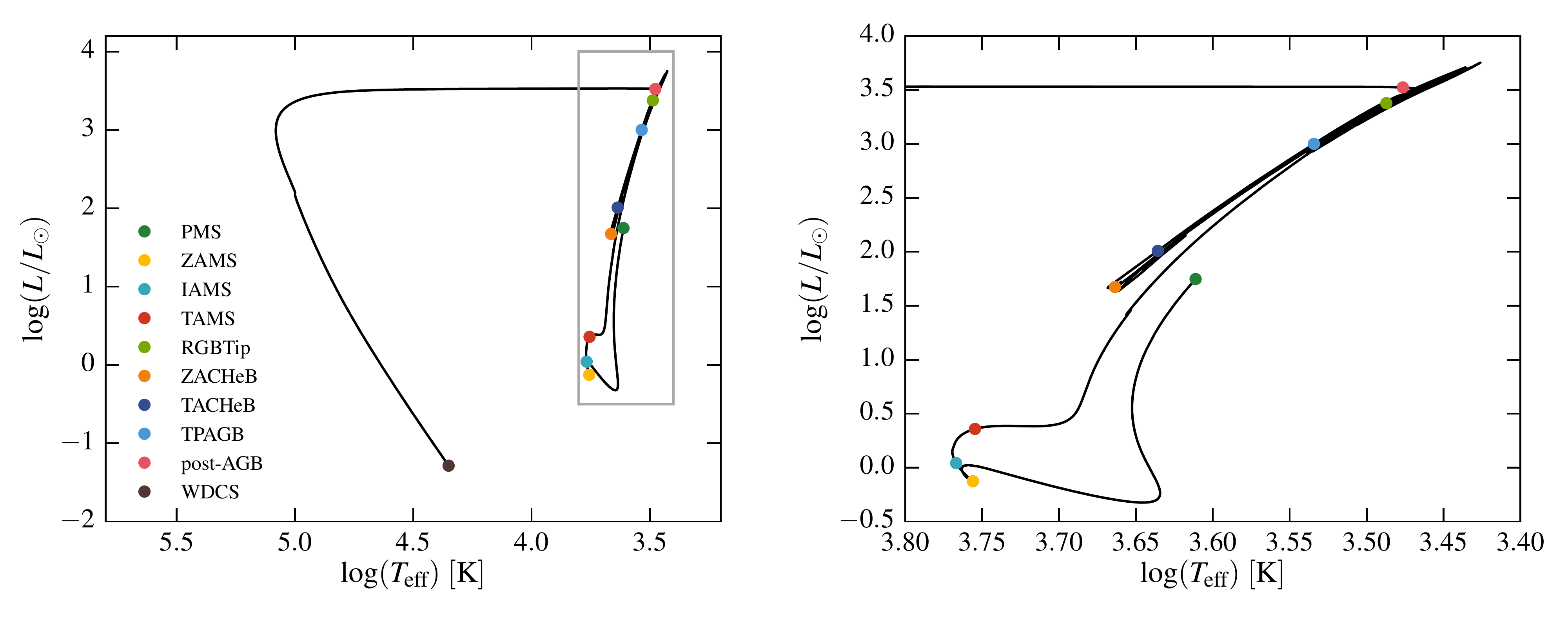} 
\caption{Left: an example $1~\msun$ evolutionary track in the equivalent evolutionary point (EEP) format, with the locations of the primary EEP points marked by colored circles. The gray box marks the zoomed-in region shown in the right panel. Right: a zoomed-in view of the track.}
\label{fig:EEPs_lowM}
\end{figure*}

\begin{figure}
\centering
\includegraphics[width=0.45\textwidth]{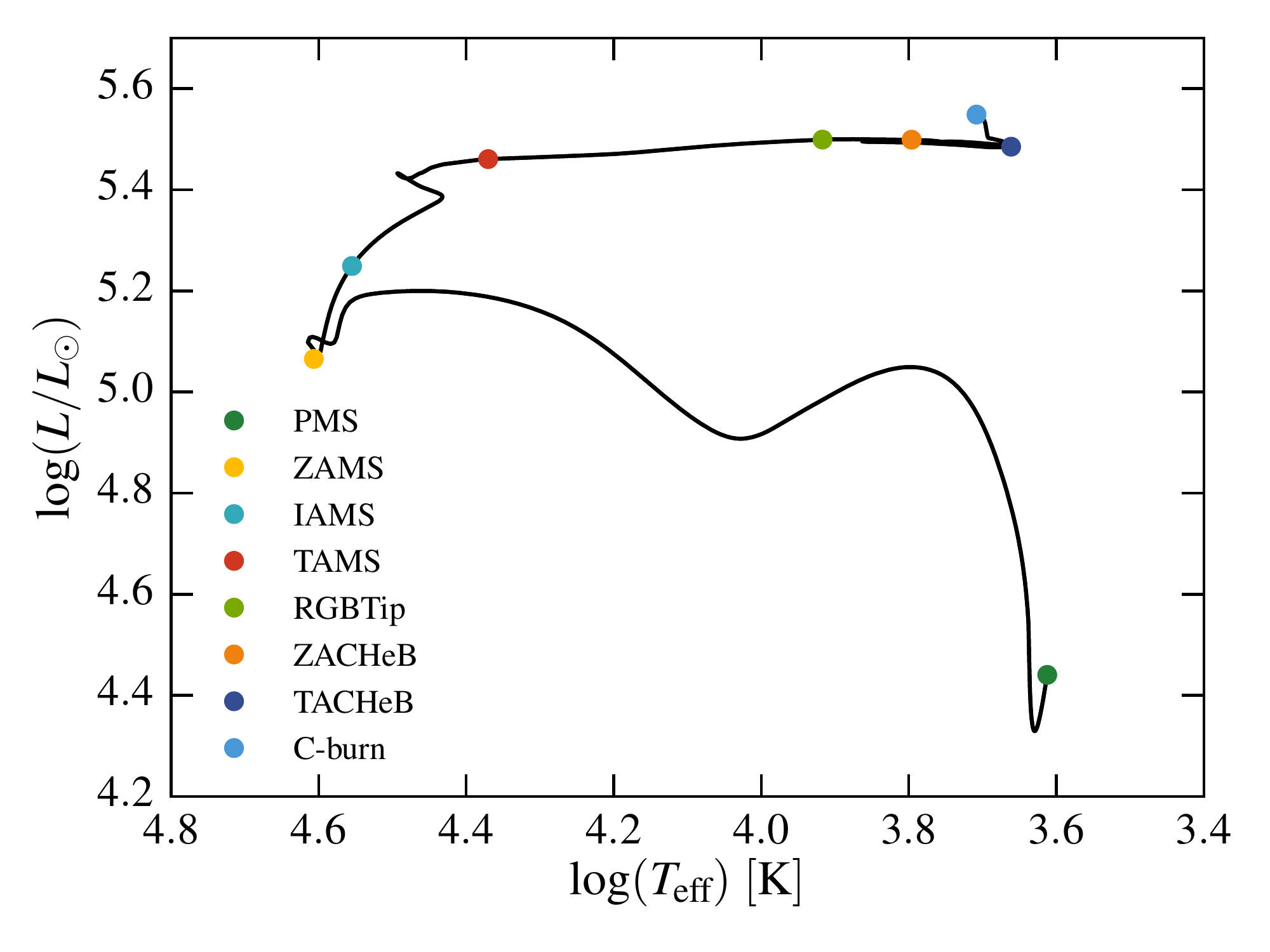} 
\caption{Same as Figure~\ref{fig:EEPs_lowM} but now for a $30~\msun$ evolutionary track. Note the different primary EEP point following the core helium burning phase. Although ``RGBTip'' does not have the same morphological significance in high-mass stars as in low-mass stars since high-mass stars ignite helium under non-degenerate conditions, we retain this terminology for consistency reasons.}
\label{fig:EEPs_highM}
\end{figure}

\subsection{Ages, Masses, Phases, Metallicities}
\label{section:ages_masses_phases_metallicities}
One of the main goals of the MIST project is to produce extensive grids of stellar evolutionary tracks and isochrones that cover a wide range in stellar masses, ages, evolutionary phases, and metallicities. The stellar mass of evolutionary tracks ranges from $0.1~\msun$ to $300~\msun$ for a total of $\gtrsim100$ models, and the ages of isochrones cover $\log \textrm{Age} = 5$ to $\log \textrm{Age} = 10.3$ in 0.05~dex steps. For $M_{\rm i}\leq0.7~\msun$, the models are terminated at TAMS, i.e., central $^{1}$H abundance drops to $10^{-4}$. For a $0.7~\msun$ star at $Z_{\odot}$, this limit is typically reached at an age > 35~Gyr. For $M_{\rm i}>0.7~\msun$, the models are either evolved through the WD cooling phase (``low-mass'' type) or the end of carbon burning (``high-mass'' type), depending on which criterion is satisfied first. We adopt this flexible approach to take into account the blurry boundary---further complicated by its metallicity dependence---between low- and intermediate-mass stars that end their lives as WDs and high-mass stars that continue to advanced stages of burning. In particular, the ``low-mass'' type models are terminated when $\Gamma$, the central plasma interaction parameter, also known as the Coulomb coupling parameter, exceeds 20. $\Gamma$ is defined to be $\bar{Z}^2e^2/a_ik_bT$, where $\bar{Z}$ is the average ion charge, $e$ is the electron charge, $a_i$ is the mean ion spacing, $k_b$ is the Boltzmann constant, and $T$ is the temperature. A large $\Gamma$ corresponds to a departure from the ideal gas limit toward solidification (crystallization of a pure oxygen WD occurs at $\Gamma\approx175$; \citealt{Paxton2011}). The ``high-mass'' type models---stars that are sufficiently massive to burn carbon---are stopped when the central $^{12}$C abundance drops to $10^{-2}$. The metallicity ranges from $\rm [Fe/H] = -2.0$ to $+0.5$, with $0.25$~dex spacing. We also provide an additional set of models evolved from the PMS to the end of core helium burning for $-4.0 \leq \rm [Z/H] < -2.0$ for modeling ancient, metal-poor populations. We provide a limited set at low metallicities at this time due to computational difficulties we have encountered. In particular, mixing between convective boundaries during the thermally pulsating AGB phase results in the ingestion of protons into a burning region, resulting in dramatically higher nuclear burning luminosities \cite[e.g.,][]{Lau2009, Stancliffe2011, Woodward2015}. Non-solar-scaled abundance grids will be presented in Paper II.

We note that in any grid, there is a subset of models that does not run to completion due to convergence issues. This is not generally problematic because the mass sampling is sufficiently fine such that there are enough models to smoothly interpolate a new EEP track and/or construct isochrones. We also note that there are interesting features in the tracks and isochrones that may appear to be numerical issues at first glance, but in fact a number of them are real phenomena captured in the MESA calculations. We refer the reader to the Appendix for a discussion of these features. Of these, an example feature that may be a numerical artifact is an extremely short-lived glitch that appears during the early post-AGB phase for a subset of the models. Since it has zero bearing on the evolution of the star, we post-process this feature out of the final evolutionary tracks in order to facilitate the construction of smooth isochrones.

\begin{figure*}
\centering
\includegraphics[width=0.95\textwidth]{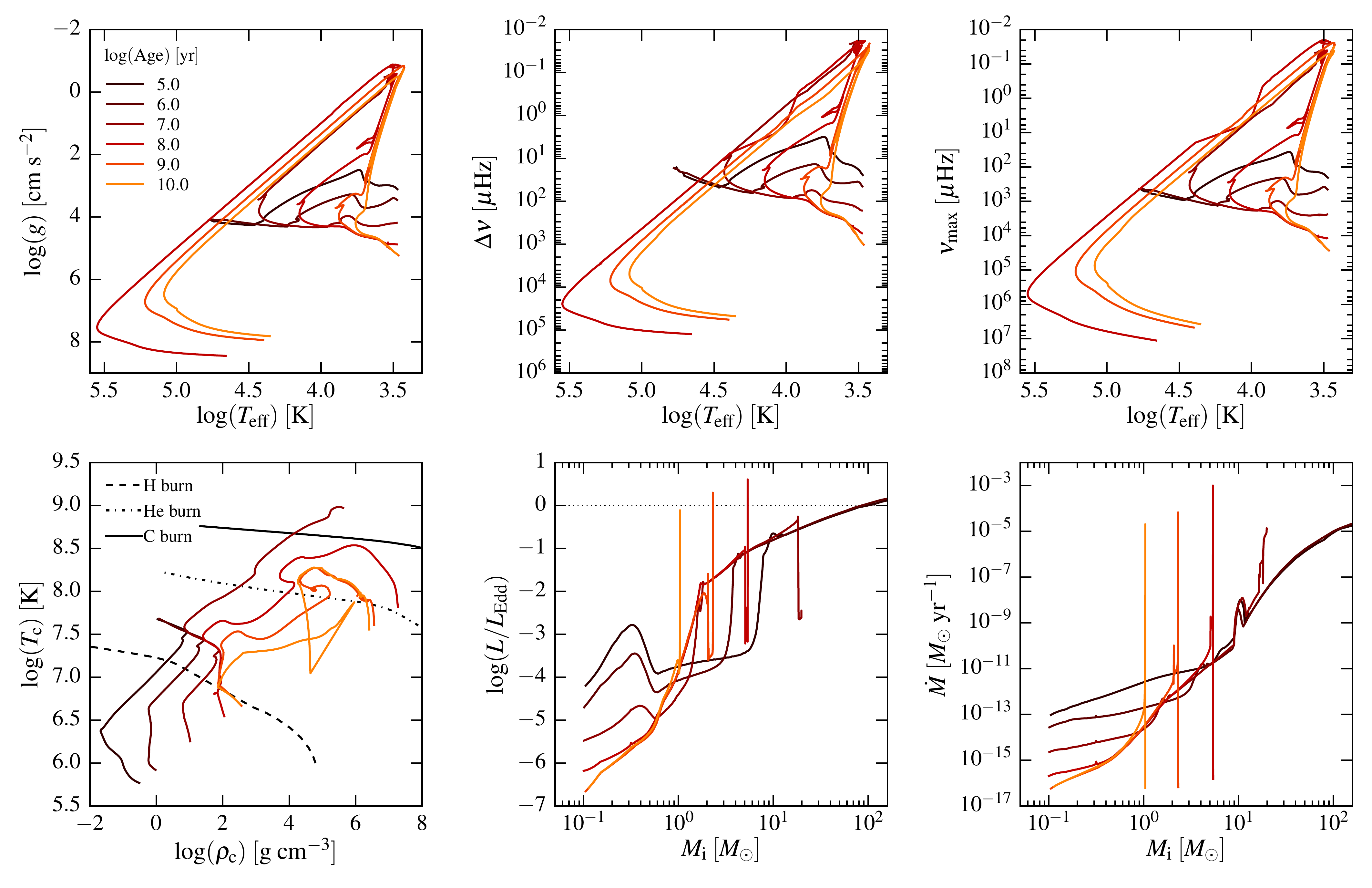}
\caption{Isochrones from $\logage = 5\textrm{--}10$ in various physical quantities. Top left: isochrones in the $\log (g)-\log (\teff)$ plane. Top middle: asteroseismic isochrones in the $\Delta \nu-\log (\teff)$ plane, where $\Delta \nu$ corresponds to the large frequency separation for p-modes. Note that $\Delta \nu$ is computed from the scaling relations. Top right: asteroseismic isochrones in the $\nu_{\rm max} -\log (\teff)$ plane, where $\nu_{\rm max}$ corresponds to the frequency of maximum power. Note that $\nu_{\rm max}$ is computed from the scaling relations. Bottom left: isochrones in the $\log (T_{c})$--$\log (\rho_{c})$ plane with the dashed, dotted--dashed, and solid lines showing thresholds for hydrogen, helium, and carbon ignition. The discontinuous feature at $\logage \ge 10$ is due to the ignition of helium in a degenerate core, i.e, helium core flash at the tip of the RGB. Bottom middle: isochrones in the $\log(L/L_{\rm Edd})$--$M_{\rm i}$ plane, where $L_{\rm Edd}$ is the Eddington luminosity. Bottom right: isochrones in the $\dot{M}$--$M_{\rm i}$ plane, where $\dot{M}$ is the mass loss rate. In the last two figures, isochrones become steadily truncated at the high-mass end as more stars evolve and die.}
\label{fig:fun_isos}
\end{figure*}

\subsection{Isochrone Construction}
\label{section:isochrone_construction}
We briefly describe the isochrone construction process and defer a detailed discussion of this topic to Paper 0 \citep{Dotter2016}. In each evolutionary track, we identify a set of the so-called primary equivalent evolutionary points (primary EEPs). These correspond to specific stages of evolution defined by a set of physical conditions, such as the terminal age main sequence (TAMS; central hydrogen exhaustion), and the tip of the RGB (RGBTip; a combination of lower limits on the central helium abundance and luminosity). Next, each segment between adjacent primary EEPs is further divided into so-called ``secondary-EEPs'' according to a distance metric that evenly samples the tracks in certain relevant variables, such as $\teff$ and $L$. Put another way, primary EEPs serve as physically meaningful reference locations along the evolutionary track and secondary EEPs finely sample the track between primary EEPs. This method maps a set of evolutionary tracks from ordinary time coordinates onto uniform EEP coordinates. The primary EEPs and corresponding evolutionary phases are listed in Table~\ref{tab:EEPs}. In Figures~\ref{fig:EEPs_lowM} and \ref{fig:EEPs_highM}, we show example $1~\msun$ and $30~\msun$ evolutionary tracks in the EEP format, with colored dots marking the locations of the primary EEPs. Note that we require both IAMS and TAMS points in order to properly resolve the MSTO for stars that burn hydrogen convectively in their cores during the MS, i.e., the Henyey hook. Also note that although ``RGBTip'' does not have the same morphological significance in high-mass stars as in low-mass stars since high-mass stars ignite helium under non-degenerate conditions, we retain this terminology for consistency reasons.

As described in Section~\ref{section:boundary_conditions}, we use three different boundary conditions depending on the initial mass of the model ($\tau=100$ tables for $0.1\textrm{--}0.3~\msun$, photosphere tables for $0.6\textrm{--}10~\msun$, and simple photosphere for $16\textrm{--}300~\msun$). To facilitate a smooth transition from one regime to another, we run both $\tau=100$ and photosphere tables for $0.3\textrm{--}0.6~\msun$ and photosphere tables and simple photosphere for $10\textrm{--}16~\msun$. For every mass in the transition regime, the two EEP tracks are blended with a smooth weighting function to create a hybrid EEP track:

\begin{equation}
w = 0.5\left [1 - \cos \left( \pi \frac{M_* - M_1}{M_2-M_1} \right ) \right]\;\;,
\end{equation}
where $M_1$ and $M_2$ are the transition masses (e.g., $M_1=0.3~\msun$ and $M_2=0.6~\msun$ for the transition from $\tau=100$ to photosphere tables).

To generate an isochrone at age $t_0$ with all of the EEP tracks now in hand, we first cycle through all masses and construct a piecewise monotonic function between $M_{\rm i}$ and $t$ for each EEP point.\footnote{Note that we no longer distinguish between primary and secondary EEPs for the purposes of isochrone construction.} Next, we interpolate to obtain $M_{\rm i}(t_0)$. Once we have $M_{\rm i}(t_0)$ for every EEP, we can now construct an isochrone for any parameter, e.g., $L$, by interpolating that parameter as a function of $M_{\rm i}$, e.g., $L(M_{\rm i}(t_0))$, at every EEP. The EEP framework is superior to a direct interpolation scheme in time coordinates as it can properly treat evolutionary phases with short timescales (e.g., post-AGB) or complex trajectories (e.g., thermally pulsating AGB; TPAGB).

A sensible approach is to construct a monotonic relationship between mass and age assuming that ``increasing mass = decreasing phase lifetime'' is always true. However, interesting non-monotonic behaviors begin to appear in certain evolutionary phases over a narrow age (or equivalently, mass) interval if the mass resolution is sufficiently high (see Figure~\ref{fig:phase_lifetimes_cumul} and also \citealt{Girardi2013}). Put another way, two stars of different initial masses are at the same evolutionary stage in terms of their EEPs over a special narrow age interval. This effect will be explored in future work but for this work, we enforce monotonicity in the age-mass relationship.

\subsection{Available Model Outputs}
The published tracks and isochrones include a wealth of information, ranging from basic parameters such as $\log (L)$, $\log (\teff)$, $\log (g)$, and surface abundances of 19 elements to more detailed quantities such as $\beta \equiv P_{\rm gas}/P_{\rm total}$ and asteroseismic parameters (the full list of available parameters is available on the project website). To highlight this fact, we show examples of isochrones in several different projections in Figure~\ref{fig:fun_isos}. From left to right, the top three panels feature isochrones in the $\log (g)-\log (\teff)$, $\Delta \nu-\log (\teff)$, and $\nu_{\rm max}-\log (\teff)$ planes. $\Delta \nu$ and $\nu_{\rm max}$ are asteroseismic quantities that correspond to the large frequency separation for p-modes and the frequency of maximum power, respectively, which can be readily obtained from power spectra of e.g., {\it Kepler} light curves. We clarify that $\Delta \nu$ and $\nu_{\rm max}$ in the MIST models are computed from simple scaling relations \citep[e.g.,][]{Ulrich1986, Brown1991, Kjeldsen1995} and not from full pulsation analysis, though the pulsation code GYRE \citep{Townsend2013} is integrated into MESA. The bottom left panel shows isochrones in the $\log (T_{c})$--$\log (\rho_{c})$ plane, where the dashed, dotted--dashed, and solid lines show thresholds for hydrogen, helium, and carbon ignition. The 10~Myr isochrone contains massive stars that are able to ignite carbon whereas at the older ages, the isochrones cannot reach sufficiently high central densities and temperatures. At $10$~Gyr, only the low-mass stars remain and helium core flash at RGBTip shows up as a discontinuous sharp feature. The bottom middle and right panels show $\log (L/L_{\rm Edd})$, the ratio of total luminosity to Eddington luminosity, and $\dot{M}$, the mass loss rate, as a function of initial mass. As expected, both quantities generally increase as the initial mass increases. A very prominent increase immediately followed by a sharp decrease at intermediate ages (0.1 and 1~Gyr) in both panels is due to the TPAGB phase, followed by the post-AGB and WD cooling phases.

\begin{figure}
\centering
\includegraphics[width=0.45\textwidth]{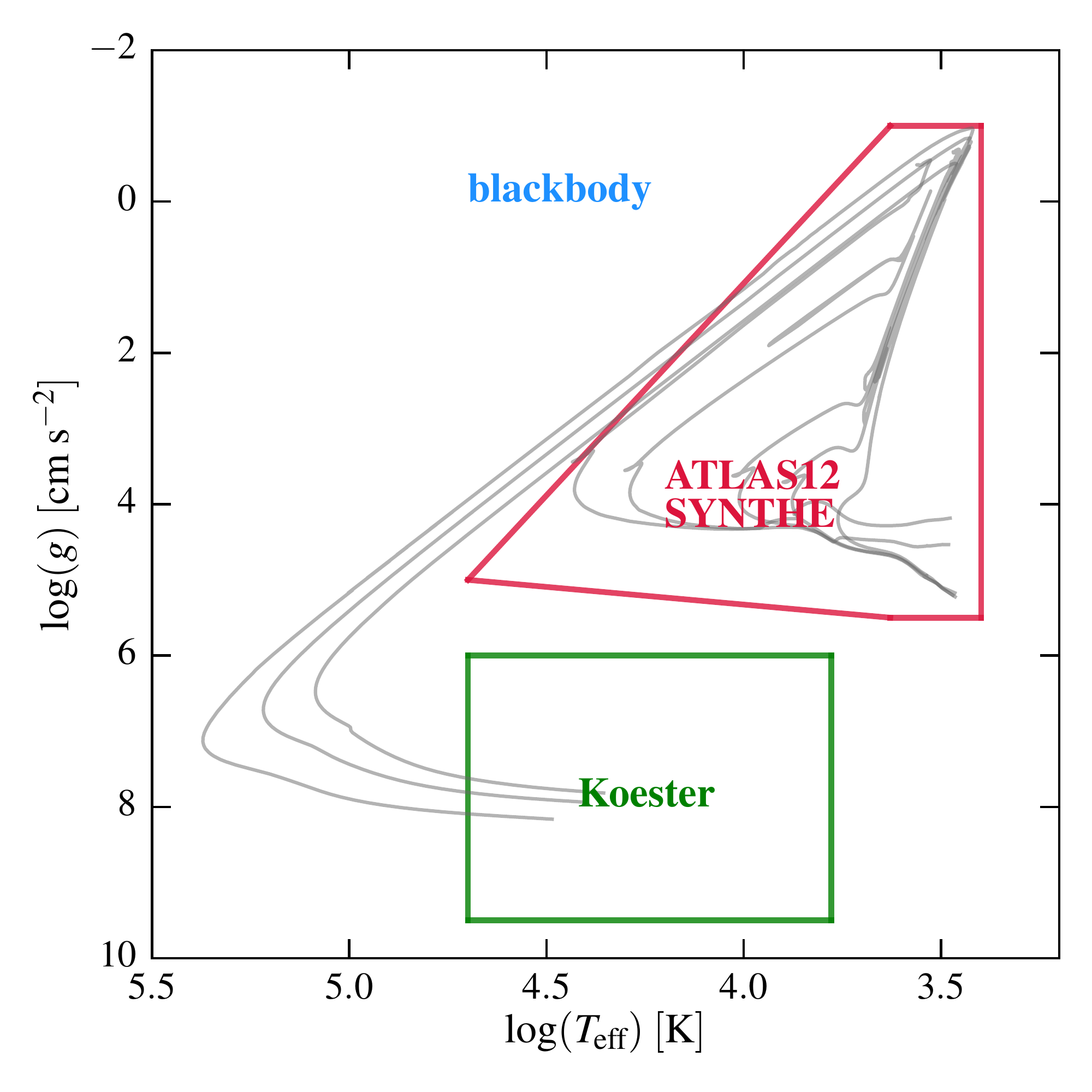}
\caption{Coverage of the synthetic spectra grids used to derive bolometric corrections. Example isochrones at $\logage = 7.0$, 7.5, 8.5, 9.0, and 10.0 are overplotted in gray for reference. The coverage in $\log g$ and $\log \teff$ is the same for all metallicities.}
\label{fig:spectra}
\end{figure}

\begin{table*}
\centering
  \begin{longtable}{ll} \\
  \caption{Current list of photometric systems.} \\
    \hline \hline \noalign{\smallskip}
    System & Reference \\
    \noalign{\smallskip}
    \hline
    \noalign{\smallskip}
    UBVRI & \citet{Bessell2012}  \\
    Str\"omgren & \citet{Bessell2011} \\
    Washington & \citet{Bessell2001}  \\
    DDO51 & www.noao.edu/kpno/mosaic/filters/ \\
    SDSS & classic.sdss.org/dr7/instruments/imager/index.html \\
    CFHT/MegaCam & www.cfht.hawaii.edu/Instruments/Imaging/Megacam/specsinformation.html \\
    PanSTARRS & \citet{Tonry2012}  \\
    DECam & www.ctio.noao.edu/noao/sites/default/files/DECam/DECam\_filters.xlsx \\
    SkyMapper & \citet{Bessell2011B} \\
    Kepler & keplergo.arc.nasa.gov/CalibrationResponse.shtml \\
    HST/ACS & www.stsci.edu/hst/acs/analysis/throughputs \\
    HST/WFPC2 & \citet{Holtzman1995} \\
    HST/WFC3 & www.stsci.edu/hst/wfc3/ins\_performance/filters/  \\
    2MASS & \citet{Cohen2003} \\
    UKIDSS & \citet{Hewett2006}  \\
    Spitzer/IRAC & \citet{Fazio2004}  \\
    WISE & \citet{Wright2010} \\
    GALEX & http://asd.gsfc.nasa.gov/archive/galex/Documents/PostLaunchResponseCurveData.html \\
    Swift & http://swift.gsfc.nasa.gov/proposals/swift\_responses.html \\
    \noalign{\smallskip}
    \hline \hline
\label{tab:photo}
  \end{longtable}  
\end{table*}

\subsection{Bolometric Corrections}
Bolometric corrections are necessary to transform theoretical isochrones into magnitudes that allow for direct comparisons with observations. The bolometric corrections are largely based on a new grid of stellar atmosphere and synthetic spectra created with the ATLAS12 and SYNTHE codes (Conroy et al.\ in preparation). These same models are used for the surface boundary conditions discussed in Section~\ref{section:boundary_conditions}. They include the latest atomic line list from R.\ Kurucz (including both laboratory and predicted lines) and many molecules including CH, CN, TiO, H$_2$, H$_2$O, SiO, C$_2$, SiH, MgH, CrH, CaH, FeH, CO, NH, VO, and OH. We have also computed model atmosphere and spectra for carbon stars with $\rm C/O=1.05$ over the range $2400<\teff<4700$~K and $-1.0<\log g~[\rm g\;cm^{-3}]<0.5$. Our carbon star spectra agree well with the models of \cite{Aringer2009}. The primary differences arise at $\rm >2~{\mu}m$ as our models do not currently include the important molecules C$_3$, HCN, and C$_2$H$_2$. We chose to create our own carbon star spectra in order to have models covering the full wavelength range and at the same resolution as our main spectral library. The ATLAS12/SYNTHE spectra are combined with synthetic spectra for H-rich WDs with $6,000 \leq \teff \leq 50,000$ K from \citet{Koester2010}. These are supplemented by a set of blackbody spectra with $ 200,000 \leq \teff \leq 1$ million K. The coverage of the different synthetic libraries is shown in Figure \ref{fig:spectra}. Example isochrones at $\logage = 7.0$, 7.5, 8.5, 9.0, and 10.0 are overplotted for reference.

\begin{figure*}
\centering
\includegraphics[width=0.95\textwidth]{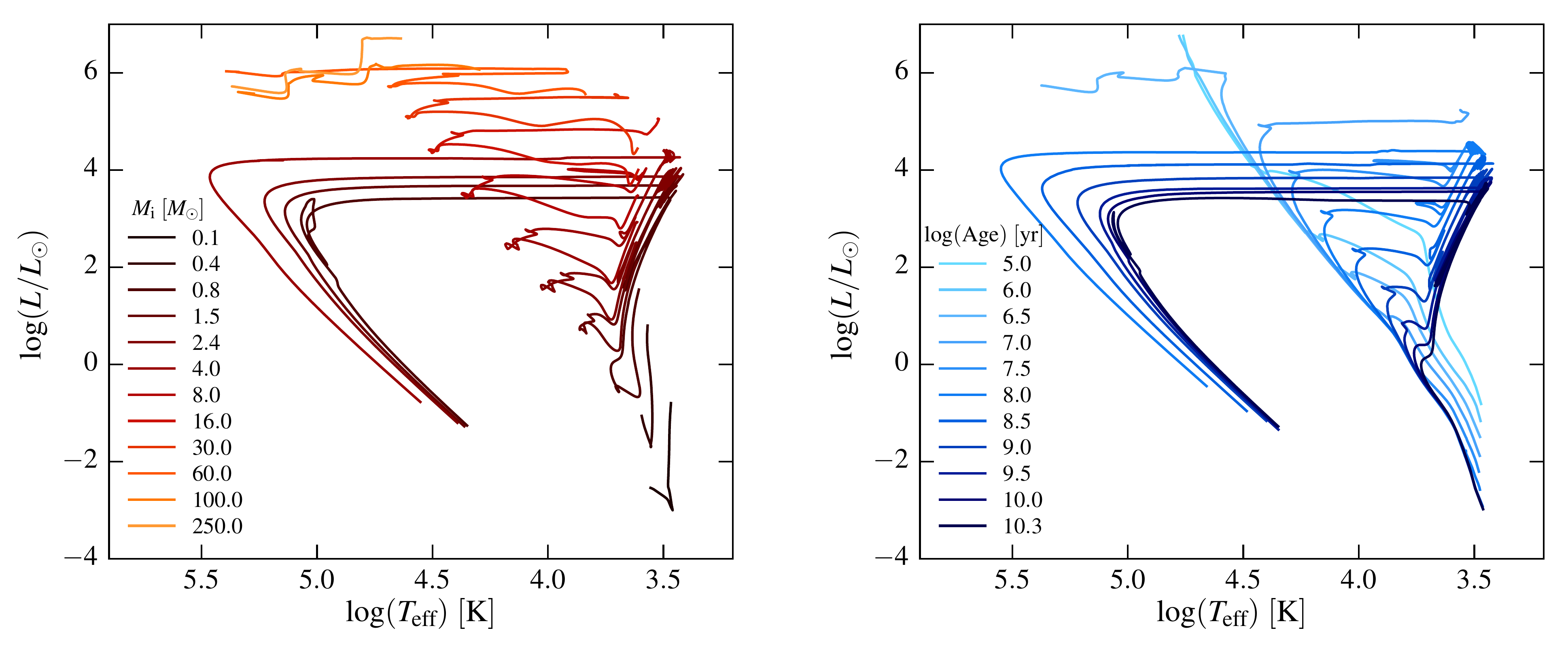}
\caption{An example solar metallicity grid of stellar evolutionary tracks (left) and isochrones (right) covering a wide range of stellar masses, ages, and evolutionary phases.}
\label{fig:tracks_iso_wholerange}
\end{figure*}

\begin{figure*}
\centering
\includegraphics[width=0.95\textwidth]{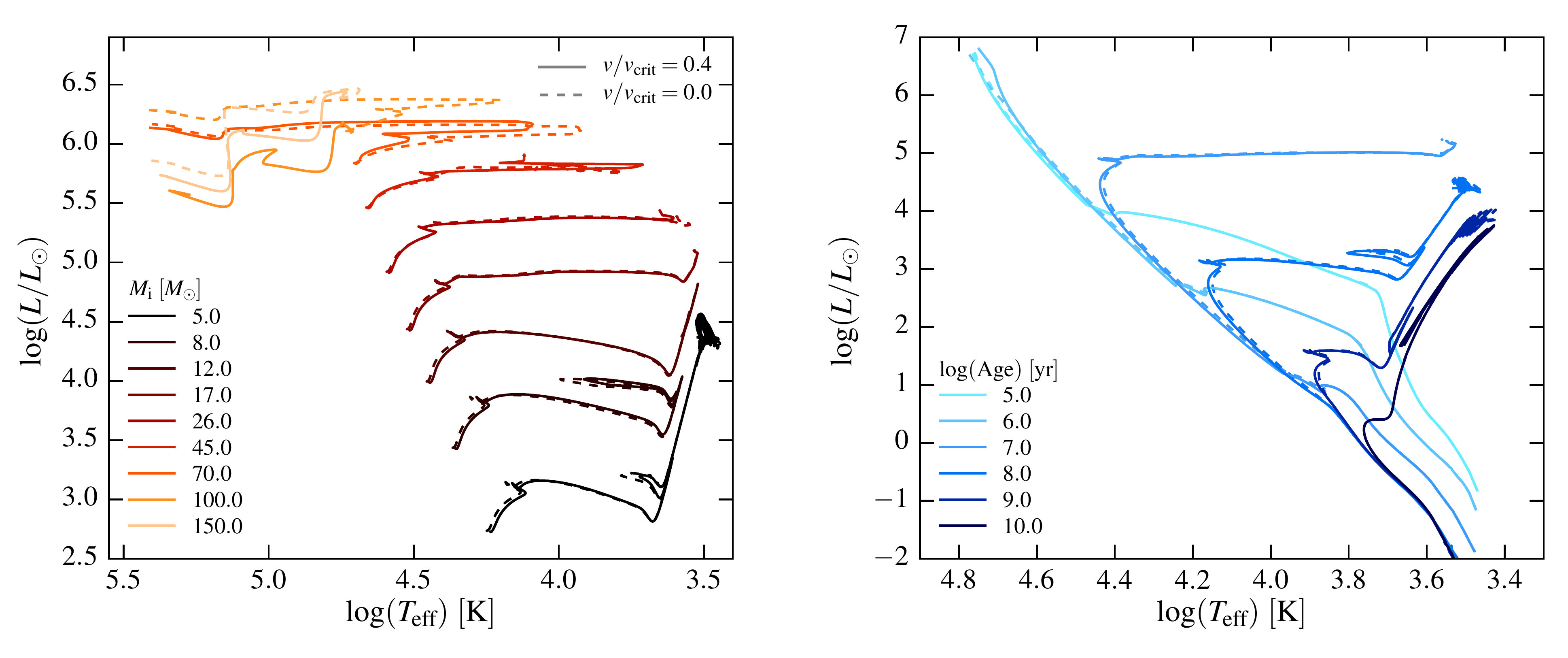}
\caption{Similar to Figure~\ref{fig:tracks_iso_wholerange} but now showing the effect of rotation on both the evolutionary tracks (left) and isochrones (right). Models with and without rotation are shown in solid and dashed lines, respectively. For rotating models, solid-body rotation with $v_{\rm ZAMS}/v_{\rm crit} = \Omega_{\rm ZAMS}/\Omega_{\rm crit} = 0.4$ is initialized at ZAMS. The PMS phase is not shown in the left panel for display purposes.}
\label{fig:tracks_iso_rot_vs_nonrot}
\end{figure*}

\begin{figure}
\centering
\includegraphics[width=0.48\textwidth]{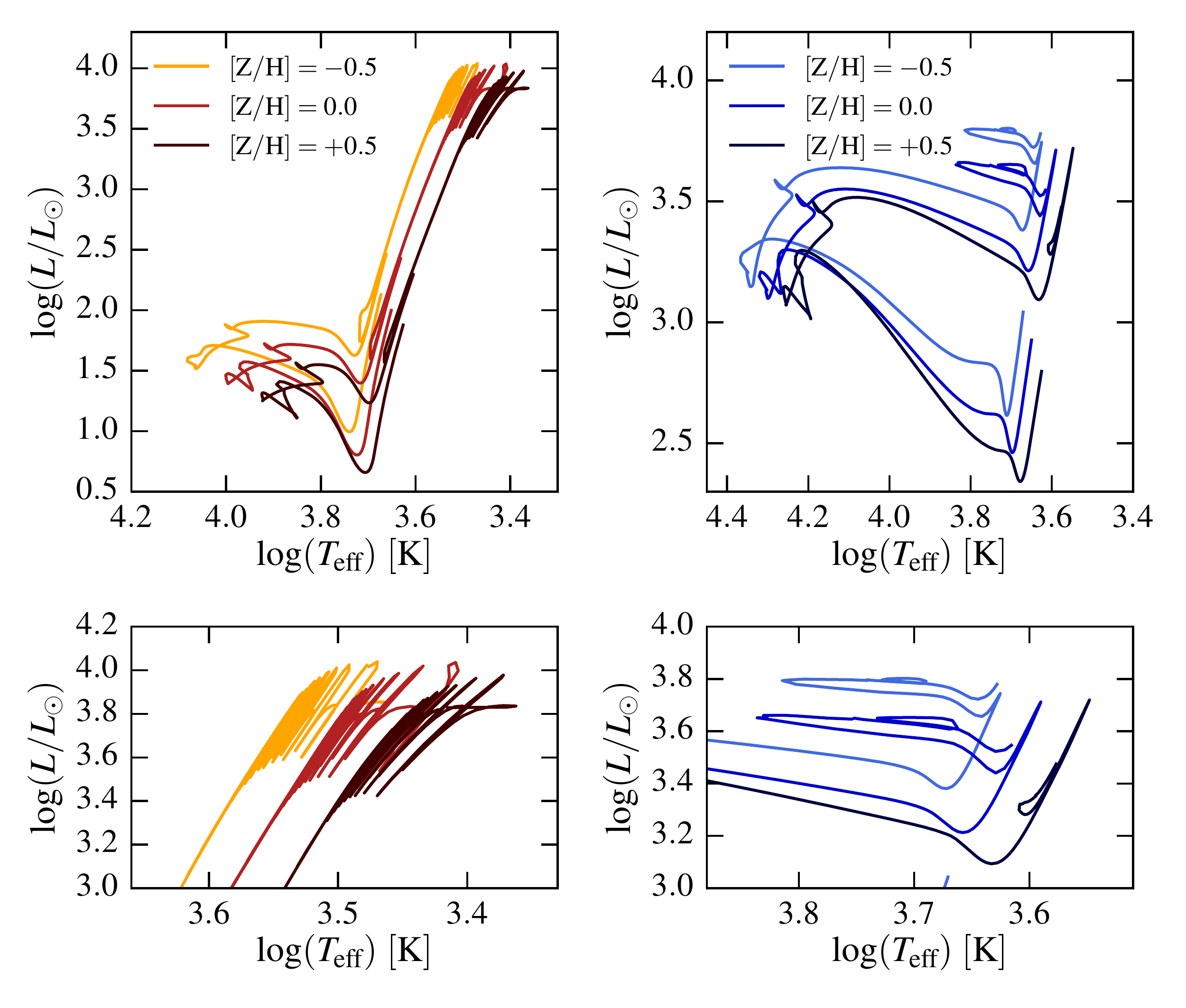}
\caption{A series of $2.2~\msun$ (left) and $6.4~\msun$ (right) evolutionary tracks for three metallicities. The tracks become hotter and more luminous with decreasing metallicity due to a lower Rosseland mean opacity. In the bottom panels, we zoom in on the TPAGB (left) and CHeB (right) to further highlight the effects of metallicity on these evolutionary phases.}
\label{fig:tracks_zoom_show_phases}
\end{figure}

Bolometric corrections are computed from the synthetic spectra following Equation~1 of \citet{Girardi2008}. As noted in the Introduction, we adopt as a zeropoint $L_{\circ}=3.0128\times10^{35}$~erg\;s$^{-1}$ to define $\mbol=0$. This is equivalent to adopting solar values of $M_{\rm bol,\odot}=4.74$ and $\lsun=3.828\times10^{33}$ erg s$^{-1}$. The bolometric corrections include a range of extinction values, as characterized by both $A_V$ and $R_V$, following the extinction curve of \citet{Cardelli1989}. We provide $A_V = 0$ to 6 with $R_V=3.1$, though other $R_V$ values can be made upon request. We emphasize that $Z$ of the bolometric correction is matched to the surface $Z$ (and surface $\rm C/O$ ratio where relevant) for each star along the isochrone.

The photometric systems included in the initial MIST release are summarized in Table \ref{tab:photo}. This is only an initial set and will expand over time. Photometric systems define their magnitude scales according to a flux standard.

\section{Overview and Basic Properties of the Models}
\subsection{Tracks and Isochrones}
\label{section:tracks_and_isochrones}
As described in detail in Section~\ref{section:ages_masses_phases_metallicities}, the MIST models cover a wide range in stellar masses, ages, metallicities, and evolutionary phases. The stellar mass of evolutionary tracks ranges from $0.1~\msun$ to $300~\msun$, the ages of isochrones cover $\log \textrm{Age} = 5$ to $\log \textrm{Age} = 10.3$, and the metallicity ranges from $\rm [Fe/H] = -4.0$ to $+0.5$. The evolution is continuously computed from the PMS phase to the end of hydrogen burning, WD cooling phase, or the end of carbon burning, depending on the initial stellar mass and metallicity.

Figure~\ref{fig:tracks_iso_wholerange} illustrates the range of stellar masses, ages, and evolutionary phases, showing the evolutionary tracks and isochrones in the left and right panels, respectively. As noted in Section~\ref{section:adopted_physics}, the models include rotation with $v/v_{\rm crit}=0.4$ by default. Figure~\ref{fig:tracks_iso_rot_vs_nonrot} shows the effect of rotation on both the evolutionary tracks and isochrones, where the rotating and non-rotating models are shown in solid and dashed lines, respectively. For display purposes, we omit the post-AGB phase where relevant. Rotation makes stars more luminous during the MS because rotational mixing (see Section~\ref{section:rot_mixing}) promotes core growth. It makes the star appear hotter or cooler depending on the efficiency of rotational mixing in the envelope: if rotational mixing introduces a sufficient amount of helium into the envelope and increases the mean molecular weight, the star becomes more compact and hotter. However, in the absence of efficient rotational mixing, the centrifugal effect dominates, making the star appear cooler and more extended. This is because in a rotating system, ordinary gravitational acceleration $g$ is replaced by $g_{\rm eff}$ that includes both the gravitational and centrifugal terms. Since $\teff \propto g_{\rm eff}^{1/4}$ for a star with a radiative envelope and $|g_{\rm eff}| < |g|$, $\teff$ is thus lower in a rotating system. Generally speaking, rotation increases $\teff$ in massive stars where rotational mixing operates efficiently, and it decreases $\teff$ in low-mass stars as well as at all ZAMS locations where the centrifugal effect dominates.

\begin{figure}
\centering
\includegraphics[width=0.45\textwidth]{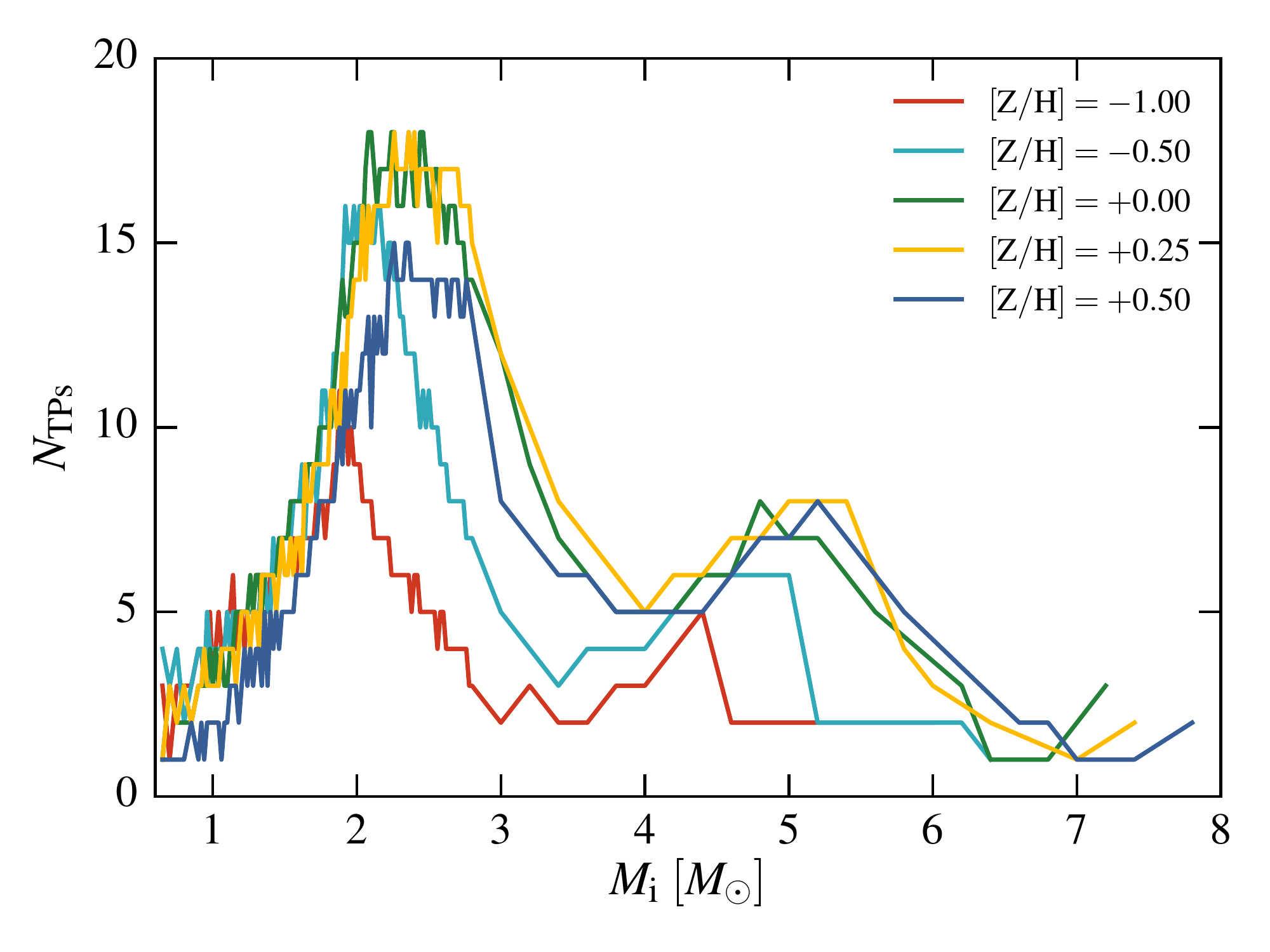} 
\caption{The number of thermal pulses (TPs) executed by each model as a function of mass at a number of metallicities. There are two notable features: first, the maximum occurs at around $2~\msun$ independent of metallicity, and two, the number of thermal pulses increases with $\rm [Z/H]$ from -1.0 to 0.0, and there is a hint that the trend reverses for $\rm [Z/H]>0$.}
\label{fig:NTP}
\end{figure}

\begin{figure}
\centering
\includegraphics[width=0.45\textwidth]{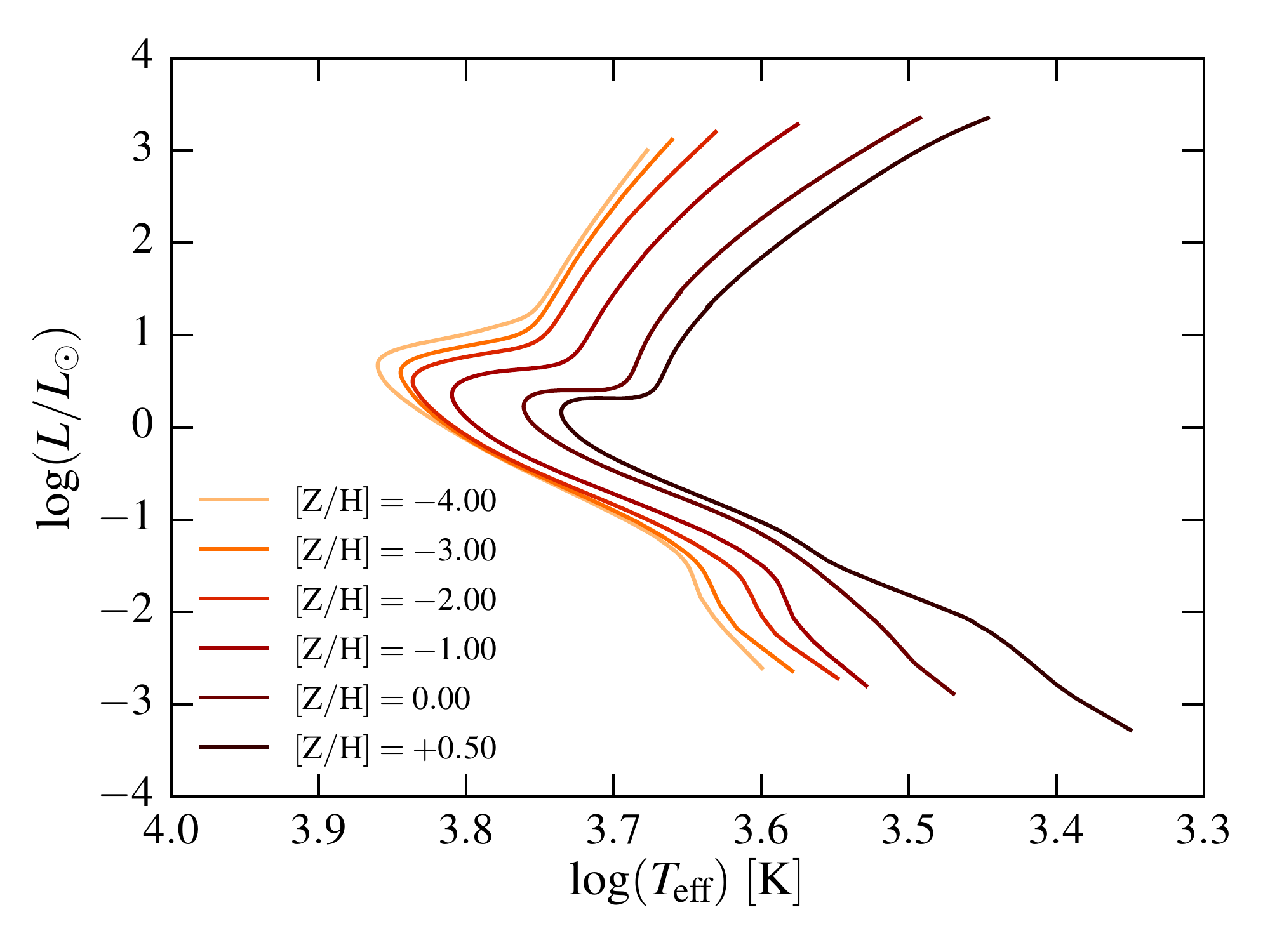}
\caption{Isochrones at 10 Gyr over a wide range of metallicities. For display purposes, we omit the phases beyond RGBTip. As the metallicity decreases, the MSTO becomes hotter and more luminous (and the MSTO mass decreases) and the RGBTip becomes fainter due to the helium ignition occurring at lower core masses. Note that the isochrone changes more subtly with metallicity in the very metal-poor regime, i.e., $\rm [Z/H]\lesssim -2$.}
\label{fig:Z_range}
\end{figure}

\begin{figure}
\centering
\includegraphics[width=0.45\textwidth]{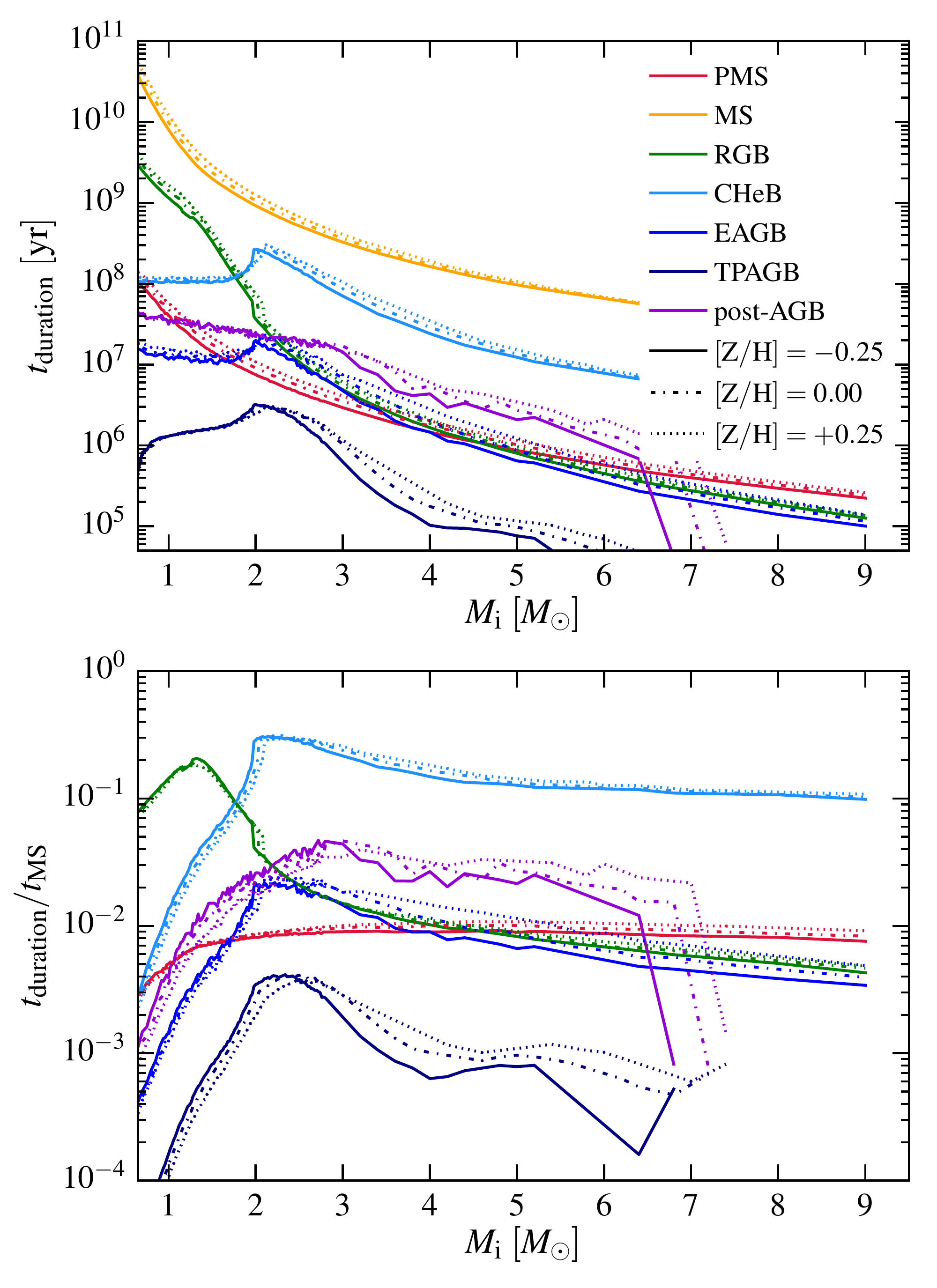}
\caption{Top: phase lifetimes as a function of initial mass for $\rm [Z/H] = -0.25$, 0.0 and $+0.25$. High mass stars are not included because they do not go through the same set of evolutionary phases featured here. Note that the ``RGB'' label refers to the phase between TAMS and helium ignition, which includes the short subgiant branch (SGB) evolution, and ``post-AGB'' includes the white dwarf cooling phase up to $\Gamma=20$. Bottom: same as above but now showing the ratio of phase lifetimes to the MS lifetime instead.}
\label{fig:phase_lifetimes}
\end{figure}

\begin{figure*}
\centering
\includegraphics[width=0.9\textwidth]{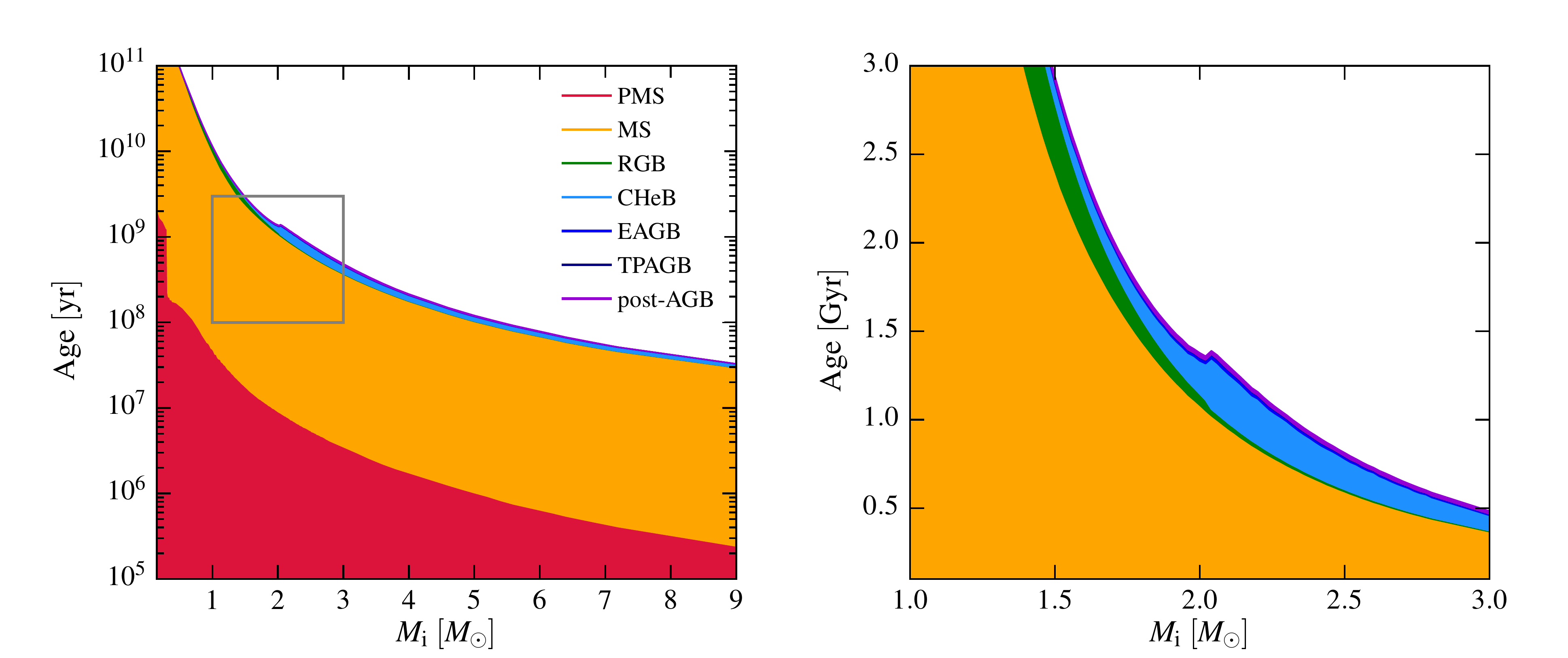}
\caption{Left: similar to Figure~\ref{fig:phase_lifetimes} but now showing the cumulative age as a function of mass for $\rm [Z/H] = 0.0$. The gray box marks the zoomed-in region shown in the right panel. Right: zooming in on a mass range that shows a critical transition from degenerate helium ignition to quiescent helium ignition. This produce a non-monotonic mass-age relationship for these phase. Note the linear scale in $y$. This figure is adapted from Figure 1 of \cite{Girardi2013}.}
\label{fig:phase_lifetimes_cumul}
\end{figure*}

The top two panels of Figure~\ref{fig:tracks_zoom_show_phases} show a series of $2.2~\msun$ (left) and $6.4~\msun$ (right) evolutionary tracks for three metallicities. A portion of the PMS phase in both panels and the evolution following the CHeB phase in the right panel are omitted for display purposes. The tracks become hotter and more luminous with decreasing metal content due to a lower Rosseland mean opacity. In the bottom panels, we zoom in on the TPAGB and CHeB to further highlight the effects of metallicity on these evolutionary phases. The blue loops become hotter and more prominent with decreasing metallicity and the entire TPAGB phase also shifts to hotter temperatures as metallicity decreases. To scrutinize the effect of metallicity on the TPAGB phase in more detail, we plot the number of TPs executed by each model as a function of mass at a number of metallicities in Figure~\ref{fig:NTP}. There are two notable features: the maximum number occurs at around $2~\msun$ independent of metallicity; and the number of thermal pulses increases with $\rm [Z/H]$ from $-1.0$ to 0.0, and there is a hint that the trend reverses for $\rm [Z/H]>0$. We leave a more detailed discussion of the number of TPs as a function of uncertain physical parameters (e.g., mixing, mass loss) and comparisons to other databases for future work.

In order to illustrate the range of metallicities available in the current MIST models, we show 10~Gyr isochrones at $\rm [Z/H] = -4$, $-3$, $-2$, $-1$, 0, and $0.5$ in Figure~\ref{fig:Z_range}. For clarity, we only show phases up to the RGBTip. As the metallicity decreases, the MSTO becomes hotter and more luminous (and the MSTO mass decreases) and the RGBTip becomes fainter due to the helium ignition occurring at lower core masses. Note that the isochrone changes more subtly with metallicity in the very metal-poor regime, i.e., $\rm [Z/H]\lesssim -2$.

In the top panel of Figure~\ref{fig:phase_lifetimes} we show phase lifetimes as a function of initial mass for $\rm [Z/H] = -0.25$, 0.0 and $+0.25$ in solid, dotted--dashed, and dotted lines, respectively. The bottom panel shows the ratio of phase lifetimes to the MS lifetime. Note that the ``RGB'' label refers to the phase between TAMS and helium ignition, which includes the short subgiant branch (SGB) evolution, and ``post-AGB'' includes the white dwarf cooling phase up to $\Gamma=20$. The post-AGB timescales in the MIST models (adopting the definition from \citealt{MillerBertolami2016}) are consistent with those reported by \cite{MillerBertolami2016} and \cite{Weiss2009}, which are a factor of $3\textrm{--}10$ shorter compared to the older post-AGB stellar evolution models \citep{Vassiliadis1994, Bloecker1995}. High mass stars are not included because they do not go through the same set of evolutionary phases featured here. The TPAGB and post-AGB phases are not shown for a subset of the models that do not completely evolve through those evolutionary stages. Unsurprisingly, the lifetimes generally decrease with increasing mass, though there are some notable exceptions including the peak in CHeB and AGB lifetimes at $\sim2~\msun$ (see the discussion below).

The left panel of Figure~\ref{fig:phase_lifetimes_cumul} is a slight variation of the previous plot, where we now show the cumulative age as a function of mass for $\rm [Z/H] = 0.0$. In the right panel, we zoom in on a particularly interesting mass range around $2~\msun$, where there a noticeable increase in the CHeB lifetime. This effect, explored in detail in \cite{Girardi2013}, is due to the transition from the explosive ignition of helium in degenerate cores of low-mass stars, i.e., helium core flash at RGBTip, to quiescent ignition of helium in more massive stars. This is because more massive stars start burning helium at lower core masses and therefore have smaller initial helium-burning luminosities compared to those that undergo the helium core flash (see also \citealt{Girardi1999}). Note that this non-monotonicity in the CHeB lifetime is present even in models computed without RGB mass loss. Thus, the simple ``increasing mass = decreasing MS lifetime'' rule of thumb is violated over a very narrow mass range. In other words, there is a special age at which two stars of different masses are at the same evolutionary stage in terms of their EEPs. This notion of ``double EEPs'' is discussed in more detail in Paper 0 \citep{Dotter2016} and will be explored in a future paper.

\begin{figure*}
\centering
\includegraphics[width=0.7\textwidth]{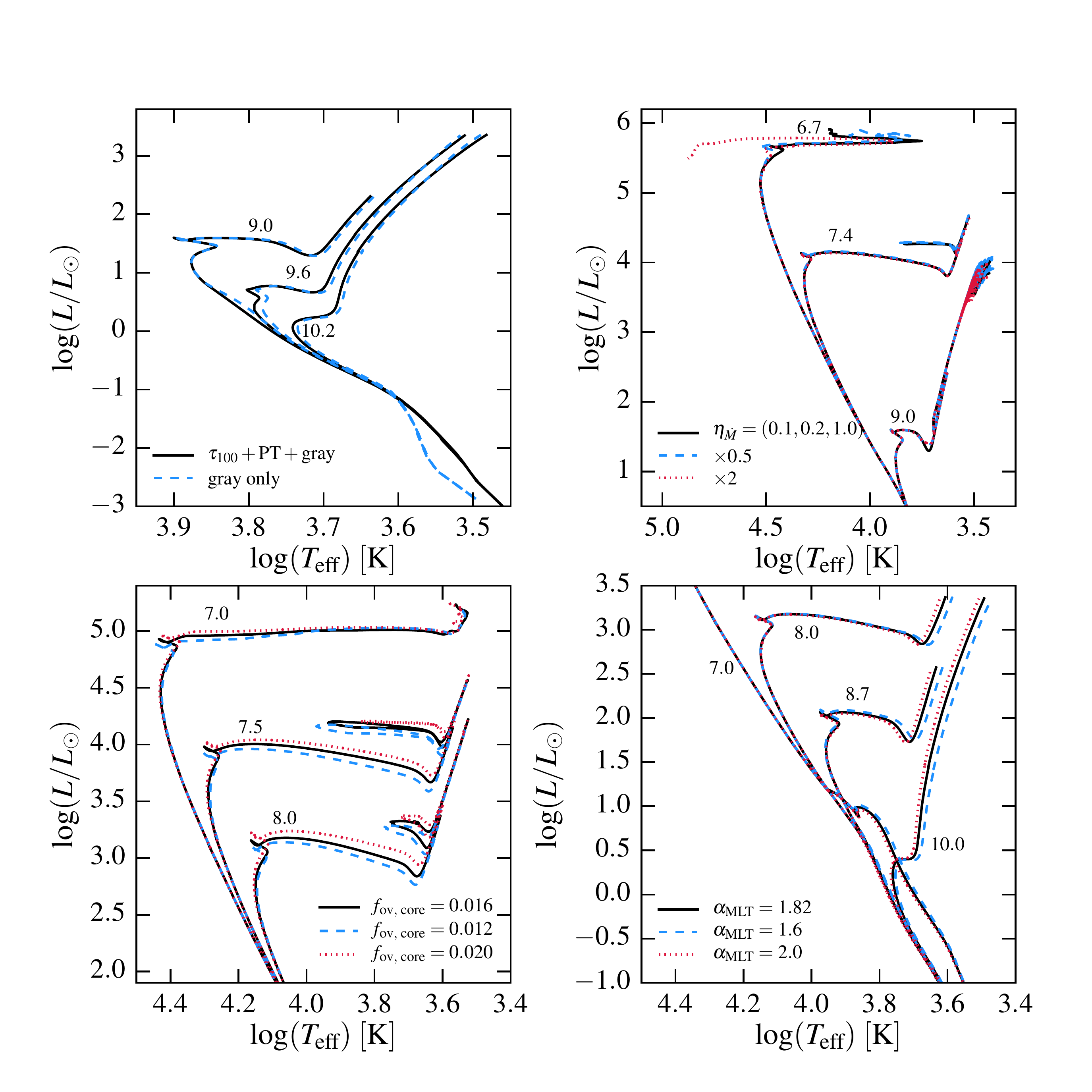}
\caption{Solar metallicity isochrones at a number of ages showing the effects of varying input physics. The fiducial model is shown in black solid lines in all four panels. Note different $x$ and $y$ axes in each panel. Top left: variations in the adopted boundary conditions for $\logage = 9.0, 9.6$, and 10.2. Top right: variations in the efficiency of mass loss for $\logage = 6.7, 7.4$, and $9.0$. The set of three numbers corresponds to $\eta_{\rm R}$, $\eta_{\rm B}$, and $\eta_{\rm Dutch}$. Bottom left: variations in the efficiency of overshoot in the core parameterized by $f_{\rm ov,\;core}$ for $\logage = 7.0, 7.5$, and 8.0. Bottom right: variations in the efficiency of convection parameterized by $\amlt$ for $\logage = 7.0, 8.0, 8.7$, and 10.0. }
\label{fig:varying_params}
\end{figure*}

\subsection{Non-standard Models}
\label{section:nonstandard_models}
To explore the effects of uncertain input physics and choices of free parameters on the resulting tracks and isochrones, we ran several sets of models varying one ingredient at a time. In Figure~\ref{fig:varying_params} we present four such variations. The black solid lines represent solar metallicity models with fiducial parameters listed in Table~\ref{tab:param_summary}. We note that several of these parameters affect the lifetimes of different evolutionary phases, though this is not explicitly shown in the figures.

In the top left panel, we show two sets of isochrones which are identical except for their surface boundary conditions. The three sets of ages correspond to $\logage = 9.0, 9.6$, and 10.2. The solid lines are isochrones with our default implementation of boundary conditions; $\tau=100$ and photosphere tables from the ATLAS12/SYNTHE atmosphere models plus the simple Eddington gray $T(\tau$) approximation for the hottest stars. The dashed lines are models with the Eddington gray $T(\tau$) relation applied across the entire mass range. For clarity, we only plot the isochrones up to the RGBTip. As expected, the choice of boundary conditions has the largest effect on the lower MS populated by cool, compact dwarfs. The shift in $\teff$ on the RGB is smaller but important, amounting $\sim50\text{--}60$~K. Differences both on the SGB and RGB and on the lower MS are larger at non-solar metallicities. In particular, the isochrones become more discrepant on the SGB and RGB at $\rm [Z/H] = -1$ and on the lower MS at $\rm [Z/H] = +0.3$.

In the top right panel, we show three sets of isochrones at $\logage = 6.7, 7.4$, and 9.0 to illustrate the impact mass loss has on various stages of evolution. The solid lines correspond to isochrones with $\eta_{\rm R}=0.1,\;\eta_{\rm B}=0.2$, and $\eta_{\rm Dutch}=1.0$, whereas the dashed lines and dotted lines represent mass loss rates that are twice and half as efficient, respectively. For display purposes, we omit the PMS and post-AGB phases. The temperature evolution is especially sensitive to the mass loss rates at the youngest ages because massive star evolution is strongly affected by the choice of input physics. The morphology of the CHeB is also directly influenced by mass loss rates: lower and higher $\dot{M}$ values yield hotter and cooler CHeB, respectively. The morphology and lifetime of the notorious TPAGB phase are also affected mass loss rates, with more efficient winds resulting in fainter and fewer TPs as expected. The mass loss efficiency, parameterized by the $\eta$ parameter, is calibrated empirically to match various observational constraints, including the number ratios of different stellar types for the high-mass stars (Section~\ref{section:highM_ratios}) and the AGB luminosity functions for the low-mass stars (Section~\ref{section:agbcs_lf}).

In the bottom left panel, we highlight the importance of core convective overshoot. The three sets of ages shown are $\logage = 7.0, 7.5$, and 8.0. The solid lines are isochrones with the default core overshoot parameter $f_{\rm ov,\;core}=0.016$ in the exponential diffusive overshoot formalism. The dashed and dotted lines represent decreased and increased ($f_{\rm ov,\;core}=0.012,\;0.020$) overshoot efficiency, respectively. For clarity, we plot the evolution up through helium burning only. Since convective overshoot enhances the mixing of fresh fuel into the core, a higher overshoot efficiency results in longer MS lifetimes and systematically higher MSTO masses and luminosities. Likewise, SGB and CHeB luminosities are higher for more efficient overshoot due to the larger resulting core masses. Note that the overshoot efficiency in the core is constrained by matching the MSTO in M67, and the overshoot efficiency in the envelope is determined during solar calibration.

In the bottom right panel, we show isochrones with different values of the mixing length parameter $\amlt$. The four ages shown are $\logage = 7.0, 8.0, 8.7$, and 10.0. The solid lines are isochrones with the solar-calibrated value $\amlt=1.82$. The dashed and dotted lines correspond to isochrones with $\amlt=1.6$ and $\amlt=2.0$, respectively. For display purposes, we only show the evolution up through the RGBTip. The physical interpretation of a small value of $\amlt$ is a fluid parcel traveling a short radial distance (in units of $H_{\rm P}$) before it deposits its internal energy and blends into the surrounding medium. The net effect is reduced convective efficiency, thus cooler temperatures and more inflated radii. For this reason, $\amlt$ is used to mimic the effects of physical ingredients, e.g., the inhibition of convection by a magnetic field, that are missing from the majority of current stellar evolution models (but see \citealt{Feiden2013}). For example, $\amlt$ that is smaller compared to the solar-calibrated value is commonly used to bring models into agreement with observations of inflated radii in low-mass stars (see Section~\ref{section:lmrt_relation}).

\section{Comparisons with Existing Databases}
In this section we compare the MIST models to several popular stellar evolution databases in the literature. Due to differences in the choice of input physics and their implementations in the codes, an apples-to-apples comparison is challenging. We stress that this is neither a comprehensive review of all published models in the literature nor a thorough and detailed comparison between different databases. Instead, we aim to provide the reader with a general impression of how the new MIST models compare to several widely used models. We refer the reader to the MESA instrument papers \citep{Paxton2011, Paxton2013, Paxton2015} for closer comparisons at the level of the codes themselves and their evolutionary track outputs. Our goal here is to compare at the level of databases, which reflects the net effect of many different choices for input physics.

\begin{figure}
\centering
\includegraphics[width=0.47\textwidth]{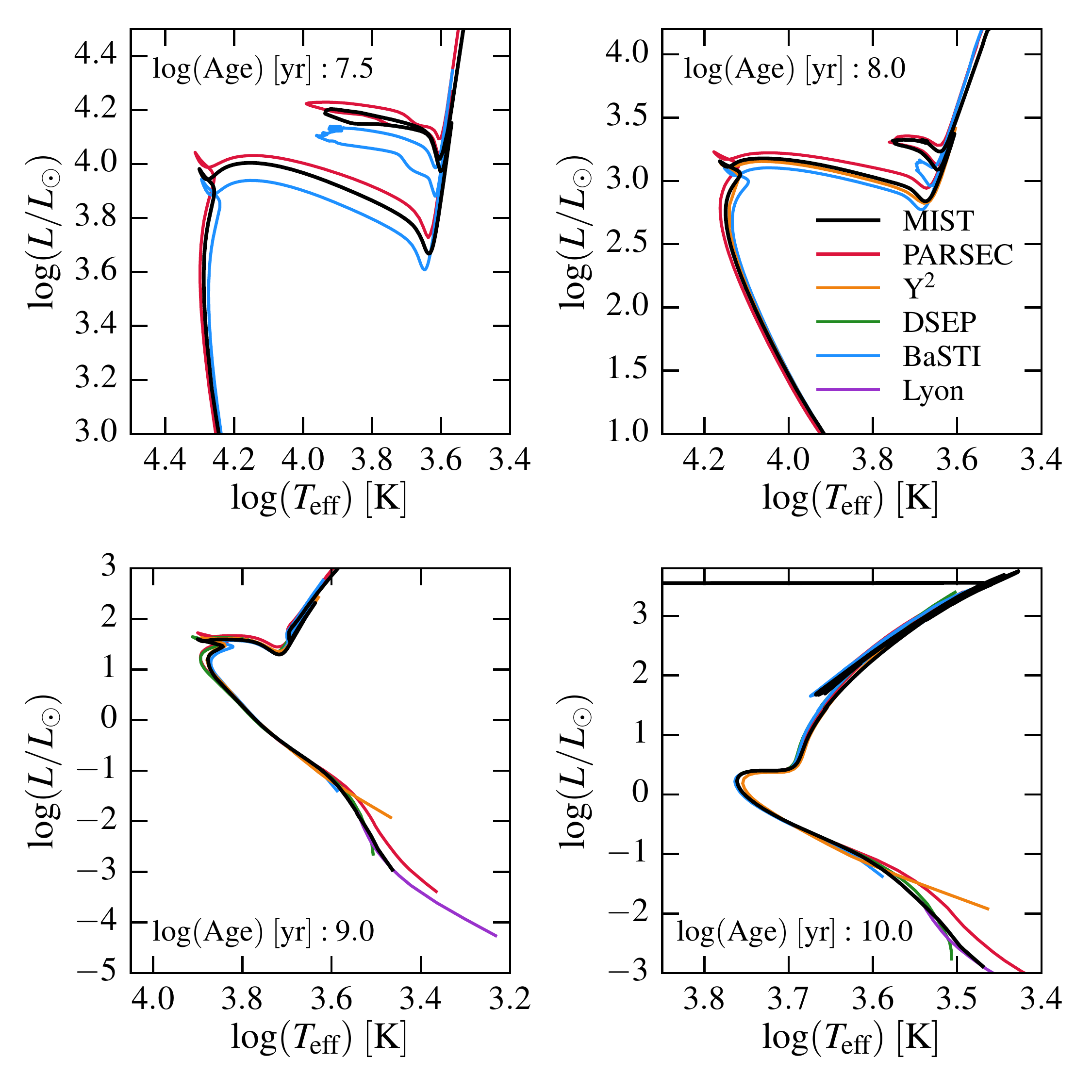}
\caption{A comparison between MIST (black; this work), PARSEC v1.2S \citep[red;][]{Bressan2012, Chen2014, Tang2014}, Y$^{2}$ \citep[orange;][]{Demarque2004}, DSEP \citep[green;][]{Dotter2008}, BaSTI ``non-canonical'' \citep[turquoise;][]{Pietrinferni2004}, and Lyon \citep[navy;][]{Baraffe1998, Baraffe2003, Baraffe2015} isochrones at $Z=Z_{\odot}$ as defined by each model.}
\label{fig:compare_isos_solar}
\end{figure}

\vspace{1cm}
\subsection{Evolutionary Tracks and Isochrones}
We first compare isochrones at $Z_{\odot}$ adopted by each model. In Figure~\ref{fig:compare_isos_solar}, we compare MIST (black; this work), PARSEC v1.2S \citep[red;][]{Bressan2012, Chen2014, Tang2014}, Y$^2$ \citep[orange;][]{Demarque2004}, DSEP \citep[green;][]{Dotter2008}, BaSTI ``non-canonical'' with $\eta=0.4$ \citep[turquoise;][]{Pietrinferni2004}, and Lyon \citep[navy;][]{Baraffe1998, Baraffe2003, Baraffe2015} for $\logage = 7.5$, 8.0, 9.0, and 10.0. The DSEP and Lyon models are not included in the top two panels because they are not available at young ages. Note that of the models featured here, only the MIST models include the effects of rotation.\footnote{There is a version of Y$^2$ models with rotation for $M<1.25~\msun$. See \cite{Spada2013}.} We choose MIST models with rotation instead of those without to make the comparison because the fiducial models include rotation and all calibrations are performed on this set. Overall, the MIST isochrones are in broad agreement with other isochrones. Although the absolute metal contents differ by as much as $30\%$ between various models due to differences in the preferred definition of $Z_{\odot}$, the isochrones are less discrepant than one might imagine because they have been calibrated to match the properties of the Sun. There is more noticeable discrepancy at the young ages due to the complex and uncertain physics---such as core convective overshoot---governing the evolution of massive stars. In particular, the CHeB phase (e.g., the development of the blue loop) is notoriously sensitive to the details of input physics \citep[e.g.,][]{McQuinn2011} though there are ongoing efforts to address this issue \citep[e.g.,][]{Tang2016}. Moreover, the PARSEC isochrones depart notably from the rest of the models on the lower MS due to their recent implementation of $T-\tau$ boundary conditions that have been empirically calibrated to match the observed mass-radius relations for cool dwarfs \citep{Chen2014}.

\begin{figure}
\centering
\includegraphics[width=0.45\textwidth]{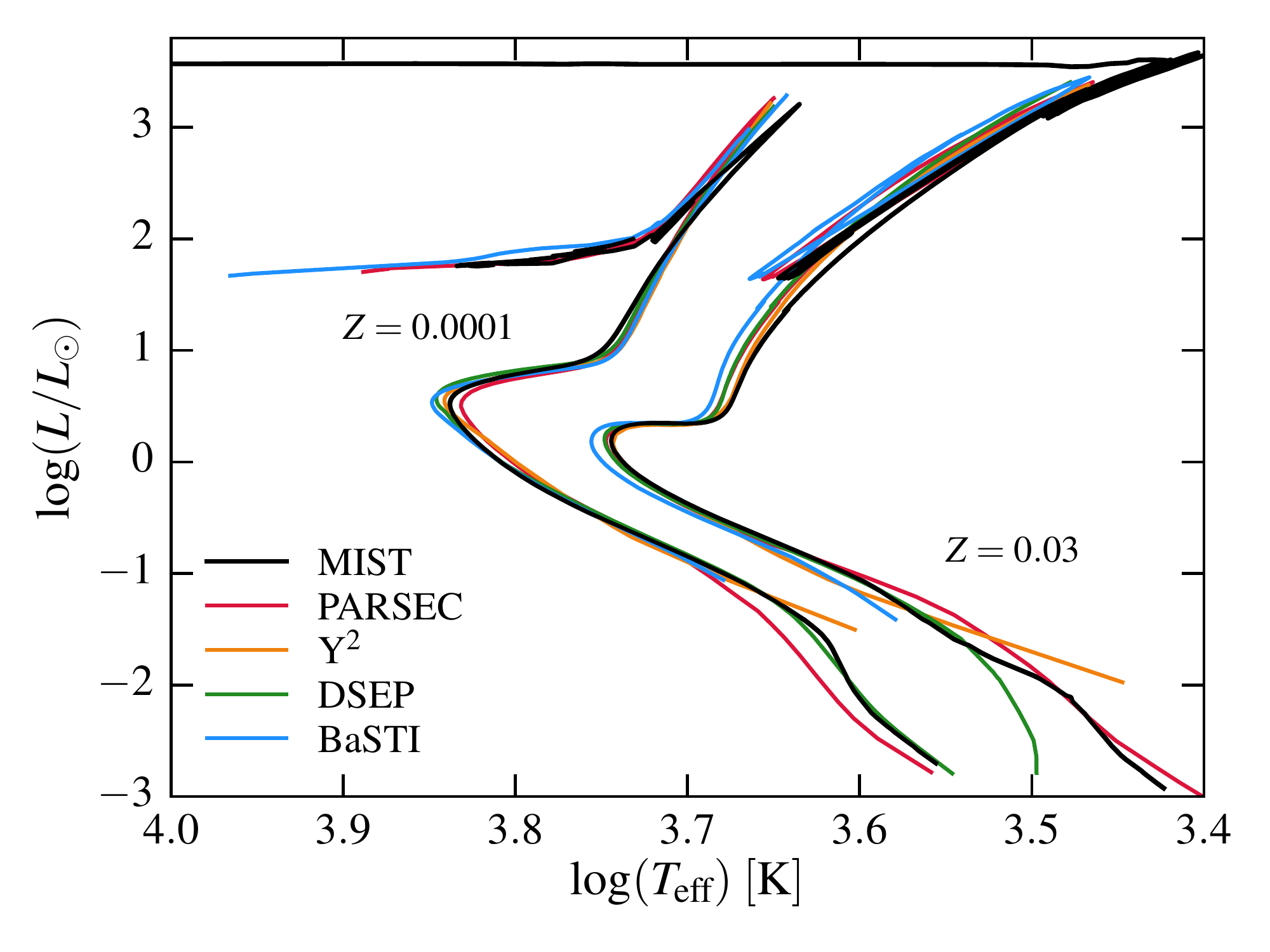}
\caption{The same as Fig~\ref{fig:compare_isos_solar}, except now at $Z=0.0001$ and $Z=0.03$ for $\log \rm Age = 10.0$.}
\label{fig:compare_isos_fixz}
\end{figure}

In Figure~\ref{fig:compare_isos_fixz}, we now compare isochrones at fixed $Z$. Note that although $Z$ is the same, there are still element-to-element variations due to the different solar abundance scales adopted by each group. We plot $\logage=10.0$ isochrones at $Z=0.0001$ and $Z=0.03$ for PARSEC, Y$^2$, DSEP, BaSTI, and MIST. Note that only the MIST models follow the evolution continuously from the helium ignition in the degenerate core (RGBTip) to the CHeB through a series of helium flashes (see also Appendix~\ref{section:appendix_features}). The models are broadly in agreement, though there are some differences in the lower MS and the extent of the CHeB. The former is likely mostly due to differences in the adopted boundary conditions in the models, and the latter is possibly due to differences in the adopted Reimers mass loss efficiency (BaSTI, PARSEC, and MIST adopt $\eta_{\rm R} = 0.4$, 0.2, and 0.1, respectively, while Y$^2$ does not include mass loss).

\begin{figure}
\centering
\includegraphics[width=0.45\textwidth]{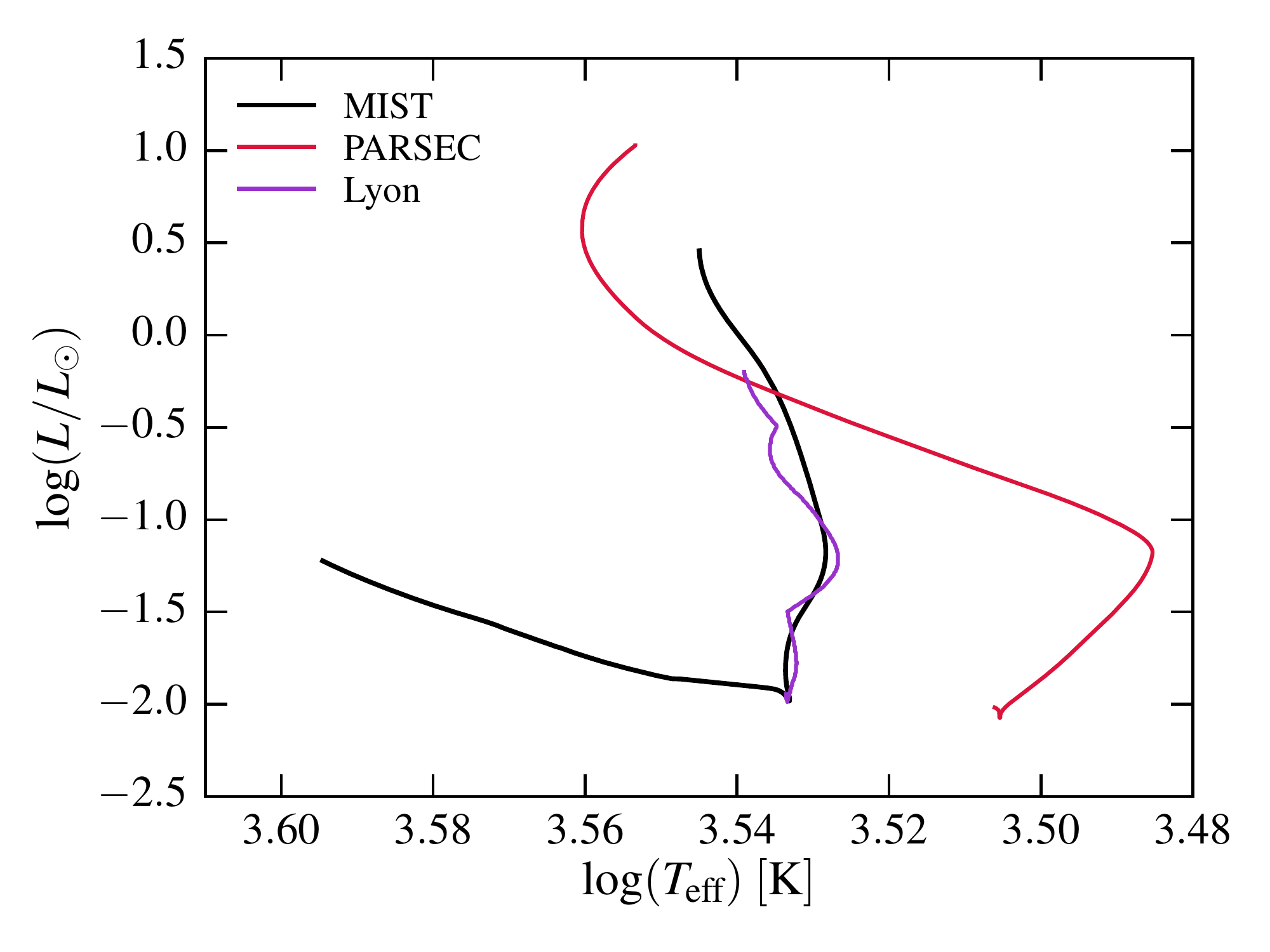}
\caption{The evolutionary tracks for a $0.3~\msun$ star from MIST, PARSEC, and Lyon models at $Z=Z_{\odot}$.}
\label{fig:lowM_tracks}
\end{figure}

Figure~\ref{fig:lowM_tracks} presents a comparison between the MIST, PARSEC, and Lyon models for a $0.3~\msun$ evolutionary track.  As noted above, the PARSEC models now adopt a modified $T-\tau$ relation for low-mass stars, which likely explains the relatively large difference between that model and the others. The Lyon and MIST models largely agree, although several sharp features are noticeable in the Lyon models during the PMS phase that do not appear in the MIST models. The origin of these small differences is unclear to us.

\begin{figure}
\centering
\includegraphics[width=0.45\textwidth]{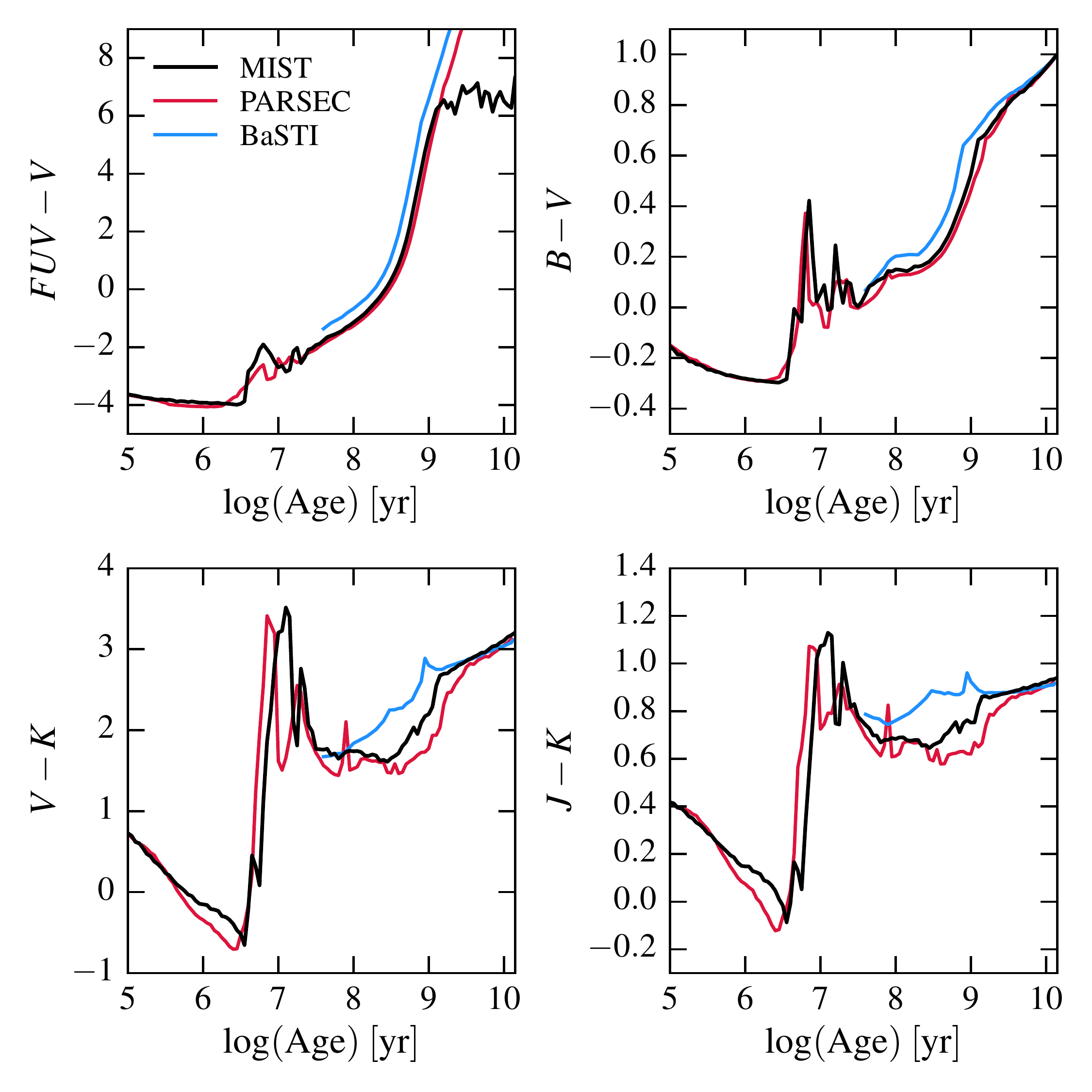}
\caption{The evolution of integrated colors of a simple stellar population for MIST (black), PARSEC/COLIBRI (red), and BaSTI (blue) isochrones at solar metallicity. The colors were computed with the Flexible Stellar Population Synthesis code \citep[FSPS, v2.6;][]{Conroy2009, Conroy2010} assuming a \cite{Kroupa2001} IMF and AGB circumstellar dust turned off. Note that we used the same bolometric corrections for all three cases so any variation in color is purely due to differences in the isochrones.}
\label{fig:ssp_colors}
\end{figure}

\subsection{Simple Stellar Population Colors}
\label{section:ssp_colors}
In Figure~\ref{fig:ssp_colors} we show the evolution of integrated colors of a simple stellar population for MIST (black), PARSEC/COLIBRI \citep[red;][]{Bressan2012, Marigo2013, Rosenfield2014}, and BaSTI \citep[blue;][]{Pietrinferni2004} isochrones at solar metallicity. The colors are computed by integrating along the isochrone at a given age with weights provided by the Kroupa IMF \citep{Kroupa2001}. They are calculated using the Python bindings\footnote{https://github.com/dfm/python-fsps} to the Flexible Stellar Population Synthesis code \citep[FSPS, v2.6;][]{Conroy2009, Conroy2010}. We turn the AGB circumstellar dust option off to enable a more direct comparison between the three models. There are no predictions from the BaSTI models at $\logage < 7.5$ because they only go up to $10~\msun$ in mass. Note that we used the same bolometric corrections for all three cases so any variation in color is purely due to differences in the isochrones.

Overall, the models are in good agreement with each other, especially in $B-V$, though there are some noticeable differences between the models in other colors. At $\logage\gtrsim9$ in $FUV-V$, the MIST prediction turns over toward bluer colors while the PARSEC/COLIBRI and BaSTI predictions continue to get redder. This qualitative difference is due to the inclusion of the post-AGB and WD phases in the MIST models. In $V-K$ and $J-K$, the large spikes at young ages ($\logage\sim7$) are due to the onset of the RSG phase from massive stars. This feature appears at a slightly later time in MIST compared to in PARSEC/COLIBRl, which points to the differences in the lifetimes of massive stars in the two databases. The inclusion of rotational mixing in the MIST models may explain the longer MS lifetimes. Finally, significant differences between the three models occur at intermediate ages in redder colors, where TPAGB stars are expected to contribute a significant fraction of the total luminosity. The MIST color predictions fall between the BaSTI and PARSEC predictions. We note that the full luminosity and temperature variations---the actual thermal pulses---are included in the MIST isochrones.

\begin{figure}
\centering
\includegraphics[width=0.45\textwidth]{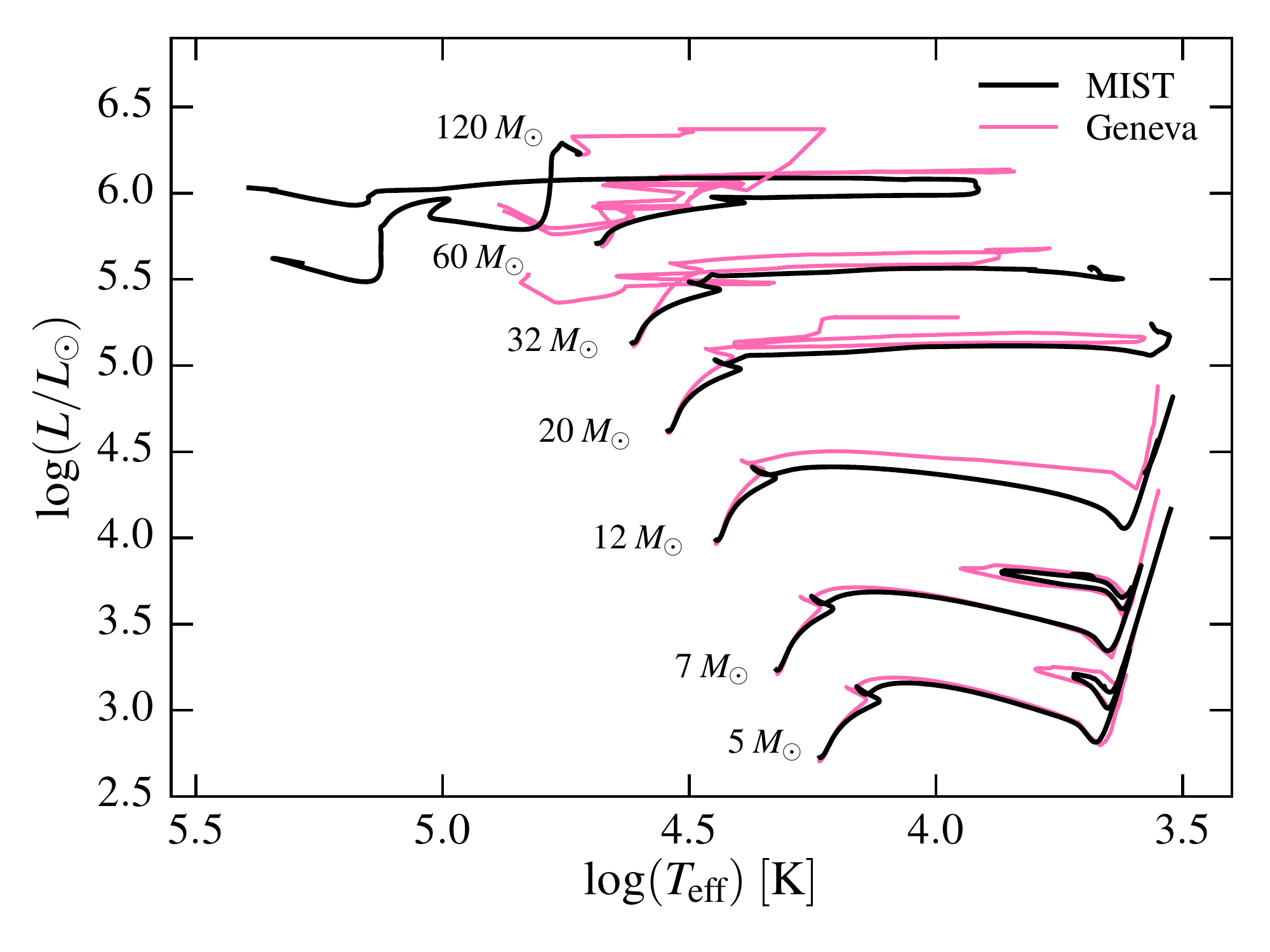}
\caption{A comparison of MIST and Geneva \citep{Ekstrom2012} evolutionary tracks with rotation in black and pink, respectively.}
\label{fig:compare_rotating_tracks_diff_models}
\end{figure}

\begin{figure*}
\centering
\includegraphics[width=0.7\textwidth]{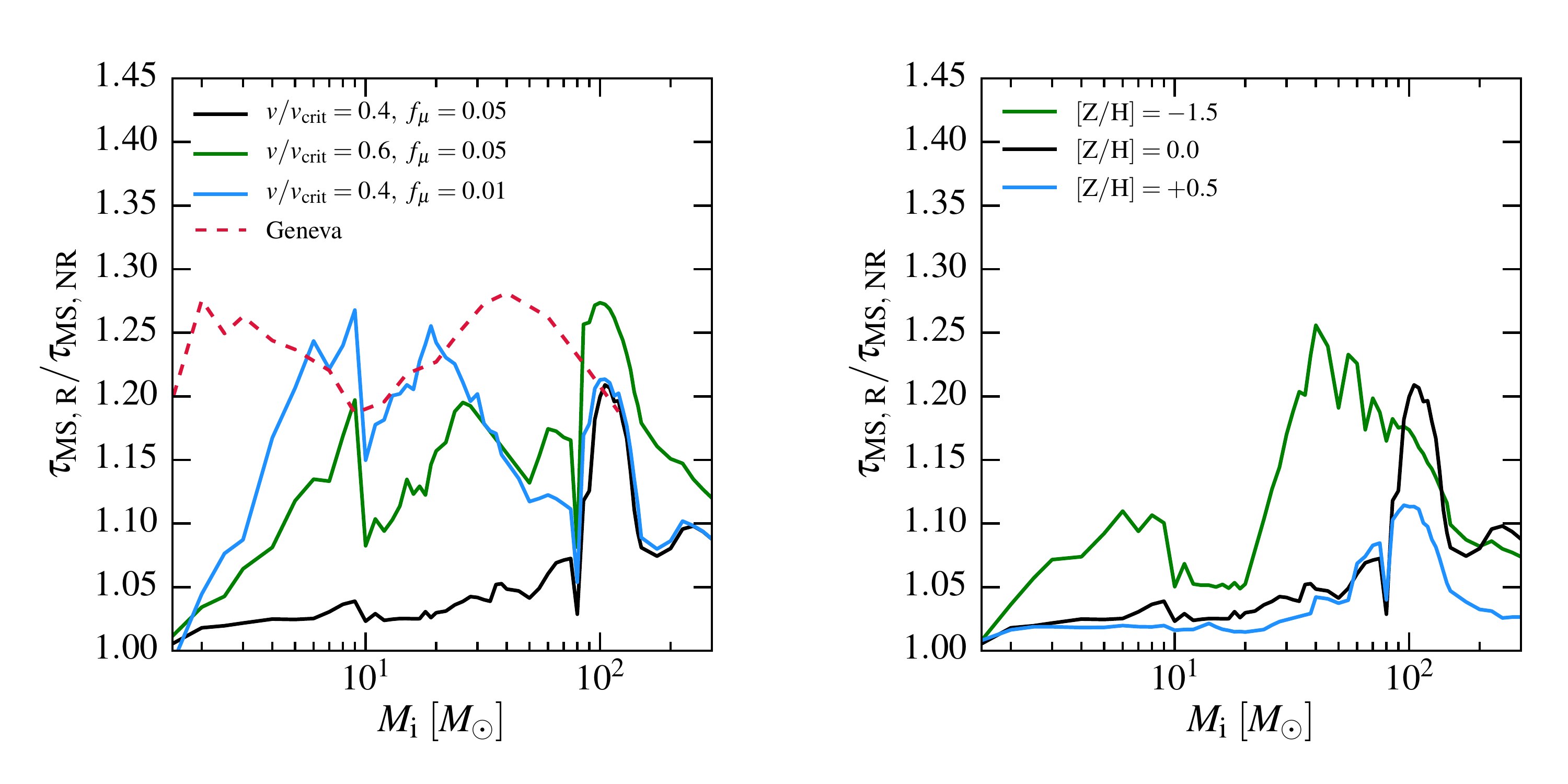}
\caption{The ratio of MS lifetimes for rotating and non-rotating models as a function of initial mass. Left: comparison at $Z_{\odot}$. The solid black, green, blue, and dashed pink lines correspond to the default model with $v/v_{\rm crit}=0.4$ and $f_\mu=0.05$, model with $v/v_{\rm crit}=0.6$, model with $f_\mu=0.01$, and the Geneva model, respectively. The default MIST model shows a modest $\sim5\%$ enhancement in the MS lifetime due to rotational mixing, whereas the Geneva model experiences a $\sim20\%$ increase due to more efficient rotational mixing. Right: comparison among MIST models at $\rm [Z/H] = -1.0$, 0.0, and $+0.5$ with default parameters $v/v_{\rm crit}=0.4$ and $f_\mu=0.05$. The efficiency of rotational mixing is larger in more metal-poor stars because line-driven mass loss---thus angular momentum loss efficiency---is lower.}
\label{fig:compare_rot_nonrot_mslifetime}
\end{figure*}

\subsection{The Effects of Rotation}
\label{section:effectofrotation}
We compare MIST and Geneva \citep{Ekstrom2012} evolutionary tracks for a wide range of masses in Figure~\ref{fig:compare_rotating_tracks_diff_models}. The Geneva models, shown in pink, also include rotation with $v/v_{\rm crit}=0.4$ and adopt a similarly low-metallicity solar abundance scale---$Z_{\odot}=0.014$ to be exact---with the elemental mixture from \cite{Asplund2005} combined with the Ne abundance from \cite{Cunha2006}. At fixed stellar mass, the Geneva models are hotter and more luminous at TAMS, which implies that rotational mixing is more efficient in their models compared to that in the MIST models. As discussed in Section~\ref{section:tracks_and_isochrones}, efficient rotational mixing gives rise to hotter temperatures and higher luminosities due to larger core sizes and increased $\mu$ in the envelope.

Some MIST models, such as the $120~\msun$ star in Figure~\ref{fig:compare_rotating_tracks_diff_models}, lose their hydrogen-rich envelope very promptly as they reach the so-called $\Omega\Gamma$-limit \citep{Maeder2000}. As evident from Equations~\ref{equation:rotboost} and \ref{equation:omegacrit}, massive stars with $\Gamma=L/L_{\rm Edd}\rightarrow1$ only require the smallest amount of rotation $\Omega$ to receive a large boost in mass loss rates. As the star evolves, its surface metallicity increases due to a combination of mixing processes and mass loss, and as a result, its surface Rosseland mean opacity increases. This in turn decreases $L_{\rm Edd}$, which makes it easier for a star to experience a large rotational boost. The star then may enter a positive feedback loop where mass loss leads to even more efficient mass loss until it removes all of its envelope and becomes a very compact star almost completely devoid of angular momentum. At the moment, it is not clear whether nature produces such stars, perhaps because of more complex behavior not included in the current 1D models.

Differences in the efficiency of rotational mixing between the MIST and Geneva models is further explored in the left panel of Figure~\ref{fig:compare_rot_nonrot_mslifetime}, which shows the ratio of MS lifetimes for rotating and non-rotating models as a function of initial mass. This ratio is expected to be greater than unity since rotational mixing channels additional fuel into the core. The solid black, green, blue, and dashed pink lines correspond to the default model at solar metallicity with $v/v_{\rm crit}=0.4$ and $f_\mu=0.05$, solar metallicity model with $v/v_{\rm crit}=0.6$ and $f_\mu=0.05$, solar metallicity model with $v/v_{\rm crit}=0.4$ and $f_\mu=0.01$, and the Geneva model, respectively. Recall that $f_\mu$ is the parameter that captures the sensitivity of rotational mixing to the mean molecular weight gradient, such that a small $f_\mu$ corresponds to efficient mixing even in the presence of a stabilizing composition gradient.\footnote{This is at a fixed value of $f_c$, the ratio of the diffusion coefficient and the turbulent viscosity. See Section 3 of \cite{Heger2000} for more details.} The default MIST model experiences only a modest enhancement in the MS lifetime. In contrast, the Geneva model experiences an overall $\sim25\%$ increase in the MS lifetime for stars more massive than $2~\msun$ \citep{Ekstrom2012, Georgy2013}. 

Although the MIST and Geneva models experience quantitively different amounts of MS lifetime boost, this is not entirely surprising given their different implementations of rotational mixing. Moreover, massive star evolution, regardless of the inclusion of rotation, is highly uncertain and very sensitive to small changes in the input physics. At fixed $v/v_{\rm crit}$, the efficiency of rotational mixing depends sensitively on $f_\mu$. As expected, MS lifetime boost in the MIST models is increased for a higher rotational mixing efficiency via increased rotation velocity or decreased $f_\mu$. The default values $f_c=1/30$ and $f_\mu=0.05$ are adopted from \cite{Heger2000}. This combination, though not unique, is able to reproduce many of the observational constraints such as the high-mass star ratios (see Section~\ref{section:highM_ratios}) and observed surface nitrogen enrichment (see Section~\ref{section:n_abun}). The fact that both MIST and Geneva models broadly reproduce observational constraints in spite of the different lifetime enhancements implies that current observations are not uniquely constraining. For reference, we note that the MS lifetimes for the rotating models in Geneva and MIST agree to within $10\textrm{--}15\%$ at solar metallicity: for low-mass stars ($\lesssim1.5~\msun$), the MS lifetime is shorter in the Geneva models, whereas for higher mass stars, the MIST models have MS lifetimes that fall between those of non-rotating and rotating Geneva models.

In the right panel, we compare the ratio of MS lifetimes among MIST models with different metallicities. Since the primary mass loss mechanism for massive stars is strongly metallicity-dependent line-driven winds, rotational mixing becomes more important at low metallicities due to the lowered efficiency of angular momentum loss, as expected.

\begin{figure}
\centering
\includegraphics[width=0.5\textwidth]{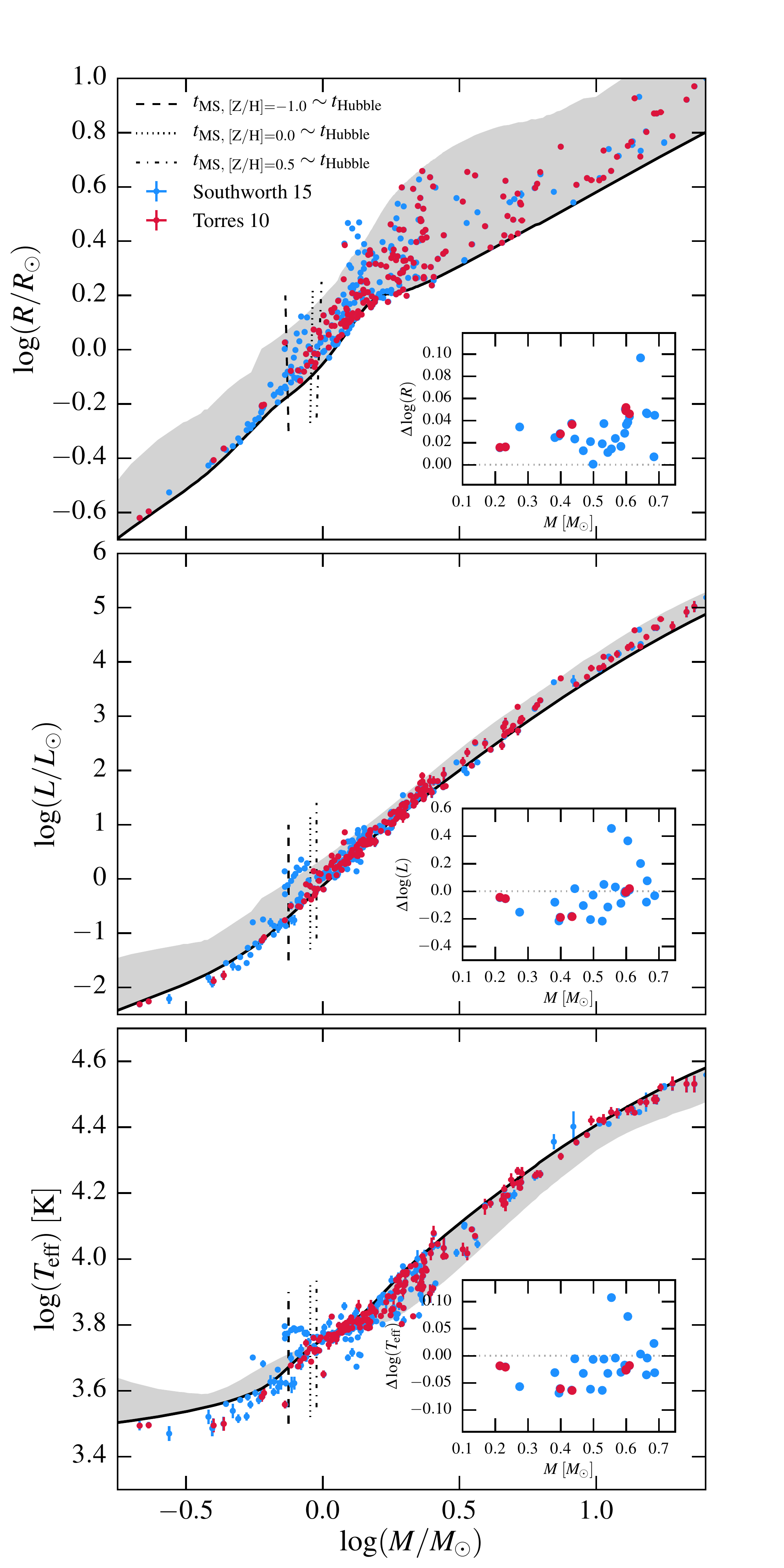}
\caption{$\log (R)$, $\log (L)$, and $\log (\teff)$ as a function of stellar mass measured for MS stars in detached eclipsing binaries (DEB). The \citet{Southworth2015} sample in blue comes from DEBCat, an online catalog of DEBs with well-measured parameters gathered from the literature. The red points correspond to a sample of DEBs selected from the literature that was homogeneously reanalyzed by \citep{Torres2010}. Note that the \cite{Torres2010} sample appears to show smaller scatter, especially around $\sim1~\msun$. The predicted ZAMS relations for solar metallicity are shown as black solid lines in each of the panels. Since the ages of the stars are unknown, we also show the full range of possible MS values as the gray shaded region. The vertical dashed, dotted, and dotted--dashed lines demarcate the initial masses for which $t_{\rm MS} \sim t_{\rm Hubble}$ for $\rm [Z/H]=-1.0$, 0.0, and $+0.5$, respectively. The insets highlight the well-known discrepancy for the low-mass stars ($<0.7~\msun$).}
\label{fig:mrlt_relation}
\end{figure}

\section{Comparisons with Data. I: Low Mass Stars}
\subsection{Luminosity--Mass--Radius--Temperature Relations}
\label{section:lmrt_relation}
Relations between mass, radius, luminosity, and temperature provide powerful and fundamental tests of stellar evolution models. In the past two decades, there have been enormous improvements in measuring these quantities to high precision from a variety of techniques, including eclipsing binaries and interferometry (see \citealt{Torres2010} for a recent review on this topic).

In Figure~\ref{fig:mrlt_relation}, we plot $\log (R)$, $\log (L)$, and $\log (\teff)$ as a function of stellar mass for the DEBCat\footnote{http://www.astro.keele.ac.uk/jkt/debcat/} sample, an online catalog of detached eclipsing binaries (DEBs) with well-measured parameters compiled from the literature \citep{Southworth2015}, and a sample of DEBs selected from the literature that was homogeneously reanalyzed by \cite{Torres2010}. Note that the \cite{Torres2010} sample appears to show smaller scatter, especially around $\sim1~\msun$. We applied a $\log (g)$ cut---$\log (g) > 4.1~\rm cm\;s^{-2}$ and $3.4~\rm cm\;s^{-2}$ for $M_{\rm i}>1.2~\msun$ and $<1.2~\msun$, respectively, as estimated from our model isochrones---to remove evolved stars from the sample of likely MS stars. Furthermore, we removed from the final sample a few conspicuous outliers identified as PMS or RGB stars in the literature. The predicted ZAMS relations for solar metallicity are shown as black solid lines in each of the panels. Since the ages of the stars are unknown, we also show the full range of possible MS values as the gray shaded region. The vertical dashed, dotted, and dotted--dashed lines demarcate the initial masses for which $t_{\rm MS} \sim t_{\rm Hubble}$ for $\rm [Z/H]=-1.0$, 0.0, and $+0.5$. Below these masses, we do not expect stars to have evolved off the ZAMS relations. Overall, the observed points fall comfortably within the region bounded by the ZAMS and TAMS relations. However, the observed stars start to deviate from the predicted relations below $M_{\rm i}\lesssim0.7~\msun$. In the insets, we zoom in on the low-mass range to show that the models systematically underpredict radii by $\sim0.03$~dex and overpredict temperatures by $\sim0.05$~dex, for a total deficit of $\sim0.2$~dex for the predicted luminosity. 

There is a well-known discrepancy between observed and predicted effective temperature, radius, and luminosity relations for stars with appreciable convective layers, most notably M dwarfs \cite[e.g.,][]{Casagrande2008, Torres2010, Kraus2011, Feiden2013, Spada2013, Torres2013, Chen2014}. At fixed stellar mass, models tend to predict stars that are $5\textrm{--}10\%$ hotter and $10\textrm{--}20\%$ smaller in radius compared to observations. This disagreement is present in both field stars and DEBs, suggesting that this is an effect intrinsic to dwarfs \citep{Boyajian2012, Spada2013}. However, there are systematic errors of a few percent expected from DEB light curve analysis due to variations in the spot size and coverage \citep{Morales2010}. A proposed explanation for this mismatch invokes magnetic activity and rotation effects that are not currently modeled accurately \citep{Spruit1986, Morales2008, Morales2010, Irwin2011, Kraus2011, Feiden2012, MacDonald2014, Jackson2014}. Large-scale magnetic fields are thought to both inhibit the upwelling of hot convective bubbles and generate more starspots on the surface \citep[e.g.,][]{Feiden2012}. In order to conserve flux, the stellar radius is inflated, causing a subsequent decrease in the surface temperature. Rotation may play a role since it is believed to generate a dynamo and has been linked to magnetic activity (see Section~\ref{section:magnetic}). Furthermore, the choice of surface boundary conditions in stellar models has a non-trivial effect on the mass--radius relation and the optical color--magnitude diagrams (CMDs) at the lowest masses \citep[e.g.,][]{Baraffe1995, Chabrier1996, Baraffe1997, Spada2013, Chen2014}. When it comes to modeling cool dwarfs, it is especially important to use accurate boundary conditions---such as those computed from atmosphere models, e.g., PHOENIX \citep{Hauschildt1999a} and ATLAS12/SYNTHE \citep{Kurucz1970, Kurucz1993}---in place of simple models that assume gray atmospheres  (see Section~\ref{section:boundary_conditions}).

\subsection{Initial--Final Mass Relation}
\label{section:ifmr}
Low- and intermediate-mass stars shed a nontrivial fraction of their mass via winds during the course of their lifetime, eventually terminating their lives as WDs. Total mass loss integrated over the lifetime directly connects the initial mass to the remnant mass through the IFMS \citep[e.g.,][]{Reimers1975, Weidemann1977, Renzini1988, Weidemann2000}. It is an important diagnostic for the cumulative effect of mass loss occurring at various stages of evolution. The expectation is that stars with higher initial masses produce more massive WD remnants \citep[e.g.,][]{Claver2001, Dobbie2004, Williams2007, Catalan2008a, Kalirai2008, Williams2009, Zhao2012}. 

\begin{figure}
\centering
\includegraphics[width=0.45\textwidth]{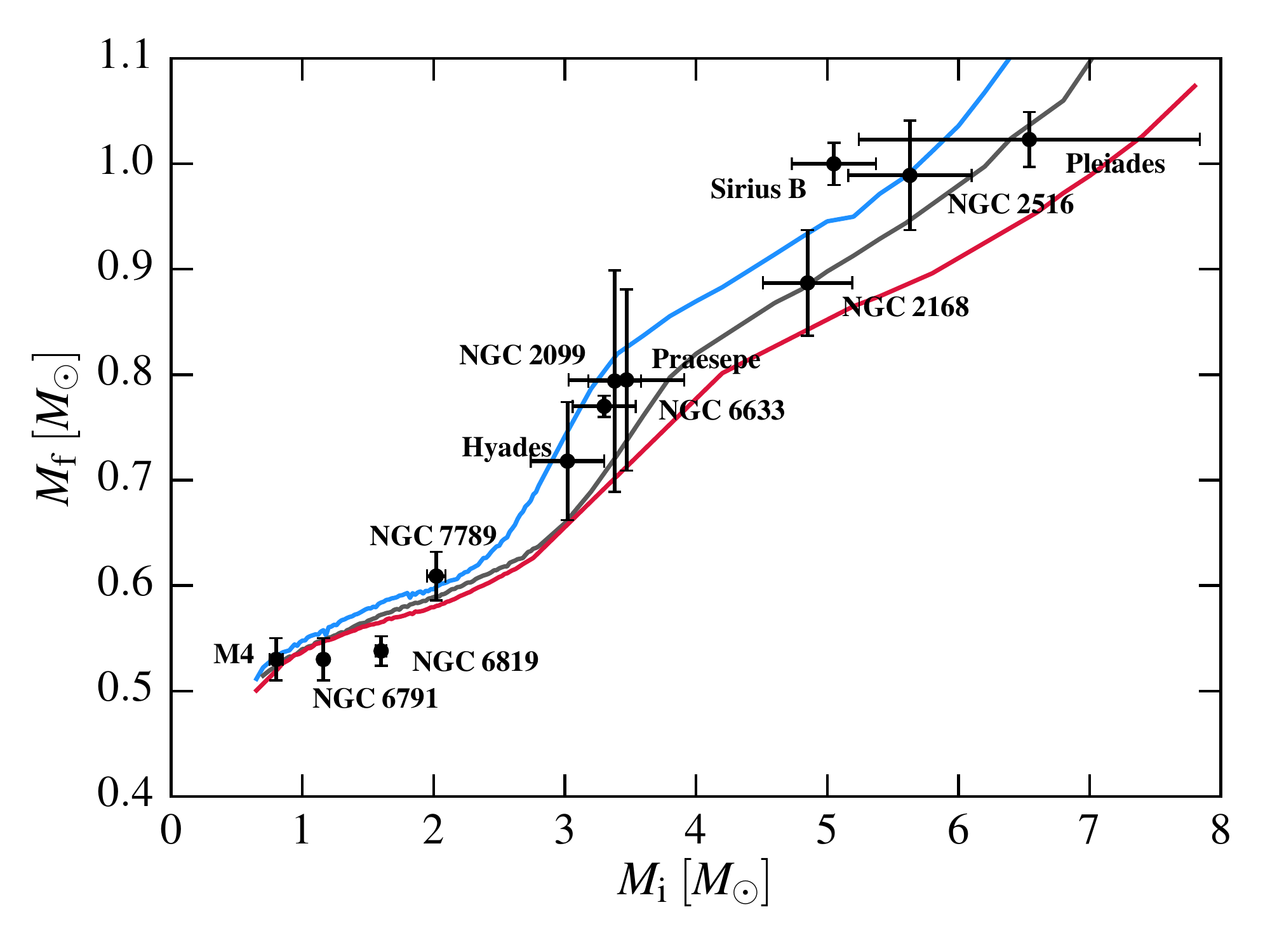}
\caption{The initial--final mass relation constructed using binned cluster data and Sirius B from \cite{Ferrario2005} (see references therein) and \cite{Kalirai2008}. The predicted relations for $\rm [Z/H]=-0.5$, 0.0, and $+0.5$ are shown in blue, gray, and red. These metallicities roughly bracket the metallicity range of the systems in the sample.}
\label{fig:ifm_relation}
\end{figure}

\begin{figure*}
\centering
\includegraphics[width=0.9\textwidth]{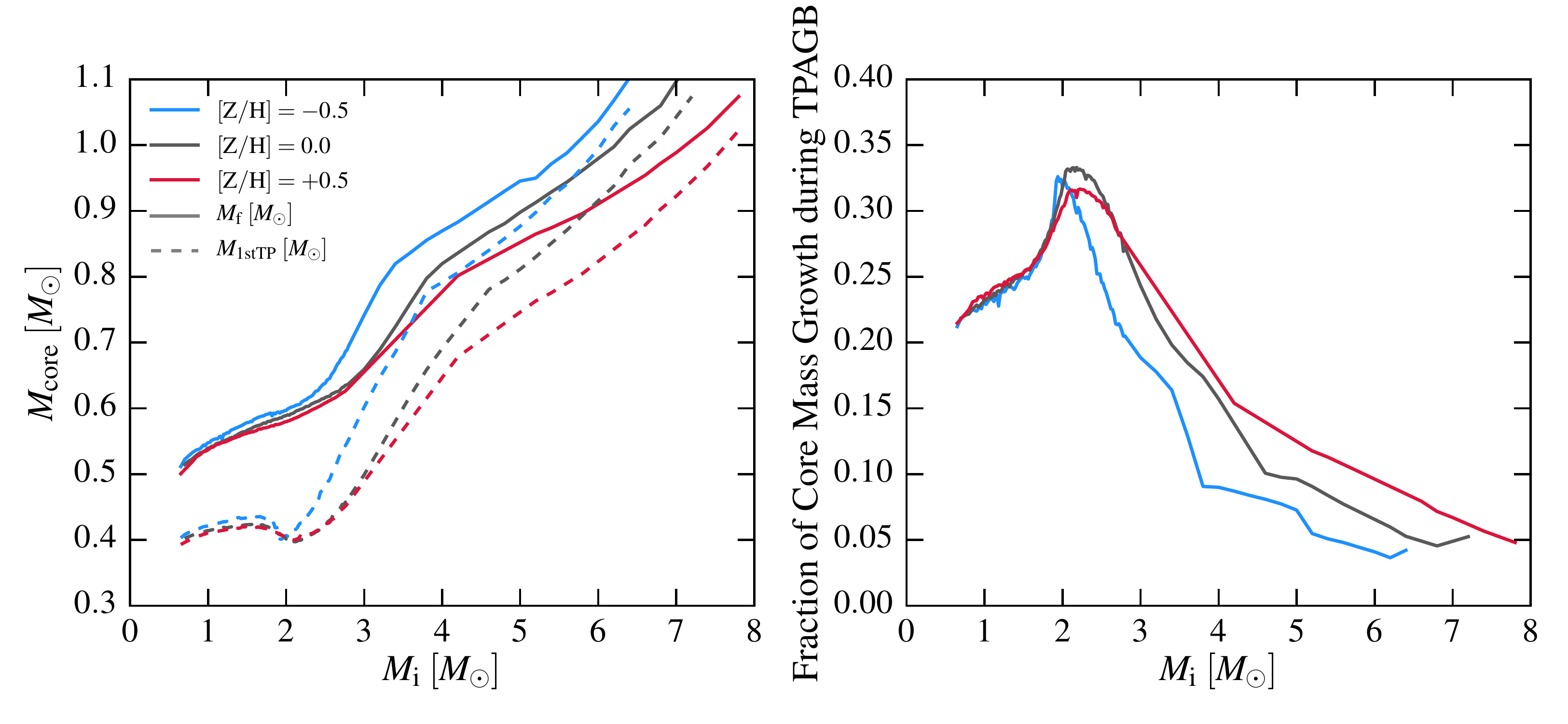}
\caption{Left: a comparison highlighting the difference between the final mass (solid) from Figure~\ref{fig:ifm_relation} and the core mass at first thermal pulse (dashed) as a function of initial mass for three metallicities. The latter is devoid of large uncertainties in mass loss, third dredge-up, and hot bottom burning that strongly influence the TPAGB evolution, and consequently, the final remnant mass. Right: fractional growth in core mass during the TPAGB phase for the same three metallicities.}
\label{fig:coremass_growth}
\end{figure*}

We compare the predicted IFMR to a sample of eight young open clusters, three older open clusters with ages $>1$~Gyr, and Sirius B (see \cite{Ferrario2005} and \cite{Kalirai2008} for references therein). It is useful to study clusters of a variety of ages because it allows us to probe a large range of initial masses. Observed initial and final masses for each WD in a cluster are inferred using the following method (see e.g., \citealt{Kalirai2008} for details). A combination of WD spectral analysis and modeling yields both the WD mass (final mass) and cooling age (age since the end of shell helium burning on the TPAGB). The WD progenitor age up to the end of the TPAGB is simply the difference between the total age of the system as estimated from the cluster turn-off and the WD cooling age from the previous step. Finally, stellar evolution models provide the progenitor mass (initial mass) corresponding to the WD progenitor age. Note that initial and final masses are not directly observed but instead are inferred from modeling: the final mass comes from the spectral analysis while the initial mass depends on stellar evolution theory and CMD analysis.

In Figure~\ref{fig:ifm_relation} we plot the predicted IFMRs for three values of [Z/H] that altogether encompass the metallicities of the systems represented here. Individual measurements within a single cluster have been binned to represent a weighted mean. Overall, the models are in excellent agreement with the data, though there are some notable outliers like NGC~6819 and Sirius B. The low-mass plateau at $M_{\rm i}\lesssim2~\msun$ ($M_{\rm f}\sim0.6~\msun$) is marginally consistent with the peak of the galactic disk WD mass function near $\sim0.6~\msun$ \citep{Liebert2005, Kleinman2013, Kepler2015}, although a full model that folds in the age and metallicity distributions as well as the IMF weights will be required for a robust comparison against the observed mass function \citep[see also][]{Catalan2008b}. 

\begin{figure*}
\centering
\includegraphics[width=\textwidth]{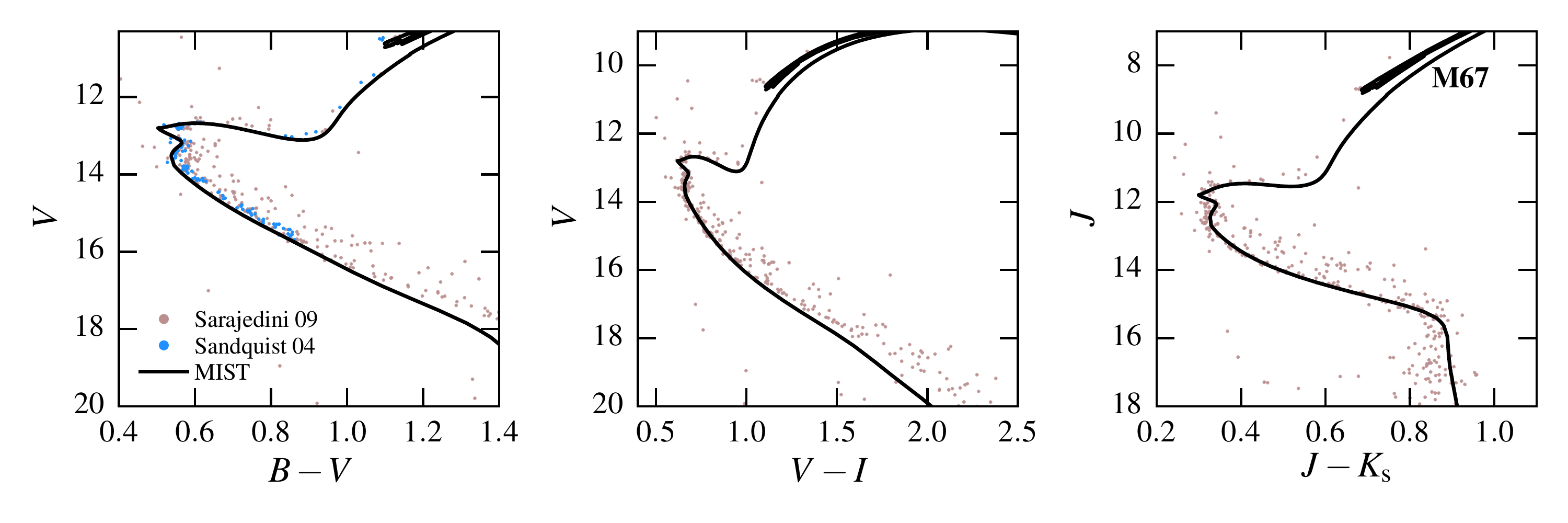} 
\caption{CMDs for M67 from {\it 2MASS} and $BVI$ photometry \citep{Sandquist2004, Sarajedini2009}. The \cite{Sandquist2004} $BVI$ sample (blue points) was carefully selected to exclude likely binaries, and the \cite{Sarajedini2009} sample (mauve points) reflects a membership probability cut of $>20\%$. MIST isochrones with $\rm [Z/H]=0.0$, $\logage = 9.58$ (3.80~Gyr), $A_V=0.2$, and $\mu=9.7$ are shown in black.}
\label{fig:M67}
\end{figure*}

Furthermore, the steep slope predicted around $M_{\rm i}\sim3~\msun$ to $4~\msun$ ($\rm slope\sim0.16$) is consistent with the empirical estimate from \cite{Cummings2015}, who found a slope of $M_{\rm f} = (0.163\pm0.022)M_{\rm i}+(0.238\pm0.071)~\msun$ from a combined sample of newly identified WDs in NGC 2099 and the WDs in the Hyades and Praesepe from \cite{Kalirai2014}. Interestingly, in agreement with \cite{Romero2015} but in contrast with \cite{Marigo2007}, our theoretical relations show a clear systematic trend with metallicity above $3~\msun$. As discussed in \cite{Marigo2007}, the core can grow/erode through shell-burning/third dredge-up (TDU) during the TPAGB, while mass loss more or less determines when the TPAGB phase terminates. Metallicity is predicted to affect all these processes: at low metallicities, TDU and hot bottom burning efficiencies, as well as the core mass at the onset of the first thermal pulse, increase, but mass loss efficiency is believed to decrease. The measurement of the IFMR as a function of metallicity therefore has great potential for constraining these uncertain evolutionary phases.

It is worth noting that calculations of $M_{\rm i}$ and $M_{\rm f}$ which rely on, e.g., WD spectral analysis and isochrone fitting, and the quality of the data vary between cluster to cluster. As noted in \cite{Kalirai2008}, the sample is likely affected by small systematic offsets and nonzero field contamination. In particular, precision measurements of ages with the MSTO method in young systems is particularly challenging due to the vertical placement of the MSTO in an optical CMD (note the larger error bars toward younger ages and higher $M_{\rm i}$). Moreover, the thickness of the hydrogen and helium layer assumed in the WD spectral model can also have a non-negligible effect on the inferred cooling age and thus the initial mass \citep{Prada2002, Catalan2008b}. \cite{Catalan2008b} found that reasonable variations in the envelope thickness can lead to differences as large as $1~\msun$ for progenitors with $M_{\rm i}\gtrsim5~\msun$ but a more modest $\sim0.1~\msun$ difference for the lower masses. Other model uncertainties include the assumed core composition. A stringent test of the IFMR should be entirely self-consistent; a single set of isochrones should be used to estimate the ``observed'' masses $M_{\rm i}$ and $M_{\rm f}$. 

\begin{figure*}
\centering
\includegraphics[width=\textwidth]{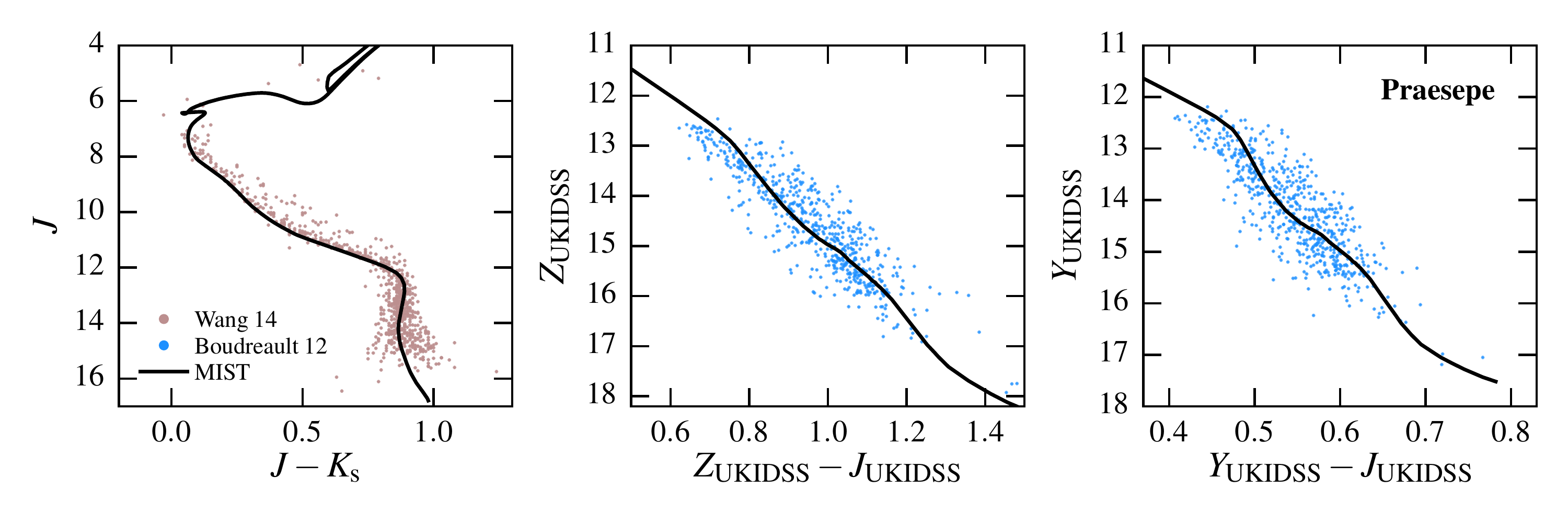} 
\caption{CMDs for Praesepe from {\it 2MASS} \citep{Wang2014} and {\it UKIDSS} Galactic Clusters Survey photometry \citep{Boudreault2012}. The \cite{Wang2014} sample (mauve points) consists of proper-motion-selected cluster members that are likely to be single stars and the \cite{Boudreault2012} sample (blue points) was identified with an astrometric and five-band photometric cut. MIST isochrones with $\rm [Z/H]=+0.15$, $\logage = 8.80$ (630~Myr), $A_V = 0.08$ and $\mu=6.26$ are shown in black.}
\label{fig:praesepe}
\end{figure*}

We conclude this section with a comparison between the final mass and the core mass at the first thermal pulse in Figure~\ref{fig:coremass_growth}. It is useful to consider the core mass before the star experiences significant core growth from the subsequent thermal pulses, because this comparison is devoid of large uncertainties in mass loss, TDU, and hot bottom burning that strongly influence the TPAGB phase \citep[e.g.,][]{Wagenhuber1998, Weidemann2000}. In the left panel, we show the predicted IFMR from Figure~\ref{fig:ifm_relation} and the core mass at the beginning of the TPAGB phase as a function of initial mass in solid and dashed lines, respectively. The right panel shows the fractional growth in core mass during the TPAGB phase ($M_{\rm f}-M_{\rm 1stTP}/M_{\rm f}$) as a function of initial mass. Overall, there is considerable growth in core mass occurring during this phase, with a broad peak over $2\textrm{--}3~\msun$. The location of maximum growth coincides with the peak in TPAGB lifetime as shown in Figure~\ref{fig:phase_lifetimes_cumul}. This result is in good agreement with what \cite{Bird2011} and \cite{Kalirai2014} found using the \cite{Pietrinferni2004} and PARSEC/COLIBRI \citep{Bressan2012, Marigo2013} models, respectively.

\subsection{Optical and NIR Color Magnitude Diagrams of Clusters}
\subsubsection{Star Clusters}
In this section we present comparisons with observed CMDs of star clusters. Models shown here include a reddening correction according to the standard $R_{\rm V} \equiv A_{\rm V}/E(B-V)=3.1$ reddening law from \cite{Cardelli1989}. The majority of the systems presented here are metal-rich and we will provide comparisons with metal-poor clusters with non-solar-scaled abundance patterns in Paper II. The CMD comparisons here are by-eye fits to check that the models yield a reasonable agreement. We plan to perform more robust CMD fitting with MATCH \citep{Dolphin2002} in future work.

\label{section:star_clusters}
{\it M67 (NGC 2682)}, an intermediate-age ($4$~Gyr) solar metallicity open cluster at a distance of $\sim800$~pc, is a benchmark system for stellar evolution models \citep{Taylor2007, Sarajedini2009}. In particular, its well-developed Henyey hook on the MSTO is used to calibrate core convective overshoot in low- and intermediate-mass stars \citep[e.g.,][]{Michaud2004, VandenBerg2006, Magic2010, Bressan2012}. 

Figure~\ref{fig:M67} shows optical and near-infrared (NIR) CMDs and $\logage=9.58$ (3.80~Gyr) isochrones with $\rm [Z/H]=0.0$, $A_V=0.2$, and $\mu=9.7$, where $\mu$ is the distance modulus. The \cite{Sandquist2004} $BV$ sample (blue points) was carefully selected using proper motion, radial velocity, variability, and CMD-location information to yield cluster members that are most likely to be single stars. The \cite{Sarajedini2009} {\it 2MASS} sample (mauve points) reflects a membership probability cut of $>20\%$.

\begin{figure*}
\centering
\includegraphics[width=\textwidth]{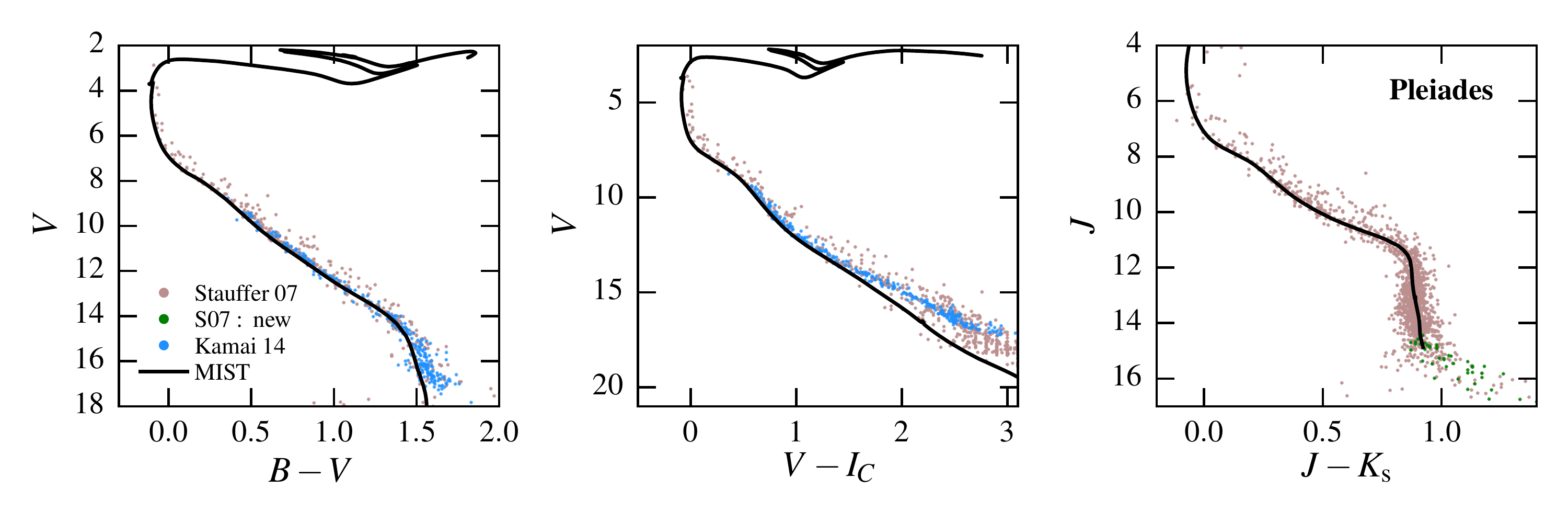} 
\caption{CMDs for Pleiades from {\it 2MASS} and $BVI_C$ photometry \cite{Stauffer2007, Kamai2014}. The mauve points represent the \cite{Stauffer2007} compilation of {\it 2MASS} and $BVI_C$ photometry from the literature, the green points correspond to a sample of newly identified {\it 2MASS} candidates by \cite{Stauffer2007}, and the blue points are the \cite{Kamai2014} sample of proper motion members from the Stauffer catalog with updated $BVI_C$ photometry. MIST isochrones with $\rm [Z/H]=0.0$, $\logage = 8.0$ (100~Myr), $A_V=0.1$, and $\mu=5.62$ are shown in black. We omit the TPAGB and post-AGB phases for display purposes.}
\label{fig:pleiades}
\end{figure*}

\begin{figure*}
\centering
\includegraphics[width=\textwidth]{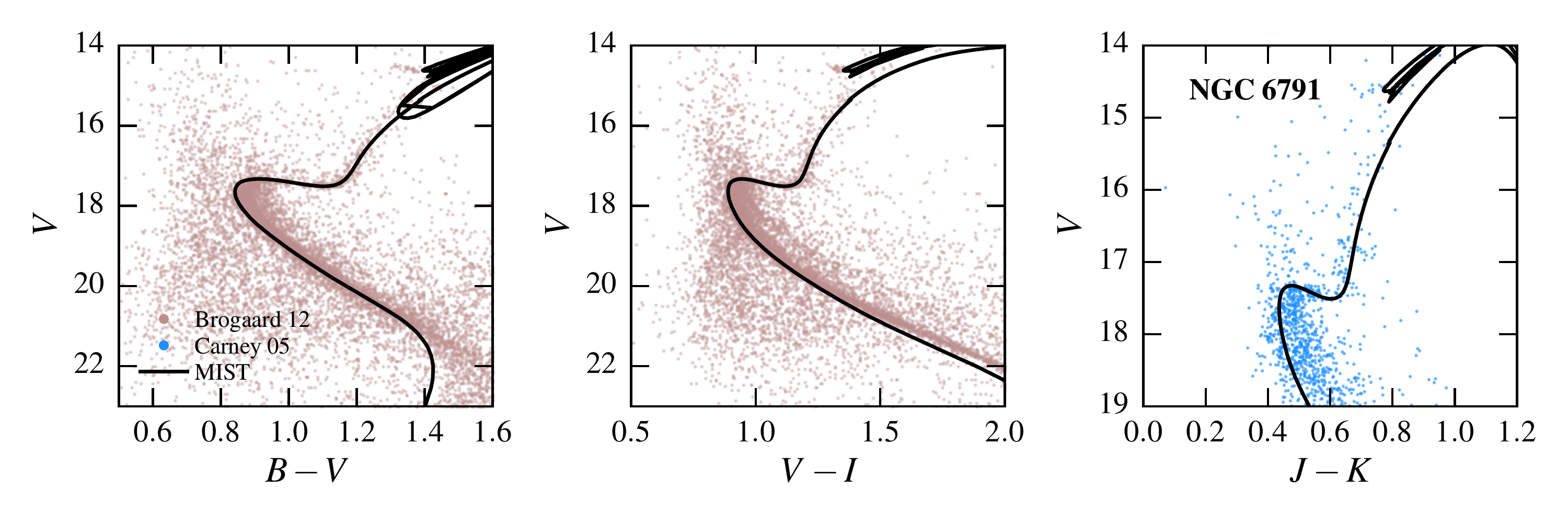} 
\caption{CMDs for NGC 6791 in $B\text{--}V$ and $V\text{--}I$ \citep[mauve points;][]{Brogaard2012} and in $J\text{--}K$ \citep[blue points;][]{Carney2005}. The \cite{Brogaard2012} sample consists of photometry from \cite{Stetson2003} that has been empirically corrected for differential reddening effects. MIST isochrones with $\rm [Z/H]=+0.47$, $\logage = 9.93$ (8.5~Gyr), $A_V = 0.32$ and $\mu=13.1$ are shown in black.}
\label{fig:ngc6791}
\end{figure*}

The MS, MSTO morphology, as well as the RC luminosity are well-matched in all three colors. However, the isochrone begins to peel away from the MS ridge line at fainter than $V\sim17$ which corresponds to $M_{\rm i}\sim0.7~\msun$. This is a well-known issue: models that successfully reproduce the NIR colors \cite[e.g.,][]{Sarajedini2009} predict MS colors that are too blue in the optical \cite[e.g.,][]{An2009, Chen2014}. One of the goals of future work is to revisit and address this problem.

{\it Praesepe (M44; NGC 2632)} is a young ($\sim757$~Myr) and moderately super-solar cluster at a distance of $180$~pc, making it one of the nearest open clusters to the Sun \citep{Taylor2006, Gaspar2009, Carrera2011}. A combination of its rich cluster membership \citep[$N\gtrsim1000$;][]{Kraus2007}, large proper motion, and proximity makes Praesepe a favorable target for stellar population studies.

In Figure~\ref{fig:praesepe}, we show {\it 2MASS} (mauve points) and {\it UKIDSS} Galactic Clusters Survey photometry data (blue points) from \cite{Wang2014} and \cite{Boudreault2012}, respectively. The \cite{Wang2014} sample consists of proper-motion-selected cluster members that are likely to be single stars and includes stars with masses as low as $\sim0.15~\msun$. The \cite{Boudreault2012} sample was obtained using astrometric and five-band photometric selection criteria and consists mostly of low-mass stars with $M_{\rm i}\lesssim 0.8~\msun$. We applied an additional cut to remove objects that were flagged as variable stars and/or had $<50\%$ membership probability. We overplot $\logage = 8.8$ (630~Myr) isochrones with $\rm [Z/H] = +0.15$, $A_V = 0.08$, and $\mu=6.26$. Overall, the MIST isochrones provide good fits in all three colors.

{\it Pleiades (M45)} is a young ($\sim100$~Myr) solar metallicity open cluster at a distance of only $133$~pc \citep{Soderblom2005, Soderblom2009, Melis2014}. Like Praesepe, its richness ($N\sim1400$) and proximity make it a popular choice for testing stellar evolution models. In fact, the tension between observed and predicted CMD locations of K and M dwarfs dates back to \cite{Herbig1962} who proposed non-coeval evolution, i.e., age spread, as a solution to the discrepancy. Theoretical isochrones were systematically offset toward higher luminosities and cooler temperatures in $B-V$ but predicted fainter and hotter stars in redder colors such as $V-I$ \citep{Stauffer2003, Kamai2014}. A more recent proposed explanation attributes this discrepancy to magnetic activity (e.g., spots) and/or rotation \citep[e.g.,][]{Stauffer2003}.

In Figure~\ref{fig:pleiades}, we show optical and NIR CMDs constructed from the \cite{Stauffer2007} compilation of {\it 2MASS} and $BVI_C$ photometry from the literature (mauve points), a sample of newly identified {\it 2MASS} candidates by \cite{Stauffer2007} (green points), and the \cite{Kamai2014} sample of proper motion members from the Stauffer catalog with updated $BVI_C$ photometry (blue points). We overplot $\logage = 8.0$ (100~Myr) isochrones with $\rm [Z/H]=0.0$, $A_V=0.1$, and $\mu=5.62$. Overall, the isochrones are well-matched to the observed CMDs in all filters. However, as seen in M67, the isochrones depart blueward from the MS ridge line in $V-I_C$ and $B-V$ at fainter than $V\sim14$, corresponding to $M_{\rm i}\sim0.6~\msun$.

{\it NGC 6791} is one of the most well-known and well-studied open clusters in the Milky Way. Its unusually old age ($\sim 8~\rm Gyr$) and high metallicity ($\rm [Fe/H] \sim 0.3\text{--}0.5$) combined with its rich membership make it a unique system for studying extreme stellar populations and their chemical evolution \citep[e.g.,][]{Bedin2005, Carney2005, Origlia2006, Kalirai2007, Brogaard2012}. NGC 6791 is particularly suitable for testing the present MIST models given its solar-scaled [$\alpha$/Fe] abundances.

In Figure~\ref{fig:ngc6791}, we show optical and NIR CMDs in $B\text{--}V$ and $V\text{--}I$ \citep[mauve points;][]{Brogaard2012} and in $J\text{--}K$ \citep[blue points;][]{Carney2005}. The \cite{Brogaard2012} sample consists of photometry from \cite{Stetson2003} that has been empirically corrected for differential reddening effects. We overplot $\logage = 9.93$ (8.5~Gyr) isochrones with $\rm [Z/H]=+0.47$, $A_V = 0.32$ and $\mu=13.1$.

The agreement between data and MIST isochrones is generally good, in particular the MSTO and SGB morphologies and the CHeB (red clump) luminosity in all three colors. However, we see the same issue in $B-V$ and $V-I$ as in Figures~\ref{fig:M67} and \ref{fig:pleiades}: MIST isochrones predict colors that are too blue for stars fainter than $V\sim21$. We return to this point at the end of this section.

\begin{figure}
\centering
\includegraphics[width=0.45\textwidth]{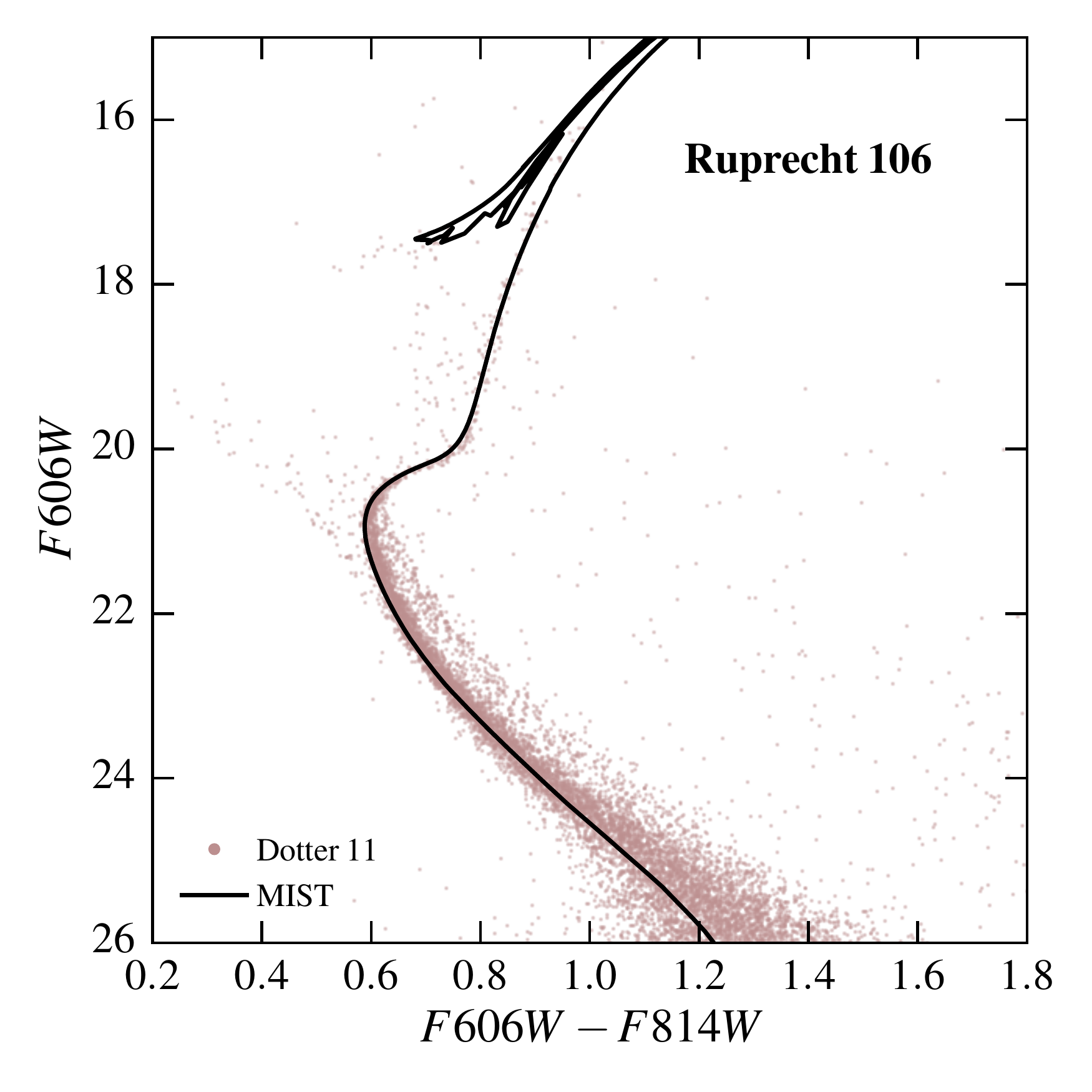} 
\caption{CMD from the {\it HST} ACS observations in the $F606W$ and $F814$ broadband filters \cite{Dotter2011}. We overplot a $\rm [Z/H]=-1.50$, $\logage=10.08$ (12.0~Gyr) isochrone with $\mu=16.7$ and $A_V=0.55$.}
\label{fig:Ruprecht106}
\end{figure}

{\it Ruprecht 106} is a relatively low-mass ($M\sim 10^{4.8}~\msun$) globular cluster with a metallicity of $\rm[Fe/H]\sim-1.5$ \citep{Kaluzny1995, Brown1997, Francois1997, Dotter2011, Villanova2013}. Its peculiar properties include the lack of $\alpha$-enhancement and the absence of abundance spread in light elements, e.g., Na-O anti-correlation, both of which are typical of globular clusters \citep{Carretta2009}. The spread in abundances found in globular clusters has been proposed to be a signature of self-enrichment through multiple generations of stellar populations \citep{Kraft1994, Gratton2004, D'Antona2008, Piotto2012, Gratton2012}. Thus the modest mass of Ruprecht 106 combined with its ``stubby'' HB morphology (consistent with the lack of helium variation via self-pollution) suggest that it is an archetypical single-population globular cluster \citep{Caloi2011, Villanova2013}. It is an excellent choice for testing our low-metallicity models because we can bypass the issue of $\alpha$-enhancement. We plan to perform many more tests against a large sample of globular clusters with our $\alpha$-enhanced models in Paper II.

Figure \ref{fig:Ruprecht106} shows the optical CMD from the {\it HST} ACS observations in the $F606W$ and $F814$ broadband filters \citep{Dotter2011}. We plot a $\logage=10.08$ (12.0~Gyr) isochrone with $\mu=16.7$, $A_V=0.55$, and $\rm [Z/H]=-1.50$ in black. The lower MS, SGB, and RGB are very well-matched though the model fails to accurately predict the extreme blue extent of the HB.

\begin{figure}
\centering
\includegraphics[width=0.45\textwidth]{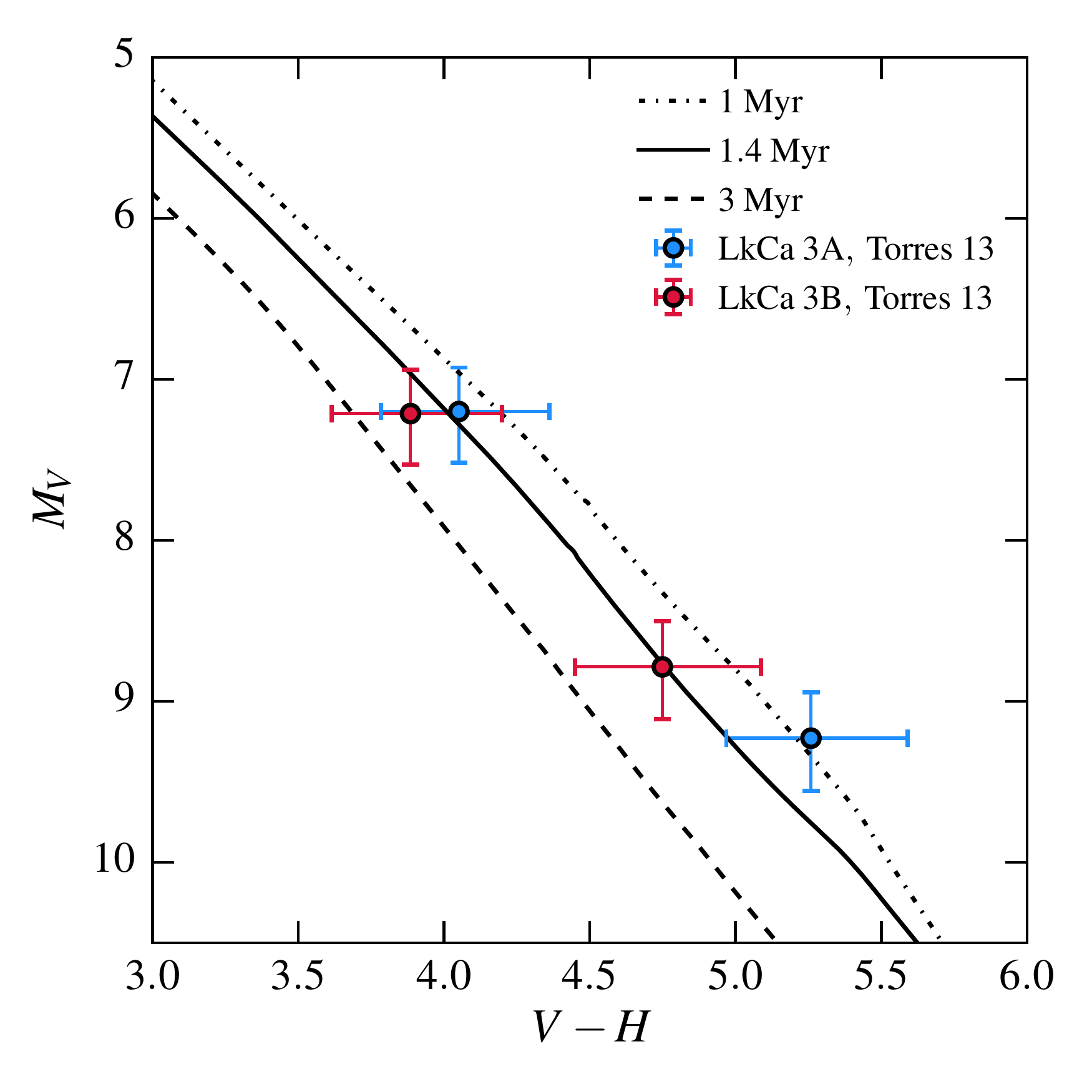}
\caption{LkCa 3, a quadruple system of PMS stars in the Taurus--Auriga star-forming region \citep{Torres2013b} is compared to 1, 1.4, and 3~Myr isochrones at solar metallicity.}
\label{fig:lkca3}
\end{figure}

We conclude this section by addressing a recurring issue raised from the CMD comparisons. Our models predict $V-I$ and $B-V$ colors that are too blue for stars below $\lesssim 0.6\textrm{--}0.7~\msun$. These discrepancies are due to missing and/or inaccurate atomic and molecular line opacity data used in our bolometric corrections. Efforts are ongoing to address these shortcomings.

\subsubsection{The Quadruple System LkCa 3} 
LkCa 3, a quadruple system of PMS stars in the Taurus--Auriga star-forming region \citep{Torres2013b}, offers an excellent opportunity to test PMS evolution models. Operating under the assumption that the system is coeval, we expect all four objects to fall on a single-age isochrone. However, there is evidence that early accretion episodes affect the location and evolution of PMS stars on the HR diagram \citep[e.g.,][]{Baraffe2009, Hosokawa2011}, possibly complicating the comparison to PMS at these young ages. \cite{Torres2013b} concluded that the predicted CMD locations of the four components in the Dartmouth models \citep{Dotter2008} are better matched than those from the Lyon models \citep{Baraffe2003}, though their recently updated models \citep{Baraffe2015} show good agreement with the observations as well. Their updates include a new solar abundance scale (a combination of \citealt{Asplund2009} and \citealt{Caffau2011}), improved linelists, and recalibrated mixing length parameter for the treatment of convection. 

In Figure~\ref{fig:lkca3}, we show the observed LkCa 3 stars from \cite{Torres2013b} with our 1, 1.4, and 3~Myr solar metallicity isochrones in the $V-H$ vs. $M_{V}$ plane. The observations as reported by \cite{Torres2013b} were already corrected for interstellar extinction assuming $A_{\rm V} = 0.31$. The best-fitting isochrone shown in solid black line indicates that the age of the LkCa 3 system is $\sim1.4$~Myr, consistent with previous estimates.

\subsection{The Age Discrepancy in Upper Scorpius}
Upper Scorpius (Upper Sco) is one of three subgroups (Upper Centaurus Lupus and Lower Centaurus Crux) in Scorpius-Centaurus (Sco-Cen), the nearest OB association from the Sun with $d\sim145~\rm pc$ \citep{deZeeuw1999, Preibisch2002}. The three subgroups altogether constitute a rich environment to study the formation and evolution of massive stars, circumstellar disks, low-mass stars, and brown dwarfs \citep[e.g.,][]{Preibisch2002, Chen2011, Lodieu2013}. There has been some recent tension in the literature over the ages of these subgroups \citep{Song2012}. In particular, the age of Upper Sco is quoted to be either $\sim5$ or $\sim11$ Myr depending on the analysis method and the spectral types of stars used to estimate the age \citep{deGeus1989, Preibisch2002, Lodieu2008, Slesnick2008, Pecaut2012, Herczeg2015}. This discrepancy poses a problem for other studies that rely on accurate age measurements, e.g., inferred mass functions \citep{Preibisch2002, Lafreniere2008, Lodieu2013}.

\cite{Pecaut2012} used isochrones to determine the age of Upper Sco from a kinematic sample of PMS F-type stars. The authors compared luminosities and temperatures derived from {\it Hipparcos} and {\it 2MASS} photometry to four different sets of models \citep{D'Antona1997, Siess2000, Demarque2004, Dotter2008}. They inferred an age of $\sim13$ Myr---much older than the previous estimates of $\sim5$ Myr---regardless of the model used in the analysis. This intriguing result prompted the authors to repeat their analysis using a number of other datasets from the literature, namely the MSTO B-type stars, M supergiant Antares ($\alpha$ Sco), MS A-type stars, and PMS G-type stars. The resulting ages were all consistently older than 5~Myr albeit with a large scatter, and the authors concluded that Upper Sco has a median age of $11\pm1\pm2$~(statistical, systematic)~Myr. 

\begin{figure}
\centering
\includegraphics[width=0.45\textwidth]{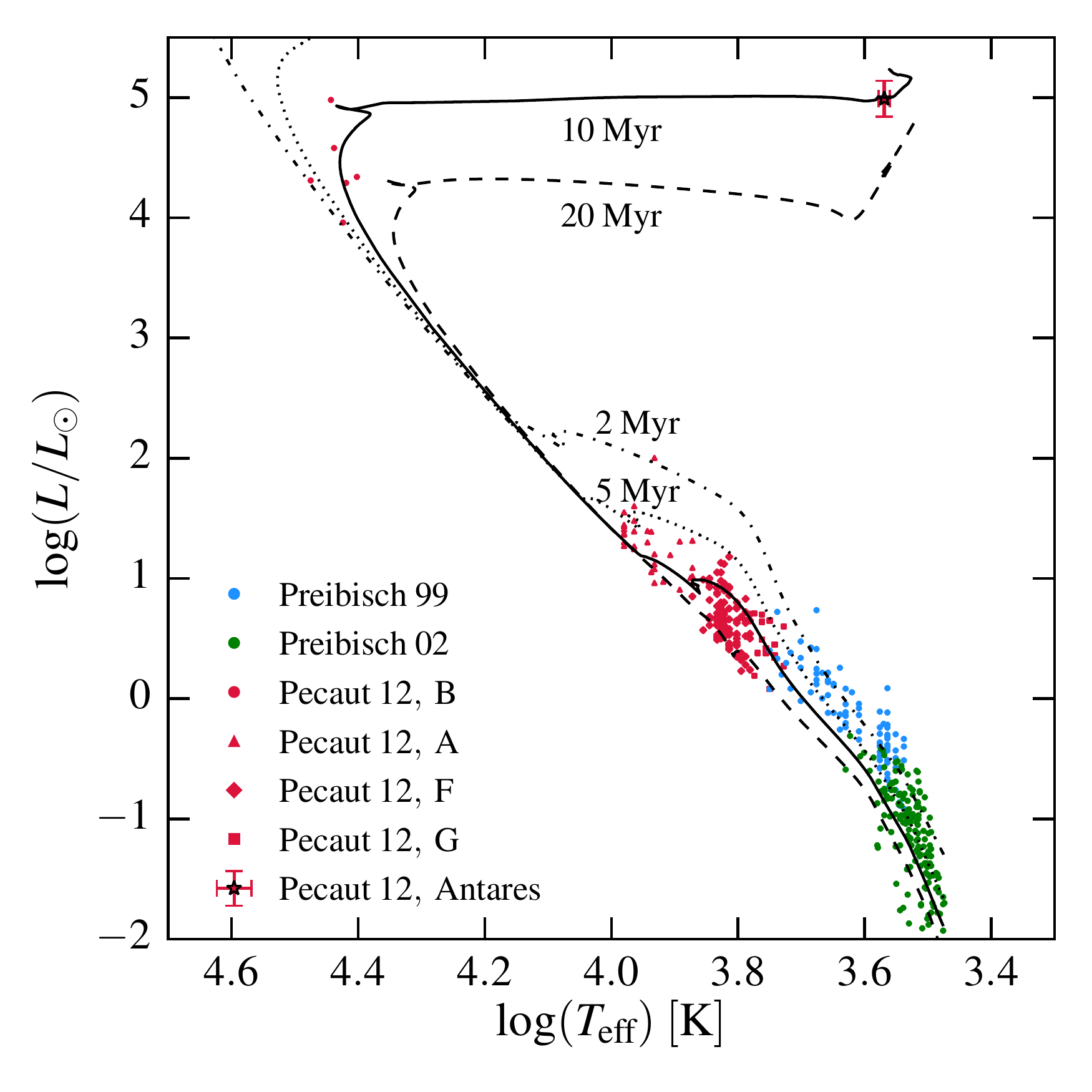}
\caption{HR diagram of Upper Sco stars from \cite{Preibisch1999}, \cite{Preibisch2002}, and \cite{Pecaut2012}, in blue, green, and red symbols, respectively. Solar metallicity isochrones with ages of 2, 5, 10, and 20 Myr are shown in black. There is a well-known discrepancy in the ages inferred from stars with spectral types earlier than G-type versus those later than G-type (with the exception of Antares which has evolved off the MS).}
\label{fig:uppersco}
\end{figure}

Figure~\ref{fig:uppersco} compares 2, 5, 10, and 20 Myr MIST isochrones at solar metallicity with observations from \cite{Preibisch1999}, \cite{Preibisch2002}, and \cite{Pecaut2012}, in blue, green, and red symbols, respectively. For spectral types later than G (with the exception of Antares, a RSG), the models yield an age of $\lesssim5$~Myr which is consistent with the earlier results \cite[e.g.,][]{deGeus1989, Preibisch2002}. For the hotter stars, however, the best-fit model has an older age of $\sim10$~Myr, consistent with the conclusion from \cite{Pecaut2012}. 

Recent work by \cite{Kraus2015} on UScoCTIO 5, a low-mass spectroscopic binary consisting of nearly identical stars that was recently observed by {\it Kepler} as part of the K2 mission \citep{Howell2014}, offered yet another perspective on this issue. Thanks to direct mass and radius measurements, the authors were able to avoid making a comparison in $\teff$ and therefore exclude potential problems with temperature scales as the culprit of this discrepancy. They found that none of the considered models---Lyon \citep{Baraffe2015}, Dartmouth \citep{Dotter2008}, Pisa \citep{Tognelli2011}, Siess \citep{Siess2000}, and PARSEC \citep{Chen2014}---predicted an age around $\sim11$~Myr given the system's stellar properties.\footnote{The Lyon, Dartmouth, Pisa, and Siess models underpredict the radius, which is a well-known problem (see Section~\ref{section:lmrt_relation}). On the other hand, the PARSEC model overpredicts the radius, which \cite{Kraus2015} attributes to a likely overcorrection in their new boundary conditions.} A simple exercise where they horizontally shifted the evolutionary tracks in the $\log L\textrm{--}\log \teff$ plane to bring the predicted $\teff$ into agreement with the observations yielded an age of $\sim11$~Myr, consistent with the age derived from hotter stars. As a result, the authors concluded that the low age predicted from low-mass stars is likely problematic and recommended $\sim11$~Myr instead of $\sim5$~Myr as the probable correct age for Upper Sco. This again emphasizes the imperative need for more robust and detailed modeling of the PMS and low MS stars, which we plan to explore in future work.

\subsection{Asymptotic Giant Branch Stars}
\label{section:agbstars}
Low- and intermediate-mass stars ($1~\msun \lesssim M_{\rm i} \lesssim 8~\msun$, depending on metallicity) ascend the Hayashi track for the second time and enter the AGB phase after they exhaust their central helium supply. During the first part of the AGB phase (the Early AGB; EAGB), the star contains at its center a degenerate carbon and oxygen core surrounded by helium-burning and hydrogen-burning shells. As the star ascends the AGB, the helium-burning shell moves outward until it reaches the hydrogen-rich zone and shuts off. Meanwhile, a thin helium-rich shell starts to grow in mass due to the helium ash raining down from the hydrogen-burning shell, which now dominates the total energy output of the star. The TPAGB phase begins when the helium shell reaches a critical mass and ignites in a thermonuclear runaway as a consequence of thin shell instability \citep{Schwarzschild1965}. The resulting expansion of the overlying material quenches the hydrogen-burning shell while the helium-burning shell settles into a period of quiescent burning. Next, the outer envelope begins to contract, causing the bottom of the hydrogen-rich shell to heat up and ignite. The helium-burning shell moves outward in mass until it eventually becomes extinguished. The entire cycle repeats as a series of thermal pulses until the star sheds its envelope and becomes a post-AGB star (observationally, a planetary nebula). The entire TPAGB phase lasts approximately $10^5$--$10^6$ years depending on mass and metallicity (see Figure~\ref{fig:phase_lifetimes}), and the majority of that time is spent in the quiescent ``interpulse'' state.

The AGB phase plays a significant role in the chemical evolution of galaxies due to its rich nucleosynthetic processes coupled with its high typical mass loss rate, which can be as large as $\sim10^{-3}~\msun \; \rm year^{-1}$ during the ``superwind'' phase \citep{Willson2000, Herwig2005}. In particular, mixing through repeated dredge-up episodes can enrich the surfaces of AGB stars with heavy elements formed from the slow neutron capture process ($s$-process). In massive ($M_{\rm i}\gtrsim4-8~\msun$) super-AGB stars, products of hot bottom burning through the CNO, NeNa, and MgAl cycles are transported up to the surfaces as well. Super-AGB stars are interesting in their own right because they occupy the blurry mass boundary within which ONe and ONeMg WDs can form as a consequence of advanced burning in the core \citep[e.g.,][]{Doherty2015}. From a stellar population synthesis perspective, the combination of high luminosities and relatively long lifetimes of AGB stars implies that stars in the AGB phase contribute a large fraction to the integrated light in intermediate-age (a few Gyr) galaxies \citep{Frogel1990, Maraston2005, Henriques2011, Melbourne2012, Conroy2013, Noel2013}. 

Given its importance, it is therefore disconcerting that the AGB phase is still one of the most poorly understood stellar evolutionary stages. This is because a number of very uncertain and complex physical processes such as mixing and mass loss operate simultaneously and contribute significantly to the evolution (see \citealt{Lattanzio2007} for an excellent short overview of these issues). As succinctly summarized in \cite{Cassisi2013}, ``AGB stars are fascinating objects, where a complicated interplay between physical and chemical processes takes place; an occurrence that still makes computing reliable stellar models for this evolutionary phase a challenge.'' Nevertheless, there has been steady progress toward improving our understanding of AGB stars by calibrating the models against optical and NIR photometry and spectroscopy of the AGB population in nearby galaxies, including the Magnellanic Clouds, dwarf spheroidals, and spirals \citep[e.g.,][]{Marigo2001, Marigo2007, Girardi2007, Girardi2010, Boyer2011, Rosenfield2014}. Since the duration of this phase strongly influences the time evolution of the rest-frame optical and NIR integrated light in intermediate-age stellar populations, it is important to carefully calibrate the models against observations in the local Universe so that integrated light from unresolved systems, e.g., high redshift galaxies, can be accurately interpreted.

\begin{figure}
\centering
\includegraphics[width=0.5\textwidth]{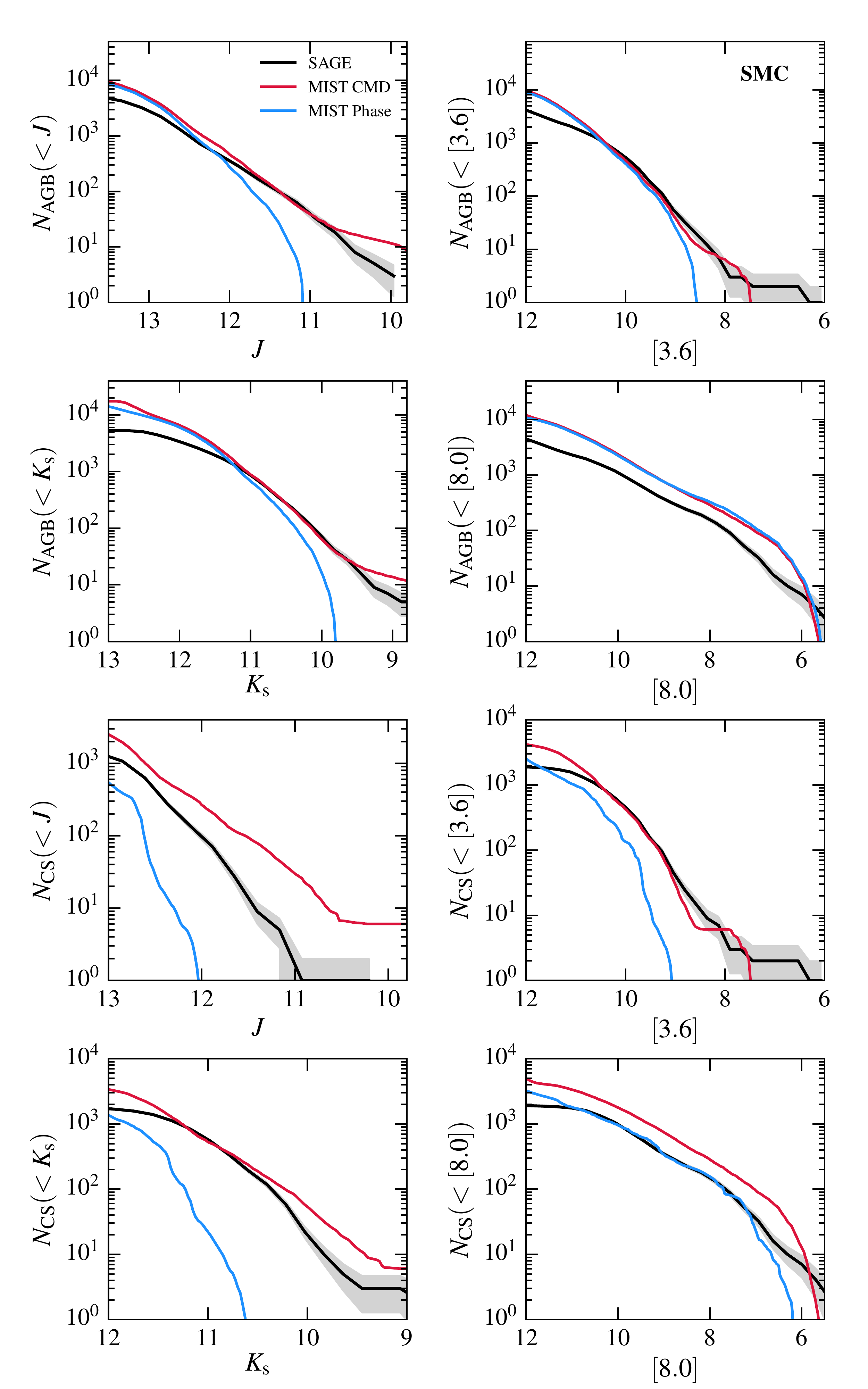}
\caption{Cumulative luminosity functions of AGB and C stars in the SMC. The observed LFs in black lines are constructed from the SAGE-SMC survey photometric catalogs (private communication, M. Boyer; see also \citealt{Gordon2011, Boyer2011} for more details) and the gray bands reflect Poisson uncertainties. The predicted LFs are computed by convolving the isochrones with star formation and metallicity histories from \cite{Harris2004} to create composite CMDs in different filters. The stars are then selected using two methods: ``CMD cut'' uses the same CMD-based criteria that were applied to the SAGE observed sample and ``phase cut'' selects all stars that are phase-tagged as ``TPAGB'' stars in the isochrones. The differences between the two methods emphasize the need for a careful analysis when comparing populations in the CMD.} 
\label{fig:smc_agbcslf}
\end{figure}

\begin{figure}
\centering
\includegraphics[width=0.5\textwidth]{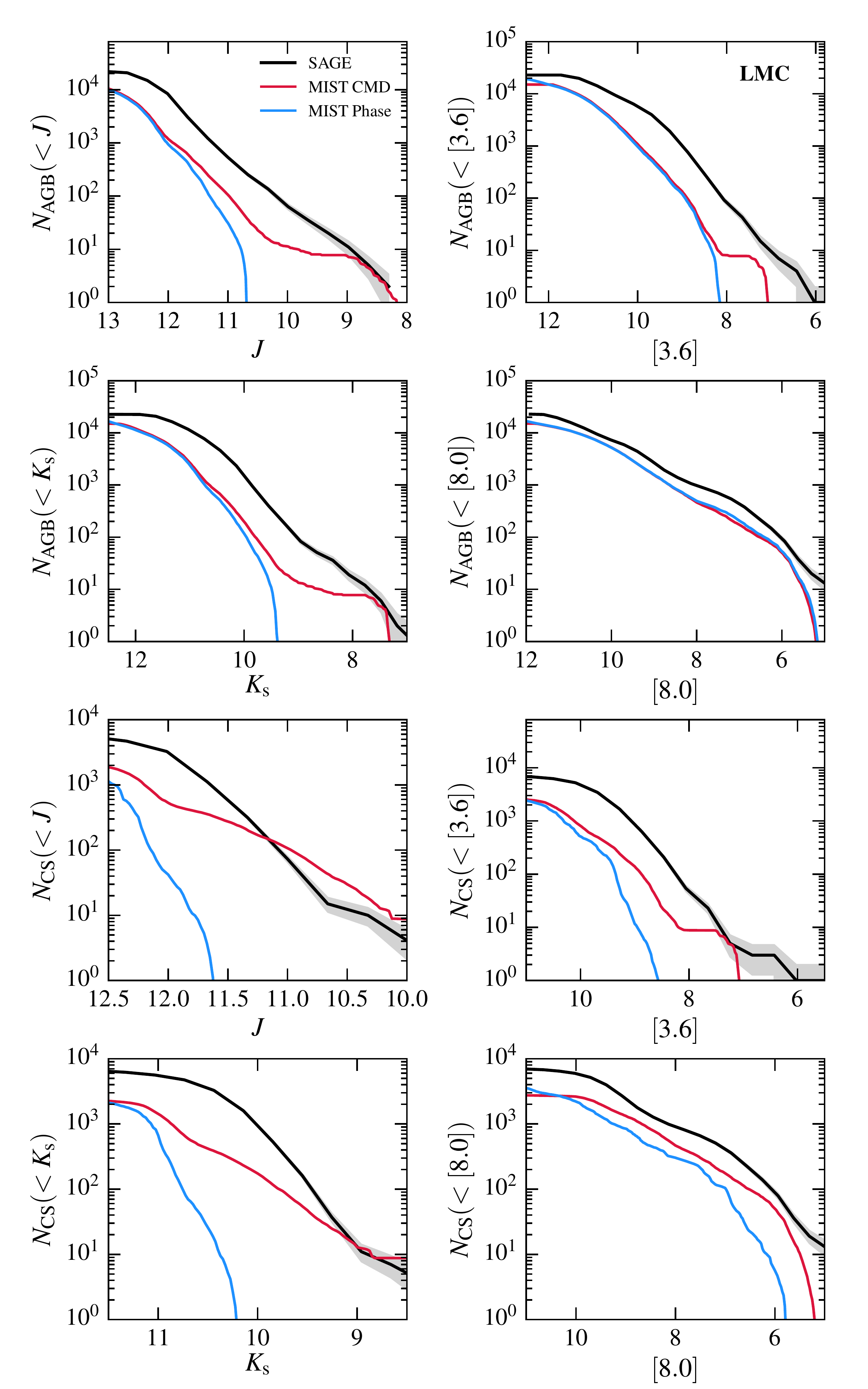}
\caption{The same as Figure~\ref{fig:smc_agbcslf}, now for the LMC. The data come from the SAGE-LMC survey (private communication, M. Boyer; see also \citealt{Meixner2006} for more details) and the star formation and metallicity histories are from \cite{Harris2009}.}
\label{fig:lmc_agbcslf}
\end{figure}

\begin{figure}
\centering
\includegraphics[width=0.5\textwidth]{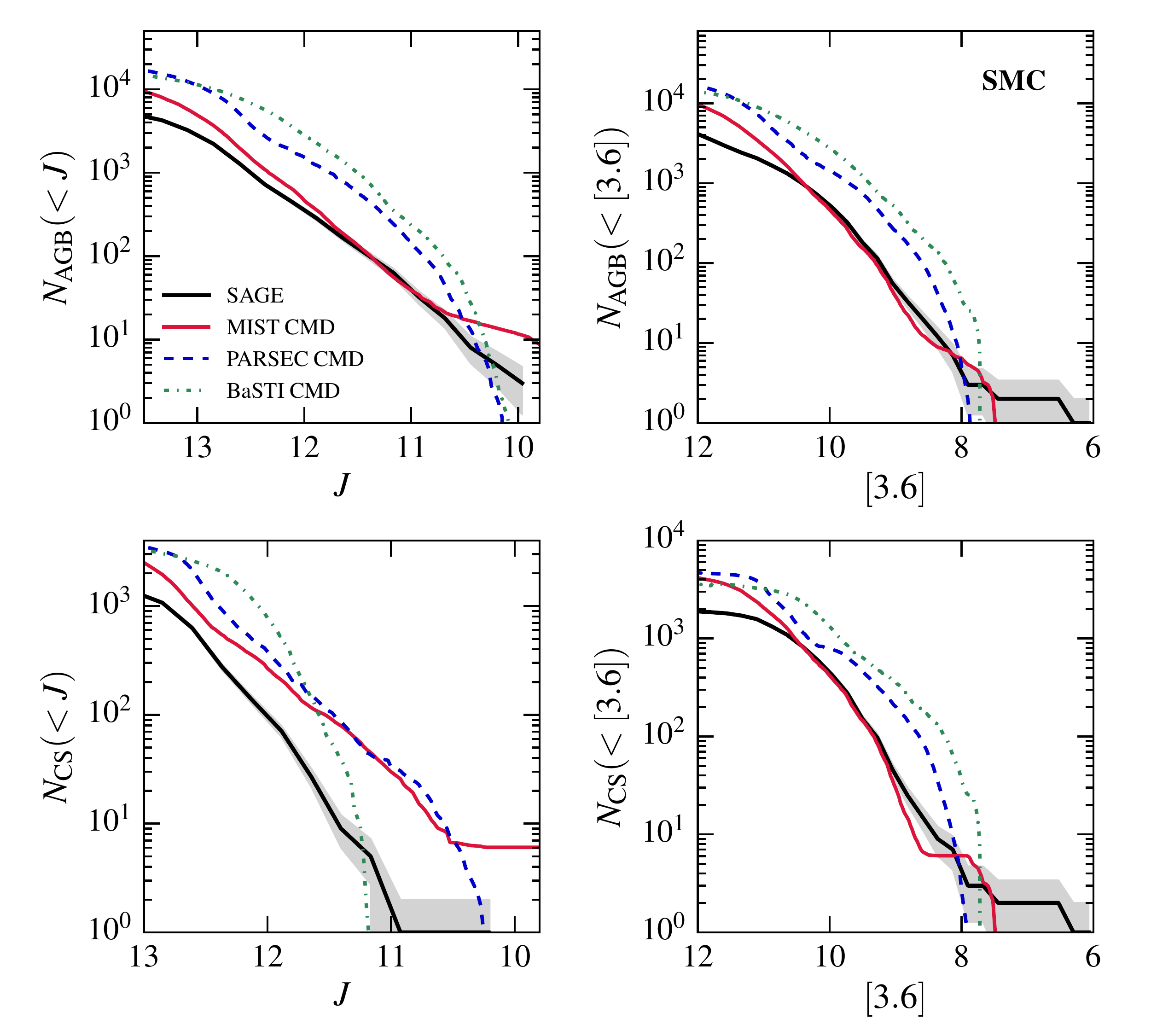}
\caption{The same as Figure~\ref{fig:smc_agbcslf}, but now comparing MIST predictions to the PARSEC/COLIBRI \citep{Bressan2012, Marigo2013, Rosenfield2014} and BaSTI \citep{Pietrinferni2004} predictions in $J$ and [3.6]. The predicted composite CMDs used to generate the LFs displayed here were selected with a CMD cut.}
\label{fig:smc_compare_agbcslf}
\end{figure}

\begin{figure}
\centering
\includegraphics[width=0.5\textwidth]{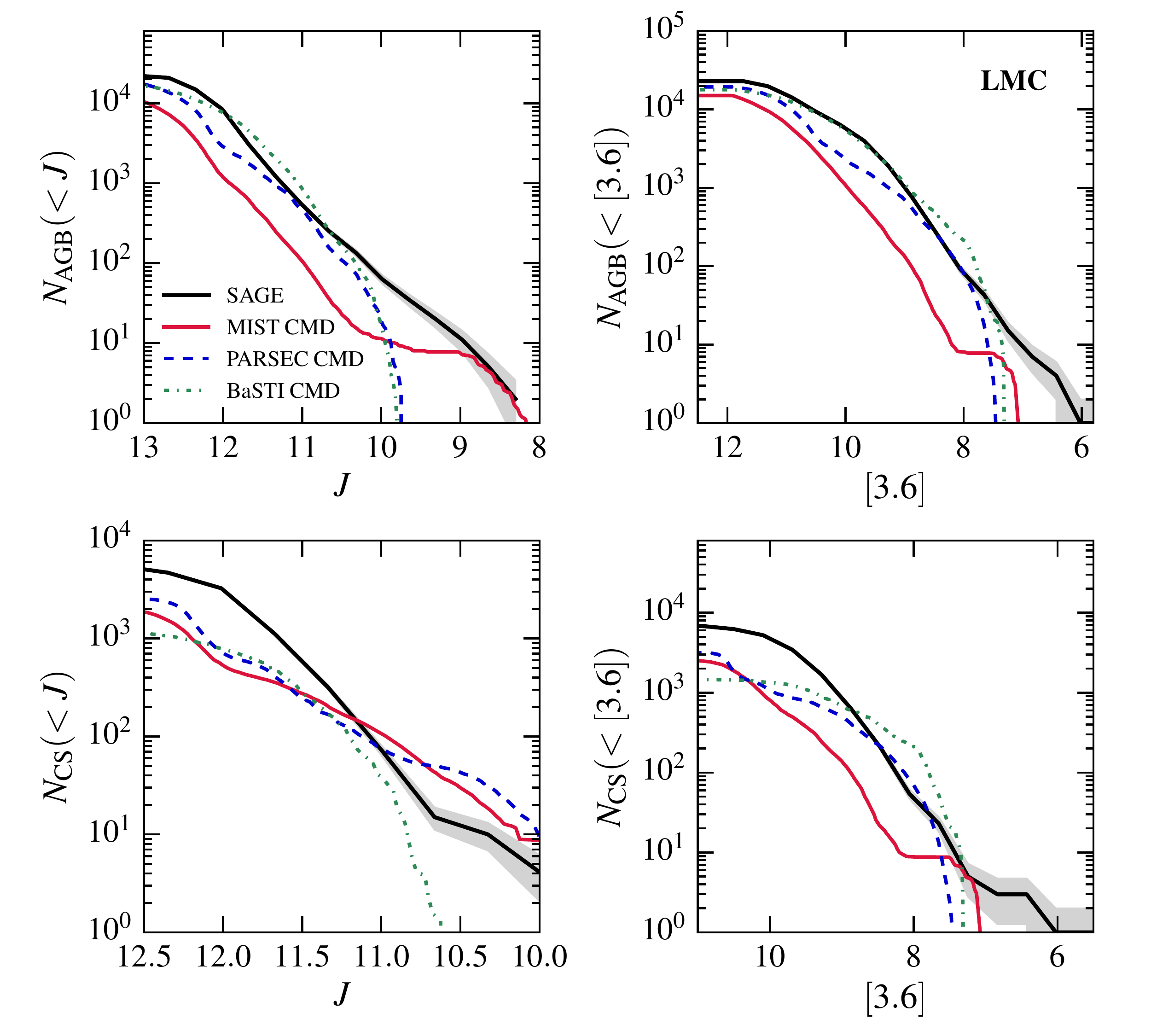}
\caption{The same as Figure~\ref{fig:smc_compare_agbcslf}, now for the LMC.}
\label{fig:lmc_compare_agbcslf}
\end{figure}

\subsubsection{AGB Luminosity Functions}
\label{section:agbcs_lf}
We calibrate the AGB phase in our models via Carbon star (C star) and total AGB luminosity functions (LFs) in the Magellanic Clouds. C stars, the most evolved subset of AGB stars, are formed when the atmosphere becomes carbon-rich ($\rm C/O>1$) as a consequence of recurrent TDU episodes \citep{Iben1983b}. The observed C star LF is a popular diagnostic used to calibrate the TDU efficiency \citep{Groenewegen1993, Marigo1999, Marigo2007}. The faint cutoff contains information about the minimum initial mass that experiences TDU, the maximum near $M_{\rm bol}\sim-5.0$ is sensitive to the efficiency of mass loss as well as TDU, and the bright end reflects the decreasing numbers both due to mass loss leading to the termination of the TPAGB phase and hot bottom burning that converts carbon to nitrogen in intermediate-mass AGB stars \citep{Cassisi2013}.

The observed LFs are constructed using photometric catalogs of cool, evolved stars in the Magellanic Clouds from the Surveying the Agents of Galaxy Evolution (SAGE)-SMC and LMC surveys (private communication, M. Boyer; see also \citealt{Meixner2006, Gordon2011, Boyer2011} for more details). The very wide baseline in wavelength---optical $UBVI$ from the Magellanic Clouds Photometric Survey \citep[MCPS;][]{Zaritsky2002} all the way out to $160~\mu$m from {\it Spitzer}/MIPS---allows for accurate photometric classification of AGB subtypes and identification of contaminants such as unresolved background galaxies, compact \ion{H}{2}  regions, and young stellar objects. From the entire catalog, we select only those located within the MCPS footprint to ensure consistency with the model prediction which folds in star formation histories derived from the MCPS observations. Next, we select AGB (x-, O-, C-, aO-AGB according to the scheme introduced in \citealt{Boyer2011}) and C stars (x-, C-AGB) to construct cumulative LFs.

To create the predicted LFs, we utilize the Flexible Stellar Population Synthesis code \citep[FSPS, v2.6;][]{Conroy2009, Conroy2010}. We first convolve the isochrones with star formation and metallicity histories (SMC; \citealt{Harris2004}, LMC; \citealt{Harris2009}) to generate composite CMDs in various filters, including the effects of circumstellar dust around AGB stars that can strongly influence the flux at longer wavelengths \citep{Villaume2015}. Next, we select the AGB and C stars using the same CMD cuts that were applied to the SAGE observed sample and construct the LFs assuming a Kroupa IMF. As a consistency check, we ensured that the convolution of the adopted star formation histories with the integrated luminosities and stellar masses reproduces the observed integrated light in the NIR and total stellar mass quoted in \cite{Harris2004} and \cite{Harris2009} to within a factor of two. 

In Figures~\ref{fig:smc_agbcslf} and \ref{fig:lmc_agbcslf}, we plot the observed cumulative AGB and C star LFs in four different bands for the SMC and LMC, respectively, in thick black lines and the associated Poisson uncertainties in gray bands. We overplot in red the MIST prediction assuming the same CMD cuts as applied to the data. Overall, the MIST models do a reasonably good job, but the models slightly overpredict and underpredict the numbers for the SMC and LMC, respectively. Originally, our goal was to use these LF comparisons as a means of calibrating various input parameters, e,g., mass loss and TDU efficiencies, that influence AGB star lifetimes, luminosities, and the formation of C-stars. Along the way, we encountered several factors that have rendered this task more challenging than initially expected.

First, there are uncertainties in the adopted star formation and metallicity histories from \cite{Harris2004} and \cite{Harris2009}. These were derived using a different set of isochrones (Padova; \citealt{Girardi2002}, to be exact) compared to the MIST isochrones used for the AGB LF predictions. There is thus a fundamental inconsistency that can be addressed by reconstructing the LMC and SMC SFHs with MIST isochrones. Moreover, the recovered star formation histories are sensitive to the adopted dust attenuation model and to crowding, which affect the completeness in high density areas like 30 Doradus. Also important is the recovered metallicity history, which is far more imprecise than the star formation history itself, since the predicted AGB colors and evolution are extremely metallicity sensitive.

Second, since predicted AGB and C stars are selected using the observed CMD cuts, a small mismatch in the locations of the isochrones, especially in color, may strongly influence the comparison with the data, e.g., separation of C-stars from O-stars ($\rm C/O<1$). Undertaking the comparison in multiple bands makes this a particularly demanding task because obtaining good agreement across all wavelengths leaves little room for error in each component: star formation and metallicity histories, stellar evolutionary models, bolometric corrections, and AGB circumstellar dust models. A perhaps more straightforward test is to compare the predicted and observed CMDs directly, though the same uncertainties will make this a challenging task as well.

Finally, although \cite{Boyer2011} took great care to ensure a high-fidelity sample, there is most likely a nonzero amount of contamination in the final sample of AGB stars by e.g., RSGs. By adopting the observed CMD cuts as the selection criteria, we can, to some degree, account for the possibility of contamination from RSGs in the AGB sample and O stars in the C star sample. 

To illustrate the challenges of comparing samples based on CMD cuts, we have also computed LFs by identifying AGB and C stars directly in the isochrone according to their evolutionary stages. It is trivial to tag the phase of every star in the predicted CMD because we have all of the necessary evolutionary information, e.g., stellar mass and surface C/O abundance. The MIST model predictions are shown in blue in Figures~\ref{fig:smc_agbcslf} and \ref{fig:lmc_agbcslf}. There are a few interesting and revealing differences. Interestingly, objects at the bright end of the $J$- and $K_{\rm s}$-band LFs are absent in the phase-selected MIST models, which suggests that RSG contamination may be important for the bright end of the AGB LFs. The C star LFs show a more dramatic difference. Possible reasons for the discrepancy include inaccurate modeling of surface abundance enrichment through TDU and winds, the current lack of C/O-variable molecular opacities in the envelope in the models, as well as any deficiencies in the AGB circumstellar dust models. The implementation of low temperature molecular opacities in MESA, which have been shown to play an important role in AGB evolution \citep{Marigo2002, Marigo2003}, is a high priority for the MIST project.

We conclude this section by comparing the predicted LFs from the MIST isochrones to those from other widely used isochrones. In Figures~\ref{fig:smc_compare_agbcslf} and ~\ref{fig:lmc_compare_agbcslf}, we plot MIST, PARSEC/COLIBRI \citep{Bressan2012, Marigo2013, Rosenfield2014}, and BaSTI \citep{Pietrinferni2004} predictions in red, dark blue, and green, respectively, for $J$ and [3.6]. The LFs were computed with stars selected from CMD cuts. The $\dot{M}$ required for the computation of AGB circumstellar dust effects came directly from the isochrone files for MIST and PARSEC/COLIBRI, while for BaSTI $\dot{M}$ was computed using the \cite{Vassiliadis1993} AGB mass loss prescription. We emphasize that all the rest of the input ingredients for constructing the LFs, including the bolometric corrections in FSPS, are identical for the three isochrones showcased here in order to isolate the effects of varying the isochrones alone. Overall, the models are in broad agreement with each other and the observations.

\section{Comparisons with Data. II: High Mass Stars}
\subsection{Width of the MS}
In Section~\ref{section:star_clusters} we used the MSTO morphology of M67 to calibrate the efficiency of core overshoot. Another popular calibration option is to match the observed width of the MS \citep[e.g.,][]{Ekstrom2012}. We check to see if the MSTO-calibrated overshoot efficiency predicts MS width that is consistent with the observed MS width reported by \cite{Castro2014}. Following \cite{Langer2014}, the authors performed their analysis on a so-called ``spectroscopic HR diagram.'' It differs from an ordinary HR diagram in that it still has $\log (\teff)$ on the $x$-axis but a new quantity $\mathscr{L}\equiv \teff^4/g$ on the $y$-axis. The main advantage of a spectroscopic HR diagram is that all of the relevant quantities can be obtained from spectroscopic analysis without having to worry about ambiguities in distance or extinction.

\begin{figure}
\centering
\includegraphics[width=0.45\textwidth]{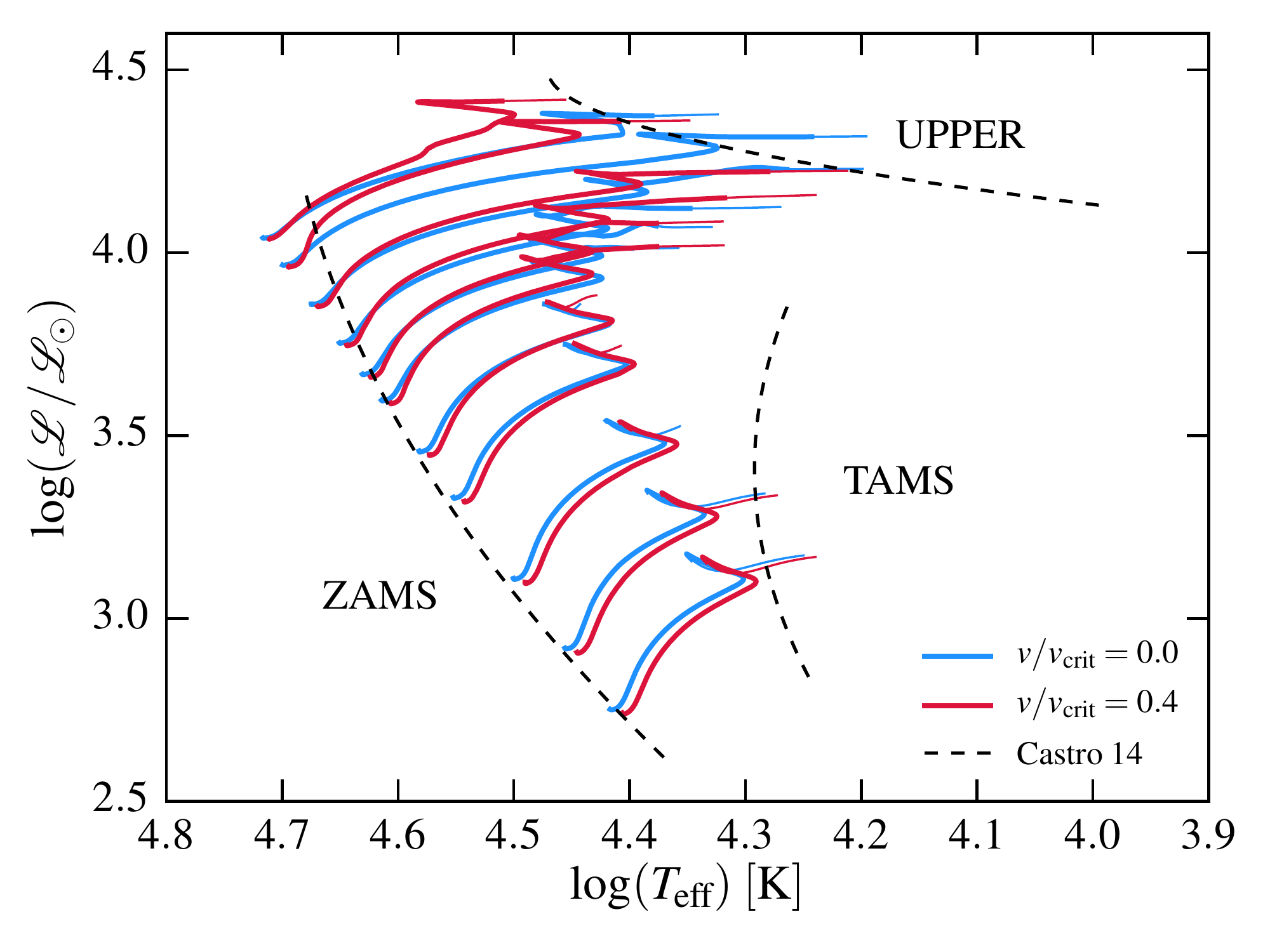}
\caption{Spectroscopic HR diagram, with $\log \teff$ on the $x$-axis and the quantity $\mathscr{L}\equiv \teff^4/g$ on the $y$-axis. Black dashed lines correspond to empirical fits to the probability density distribution constructed from a sample of more than 600 stars \citep{Castro2014}. The red and blue tracks show the MS in the solar metallicity MIST evolutionary tracks with and without rotation, respectively.}
\label{fig:sHRD}
\end{figure}

In Figure~\ref{fig:sHRD}, we plot three lines demarcating the MS region (ZAMS, TAMS, UPPER) from \cite{Castro2014}, which are empirical fits to the probability density distribution constructed from a sample of more than 600 stars. The top portion of the TAMS line is missing because there is no clean break from the MS to cooler temperatures in the observed distribution of stars for $\log \mathscr{L}/\mathscr{L}_{\odot}\gtrsim4$. This continuous distribution could be explained by the inflation of stars approaching the Eddington limit or the presence of helium burning stars. \cite{Castro2014} compared evolutionary tracks from \cite{Ekstrom2012} and \cite{Brott2011} both with and without rotation and found that the ZAMS loci are generally well-reproduced by all models except at the highest masses. However, these massive objects could be missing from the observed sample simply due to their rarity or obscuration by their birth clouds. The authors concluded that no model can reproduce the broad MS width at high $\mathscr{L}$ and suggested that core overshoot efficiency may be mass-dependent. 

We overplot a series of solar-metallicity MIST evolutionary tracks with masses ranging from 10 to $80~\msun$ with (solid red) and without (solid blue) rotation in Figure~\ref{fig:sHRD}. The MIST models also correctly predict the ZAMS line but fail to reproduce the broad MS width at the highest masses. These models suggest that rotation alone cannot explain the full extent of the MS width. A mass-dependent core overshoot efficiency is a topic we plan to explore in future work.

\subsection{Locations of Red Supergiants on the HR Diagram}
When a high-mass star runs out of hydrogen in its core, it may migrate toward the Hayashi track and become a RSG. For a long time, the observed RSGs were found to be too luminous and cool compared to the predicted RSGs \citep{Massey2003}, which was problematic as the region redward of the theoretical Hayashi track represents a ``forbidden zone.'' At fixed metallicity, each point along the Hayashi line corresponds to the coolest possible model in hydrostatic equilibrium.

\cite{Levesque2005} resolved this problem for Galactic stars, demonstrating that the new effective temperature scale computed from improved MARCS atmospheric models yielded much better agreement between the Geneva evolutionary tracks and the observed HR diagram locations. Shortly after, \cite{Levesque2006} utilized these new models to analyze a sample of RSGs in the Magellanic Clouds and confirmed previous results from \cite{Elias1985} that these RSGs belong to earlier spectral subtypes compared to their galactic counterparts. This is also consistent with the theoretical expectation that the Hayashi line should shift toward warmer temperatures at lower metallicities. \cite{Levesque2006} also found that RSGs in the SMC span a wide range of $\teff$ for a given $\mbol$, possibly due to the increased importance of rotational mixing at lower metallicities (see Section~\ref{section:effectofrotation} for more details).

\begin{figure}
\centering
\includegraphics[width=0.5\textwidth]{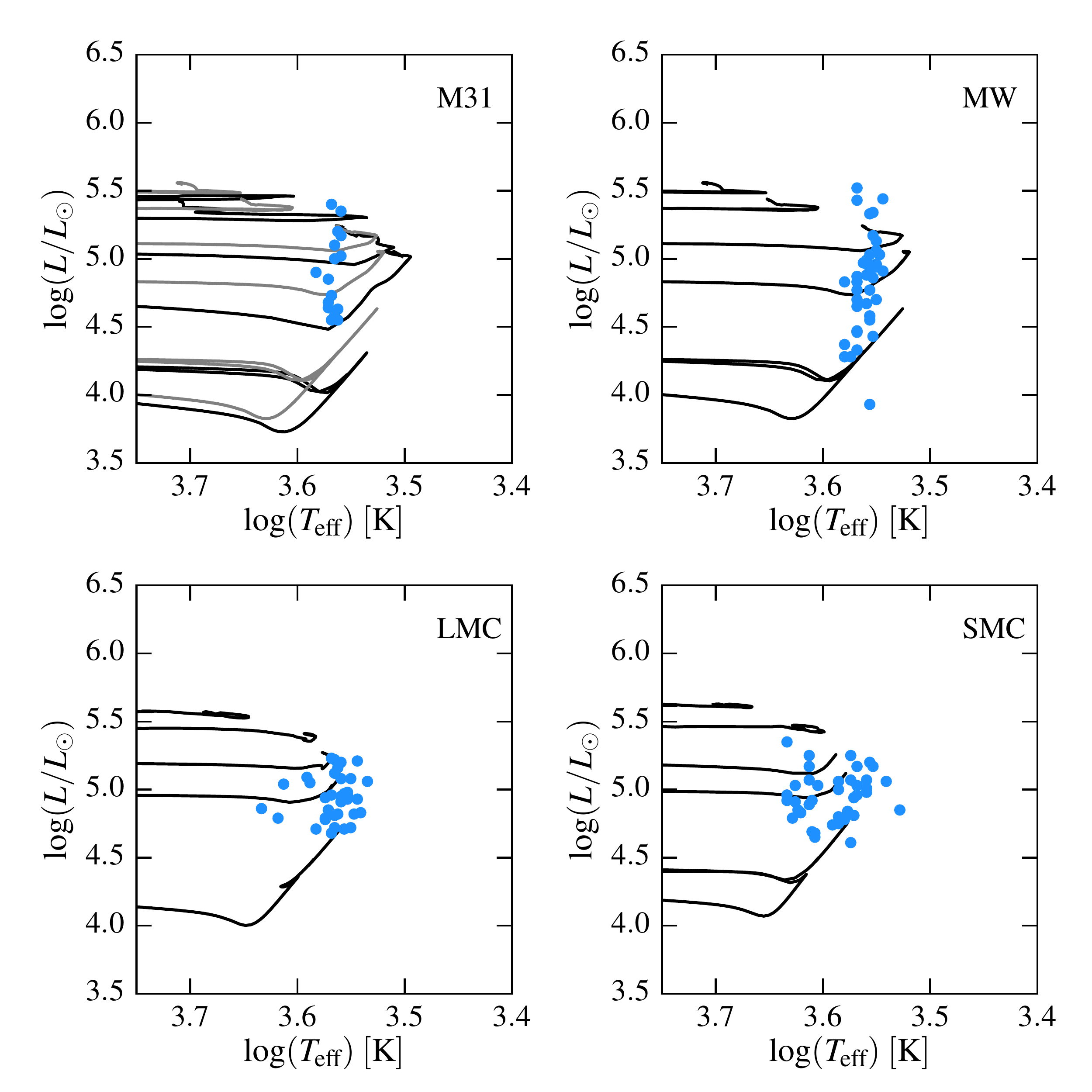} 
\caption{A comparison between the MIST evolutionary tracks with rotation and observed RSGs \citep{Levesque2005, Levesque2006, Massey2009}. The masses of the tracks displayed are 10, 16, 20, 26, and $30~\msun$. The observed $L_{\rm bol}$ is calculated from $K$-band photometry. Top left: M31 ($\rm [Z/H]=+0.3$ in black; $\rm [Z/H]=0.0$ in gray). {\it Top right}: Milky Way ($\rm [Z/H]=0.0$). Bottom left: LMC ($\rm [Z/H]=-0.5$). Bottom right: SMC ($\rm [Z/H]=-0.75$).}
\label{fig:rsg_locations}
\end{figure}

\begin{figure*}
\centering
\includegraphics[width=\textwidth]{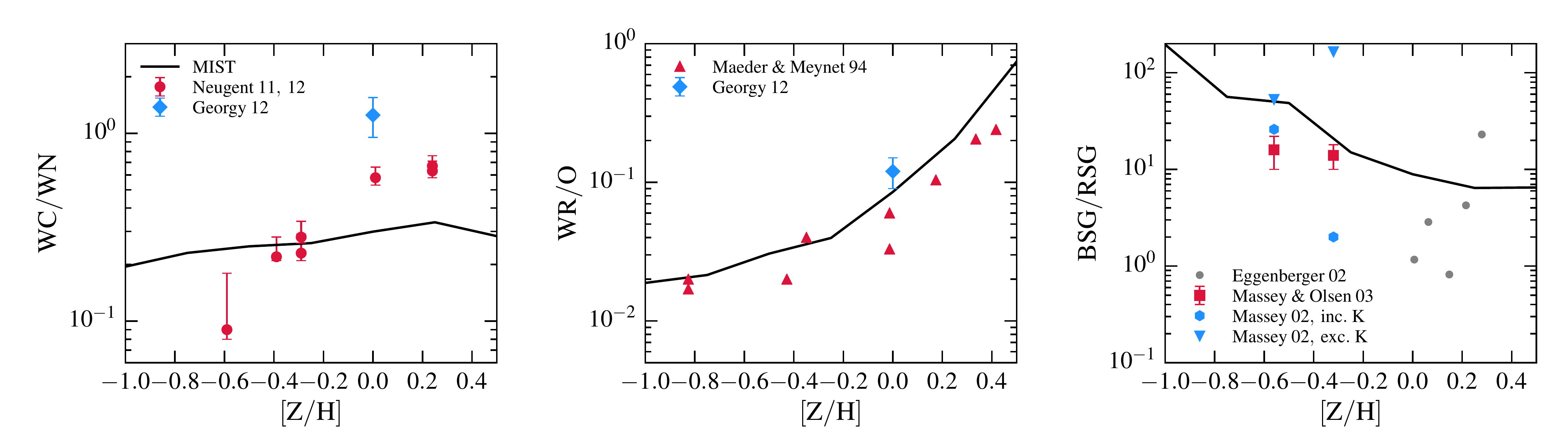} 
\caption{Left: the predicted WC to WN ratio as a function of metallicity. The red circles are observations from \cite{Neugent2011} and \cite{Neugent2012_wr}, and the blue diamond point is the observed ratio from \cite{Georgy2012}, estimated in the 3~kpc radius volume around the solar neighborhood. Middle: the same as the left panel except now showing the WR to O ratio. The red triangles are observed number ratios from Table 6 in \cite{Maeder1994b} and the blue diamond point is the observed ratio from \cite{Georgy2012}, estimated in the 2.5~kpc radius volume around the solar neighborhood. Right: the same as the left panel except now showing the BSG to RSG ratio. The gray points are observed ratios from young star clusters in the Milky Way and the Magellanic Clouds \citep{Eggenberger2002}. The blue hexagons and triangles are observed ratios computed with and without K-type stars in the RSG category \citep{Massey2002}. The red squares are observed ratios computed from spectroscopically confirmed RSG stars \citep{Massey2003}.}
\label{fig:highmass_ratios}
\end{figure*}

In Figure \ref{fig:rsg_locations}, we compare the MIST evolutionary tracks in black and the sample of observed RSGs from \cite{Massey2009}. Only for the top left panel (M31), we show additional tracks at $\rm [Z/H] = 0.0$ in gray since the metallicity of M31 is still under debate \citep{Venn2000, Trundle2002, Sanders2012, Zurita2012}. The observed $L_{\rm bol}$ shown here is derived from $M_K$ rather than $M_V$ since the former is less sensitive to extinction. The typical measurement uncertainty in $\teff$ ranges from $\sim100$~K for the warmest stars to $\sim20$~K for the coolest stars, and the uncertainty in $\log (L) $ is negligible ($\sim0.05$~dex). We expect the high density of observed stars to coincide with the location of the Hayashi line \citep{Drout2012}. For the LMC and the SMC, $\teff$ and maximum luminosity are both reproduced by the models. For M31 and the MW, the predicted slopes of the RGB tracks are too shallow compared to the observations, but it is still encouraging that no observed RSGs fall in the forbidden zone. We plan to investigate this further in the future.

\subsection{Relative Lifetimes}
\label{section:highM_ratios} 
One of the most popular tests of massive star models is to compare the observed and predicted ratios of stars belonging to different evolutionary stages as a function of metallicity. The ratio of the IMF-weighted sums of phase $A$ and $B$ lifetimes serves as a proxy for the observed number ratio of stars in phases $A$ and $B$, $N_{\rm obs,\;A}/N_{\rm obs,\;B}$:
\begin{equation}
\frac{\int t_{\rm A}(M)\; \phi(M) \; dM}{\int t_{\rm B}(M) \; \phi(M) \; dM} = \frac{N_{\rm obs,\; A}}{N_{\rm obs,\; B}}\;\;,
\end{equation}
where $t$ is the phase lifetime and $\phi$ is the IMF weight. This implicitly assumes that the star formation history is constant over the range of ages considered, which is likely a reasonable approximation for massive stars with $M_{\rm i}\gtrsim10~\msun$ and MS lifetimes $\lesssim20~\rm Myr$ (but see also e.g., \citealt{DohmPalmer2002} where the authors examined the ratio of blue to red supergiants as a function of age in Sextans A).

Here we present three such tests: the ratio of WR subtypes WC to WN, the ratio of WR to O-type stars, and the ratio of blue supergiants (BSGs) to RSGs. We convert metallicities reported in the literature---$\log (\rm O/H)+12$, $Z$, and [Fe/H]---to a common scale in [Z/H] to enable comparison with the models. There is an estimated $\sim0.1$~dex uncertainty in our converted [Z/H] values since there are spatial metallicity gradients within the galaxies and variations in the solar abundances adopted by different groups. Note that these models do not include the effects of binary evolution (see \citealt{Eldridge2008}).

\subsubsection{WC/WN}
A WR star is an evolved massive star with little to no hydrogen in its outer layers. It is formed once the star sheds its hydrogen-rich envelope through mass loss, revealing hydrogen-burning products such as helium and nitrogen (WN subtype) and helium-burning products such as carbon and oxygen (WC subtype). Since the predominant mass loss mechanism in hot massive stars is likely radiative momentum transfer onto metal ions in the atmosphere, mass loss is predicted to increase with metallicity \citep{Vink2005}. As a result, the ratio of WC to WN subtypes is expected to increase with increasing metallicity of the environment \citep{Maeder1994}, which makes this ratio a useful calibrator for metallicity-dependent mass loss in massive star evolutionary models.

There has been a long-standing mismatch between the predicted and observed WC/WN ratios, especially at high metallicities \citep{Meynet2005, Neugent2011, Neugent2012_wr}. However, it was unclear whether this disagreement was due to poor models or completeness issues with observations. On the observations front, \cite{Neugent2012_wr} completely revised the WN/WC ratio in M31 by discovering more than 100 new WR stars with an estimated completeness fraction of $\sim0.95$. A comparison between the observed WC/WN ratios in M31, M33, SMC, and LMC and the ratios predicted by non-rotating and rotating Geneva models \citep{Meynet2005, Ekstrom2012} revealed only a marginal improvement from the new rotating models. Furthermore, they concluded that additional models at different metallicities (full grids of models were only available for two metallicities at that time) were required for a more informative comparison.

We identify WR stars in our models following the classification scheme introduced in \cite{Georgy2012}. We group the WNL and WNE stars (late- and early-subtypes of WN) as part of the WN stars, exclude the ambiguous WNC stars (WN to WC transition), and include the WO subtype with the WC stars. We emphasize that this theoretical classification scheme---based on the average surface abundances---is technically not equivalent to the classification scheme used by observers who rely on the spectroscopic detection of emission lines \citep[see e.g.,][]{vanderHucht2001}. 

First we compute the phase lifetimes for each evolutionary track. For each phase, we sum up the lifetimes for all stellar masses with weights provided by the IMF \citep{Kroupa2001}. The total lifetime for a given phase is a theoretical proxy for the observed number of stars in that phase, which means that we can now take the ratio of WC to WN lifetimes and directly compare to the observations.

In the left panel of Figure~\ref{fig:highmass_ratios}, we plot the predicted WC to WN ratio as a function of metallicity in a solid black line. The red circles are observations from \cite{Neugent2011} and \cite{Neugent2012_wr}, and the blue diamond point is the observed ratio computed by \cite{Georgy2012}, estimated in the 3~kpc radius volume around the solar neighborhood (see their Section 4.4 for references therein). Although the MIST model is currently unable to reproduce the observed trend with metallicity, the predicted ratios are in agreement with the observed values to within a factor of 2 to 3. We plan to improve this further in future work.

\subsubsection{WR/O}
The predicted ratio of WR to O-type stars is computed from the models using the method outlined in the previous section and compared to observations. Here, the WR population is the sum of all WR subtypes. O-type MS stars are identified according to the \cite{Georgy2012} classification scheme.

In the middle panel of Figure~\ref{fig:highmass_ratios}, we show the predicted WR to O ratio as a function of metallicity in a solid black line. The red triangles are observed number ratios from Table 6 in \cite{Maeder1994b}\footnote{There are no uncertainties reported by the authors.} and the blue diamond point is the observed ratio computed by \cite{Georgy2012}, estimated in the 2.5~kpc radius volume around the solar neighborhood. Again, the model prediction is in qualitative agreement with the observations; WR stars become more abundant in higher metallicity environments. This is to be expected because efficient mass loss at high metallicities readily removes the hydrogen-rich outer layers and promotes the formation of WR stars.

\subsubsection{BSG/RSG}
The number ratio of blue to red supergiants (BSG/RSG) has long been known to decrease with increasing galactocentric radius in the Milky Way, the Magellanic Clouds, and M33 \citep[e.g.,][]{Walker1964, Hartwick1970, Cowley1979, Humphreys1979, Meylan1982}. This was explained by invoking radial metallicity gradients in the disks \citep{vandenBergh1968}. The BSG/RSG ratio is an excellent diagnostic tool because whether a star becomes a RSG or a BSG hinges very sensitively on, for example, the details of semiconvection and convective overshoot \citep[e.g,][]{Langer1991}.

Most stellar evolution models have struggled to reproduce this radial/metallicity trend \cite[see][for an in-depth discussion of this topic]{Langer1995}. However, \cite{Maeder2001} found that the inclusion of rotation in their models produced more RSGs at low metallicities, which dramatically lowered the predicted BSG/RSG ratio in the SMC and brought the model prediction into better agreement with the observations. \cite{Eldridge2008} computed their model predictions with and without the effects of binarity and concluded that their single-star model underpredicted the BSG/RSG ratio. A mixed population model with a two-thirds binary fraction was required to reproduce the observations.

In the right panel of Figure~\ref{fig:highmass_ratios}, we plot the observed BSG/RSG ratios in the Magellanic Clouds from \cite{Massey2002} and \cite{Massey2003}. The two \cite{Massey2002} symbols correspond to ratios computed including (blue hexagon) and excluding (blue triangle) K-type stars in the RSG category from their $UBVR$-photometry-selected sample. Their sample was limited to $\mbol < -7.5$ (roughly $\log (L/ \lsun ) > 4.9$) in order to minimize contamination by the AGB stars and improve completeness. There was an estimated $\sim10\%$ contamination by foreground red dwarfs in their photometric sample, but this was dramatically improved in \cite{Massey2003} with spectroscopic radial velocity measurements. The updated ratios from spectroscopically confirmed RSG stars \citep{Massey2003} are shown as red squares.

We also show the observed BSG/RSG ratios computed from spectroscopically identified candidates in a sample of 45 young ($6.8 < \logage < 7.5$) clusters in the Milky Way and the Magellanic Clouds \cite[gray;][]{Eggenberger2002}. The authors computed the observed BSG/RSG ratios at different metallicities by binning the stars according to their galactocentric distances and assuming a radial metallicity gradient to assign metallicities to each radius bin. Their results showed increasing BSG/RSG ratio with increasing metallicity.

We overplot our prediction as a solid black line. RSGs and BSGs were identified using the selection criteria from \cite{Eldridge2008}, which are consistent with observational cuts made by \cite{Massey2002} and \cite{Massey2003}. The model predictions are bracketed by the two \cite{Massey2002} points and marginally consistent with \cite{Massey2003}. We do not reproduce the positive trend reported by \cite{Eggenberger2002}, but \cite{Eldridge2008} raised the concern that clusters in the \cite{Eggenberger2002} sample generally have an age spread similar to the age of the cluster itself. Additional data over a wider range of environments would be valuable in assessing the quality of the massive star models.

\subsection{The Effects of Rotation on Surface Abundances}
\label{section:n_abun}
Rotation has been proposed as a viable mechanism for enriching the surfaces of stars (\citealt{Heger2000, Meynet2000}; see also Sections~\ref{section:rot_mixing} and \ref{section:rot_mass_loss}) by inducing extra mixing and enhancing mass loss.\footnote{While gravitational settling works against these processes, it is a slow process with a negligible effect in the presence of rapid rotation.} Models with rotation generally predict a surface enrichment of helium and nitrogen along with a concomitant depletion of carbon and oxygen during the MS as the products of the CNO cycle get dredged up to the surface through rotational mixing \citep{Maeder2000, Yoon2005}. For this reason, observed surface abundances have been used to calibrate the efficiency of rotational mixing in the models \citep[$f_\mu$ and $f_c$;][]{Pinsonneault1989, Heger2000, Brott2011}.  

\begin{figure}
\centering
\includegraphics[width=0.4\textwidth]{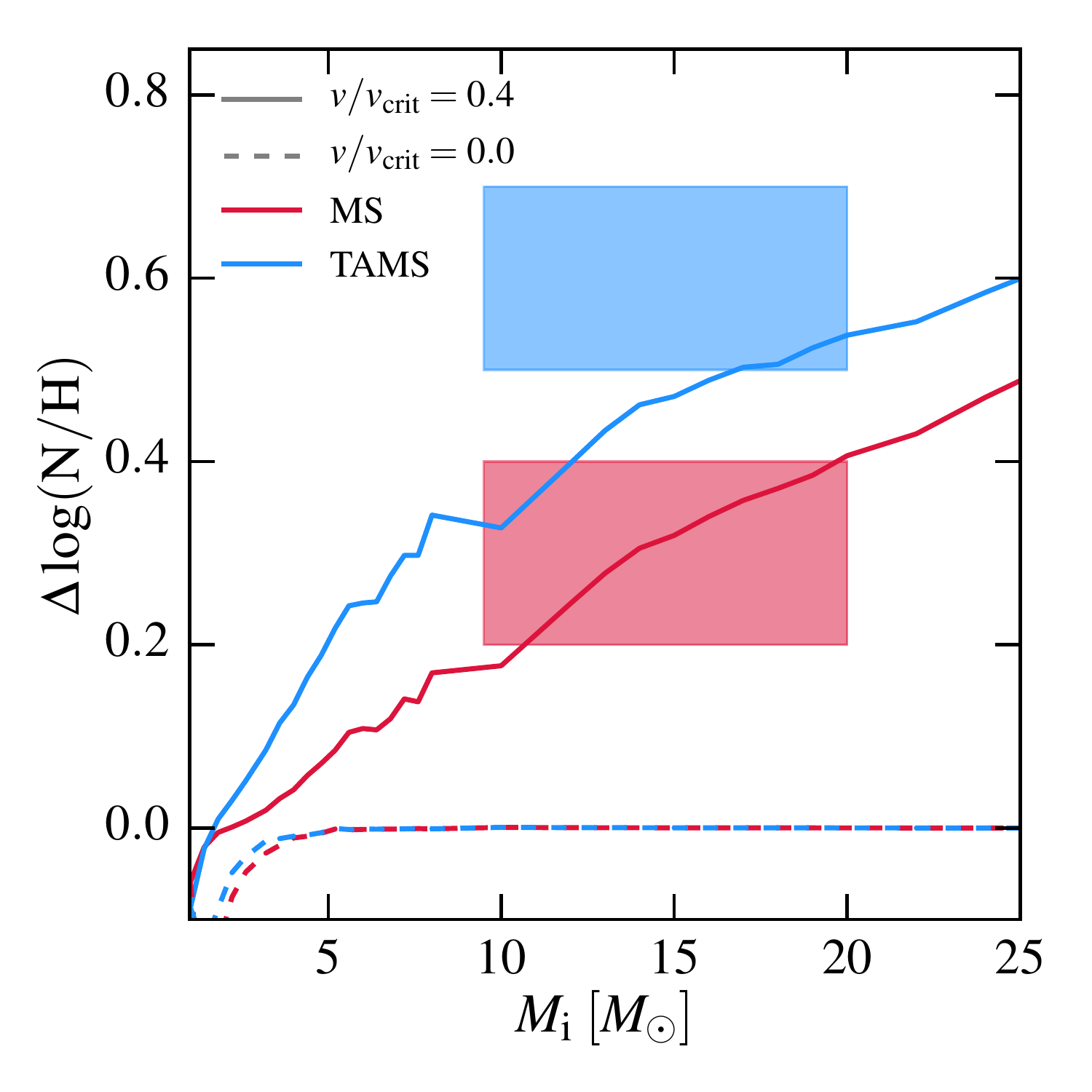} 
\caption{Surface nitrogen enrichment midway through the MS ($X_{\rm c}\sim0.5$; red) and at TAMS (blue) as a function of initial mass. The red and blue shaded boxes correspond to the average nitrogen excess observed for galactic MS B-type stars with $M_{\rm i}<20~\msun$ and the maximum observed excess \citep{Gies1992, Kilian1992, Morel2008, Hunter2009}. Without rotation, the predicted nitrogen enrichment during the MS at these masses is zero. This figure is adapted from Figure~11 of \cite{Ekstrom2012}.}
\label{fig:surf_n_abun_delta}
\end{figure}

\begin{figure*}
\centering
\includegraphics[width=\textwidth]{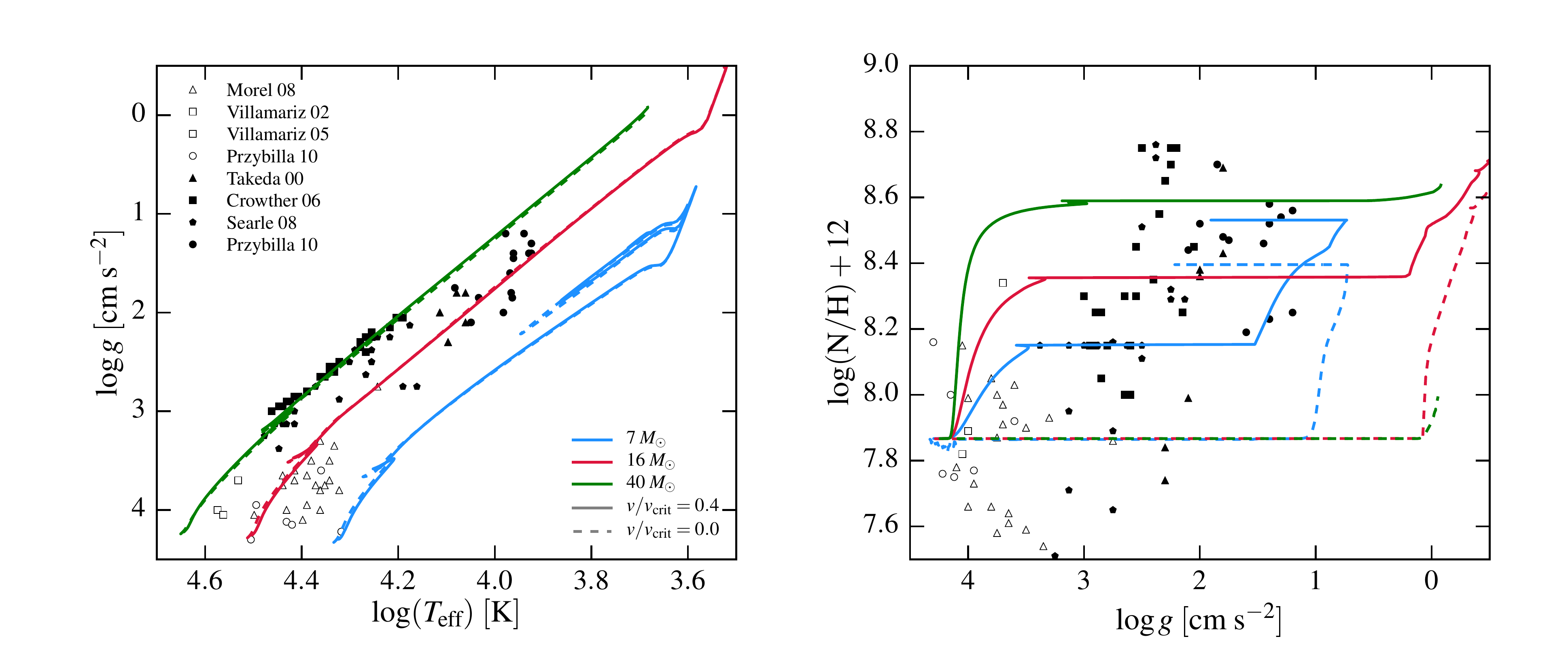} 
\caption{Left: evolutionary tracks in the $\log(g)-\log(\teff)$ plane for 7, 16, and $40~\msun$ stars at solar metallicity. Solid and dotted lines are models with and without rotation and open and filled symbols correspond to O- and B-type galactic dwarfs and supergiants, respectively. This figure is adapted from Figure~12 of \cite{Ekstrom2012}. Right: surface nitrogen abundance evolution for the same three models.} 
\label{fig:surf_n_abun_evol}
\end{figure*}

Following \cite{Ekstrom2012}, we check that our rotating models are able to match the range of observed surface nitrogen enrichment in galactic O- and B-type stars. Figure~\ref{fig:surf_n_abun_delta} shows the predicted surface nitrogen enrichment midway through the MS ($X_{\rm c}\sim0.5$; red) and at TAMS (blue) as a function of initial mass. Models with and without rotation are plotted in solid and dashed lines, respectively. The red shaded box corresponds to the mean surface nitrogen excess observed in a sample of galactic MS B-type stars with masses below $20~\msun$ and the blue shaded box on top corresponds to the maximum observed excess. These observed numbers come from Table 2 of \cite{Maeder2012}, which is a compilation of data from \cite{Gies1992}, \cite{Kilian1992}, \cite{Morel2008}, and \cite{Hunter2009}. Overall, our models are in excellent agreement with the observed range of nitrogen enrichment on the surfaces of B-type MS stars, and in marginal agreement with the maximum observed excess. They are also broadly in agreement with the predictions from the Geneva models \citep[see Figure 11 of ][]{Ekstrom2012}, though there is a noticeable difference for stars below $\sim2~\msun{}$. This is due to the inclusion of magnetic braking effects in the Geneva models; stars with masses below $\sim 1.7~\msun$ experience extra surface nitrogen enrichment due to enhanced mixing induced by large shear in the outer layers, though the authors caution that this effect may be overestimated in their current implementation. In the MIST models, we turn off rotation for stars with $M_{\rm i}<1.2~\msun$ and gradually increase the rotation rate from $v_{\rm ZAMS}/v_{\rm crit}= 0.0$ to $0.4$ over the mass range $M_{\rm i}=1.2$ to $1.8~\msun$. As discussed in Section~\ref{section:rotation}, the purpose of this ramping scheme is to compensate for the absence of magnetic braking in MESA which is important for low-mass stars with appreciable convective envelopes. At low masses where the MS lifetimes are long and rotational mixing is inefficient or non-existent, the predicted surface nitrogen abundances actually decrease during the MS due to diffusion. At higher masses without the inclusion of rotation, the predicted nitrogen enrichment during the MS is zero. Additional nitrogen enhancement measurements in stars with $M_{\rm i}<10~\msun$ would provide a valuable constraint on the models.

In the left panel of Figure~\ref{fig:surf_n_abun_evol}, we show 7, 16, and $40~\msun$ solar metallicity evolutionary tracks with (solid) and without (dashed) rotation in the $\log (g) \text{--} \log (\teff)$ plane. In the right panel, we show the evolution of surface nitrogen abundance for the same three models. The symbols correspond to O- and B-type galactic dwarfs and supergiants \citep{Takeda2000, Villamariz2002, Villamariz2005, Crowther2006, Morel2008, Searle2008, Przybilla2010}. To limit the comparison to O- and B-type stars, we exclude the A-, F-, and Cepheid stars from the \cite{Takeda2000} sample. We adopt the symbol and color scheme from Figure~12 of \cite{Ekstrom2012} to aid comparison with their figure. As in \cite{Ekstrom2012}, we do not compare surface rotation velocities because only $v\sin i$ is known for most observed stars. The left panel suggests that the observed stars have initial masses ranging roughly between 7 and $40~\msun$ and that the observed $\log (g)$ values for the dwarfs (open symbols) are in good agreement with the MS location. The right panel demonstrates that models can broadly reproduce the range of observed surface nitrogen abundances. A fair number of observed points fall below the initial N/H ratio in the models, which \cite{Ekstrom2012} suggest is due to abundance variations in the birth cloud.

We also note that for some samples consisting exclusively of slow rotators (e.g., the \cite{Morel2008} sample of B stars whose $v\sin i$ values range from 10 to $60~\rm km\;s^{-1}$), our default rotating models may be inappropriate for a direct comparison. The observed rotational distribution function is quite broad \citep{Huang2010} and also the observed stars were most likely born with a range of initial metallicities, which would increase scatter in the observed nitrogen abundances. Moreover, there are physical mechanisms that have not been taken into account in our models, such as mass loss and gain from binary mass transfer \citep[see e.g.,][]{Eldridge2008, deMink2009, deMink2013}, and typical measurement uncertainties are quite large: 1000~K for $\teff$, $0.1\text{--}0.2$~dex for $\log (g)$, and $0.2$~dex for the N/H ratio. This is meant to be a coarse-grained comparison to assess whether or not the models can {\it reasonably} reproduce the range of observed surface nitrogen abundances. Finally, we note that this is still a controversial topic: there are known slow/fast-rotators with/without surface nitrogen enrichment, and some authors even find that de-projected velocities show no statistically significant correlation with surface nitrogen abundances \citep[e.g,][]{Hunter2008, Brott2011b, RiveroGonzalez2012, Bouret2013, Aerts2014}. We defer a more detailed comparison and discussion of these issues to future work.

\section{Caveats and Future Work}
In this paper we presented an overview of the MIST models, including a comprehensive discussion of the input physics and comparisons with existing databases and observational constraints. We conclude with a discussion of some of the caveats and plans for future work.

Perhaps the most significant shortcoming common to all stellar evolutionary tracks and isochrones of this generation is that they are computed within a 1D framework despite the inherently 3D nature of stellar astrophysical phenomena, e.g., mass-loss, convective mixing, rotation, and magnetic fields. There has been recent progress in 2D and 3D simulations of stellar interiors and atmospheres, but they are limited to small spatial scales and physical time durations of order only $\sim10$ hours \citep[e.g.,][]{Woodward2015}. Although full 3D simulations of stellar evolution are much too far beyond grasp today, we are nevertheless taking small steps forward. For example, it is becoming increasingly common to map 3D simulation results to a 1D formulation in order to incorporate the valuable insight we are gaining from these sophisticated simulations into standard 1D stellar evolution models \cite[see e.g.,][]{Brown2013, Trampedach2014, Arnett2015, Magic2015}. This is an active area of development in MESA.

Another major caveat is that binary interaction is not taken into account in these models. Multiplicity is extremely common among O- and B-type stars; binary mass exchange is believed to occur for $\gtrsim70\%$ of O-type stars and about a third of those stars will ultimately form a binary merger product \citep[e.g.,][]{Chini2012, Sana2012, Sana2013, deMink2014}. These numbers have serious implications for the evolution of massive stars and their explosive final fates, including the expected frequency of different types of core-collapse supernovae \citep[see the review by][]{Smartt2009}. Since binarity dramatically expands the size of the parameter space to be explored (e.g., mass ratio, eccentricity, and separation), it is currently computationally infeasible to construct a multi-dimensional grid of binary models from full stellar evolution calculations. In order to make this a tractable problem, the standard approach has been to either couple detailed stellar evolution codes such as MESA to binary population synthesis codes \citep[e.g.,][]{Eldridge2008} or to make use of fitting formulae to approximate models of single stars \citep[e.g.,][]{Hurley2002, Izzard2006}. Although binarity is not included in the current models, some of its evolutionary consequences must be at least partially mimicked through the effects of rotation and ordinary single-star mass loss, given that the models are in broad agreement with a number of observations. Moreover, it is possible to couple MIST to available binary synthesis codes to investigate and model binary effects in detail in the future.

In Paper II, we will present models with non-solar-scaled abundance patterns for the same large range of masses, ages, metallicities, and evolutionary phases, and make extensive comparisons with the observed properties of globular clusters. Most of our intended future directions are also areas of known limitations. Nevertheless, other future projects we hope to pursue include: investigating the effects of varying $\amlt$ across the HR diagram and as a function of metallicity \citep[e.g.,][]{Trampedach2014, Magic2015}, exploring the effects of magnetic fields in low-mass and PMS stars, improving the WD cooling models for joint fitting of entire CMDs of star clusters, implementing low temperature C/O-variable molecular opacities for modeling the envelopes in AGB stars, performing more precise calibration of the AGB phase by jointly modeling the SFHs of the Magellanic Clouds and their AGB LFs, and developing non-standard and innovative isochrone construction methods.  

\acknowledgments{
We thank the referee for his/her constructive comments. We thank Martha Boyer, Jason Kalirai, and Guillermo Torres for generously sharing their data with us and Francis Timmes for his assistance with nuclear reaction networks in MESA. We thank Leo Girardi and Dan Weisz for fruitful conversations and comments on the manuscript. We thank Conny Aerts, Lars Bildsten, Jonathan Bird, Phillip Cargile, Selma de Mink, Falk Herwig, Jessica Lu, Marc Pinsonneault, Philip Rosenfield, and Victor Silva-Aguirre for sharing their expertise. We also thank the entire MESA community for making this work possible, with special thanks to the MESA Council. Finally, we thank the participants of the KITP workshop ``Galactic Archaeology and Precision Stellar Astrophysics'' for stimulating discussions. This research was supported in part by the National Science Foundation under Grant No. NSF PHY11-25915. JC acknowledges support from the National Science Foundation Graduate Research Fellowship Program. AD received support from the Australian Research Council under grant FL110100012. CC acknowledges support from NASA grant NNX13AI46G, NSF grant AST-1313280, and the Packard Foundation. This work was supported in part by NASA grant NNX15AK14G. The computations in this paper were run on the Odyssey cluster supported by the FAS Division of Science, Research Computing Group at Harvard University.}

\begin{appendix}
\section{Features in the Evolutionary Tracks and Isochrones}
\label{section:appendix_features}
\begin{figure*}
\centering
\includegraphics[width=\textwidth]{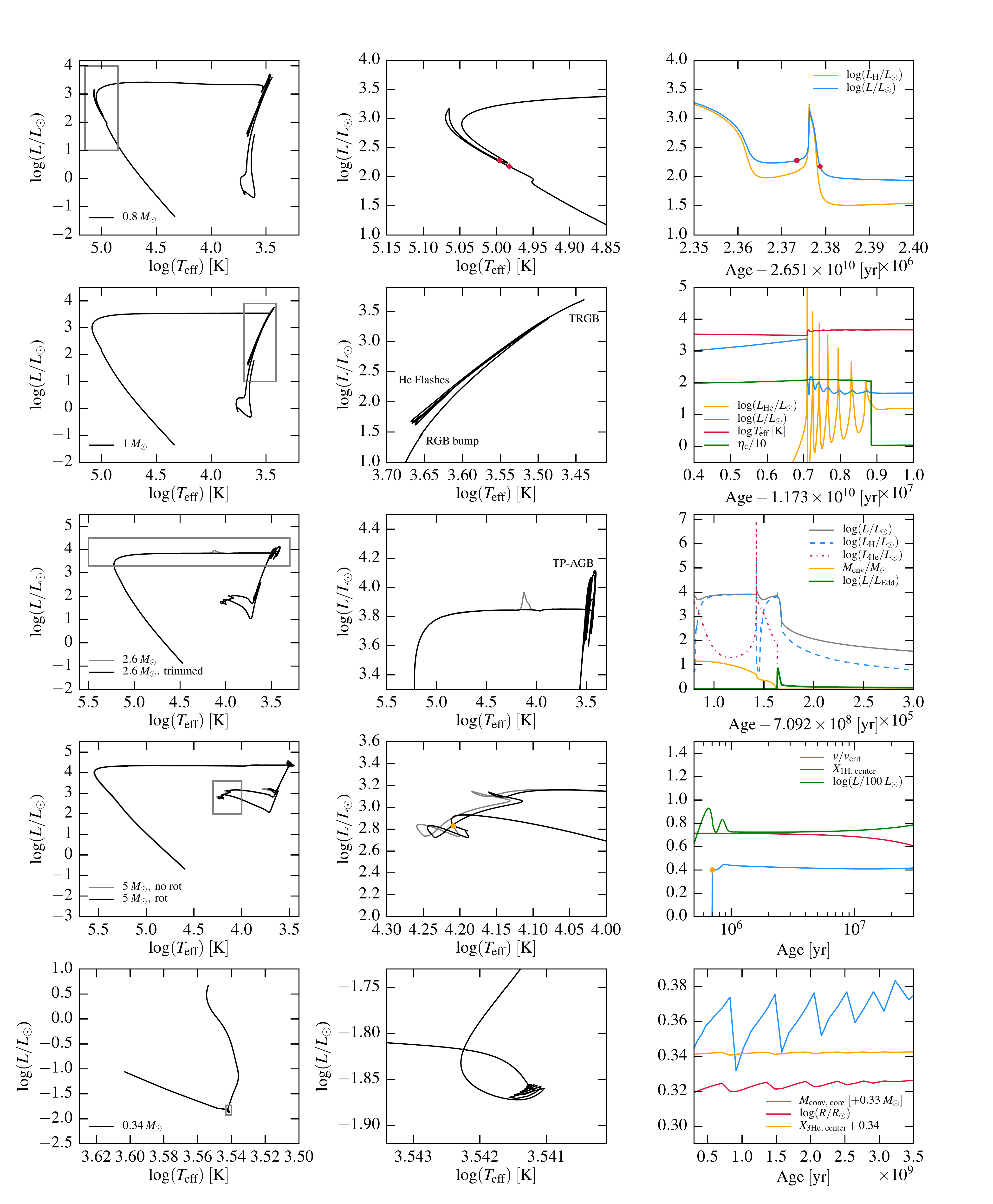} 
\caption{An illustration of interesting and unusual features in the evolutionary tracks. For each row, the left panel shows the evolutionary track in its entirety, and the gray box marks the zoomed-in region shown in the middle panel. The right panel shows the relevant physical quantities as a function of time. First row: ``born-again'' evolution during the post-AGB phase. Second row: helium flashes following helium ignition at the tip of the RGB. Third row: TPAGB phase and post-AGB bump. Fourth row: a shift in $\teff$ due to the initialization of rotation near ZAMS. Fifth row: $^3$He-driven instability near the transition from fully convective to radiative core during the MS.}
\label{fig:features}
\end{figure*}

Here we identify and discuss some of the interesting and unusual features in the evolutionary tracks and isochrones. Figure~\ref{fig:features} shows five such examples. For each row, the left panel shows the evolutionary track in its entirety, and the gray box marks the zoomed-in region shown in the middle panel. The right panel shows a series of relevant physical quantities plotted as a function of time.

The first row displays an example of ``born-again'' evolution during the post-AGB phase \citep[e.g.,][]{Schoenberner1979, Iben1982, Iben1983a}. Following the end of the TPAGB phase, the star enters the post-AGB phase and rapidly evolves toward hotter temperatures. During this short-lived phase ($\sim10^3\textrm{--}10^4$~years), the luminosity remains roughly constant, powered primarily by hydrogen shell-burning with some contribution from gravitational contraction. In the most uneventful scenario (e.g., the $1~\msun$ model in the left panel, second row), the surface hydrogen envelope mass eventually falls below the critical value and hydrogen burning shuts off as a result. Having lost its nuclear energy source, the star then begins to cool and fade away as a WD. In contrast, more interesting things can happen if the star leaves the TPAGB in the middle of its last TP cycle, in which case it may subsequently undergo a late TP either as a post-AGB star (Late Thermal Pulse; LTP) or as a WD (Very Late Thermal Pulse; VLTP) \citep{Kawaler1996}. In either case, the star loops back around toward cooler temperatures as it puffs up from the sudden injection of energy into the envelope. The middle panel zooms in on the VLPT, where the beginning and the end of this episode are marked by the red circle and diamond, respectively. As the right panel shows, the sudden ignition of hydrogen is what triggers the VLTP in the star just as it enters the WD cooling phase. The stellar bolometric luminosity and hydrogen burning luminosity are shown in blue and orange.

The second row illustrates a series of helium flashes in the degenerate core of a low-mass star as it settles into quiescent helium burning. As the star climbs up the RGB, the inert helium core becomes denser and more degenerate as it undergoes gravitational contraction, continuously growing in mass as the hydrogen-burning shell above rains down freshly fused helium. For sufficiently large central densities, neutrino cooling becomes significant and the peak in temperature actually shifts away from the center. As a result, helium ignition occurs off-center at the RGBTip \citep{Thomas1967}.\footnote{\cite{Schwarzschild1962} actually computed the evolution through helium flash prior to this but they did not account for neutrino cooling, so their models ignited helium in the center.} The initial helium flash is followed by a series of weaker helium flashes that move radially inward until helium burning reaches the center. This lifts the degeneracy of the core and the star commences quiescent helium burning. The oscillatory features in the middle panel correspond to these successive helium flashes. The right panel shows the helium burning luminosity, stellar bolometric luminosity, and temperature in orange, blue, and red, respectively. The sharp decline in stellar luminosity in concomitance with a sharp peak in helium burning luminosity marks the location of the RGBTip. The degeneracy parameter $\eta=\epsilon_F/k_BT$ in the center (shown in green, rescaled for display purposes) plummets to $\sim0$ once the helium flash reaches the innermost core.

The third row shows the TPAGB phase followed by a bump in the post-AGB track of a $2.6~\msun{}$ star. The TPAGB phase is famous both for its namesake thermal pulses triggered by alternating hydrogen and helium burning in shells (see Section~\ref{section:agbstars} for a more detailed explanation) and high mass loss rates (see \citealt{Herwig2005} for an excellent review). The TPAGB phase is terminated once the star sheds almost all of its outer hydrogen-rich envelope to reveal the hot CO core, which launches the star leftward in the HR diagram. In the right panel, the alternating hydrogen and helium burning luminosities for the last $\sim$two TPs are plotted in blue dashed and red dotted--dashed lines. The envelope mass shown in orange decreases rapidly until it reaches 0, marking the end of the TPAGB phase. Since the mass of the envelope is dramatically reduced but the stellar luminosity remains more or less the same, the Eddington ratio $L/L_{\rm Edd}$ goes up and the stellar surface becomes unstable. The star glitches over a very short timescale ($<1~\rm yr$) as it attempts to achieve hydrostatic equilibrium, which shows up as a sharp bump at around $\log (\teff) \sim 4.2$~K in the middle panel. We note that this feature only appears sometimes (generally for $>2.5~\msun$ at solar metallicity). Since it is unclear whether or not this behavior is real or a numerical artifact, and it has zero bearing on the evolution of the star due to its extremely short-lived nature, we post-process this feature out of the final evolutionary tracks in order to facilitate the construction of smooth isochrones. Specifically, we trim out any points with $|d(\log L)/dt| < 0.1$ during the post-AGB phase only. The resulting track is shown in black in the left and middle panels.

The fourth row illustrates the effect of turning on rotation at ZAMS. Once the star reaches ZAMS (defined to be $L_{\rm nuc}/L\geq0.9$ in MESA) solid-body rotation is established over 10 time steps. In the absence of efficient rotational mixing, rotation simply makes stars cooler (see Section~\ref{section:tracks_and_isochrones}). The middle panel shows that the rotating (black) and non-rotating (gray) tracks overlap up until ZAMS marked by an orange circle. Once rotation is established, the two curves diverge as the rotating model settles to a lower $\teff$. This jump in temperature is extremely short-lived and is purely a feature of our implementation of rotation; a real star is born from a birth cloud with nonzero initial angular momentum. A more realistic model of the PMS phase that includes the effects of rotation will be explored in the future. In the right panel, we plot $v/v_{\rm crit}$, central hydrogen abundance, and $\log L$ as a function of time. The small increase in velocity immediately following velocity initialization is due to the star experiencing additional contraction before it begins to steadily burn hydrogen. During the MS, $v/v_{\rm crit}$ decreases as the stellar radius and luminosity gradually increase.

The fifth and final row demonstrates the presence of a $^{3}$He-driven instability in a $0.34~\msun$ star. This stellar mass represents the transition from stars that are fully convective (lower mass) on the MS to those with radiative cores and convective envelopes (higher mass). This instability was first reported by \cite{vanSaders2012} who found that these low-mass stars develop small convective cores---normally characteristic of CNO burning in more massive stars ($\gtrsim1.1~\msun$)---due to non-equilibrium $^{3}$He  abundances at these low central temperatures. The net production of $^{3}$He in the center leads to the development of a small convective core, which steadily grows in extent and eventually makes contact with the bottom of the deep convective envelope. As a result of vigorous convective mixing, the local $^{3}$He enhancement in the center is erased as the central $^{3}$He abundance dilutes back to the bulk abundance and the cycle resets. This event occurs as the star leaves the Hayashi track, as can be seen in the left and middle panels. As the right panel shows, a sudden drop in central $^{3}$He abundance (orange) caused by the convective core coming into contact with the envelope coincides with a shrinking of the convective core (blue) and a drop in the stellar radius (red) and luminosity (not shown for display purposes).

\section{The Effects of Varying Time Step and Mesh Controls in MESA Evolution Calculations}
Here we examine the temporal and spatial convergence in our MESA stellar evolution calculations following the methodology in \cite{Paxton2011}. In MESA, the sizes of time steps and grid cells are controlled by {\tt \small varcontrol\_target} and {\tt \small mesh\_delta\_coeff}, respectively, but there is additional flexibility to adjust the resolution specifically for certain zones or evolutionary phases. {\tt \small varcontrol\_target} is the target value for relative changes in the stellar structure variables, e.g., $\rho(m)$, between two consecutive time steps and {\tt \small mesh\_delta\_coeff} is the analogous parameter for differences between adjacent grid cells.

Adopting the notations introduced in \cite{Paxton2011}, we vary temporal and spatial resolution in lockstep according to $C$, a numerical factor by which we multiply the default values for {\tt \small varcontrol\_target} and {\tt \small mesh\_delta\_coeff} simultaneously. A small $C$ demands small time steps and fine grid cells which increase the numerical accuracy but at the expense of longer computation times. It is thus advantageous to search for the largest $C$ we can afford to use without sacrificing the quality of the models. 

Here we consider five values of $C$: 1/4, 1/2, 1, 2, and 4.\footnote{We initially considered additional smaller values of $C$ but many models failed to complete successfully within a reasonable amount of time.} Note that $C=1$ corresponds to the models computed in this paper. For each value of $C$, we compute evolutionary tracks for four masses---1, 5, 20, and $100~\msun$---that are representative of the range of stellar types in the MIST models. Each $M_{\rm i}$ and $C$ combination is computed until the end of hydrogen and helium burning in the core. Convergence of a given model is quantified using the $\xi$ parameter, which is the difference between a variable at a given resolution and at the highest resolution in the study ($C=1/4$ in this case). For $100~\msun$, however, the highest resolution is $C=1/2$ because the $C=1/4$ models did not finish running in a reasonable amount of time. Table~\ref{tab:conv_test} summarizes the results.

For the low-mass stars, whose early evolution is relatively simple, the result at our default resolution $C=1$ differs at most by a few percent compared to the model calculations at four times the resolution. We were also able to successfully run $C=1/8$ and $1/16$ models for 1 and $5~\msun$ models to TAMS. Even compared to model calculations at 16 times the resolution, our default $C=1$ model is converged to within a few percent. As an additional test, we ran $C=2$ and $C=4$ models through to the WDCS to check if the final mass of the WD depends sensitively on the resolution; for example, improperly resolved convective boundaries may influence the growth of the core during the TPAGB phase. We find that decreasing the resolution changes the final mass of the WD by less than a percent, which is highly encouraging.

The behavior is less clear at higher masses (20 and $100~\msun$), but this is not unexpected since massive star evolution is much more complex and sensitive to the choice of input physics. Although the answer can change as much as $\sim12\%$ between our default resolution and $C=1/2$, this is not a huge effect given the uncertainties in the evolution calculations of massive stars. We regard agreement at the $\sim10\%$ level as satisfactory because at this level of precision nearly every other detail of the input physics matters, from MLT implementation to mass loss and rotation.

\begin{table*}
\centering
\begin{longtable}{lccccclccccc}
\caption{Convergence test results.} \\
\hline \hline \noalign{\smallskip}
\multicolumn{1}{c}{\hfill $1~\msun$} \\
$C$: end of H burn & 4 & 2 & 1 & $1/2$ & $1/4$ & $C$: end of He burn & 4 & 2 & 1 & $1/2$ & $1/4$ \\
\hline \noalign{\smallskip}
$N_{\rm cells}$ & 347 & 486 & 821 & 1593 & 3174 & $N_{\rm cells}$ & 586 & 1139 & 2208 & 3931 & 6931\\
$N_{\rm timesteps}$ &335 & 500 & 818 & 1432 & 2726 & $N_{\rm timesteps}$ &9607 & 12185 & 17826 & 32928 & 66044\\ \noalign{\smallskip}
$M/M_{\odot}$ & 1.0 & 1.0 & 1.0 & 1.0 & 1.0 & $M/M_{\odot}$ & 0.95 & 0.951 & 0.951 & 0.951 & 0.951\\
$\xi~(\times10^{-3}$) & 0.006 & 0.009 & 0.004 & 0.001 & 0.0 & $\xi~(\times10^{-3}$) & -0.615 & -0.148 & 0.063 & 0.046 & 0.0\\ \noalign{\smallskip}
$\log (L/L_{\odot})$ & 0.261 & 0.256 & 0.256 & 0.256 & 0.256 & $\log (L/L_{\odot})$ & 2.031 & 2.024 & 2.008 & 2.001 & 2.001\\
$\xi~(\times10^{-3}$) & 10.64 & 1.151 & 0.586 & 1.054 & 0.0 & $\xi~(\times10^{-3}$) & 70.953 & 53.714 & 14.967 & -0.799 & 0.0\\ \noalign{\smallskip}
$\log (T_{\rm eff})~[\rm K]$ & 3.764 & 3.766 & 3.767 & 3.767 & 3.767 & $\log (T_{\rm eff}~[\rm K])$  & 3.63 & 3.634 & 3.636 & 3.637 & 3.637\\
$\xi~(\times10^{-3}$) & -5.652 & -1.65 & -0.493 & -0.159 & 0.0 & $\xi~(\times10^{-3}$) & -14.812 & -6.501 & -1.779 & -0.06 & 0.0\\ \noalign{\smallskip}
$\log(t)$ & 9.94 & 9.939 & 9.941 & 9.942 & 9.943 & $\log(t)$ & 10.056 & 10.057 & 10.059 & 10.06 & 10.06\\
$\xi~(\times10^{-3}$) & -5.617 & -8.577 & -3.747 & -0.836 & 0.0 & $\xi~(\times10^{-3}$) & -10.478 & -7.686 & -3.576 & -1.376 & 0.0\\ \noalign{\smallskip}
$\log (T_{\rm c})~[\rm K]$ & 7.295 & 7.294 & 7.294 & 7.294 & 7.294 & $\log (T_{\rm c})~[\rm K]$ & 8.282 & 8.28 & 8.275 & 8.273 & 8.273\\
$\xi~(\times10^{-3}$) & 1.165 & -0.103 & 0.037 & 0.202 & 0.0 & $\xi~(\times10^{-3}$) & 20.975 & 16.362 & 4.178 & 0.649 & 0.0\\ \noalign{\smallskip}
$\log (\rho_{\rm c})~[\rm g\;cm^{-3}]$ & 2.805 & 2.788 & 2.786 & 2.786 & 2.784 & $\log (\rho_{\rm c})~[\rm g\;cm^{-3}]$ & 4.754 & 4.742 & 4.733 & 4.727 & 4.726\\
$\xi~(\times10^{-3}$) & 48.7 & 10.331 & 4.975 & 4.094 & 0.0 & $\xi~(\times10^{-3}$) & 66.899 & 37.961 & 15.172 & 2.266 & 0.0\\ \noalign{\smallskip}
\hline \noalign{\smallskip}
\multicolumn{1}{c}{\hfill $5~\msun$} \\
$C$: end of H burn & 4 & 2 & 1 & $1/2$ & $1/4$ & $C$: end of He burn & 4 & 2 & 1 & $1/2$ & $1/4$ \\
\hline \noalign{\smallskip}
$N_{\rm cells}$ & 345 & 547 & 1015 & 2003 & 3992 & $N_{\rm cells}$ & 603 & 1248 & 2148 & 4582 & 7283\\
$N_{\rm timesteps}$ &383 & 584 & 1019 & 1919 & 3699 & $N_{\rm timesteps}$ &644 & 1117 & 2123 & 4262 & 8454\\ \noalign{\smallskip}
$M/M_{\odot}$ & 4.997 & 4.997 & 4.997 & 4.997 & 4.997 & $M/M_{\odot}$ & 4.984 & 4.985 & 4.985 & 4.985 & 4.986\\
$\xi~(\times10^{-3}$) & -0.057 & -0.051 & -0.02 & -0.019 & 0.0 & $\xi~(\times10^{-3}$) & -0.355 & -0.124 & -0.033 & -0.07 & 0.0\\ \noalign{\smallskip}
$\log (L/L_{\odot})$ & 3.168 & 3.154 & 3.139 & 3.137 & 3.136 & $\log (L/L_{\odot})$ & 3.294 & 3.23 & 3.173 & 3.175 & 3.17\\
$\xi~(\times10^{-3}$) & 74.552 & 40.912 & 5.28 & 0.766 & 0.0 & $\xi~(\times10^{-3}$) & 330.852 & 147.317 & 7.776 & 12.006 & 0.0\\ \noalign{\smallskip}
$\log (T_{\rm eff})~[\rm K]$ & 4.159 & 4.16 & 4.16 & 4.16 & 4.164 & $\log (T_{\rm eff}~[\rm K])$  & 3.614 & 3.624 & 3.63 & 3.63 & 3.631\\
$\xi~(\times10^{-3}$) & -12.433 & -10.03 & -7.975 & -8.023 & 0.0 & $\xi~(\times10^{-3}$) & -37.609 & -15.963 & -1.504 & -1.559 & 0.0\\ \noalign{\smallskip}
$\log(t)$ & 8.025 & 8.021 & 8.01 & 8.009 & 8.008 & $\log(t)$ & 8.077 & 8.076 & 8.071 & 8.072 & 8.071\\
$\xi~(\times10^{-3}$) & 40.171 & 29.525 & 5.393 & 2.926 & 0.0 & $\xi~(\times10^{-3}$) & 13.82 & 12.067 & -0.536 & 2.725 & 0.0\\ \noalign{\smallskip}
$\log (T_{\rm c})~[\rm K]$ & 7.663 & 7.658 & 7.657 & 7.657 & 7.657 & $\log (T_{\rm c})~[\rm K]$ & 8.379 & 8.371 & 8.368 & 8.371 & 8.369\\
$\xi~(\times10^{-3}$) & 13.611 & 2.911 & 0.973 & 0.874 & 0.0 & $\xi~(\times10^{-3}$) & 21.884 & 3.362 & -2.952 & 4.452 & 0.0\\ \noalign{\smallskip}
$\log (\rho_{\rm c})~[\rm g\;cm^{-3}]$ & 2.001 & 2.001 & 2.006 & 2.007 & 2.007 & $\log (\rho_{\rm c})~[\rm g\;cm^{-3}]$ & 4.243 & 4.245 & 4.246 & 4.247 & 4.243\\
$\xi~(\times10^{-3}$) & -13.275 & -14.014 & -0.887 & 0.314 & 0.0 & $\xi~(\times10^{-3}$) & 2.19 & 6.022 & 7.495 & 9.271 & 0.0\\ \noalign{\smallskip}
\hline \noalign{\smallskip}
\multicolumn{1}{c}{\hfill $20~\msun$} \\
$C$: end of H burn & 4 & 2 & 1 & $1/2$ & $1/4$ & $C$: end of He burn & 4 & 2 & 1 & $1/2$ & $1/4$ \\
\hline \noalign{\smallskip}
$N_{\rm cells}$ & 1467 & 2683 & 5024 & 9912 & 17655 & $N_{\rm cells}$ & 2325 & 4598 & 9133 & 18303 & 36666\\
$N_{\rm timesteps}$ &1803 & 2493 & 4460 & 8438 & 16409 & $N_{\rm timesteps}$ &2951 & 3635 & 6218 & 11591 & 22501\\ \noalign{\smallskip}
$M/M_{\odot}$ & 19.279 & 19.202 & 19.184 & 19.188 & 19.212 & $M/M_{\odot}$ & 12.156 & 12.783 & 13.115 & 14.655 & 14.856\\
$\xi~(\times10^{-3}$) & 3.456 & -0.532 & -1.475 & -1.287 & 0.0 & $\xi~(\times10^{-3}$) & -181.746 & -139.532 & -117.17 & -13.479 & 0.0\\ \noalign{\smallskip}
$\log (L/L_{\odot})$ & 5.083 & 5.03 & 5.02 & 5.017 & 5.015 & $\log (L/L_{\odot})$ & 5.174 & 5.131 & 5.121 & 5.101 & 5.099\\
$\xi~(\times10^{-3}$) & 169.241 & 35.182 & 11.427 & 5.929 & 0.0 & $\xi~(\times10^{-3}$) & 190.493 & 77.659 & 53.169 & 4.741 & 0.0\\ \noalign{\smallskip}
$\log (T_{\rm eff})~[\rm K]$ & 4.468 & 4.442 & 4.437 & 4.434 & 4.435 & $\log (T_{\rm eff}~[\rm K])$  & 3.568 & 3.535 & 3.532 & 3.537 & 3.537\\
$\xi~(\times10^{-3}$) & 78.496 & 17.202 & 4.979 & -1.602 & 0.0 & $\xi~(\times10^{-3}$) & 74.031 & -4.799 & -10.998 & -0.389 & 0.0\\ \noalign{\smallskip}
$\log(t)$ & 6.998 & 6.961 & 6.953 & 6.949 & 6.947 & $\log(t)$ & 7.034 & 7.002 & 6.995 & 6.993 & 6.991\\
$\xi~(\times10^{-3}$) & 123.993 & 31.461 & 12.596 & 5.123 & 0.0 & $\xi~(\times10^{-3}$) & 103.348 & 25.481 & 9.44 & 4.182 & 0.0\\ \noalign{\smallskip}
$\log (T_{\rm c})~[\rm K]$ & 7.819 & 7.815 & 7.815 & 7.815 & 7.814 & $\log (T_{\rm c})~[\rm K]$ & 8.511 & 8.505 & 8.504 & 8.503 & 8.503\\
$\xi~(\times10^{-3}$) & 10.992 & 2.764 & 1.188 & 0.834 & 0.0 & $\xi~(\times10^{-3}$) & 17.555 & 5.504 & 2.622 & 0.795 & 0.0\\ \noalign{\smallskip}
$\log (\rho_{\rm c})~[\rm g\;cm^{-3}]$ & 1.375 & 1.393 & 1.396 & 1.397 & 1.397 & $\log (\rho_{\rm c})~[\rm g\;cm^{-3}]$ & 3.525 & 3.545 & 3.548 & 3.55 & 3.551\\
$\xi~(\times10^{-3}$) & -48.836 & -10.201 & -2.678 & -0.329 & 0.0 & $\xi~(\times10^{-3}$) & -58.531 & -13.623 & -6.043 & -1.536 & 0.0\\ \noalign{\smallskip}
\hline \noalign{\smallskip}
\multicolumn{1}{c}{\hfill $100~\msun$} \\
$C$: end of H burn & 4 & 2 & 1 & $1/2$ & $1/4$ & $C$: end of He burn & 4 & 2 & 1 & $1/2$ & $1/4$ \\
\hline \noalign{\smallskip}
$N_{\rm cells}$ & 1468 & 2727 & 5448 & 10818 & - & $N_{\rm cells}$ & 1569 & 3152 & 6486 & 12465 & - \\
$N_{\rm timesteps}$ &275 & 323 & 323 & 328 & - & $N_{\rm timesteps}$ &696 & 572 & 559 & 584 & -\\ \noalign{\smallskip}
$M/M_{\odot}$ & 65.916 & 25.712 & 24.583 & 23.752 & - &  $M/M_{\odot}$ & 33.813 & 11.914 & 11.664 & 11.454 & -\\
$\xi~(\times10^{-3}$) & 1775.107 & 82.495 & 34.981 & 0.0 & - & $\xi~(\times10^{-3}$) & 1952.055 & 40.126 & 18.353 & 0.0 & -\\ \noalign{\smallskip}
$\log (L/L_{\odot})$ & 6.274 & 5.805 & 5.774 & 5.749 & - & $\log (L/L_{\odot})$ & 6.162 & 5.504 & 5.489 & 5.475 & -\\
$\xi~(\times10^{-3}$) & 2352.624 & 139.574 & 58.767 & 0.0 & - & $\xi~(\times10^{-3}$) & 3867.197 & 67.71 & 31.926 & 0.0 & -\\ \noalign{\smallskip}
$\log (T_{\rm eff})~[\rm K]$ & 4.603 & 4.882 & 4.877 & 4.874 & - & $\log (T_{\rm eff}~[\rm K])$  & 5.228 & 5.2 & 5.199 & 5.198 & -\\
$\xi~(\times10^{-3}$) & -463.387 & 18.996 & 8.502 & 0.0 & - & $\xi~(\times10^{-3}$) & 71.311 & 4.274 & 3.703 & 0.0 & -\\ \noalign{\smallskip}
$\log(t)$ & 6.476 & 6.546 & 6.553 & 6.56 &  - & $\log(t)$ & 6.52 & 6.595 & 6.601 & 6.608 & -\\
$\xi~(\times10^{-3}$) & -174.392 & -29.716 & -15.581 & 0.0 &  - & $\xi~(\times10^{-3}$) & -183.629 & -28.855 & -14.912 & 0.0 & -\\ \noalign{\smallskip}
$\log (T_{\rm c})~[\rm K]$ & 7.894 & 7.87 & 7.868 & 7.867 &  - & $\log (T_{\rm c})~[\rm K]$ & 8.56 & 8.529 & 8.53 & 8.528 & -\\
$\xi~(\times10^{-3}$) & 65.67 & 8.314 & 3.609 & 0.0 &  - & $\xi~(\times10^{-3}$) & 78.495 & 3.91 & 4.801 & 0.0 & -\\ \noalign{\smallskip}
$\log (\rho_{\rm c})~[\rm g\;cm^{-3}]$ & 1.03 & 1.208 & 1.216 & 1.223 & - & $\log (\rho_{\rm c})~[\rm g\;cm^{-3}]$ & 3.151 & 3.383 & 3.392 & 3.392 & -\\
$\xi~(\times10^{-3}$) & -359.052 & -35.601 & -15.776 & 0.0 & - &  $\xi~(\times10^{-3}$) & -426.172 & -19.84 & -0.399 & 0.0 & -\\ \noalign{\smallskip}
\hline
\hline
\label{tab:conv_test}
\end{longtable}
\end{table*}
\end{appendix}

\newpage
\bibliographystyle{apj}
\bibliography{bibtex.bib}

\begin{thebibliography}{}
\expandafter\ifx\csname natexlab\endcsname\relax\def\natexlab#1{#1}\fi

\bibitem[{{Aerts} {et~al.}(2014){Aerts}, {Molenberghs}, {Kenward}, \&
  {Neiner}}]{Aerts2014}
{Aerts}, C., {Molenberghs}, G., {Kenward}, M.~G., \& {Neiner}, C. 2014, \apj,
  781, 88

\bibitem[{{Alastuey} \& {Jancovici}(1978)}]{Alastuey1978}
{Alastuey}, A., \& {Jancovici}, B. 1978, \apj, 226, 1034

\bibitem[{{Alecian} {et~al.}(1993){Alecian}, {Michaud}, \&
  {Tully}}]{Alecian1993}
{Alecian}, G., {Michaud}, G., \& {Tully}, J. 1993, \apj, 411, 882

\bibitem[{{An} {et~al.}(2009){An}, {Pinsonneault}, {Masseron}, {Delahaye},
  {Johnson}, {Terndrup}, {Beers}, {Ivans}, \& {Ivezi{\'c}}}]{An2009}
{An}, D., {Pinsonneault}, M.~H., {Masseron}, T., {et~al.} 2009, \apj, 700, 523

\bibitem[{{Anders} \& {Grevesse}(1989)}]{Anders1989}
{Anders}, E., \& {Grevesse}, N. 1989, \gca, 53, 197

\bibitem[{{Angulo} {et~al.}(1999){Angulo}, {Arnould}, {Rayet}, {Descouvemont},
  {Baye}, {Leclercq-Willain}, {Coc}, {Barhoumi}, {Aguer}, {Rolfs}, {Kunz},
  {Hammer}, {Mayer}, {Paradellis}, {Kossionides}, {Chronidou}, {Spyrou},
  {degl'Innocenti}, {Fiorentini}, {Ricci}, {Zavatarelli}, {Providencia},
  {Wolters}, {Soares}, {Grama}, {Rahighi}, {Shotter}, \& {Lamehi
  Rachti}}]{Angulo1999}
{Angulo}, C., {Arnould}, M., {Rayet}, M., {et~al.} 1999, Nuclear Physics A,
  656, 3

\bibitem[{{Antia} \& {Basu}(2005)}]{Antia2005}
{Antia}, H.~M., \& {Basu}, S. 2005, \apjl, 620, L129

\bibitem[{{Aringer} {et~al.}(2009){Aringer}, {Girardi}, {Nowotny}, {Marigo}, \&
  {Lederer}}]{Aringer2009}
{Aringer}, B., {Girardi}, L., {Nowotny}, W., {Marigo}, P., \& {Lederer}, M.~T.
  2009, \aap, 503, 913

\bibitem[{{Arnett} {et~al.}(2015){Arnett}, {Meakin}, {Viallet}, {Campbell},
  {Lattanzio}, \& {Moc{\'a}k}}]{Arnett2015}
{Arnett}, W.~D., {Meakin}, C., {Viallet}, M., {et~al.} 2015, \apj, 809, 30

\bibitem[{{Asplund} {et~al.}(2005){Asplund}, {Grevesse}, \&
  {Sauval}}]{Asplund2005}
{Asplund}, M., {Grevesse}, N., \& {Sauval}, A.~J. 2005, in Astronomical Society
  of the Pacific Conference Series, Vol. 336, Cosmic Abundances as Records of
  Stellar Evolution and Nucleosynthesis, ed. T.~G. {Barnes}, III \& F.~N.
  {Bash}, 25

\bibitem[{{Asplund} {et~al.}(2004){Asplund}, {Grevesse}, {Sauval}, {Allende
  Prieto}, \& {Kiselman}}]{Asplund2004}
{Asplund}, M., {Grevesse}, N., {Sauval}, A.~J., {Allende Prieto}, C., \&
  {Kiselman}, D. 2004, \aap, 417, 751

\bibitem[{{Asplund} {et~al.}(2009){Asplund}, {Grevesse}, {Sauval}, \&
  {Scott}}]{Asplund2009}
{Asplund}, M., {Grevesse}, N., {Sauval}, A.~J., \& {Scott}, P. 2009, \araa, 47,
  481

\bibitem[{{Baglin} {et~al.}(2006){Baglin}, {Auvergne}, {Barge}, {Deleuil},
  {Catala}, {Michel}, {Weiss}, \& {COROT Team}}]{Baglin2006}
{Baglin}, A., {Auvergne}, M., {Barge}, P., {et~al.} 2006, in ESA Special
  Publication, Vol. 1306, ESA Special Publication, ed. M.~{Fridlund},
  A.~{Baglin}, J.~{Lochard}, \& L.~{Conroy}, 33

\bibitem[{{Bahcall} {et~al.}(2005{\natexlab{a}}){Bahcall}, {Basu},
  {Pinsonneault}, \& {Serenelli}}]{Bahcall2005a}
{Bahcall}, J.~N., {Basu}, S., {Pinsonneault}, M., \& {Serenelli}, A.~M.
  2005{\natexlab{a}}, \apj, 618, 1049

\bibitem[{{Bahcall} {et~al.}(2005{\natexlab{b}}){Bahcall}, {Basu}, \&
  {Serenelli}}]{Bahcall2005b}
{Bahcall}, J.~N., {Basu}, S., \& {Serenelli}, A.~M. 2005{\natexlab{b}}, \apj,
  631, 1281

\bibitem[{{Bahcall} {et~al.}(2006){Bahcall}, {Serenelli}, \&
  {Basu}}]{Bahcall2006}
{Bahcall}, J.~N., {Serenelli}, A.~M., \& {Basu}, S. 2006, \apjs, 165, 400

\bibitem[{Bailey {et~al.}(2015)Bailey, Nagayama, Loisel, Rochau, Blancard,
  Colgan, Cosse, Faussurier, Fontes, Gilleron, \& et~al.}]{Bailey2015}
Bailey, J.~E., Nagayama, T., Loisel, G.~P., {et~al.} 2015, Nature, 517, 56Ð59

\bibitem[{{Baraffe} {et~al.}(1995){Baraffe}, {Chabrier}, {Allard}, \&
  {Hauschildt}}]{Baraffe1995}
{Baraffe}, I., {Chabrier}, G., {Allard}, F., \& {Hauschildt}, P.~H. 1995,
  \apjl, 446, L35

\bibitem[{{Baraffe} {et~al.}(1997){Baraffe}, {Chabrier}, {Allard}, \&
  {Hauschildt}}]{Baraffe1997}
---. 1997, \aap, 327, 1054

\bibitem[{{Baraffe} {et~al.}(1998){Baraffe}, {Chabrier}, {Allard}, \&
  {Hauschildt}}]{Baraffe1998}
---. 1998, \aap, 337, 403

\bibitem[{{Baraffe} {et~al.}(2003){Baraffe}, {Chabrier}, {Barman}, {Allard}, \&
  {Hauschildt}}]{Baraffe2003}
{Baraffe}, I., {Chabrier}, G., {Barman}, T.~S., {Allard}, F., \& {Hauschildt},
  P.~H. 2003, \aap, 402, 701

\bibitem[{{Baraffe} {et~al.}(2009){Baraffe}, {Chabrier}, \&
  {Gallardo}}]{Baraffe2009}
{Baraffe}, I., {Chabrier}, G., \& {Gallardo}, J. 2009, \apjl, 702, L27

\bibitem[{{Baraffe} {et~al.}(2015){Baraffe}, {Homeier}, {Allard}, \&
  {Chabrier}}]{Baraffe2015}
{Baraffe}, I., {Homeier}, D., {Allard}, F., \& {Chabrier}, G. 2015, \aap, 577,
  A42

\bibitem[{{Basu} \& {Antia}(2004)}]{Basu2004}
{Basu}, S., \& {Antia}, H.~M. 2004, \apjl, 606, L85

\bibitem[{{Bedin} {et~al.}(2005){Bedin}, {Salaris}, {Piotto}, {King},
  {Anderson}, {Cassisi}, \& {Momany}}]{Bedin2005}
{Bedin}, L.~R., {Salaris}, M., {Piotto}, G., {et~al.} 2005, \apjl, 624, L45

\bibitem[{{Bessell} {et~al.}(2011){Bessell}, {Bloxham}, {Schmidt}, {Keller},
  {Tisserand}, \& {Francis}}]{Bessell2011B}
{Bessell}, M., {Bloxham}, G., {Schmidt}, B., {et~al.} 2011, \pasp, 123, 789

\bibitem[{{Bessell} \& {Murphy}(2012)}]{Bessell2012}
{Bessell}, M., \& {Murphy}, S. 2012, \pasp, 124, 140

\bibitem[{{Bessell}(2001)}]{Bessell2001}
{Bessell}, M.~S. 2001, \pasp, 113, 66

\bibitem[{{Bessell}(2011)}]{Bessell2011}
---. 2011, \pasp, 123, 1442

\bibitem[{{Bird} \& {Pinsonneault}(2011)}]{Bird2011}
{Bird}, J.~C., \& {Pinsonneault}, M.~H. 2011, \apj, 733, 81

\bibitem[{{Bjorkman} \& {Cassinelli}(1993)}]{Bjorkman1993}
{Bjorkman}, J.~E., \& {Cassinelli}, J.~P. 1993, \apj, 409, 429

\bibitem[{{Bl{\"o}cker}(1995)}]{Bloecker1995}
{Bl{\"o}cker}, T. 1995, \aap, 297, 727

\bibitem[{{B{\"o}hm}(1963)}]{Bohm1963}
{B{\"o}hm}, K.-H. 1963, \apj, 138, 297

\bibitem[{{B{\"o}hm-Vitense}(1958)}]{BohmVitense1958}
{B{\"o}hm-Vitense}, E. 1958, \zap, 46, 108

\bibitem[{{Boudreault} {et~al.}(2012){Boudreault}, {Lodieu}, {Deacon}, \&
  {Hambly}}]{Boudreault2012}
{Boudreault}, S., {Lodieu}, N., {Deacon}, N.~R., \& {Hambly}, N.~C. 2012,
  \mnras, 426, 3419

\bibitem[{{Bouret} {et~al.}(2013){Bouret}, {Lanz}, {Martins}, {Marcolino},
  {Hillier}, {Depagne}, \& {Hubeny}}]{Bouret2013}
{Bouret}, J.-C., {Lanz}, T., {Martins}, F., {et~al.} 2013, \aap, 555, A1

\bibitem[{{Bouvier}(2008)}]{Bouvier2008}
{Bouvier}, J. 2008, \aap, 489, L53

\bibitem[{{Boyajian} {et~al.}(2012){Boyajian}, {von Braun}, {van Belle},
  {McAlister}, {ten Brummelaar}, {Kane}, {Muirhead}, {Jones}, {White},
  {Schaefer}, {Ciardi}, {Henry}, {L{\'o}pez-Morales}, {Ridgway}, {Gies}, {Jao},
  {Rojas-Ayala}, {Parks}, {Sturmann}, {Sturmann}, {Turner}, {Farrington},
  {Goldfinger}, \& {Berger}}]{Boyajian2012}
{Boyajian}, T.~S., {von Braun}, K., {van Belle}, G., {et~al.} 2012, \apj, 757,
  112

\bibitem[{{Boyer} {et~al.}(2011){Boyer}, {Srinivasan}, {van Loon}, {McDonald},
  {Meixner}, {Zaritsky}, {Gordon}, {Kemper}, {Babler}, {Block}, {Bracker},
  {Engelbracht}, {Hora}, {Indebetouw}, {Meade}, {Misselt}, {Robitaille},
  {Sewi{\l}o}, {Shiao}, \& {Whitney}}]{Boyer2011}
{Boyer}, M.~L., {Srinivasan}, S., {van Loon}, J.~T., {et~al.} 2011, \aj, 142,
  103

\bibitem[{{Braithwaite}(2006)}]{Braithwaite2006}
{Braithwaite}, J. 2006, \aap, 449, 451

\bibitem[{{Braithwaite} \& {Spruit}(2004)}]{Braithwaite2004}
{Braithwaite}, J., \& {Spruit}, H.~C. 2004, \nat, 431, 819

\bibitem[{{Brandenburg} \& {Subramanian}(2005)}]{Brandenburg2005}
{Brandenburg}, A., \& {Subramanian}, K. 2005, \physrep, 417, 1

\bibitem[{{Bressan} {et~al.}(2012){Bressan}, {Marigo}, {Girardi}, {Salasnich},
  {Dal Cero}, {Rubele}, \& {Nanni}}]{Bressan2012}
{Bressan}, A., {Marigo}, P., {Girardi}, L., {et~al.} 2012, \mnras, 427, 127

\bibitem[{{Bressan} {et~al.}(1981){Bressan}, {Chiosi}, \&
  {Bertelli}}]{Bressan1981}
{Bressan}, A.~G., {Chiosi}, C., \& {Bertelli}, G. 1981, \aap, 102, 25

\bibitem[{{Brogaard} {et~al.}(2012){Brogaard}, {VandenBerg}, {Bruntt},
  {Grundahl}, {Frandsen}, {Bedin}, {Milone}, {Dotter}, {Feiden}, {Stetson},
  {Sandquist}, {Miglio}, {Stello}, \& {Jessen-Hansen}}]{Brogaard2012}
{Brogaard}, K., {VandenBerg}, D.~A., {Bruntt}, H., {et~al.} 2012, \aap, 543,
  A106

\bibitem[{{Brott} {et~al.}(2011{\natexlab{a}}){Brott}, {de Mink}, {Cantiello},
  {Langer}, {de Koter}, {Evans}, {Hunter}, {Trundle}, \& {Vink}}]{Brott2011}
{Brott}, I., {de Mink}, S.~E., {Cantiello}, M., {et~al.} 2011{\natexlab{a}},
  \aap, 530, A115

\bibitem[{{Brott} {et~al.}(2011{\natexlab{b}}){Brott}, {Evans}, {Hunter}, {de
  Koter}, {Langer}, {Dufton}, {Cantiello}, {Trundle}, {Lennon}, {de Mink},
  {Yoon}, \& {Anders}}]{Brott2011b}
{Brott}, I., {Evans}, C.~J., {Hunter}, I., {et~al.} 2011{\natexlab{b}}, \aap,
  530, A116

\bibitem[{{Brown} {et~al.}(1997){Brown}, {Wallerstein}, \&
  {Zucker}}]{Brown1997}
{Brown}, J.~A., {Wallerstein}, G., \& {Zucker}, D. 1997, \aj, 114, 180

\bibitem[{{Brown} {et~al.}(2013){Brown}, {Garaud}, \& {Stellmach}}]{Brown2013}
{Brown}, J.~M., {Garaud}, P., \& {Stellmach}, S. 2013, \apj, 768, 34

\bibitem[{{Brown} {et~al.}(1989){Brown}, {Christensen-Dalsgaard},
  {Dziembowski}, {Goode}, {Gough}, \& {Morrow}}]{Brown1989}
{Brown}, T.~M., {Christensen-Dalsgaard}, J., {Dziembowski}, W.~A., {et~al.}
  1989, \apj, 343, 526

\bibitem[{{Brown} {et~al.}(1991){Brown}, {Gilliland}, {Noyes}, \&
  {Ramsey}}]{Brown1991}
{Brown}, T.~M., {Gilliland}, R.~L., {Noyes}, R.~W., \& {Ramsey}, L.~W. 1991,
  \apj, 368, 599

\bibitem[{{Burbidge} {et~al.}(1957){Burbidge}, {Burbidge}, {Fowler}, \&
  {Hoyle}}]{Burbidge1957}
{Burbidge}, E.~M., {Burbidge}, G.~R., {Fowler}, W.~A., \& {Hoyle}, F. 1957,
  Reviews of Modern Physics, 29, 547

\bibitem[{{Caffau} {et~al.}(2011){Caffau}, {Ludwig}, {Steffen}, {Freytag}, \&
  {Bonifacio}}]{Caffau2011}
{Caffau}, E., {Ludwig}, H.-G., {Steffen}, M., {Freytag}, B., \& {Bonifacio}, P.
  2011, \solphys, 268, 255

\bibitem[{{Caloi} \& {D'Antona}(2011)}]{Caloi2011}
{Caloi}, V., \& {D'Antona}, F. 2011, \mnras, 417, 228

\bibitem[{{Cantiello} \& {Langer}(2010)}]{Cantiello2010}
{Cantiello}, M., \& {Langer}, N. 2010, \aap, 521, A9

\bibitem[{{Cantiello} {et~al.}(2014){Cantiello}, {Mankovich}, {Bildsten},
  {Christensen-Dalsgaard}, \& {Paxton}}]{Cantiello2014}
{Cantiello}, M., {Mankovich}, C., {Bildsten}, L., {Christensen-Dalsgaard}, J.,
  \& {Paxton}, B. 2014, \apj, 788, 93

\bibitem[{{Canto Martins} {et~al.}(2011){Canto Martins}, {L{\`e}bre},
  {Palacios}, {de Laverny}, {Richard}, {Melo}, {Do Nascimento}, \& {de
  Medeiros}}]{CantoMartins2011}
{Canto Martins}, B.~L., {L{\`e}bre}, A., {Palacios}, A., {et~al.} 2011, \aap,
  527, A94

\bibitem[{{Cardelli} {et~al.}(1989){Cardelli}, {Clayton}, \&
  {Mathis}}]{Cardelli1989}
{Cardelli}, J.~A., {Clayton}, G.~C., \& {Mathis}, J.~S. 1989, \apj, 345, 245

\bibitem[{{Carney} {et~al.}(2005){Carney}, {Lee}, \& {Dodson}}]{Carney2005}
{Carney}, B.~W., {Lee}, J.-W., \& {Dodson}, B. 2005, \aj, 129, 656

\bibitem[{{Carrera} \& {Pancino}(2011)}]{Carrera2011}
{Carrera}, R., \& {Pancino}, E. 2011, \aap, 535, A30

\bibitem[{{Carretta} {et~al.}(2009){Carretta}, {Bragaglia}, {Gratton},
  {Lucatello}, {Catanzaro}, {Leone}, {Bellazzini}, {Claudi}, {D'Orazi},
  {Momany}, {Ortolani}, {Pancino}, {Piotto}, {Recio-Blanco}, \&
  {Sabbi}}]{Carretta2009}
{Carretta}, E., {Bragaglia}, A., {Gratton}, R.~G., {et~al.} 2009, \aap, 505,
  117

\bibitem[{{Casagrande} {et~al.}(2008){Casagrande}, {Flynn}, \&
  {Bessell}}]{Casagrande2008}
{Casagrande}, L., {Flynn}, C., \& {Bessell}, M. 2008, \mnras, 389, 585

\bibitem[{{Cassisi} {et~al.}(2007){Cassisi}, {Potekhin}, {Pietrinferni},
  {Catelan}, \& {Salaris}}]{Cassisi2007}
{Cassisi}, S., {Potekhin}, A.~Y., {Pietrinferni}, A., {Catelan}, M., \&
  {Salaris}, M. 2007, \apj, 661, 1094

\bibitem[{{Cassisi} \& {Salaris}(2013)}]{Cassisi2013}
{Cassisi}, S., \& {Salaris}, M. 2013, {Old Stellar Populations: How to Study
  the Fossil Record of Galaxy Formation}

\bibitem[{{Castelli} {et~al.}(1997){Castelli}, {Gratton}, \&
  {Kurucz}}]{Castelli1997}
{Castelli}, F., {Gratton}, R.~G., \& {Kurucz}, R.~L. 1997, \aap, 318, 841

\bibitem[{{Castor} {et~al.}(1975){Castor}, {Abbott}, \& {Klein}}]{Castor1975}
{Castor}, J.~I., {Abbott}, D.~C., \& {Klein}, R.~I. 1975, \apj, 195, 157

\bibitem[{{Castro} {et~al.}(2014){Castro}, {Fossati}, {Langer},
  {Sim{\'o}n-D{\'{\i}}az}, {Schneider}, \& {Izzard}}]{Castro2014}
{Castro}, N., {Fossati}, L., {Langer}, N., {et~al.} 2014, \aap, 570, L13

\bibitem[{{Catal{\'a}n} {et~al.}(2008{\natexlab{a}}){Catal{\'a}n}, {Isern},
  {Garc{\'{\i}}a-Berro}, \& {Ribas}}]{Catalan2008b}
{Catal{\'a}n}, S., {Isern}, J., {Garc{\'{\i}}a-Berro}, E., \& {Ribas}, I.
  2008{\natexlab{a}}, \mnras, 387, 1693

\bibitem[{{Catal{\'a}n} {et~al.}(2008{\natexlab{b}}){Catal{\'a}n}, {Isern},
  {Garc{\'{\i}}a-Berro}, {Ribas}, {Allende Prieto}, \&
  {Bonanos}}]{Catalan2008a}
{Catal{\'a}n}, S., {Isern}, J., {Garc{\'{\i}}a-Berro}, E., {et~al.}
  2008{\natexlab{b}}, \aap, 477, 213

\bibitem[{{Chaboyer} {et~al.}(2001){Chaboyer}, {Fenton}, {Nelan}, {Patnaude},
  \& {Simon}}]{Chaboyer2001}
{Chaboyer}, B., {Fenton}, W.~H., {Nelan}, J.~E., {Patnaude}, D.~J., \& {Simon},
  F.~E. 2001, \apj, 562, 521

\bibitem[{{Chaboyer} \& {Zahn}(1992)}]{Chaboyer1992}
{Chaboyer}, B., \& {Zahn}, J.-P. 1992, \aap, 253, 173

\bibitem[{{Chabrier} {et~al.}(1996){Chabrier}, {Baraffe}, \&
  {Plez}}]{Chabrier1996}
{Chabrier}, G., {Baraffe}, I., \& {Plez}, B. 1996, \apjl, 459, L91

\bibitem[{{Charbonneau} \& {MacGregor}(1993)}]{Charbonneau1993}
{Charbonneau}, P., \& {MacGregor}, K.~B. 1993, \apj, 417, 762

\bibitem[{{Charbonnel} \& {Lagarde}(2010)}]{Charbonnel2010}
{Charbonnel}, C., \& {Lagarde}, N. 2010, \aap, 522, A10

\bibitem[{{Charbonnel} \& {Talon}(2005)}]{Charbonnel2005}
{Charbonnel}, C., \& {Talon}, S. 2005, Science, 309, 2189

\bibitem[{{Charbonnel} \& {Zahn}(2007)}]{Charbonnel2007}
{Charbonnel}, C., \& {Zahn}, J.-P. 2007, \aap, 467, L15

\bibitem[{{Chen} {et~al.}(2011){Chen}, {Mamajek}, {Bitner}, {Pecaut}, {Su}, \&
  {Weinberger}}]{Chen2011}
{Chen}, C.~H., {Mamajek}, E.~E., {Bitner}, M.~A., {et~al.} 2011, \apj, 738, 122

\bibitem[{{Chen} {et~al.}(2014){Chen}, {Girardi}, {Bressan}, {Marigo},
  {Barbieri}, \& {Kong}}]{Chen2014}
{Chen}, Y., {Girardi}, L., {Bressan}, A., {et~al.} 2014, \mnras, 444, 2525

\bibitem[{{Chieffi} \& {Limongi}(2013)}]{Chieffi2013}
{Chieffi}, A., \& {Limongi}, M. 2013, \apj, 764, 21

\bibitem[{{Chini} {et~al.}(2012){Chini}, {Hoffmeister}, {Nasseri}, {Stahl}, \&
  {Zinnecker}}]{Chini2012}
{Chini}, R., {Hoffmeister}, V.~H., {Nasseri}, A., {Stahl}, O., \& {Zinnecker},
  H. 2012, \mnras, 424, 1925

\bibitem[{{Christensen-Dalsgaard} {et~al.}(2009){Christensen-Dalsgaard}, {di
  Mauro}, {Houdek}, \& {Pijpers}}]{ChristensenDalsgaard2009}
{Christensen-Dalsgaard}, J., {di Mauro}, M.~P., {Houdek}, G., \& {Pijpers}, F.
  2009, \aap, 494, 205

\bibitem[{{Claver} {et~al.}(2001){Claver}, {Liebert}, {Bergeron}, \&
  {Koester}}]{Claver2001}
{Claver}, C.~F., {Liebert}, J., {Bergeron}, P., \& {Koester}, D. 2001, \apj,
  563, 987

\bibitem[{{Cohen} {et~al.}(2003){Cohen}, {Wheaton}, \& {Megeath}}]{Cohen2003}
{Cohen}, M., {Wheaton}, W.~A., \& {Megeath}, S.~T. 2003, \aj, 126, 1090

\bibitem[{{Conroy}(2013)}]{Conroy2013}
{Conroy}, C. 2013, \araa, 51, 393

\bibitem[{{Conroy} \& {Gunn}(2010)}]{Conroy2010}
{Conroy}, C., \& {Gunn}, J.~E. 2010, \apj, 712, 833

\bibitem[{{Conroy} {et~al.}(2009){Conroy}, {Gunn}, \& {White}}]{Conroy2009}
{Conroy}, C., {Gunn}, J.~E., \& {White}, M. 2009, \apj, 699, 486

\bibitem[{{Cowley} {et~al.}(1979){Cowley}, {Dawson}, \&
  {Hartwick}}]{Cowley1979}
{Cowley}, A.~P., {Dawson}, P., \& {Hartwick}, F.~D.~A. 1979, \pasp, 91, 628

\bibitem[{{Cox} \& {Stewart}(1965)}]{Cox1965}
{Cox}, A.~N., \& {Stewart}, J.~N. 1965, \apjs, 11, 22

\bibitem[{{Cox} \& {Stewart}(1970)}]{Cox1970}
---. 1970, \apjs, 19, 243

\bibitem[{{Cox}(1968)}]{Cox1968}
{Cox}, J.~P. 1968, {Principles of stellar structure - Vol.1: Physical
  principles; Vol.2: Applications to stars}

\bibitem[{{Crowther} {et~al.}(2006){Crowther}, {Lennon}, \&
  {Walborn}}]{Crowther2006}
{Crowther}, P.~A., {Lennon}, D.~J., \& {Walborn}, N.~R. 2006, \aap, 446, 279

\bibitem[{{Cummings} {et~al.}(2015){Cummings}, {Kalirai}, {Tremblay}, \&
  {Ramirez-Ruiz}}]{Cummings2015}
{Cummings}, J.~D., {Kalirai}, J.~S., {Tremblay}, P.-E., \& {Ramirez-Ruiz}, E.
  2015, \apj, 807, 90

\bibitem[{{Cunha} {et~al.}(2006){Cunha}, {Hubeny}, \& {Lanz}}]{Cunha2006}
{Cunha}, K., {Hubeny}, I., \& {Lanz}, T. 2006, \apjl, 647, L143

\bibitem[{{Cunha} \& {Lambert}(1994)}]{Cunha1994}
{Cunha}, K., \& {Lambert}, D.~L. 1994, \apj, 426, 170

\bibitem[{{Cyburt} {et~al.}(2010){Cyburt}, {Amthor}, {Ferguson}, {Meisel},
  {Smith}, {Warren}, {Heger}, {Hoffman}, {Rauscher}, {Sakharuk}, {Schatz},
  {Thielemann}, \& {Wiescher}}]{Cyburt2010}
{Cyburt}, R.~H., {Amthor}, A.~M., {Ferguson}, R., {et~al.} 2010, \apjs, 189,
  240

\bibitem[{{Daflon} {et~al.}(2001){Daflon}, {Cunha}, {Becker}, \&
  {Smith}}]{Daflon2001}
{Daflon}, S., {Cunha}, K., {Becker}, S.~R., \& {Smith}, V.~V. 2001, \apj, 552,
  309

\bibitem[{{Dalcanton} {et~al.}(2012){Dalcanton}, {Williams}, {Lang}, {Lauer},
  {Kalirai}, {Seth}, {Dolphin}, {Rosenfield}, {Weisz}, {Bell}, {Bianchi},
  {Boyer}, {Caldwell}, {Dong}, {Dorman}, {Gilbert}, {Girardi}, {Gogarten},
  {Gordon}, {Guhathakurta}, {Hodge}, {Holtzman}, {Johnson}, {Larsen}, {Lewis},
  {Melbourne}, {Olsen}, {Rix}, {Rosema}, {Saha}, {Sarajedini}, {Skillman}, \&
  {Stanek}}]{Dalcanton2012}
{Dalcanton}, J.~J., {Williams}, B.~F., {Lang}, D., {et~al.} 2012, \apjs, 200,
  18

\bibitem[{{D'Antona} \& {Caloi}(2008)}]{D'Antona2008}
{D'Antona}, F., \& {Caloi}, V. 2008, \mnras, 390, 693

\bibitem[{{D'Antona} \& {Mazzitelli}(1997)}]{D'Antona1997}
{D'Antona}, F., \& {Mazzitelli}, I. 1997, \memsai, 68, 807

\bibitem[{{de Geus} {et~al.}(1989){de Geus}, {de Zeeuw}, \& {Lub}}]{deGeus1989}
{de Geus}, E.~J., {de Zeeuw}, P.~T., \& {Lub}, J. 1989, \aap, 216, 44

\bibitem[{{de Jager} {et~al.}(1988){de Jager}, {Nieuwenhuijzen}, \& {van der
  Hucht}}]{deJager1988}
{de Jager}, C., {Nieuwenhuijzen}, H., \& {van der Hucht}, K.~A. 1988, \aaps,
  72, 259

\bibitem[{{de Mink} {et~al.}(2010){de Mink}, {Cantiello}, {Langer}, \&
  {Pols}}]{deMink2010}
{de Mink}, S.~E., {Cantiello}, M., {Langer}, N., \& {Pols}, O.~R. 2010, in
  American Institute of Physics Conference Series, Vol. 1314, American
  Institute of Physics Conference Series, ed. V.~{Kalogera} \& M.~{van der
  Sluys}, 291--296

\bibitem[{{de Mink} {et~al.}(2009){de Mink}, {Cantiello}, {Langer}, {Pols},
  {Brott}, \& {Yoon}}]{deMink2009}
{de Mink}, S.~E., {Cantiello}, M., {Langer}, N., {et~al.} 2009, \aap, 497, 243

\bibitem[{{de Mink} {et~al.}(2013){de Mink}, {Langer}, {Izzard}, {Sana}, \& {de
  Koter}}]{deMink2013}
{de Mink}, S.~E., {Langer}, N., {Izzard}, R.~G., {Sana}, H., \& {de Koter}, A.
  2013, \apj, 764, 166

\bibitem[{{de Mink} {et~al.}(2014){de Mink}, {Sana}, {Langer}, {Izzard}, \&
  {Schneider}}]{deMink2014}
{de Mink}, S.~E., {Sana}, H., {Langer}, N., {Izzard}, R.~G., \& {Schneider},
  F.~R.~N. 2014, \apj, 782, 7

\bibitem[{{De Silva} {et~al.}(2015){De Silva}, {Freeman}, {Bland-Hawthorn},
  {Martell}, {de Boer}, {Asplund}, {Keller}, {Sharma}, {Zucker}, {Zwitter},
  {Anguiano}, {Bacigalupo}, {Bayliss}, {Beavis}, {Bergemann}, {Campbell},
  {Cannon}, {Carollo}, {Casagrande}, {Casey}, {Da Costa}, {D'Orazi}, {Dotter},
  {Duong}, {Heger}, {Ireland}, {Kafle}, {Kos}, {Lattanzio}, {Lewis}, {Lin},
  {Lind}, {Munari}, {Nataf}, {O'Toole}, {Parker}, {Reid}, {Schlesinger},
  {Sheinis}, {Simpson}, {Stello}, {Ting}, {Traven}, {Watson}, {Wittenmyer},
  {Yong}, \& {{\v Z}erjal}}]{DeSilva2015}
{De Silva}, G.~M., {Freeman}, K.~C., {Bland-Hawthorn}, J., {et~al.} 2015,
  \mnras, 449, 2604

\bibitem[{{de Zeeuw} {et~al.}(1999){de Zeeuw}, {Hoogerwerf}, {de Bruijne},
  {Brown}, \& {Blaauw}}]{deZeeuw1999}
{de Zeeuw}, P.~T., {Hoogerwerf}, R., {de Bruijne}, J.~H.~J., {Brown}, A.~G.~A.,
  \& {Blaauw}, A. 1999, \aj, 117, 354

\bibitem[{{Demarque} {et~al.}(2004){Demarque}, {Woo}, {Kim}, \&
  {Yi}}]{Demarque2004}
{Demarque}, P., {Woo}, J.-H., {Kim}, Y.-C., \& {Yi}, S.~K. 2004, \apjs, 155,
  667

\bibitem[{{Denissenkov}(2010)}]{Denissenkov2010}
{Denissenkov}, P.~A. 2010, \apj, 723, 563

\bibitem[{{Denissenkov} \& {Pinsonneault}(2007)}]{Denissenkov2007}
{Denissenkov}, P.~A., \& {Pinsonneault}, M. 2007, \apj, 655, 1157

\bibitem[{{Dobbie} {et~al.}(2004){Dobbie}, {Pinfield}, {Napiwotzki}, {Hambly},
  {Burleigh}, {Barstow}, {Jameson}, \& {Hubeny}}]{Dobbie2004}
{Dobbie}, P.~D., {Pinfield}, D.~J., {Napiwotzki}, R., {et~al.} 2004, \mnras,
  355, L39

\bibitem[{{Doherty} {et~al.}(2015){Doherty}, {Gil-Pons}, {Siess}, {Lattanzio},
  \& {Lau}}]{Doherty2015}
{Doherty}, C.~L., {Gil-Pons}, P., {Siess}, L., {Lattanzio}, J.~C., \& {Lau},
  H.~H.~B. 2015, \mnras, 446, 2599

\bibitem[{{Dohm-Palmer} \& {Skillman}(2002)}]{DohmPalmer2002}
{Dohm-Palmer}, R.~C., \& {Skillman}, E.~D. 2002, \aj, 123, 1433

\bibitem[{{Dolphin}(2002)}]{Dolphin2002}
{Dolphin}, A.~E. 2002, \mnras, 332, 91

\bibitem[{{Dotter}(2016)}]{Dotter2016}
{Dotter}, A. 2016, \apjs, 222, 8

\bibitem[{{Dotter} {et~al.}(2008){Dotter}, {Chaboyer}, {Jevremovi{\'c}},
  {Kostov}, {Baron}, \& {Ferguson}}]{Dotter2008}
{Dotter}, A., {Chaboyer}, B., {Jevremovi{\'c}}, D., {et~al.} 2008, \apjs, 178,
  89

\bibitem[{{Dotter} {et~al.}(2011){Dotter}, {Sarajedini}, \&
  {Anderson}}]{Dotter2011}
{Dotter}, A., {Sarajedini}, A., \& {Anderson}, J. 2011, \apj, 738, 74

\bibitem[{{Drake} \& {Testa}(2005)}]{Drake2005}
{Drake}, J.~J., \& {Testa}, P. 2005, \nat, 436, 525

\bibitem[{{Drout} {et~al.}(2012){Drout}, {Massey}, \& {Meynet}}]{Drout2012}
{Drout}, M.~R., {Massey}, P., \& {Meynet}, G. 2012, \apj, 750, 97

\bibitem[{{Eggenberger} {et~al.}(2005){Eggenberger}, {Maeder}, \&
  {Meynet}}]{Eggenberger2005}
{Eggenberger}, P., {Maeder}, A., \& {Meynet}, G. 2005, \aap, 440, L9

\bibitem[{{Eggenberger} {et~al.}(2002){Eggenberger}, {Meynet}, \&
  {Maeder}}]{Eggenberger2002}
{Eggenberger}, P., {Meynet}, G., \& {Maeder}, A. 2002, \aap, 386, 576

\bibitem[{{Eggenberger} {et~al.}(2008){Eggenberger}, {Meynet}, {Maeder},
  {Hirschi}, {Charbonnel}, {Talon}, \& {Ekstr{\"o}m}}]{Eggenberger2008}
{Eggenberger}, P., {Meynet}, G., {Maeder}, A., {et~al.} 2008, \apss, 316, 43

\bibitem[{{Eggenberger} {et~al.}(2012){Eggenberger}, {Montalb{\'a}n}, \&
  {Miglio}}]{Eggenberger2012}
{Eggenberger}, P., {Montalb{\'a}n}, J., \& {Miglio}, A. 2012, \aap, 544, L4

\bibitem[{{Eggleton}(1971)}]{Eggleton1971}
{Eggleton}, P.~P. 1971, \mnras, 151, 351

\bibitem[{{Eggleton} {et~al.}(2006){Eggleton}, {Dearborn}, \&
  {Lattanzio}}]{Eggleton2006}
{Eggleton}, P.~P., {Dearborn}, D.~S.~P., \& {Lattanzio}, J.~C. 2006, Science,
  314, 1580

\bibitem[{{Ekstr{\"o}m} {et~al.}(2012){Ekstr{\"o}m}, {Georgy}, {Eggenberger},
  {Meynet}, {Mowlavi}, {Wyttenbach}, {Granada}, {Decressin}, {Hirschi},
  {Frischknecht}, {Charbonnel}, \& {Maeder}}]{Ekstrom2012}
{Ekstr{\"o}m}, S., {Georgy}, C., {Eggenberger}, P., {et~al.} 2012, \aap, 537,
  A146

\bibitem[{{Eldridge} {et~al.}(2008){Eldridge}, {Izzard}, \&
  {Tout}}]{Eldridge2008}
{Eldridge}, J.~J., {Izzard}, R.~G., \& {Tout}, C.~A. 2008, \mnras, 384, 1109

\bibitem[{{Eldridge} \& {Tout}(2004)}]{Eldridge2004}
{Eldridge}, J.~J., \& {Tout}, C.~A. 2004, \mnras, 348, 201

\bibitem[{{Elias} {et~al.}(1985){Elias}, {Frogel}, \& {Humphreys}}]{Elias1985}
{Elias}, J.~H., {Frogel}, J.~A., \& {Humphreys}, R.~M. 1985, \apjs, 57, 91

\bibitem[{{Endal} \& {Sofia}(1976)}]{Endal1976}
{Endal}, A.~S., \& {Sofia}, S. 1976, \apj, 210, 184

\bibitem[{{Endal} \& {Sofia}(1978)}]{Endal1978}
---. 1978, \apj, 220, 279

\bibitem[{{Faulkner} \& {Cannon}(1973)}]{Faulkner1973}
{Faulkner}, D.~J., \& {Cannon}, R.~D. 1973, \apj, 180, 435

\bibitem[{{Fazio} {et~al.}(2004){Fazio}, {Hora}, {Allen}, {Ashby}, {Barmby},
  {Deutsch}, {Huang}, {Kleiner}, {Marengo}, {Megeath}, {Melnick}, {Pahre},
  {Patten}, {Polizotti}, {Smith}, {Taylor}, {Wang}, {Willner}, {Hoffmann},
  {Pipher}, {Forrest}, {McMurty}, {McCreight}, {McKelvey}, {McMurray}, {Koch},
  {Moseley}, {Arendt}, {Mentzell}, {Marx}, {Losch}, {Mayman}, {Eichhorn},
  {Krebs}, {Jhabvala}, {Gezari}, {Fixsen}, {Flores}, {Shakoorzadeh}, {Jungo},
  {Hakun}, {Workman}, {Karpati}, {Kichak}, {Whitley}, {Mann}, {Tollestrup},
  {Eisenhardt}, {Stern}, {Gorjian}, {Bhattacharya}, {Carey}, {Nelson},
  {Glaccum}, {Lacy}, {Lowrance}, {Laine}, {Reach}, {Stauffer}, {Surace},
  {Wilson}, {Wright}, {Hoffman}, {Domingo}, \& {Cohen}}]{Fazio2004}
{Fazio}, G.~G., {Hora}, J.~L., {Allen}, L.~E., {et~al.} 2004, \apjs, 154, 10

\bibitem[{{Feiden} \& {Chaboyer}(2012)}]{Feiden2012}
{Feiden}, G.~A., \& {Chaboyer}, B. 2012, \apj, 757, 42

\bibitem[{{Feiden} \& {Chaboyer}(2013)}]{Feiden2013}
---. 2013, \apj, 779, 183

\bibitem[{{Ferguson} {et~al.}(2005){Ferguson}, {Alexander}, {Allard}, {Barman},
  {Bodnarik}, {Hauschildt}, {Heffner-Wong}, \& {Tamanai}}]{Ferguson2005}
{Ferguson}, J.~W., {Alexander}, D.~R., {Allard}, F., {et~al.} 2005, \apj, 623,
  585

\bibitem[{{Ferrario} {et~al.}(2005){Ferrario}, {Wickramasinghe}, {Liebert}, \&
  {Williams}}]{Ferrario2005}
{Ferrario}, L., {Wickramasinghe}, D., {Liebert}, J., \& {Williams}, K.~A. 2005,
  \mnras, 361, 1131

\bibitem[{{Fontaine} {et~al.}(2008){Fontaine}, {Brassard}, {Green}, {Chayer},
  {Charpinet}, {Andersen}, \& {Portouw}}]{Fontaine2008}
{Fontaine}, G., {Brassard}, P., {Green}, E.~M., {et~al.} 2008, \aap, 486, L39

\bibitem[{{Francois} {et~al.}(1997){Francois}, {Danziger}, {Buonanno}, \&
  {Perrin}}]{Francois1997}
{Francois}, P., {Danziger}, J., {Buonanno}, R., \& {Perrin}, M.~N. 1997, \aap,
  327, 121

\bibitem[{{Freedman} {et~al.}(2008){Freedman}, {Marley}, \&
  {Lodders}}]{Freedman2008}
{Freedman}, R.~S., {Marley}, M.~S., \& {Lodders}, K. 2008, \apjs, 174, 504

\bibitem[{{Freytag} {et~al.}(1996){Freytag}, {Ludwig}, \&
  {Steffen}}]{Freytag1996}
{Freytag}, B., {Ludwig}, H.-G., \& {Steffen}, M. 1996, \aap, 313, 497

\bibitem[{{Fricke} \& {Kippenhahn}(1972)}]{Fricke1972}
{Fricke}, K.~J., \& {Kippenhahn}, R. 1972, \araa, 10, 45

\bibitem[{{Friend} \& {Abbott}(1986)}]{Friend1986}
{Friend}, D.~B., \& {Abbott}, D.~C. 1986, \apj, 311, 701

\bibitem[{{Frogel} {et~al.}(1990){Frogel}, {Mould}, \& {Blanco}}]{Frogel1990}
{Frogel}, J.~A., {Mould}, J., \& {Blanco}, V.~M. 1990, \apj, 352, 96

\bibitem[{{Frommhold} {et~al.}(2010){Frommhold}, {Abel}, {Wang}, {Gustafsson},
  {Li}, \& {Hunt}}]{Frommhold2010}
{Frommhold}, L., {Abel}, M., {Wang}, F., {et~al.} 2010, Molecular Physics, 108,
  2265

\bibitem[{{Fynbo} {et~al.}(2005){Fynbo}, {Diget}, {Bergmann}, {Borge},
  {Cederk{\"a}ll}, {Dendooven}, {Fraile}, {Franchoo}, {Fedosseev}, {Fulton},
  {Huang}, {Huikari}, {Jeppesen}, {Jokinen}, {Jones}, {Jonson}, {K{\"o}ster},
  {Langanke}, {Meister}, {Nilsson}, {Nyman}, {Prezado}, {Riisager},
  {Rinta-Antila}, {Tengblad}, {Turrion}, {Wang}, {Weissman}, {Wilhelmsen},
  {{\"A}yst{\"o}}, \& {ISOLDE Collaboration}}]{Fynbo2005}
{Fynbo}, H.~O.~U., {Diget}, C.~A., {Bergmann}, U.~C., {et~al.} 2005, \nat, 433,
  136

\bibitem[{{Gallet} \& {Bouvier}(2013)}]{Gallet2013}
{Gallet}, F., \& {Bouvier}, J. 2013, \aap, 556, A36

\bibitem[{{G{\'a}sp{\'a}r} {et~al.}(2009){G{\'a}sp{\'a}r}, {Rieke}, {Su},
  {Balog}, {Trilling}, {Muzzerole}, {Apai}, \& {Kelly}}]{Gaspar2009}
{G{\'a}sp{\'a}r}, A., {Rieke}, G.~H., {Su}, K.~Y.~L., {et~al.} 2009, \apj, 697,
  1578

\bibitem[{{Gathier} {et~al.}(1981){Gathier}, {Lamers}, \& {Snow}}]{Gathier1981}
{Gathier}, R., {Lamers}, H.~J.~G.~L.~M., \& {Snow}, T.~P. 1981, \apj, 247, 173

\bibitem[{{Georgy} {et~al.}(2012){Georgy}, {Ekstr{\"o}m}, {Meynet}, {Massey},
  {Levesque}, {Hirschi}, {Eggenberger}, \& {Maeder}}]{Georgy2012}
{Georgy}, C., {Ekstr{\"o}m}, S., {Meynet}, G., {et~al.} 2012, \aap, 542, A29

\bibitem[{{Georgy} {et~al.}(2013){Georgy}, {Ekstr{\"o}m}, {Eggenberger},
  {Meynet}, {Haemmerl{\'e}}, {Maeder}, {Granada}, {Groh}, {Hirschi}, {Mowlavi},
  {Yusof}, {Charbonnel}, {Decressin}, \& {Barblan}}]{Georgy2013}
{Georgy}, C., {Ekstr{\"o}m}, S., {Eggenberger}, P., {et~al.} 2013, \aap, 558,
  A103

\bibitem[{{Gies} \& {Lambert}(1992)}]{Gies1992}
{Gies}, D.~R., \& {Lambert}, D.~L. 1992, \apj, 387, 673

\bibitem[{{Gilliland} {et~al.}(2010){Gilliland}, {Brown},
  {Christensen-Dalsgaard}, {Kjeldsen}, {Aerts}, {Appourchaux}, {Basu},
  {Bedding}, {Chaplin}, {Cunha}, {De Cat}, {De Ridder}, {Guzik}, {Handler},
  {Kawaler}, {Kiss}, {Kolenberg}, {Kurtz}, {Metcalfe}, {Monteiro}, {Szab{\'o}},
  {Arentoft}, {Balona}, {Debosscher}, {Elsworth}, {Quirion}, {Stello},
  {Su{\'a}rez}, {Borucki}, {Jenkins}, {Koch}, {Kondo}, {Latham}, {Rowe}, \&
  {Steffen}}]{Gilliland2010}
{Gilliland}, R.~L., {Brown}, T.~M., {Christensen-Dalsgaard}, J., {et~al.} 2010,
  \pasp, 122, 131

\bibitem[{{Girardi}(1999)}]{Girardi1999}
{Girardi}, L. 1999, \mnras, 308, 818

\bibitem[{{Girardi} {et~al.}(2002){Girardi}, {Bertelli}, {Bressan}, {Chiosi},
  {Groenewegen}, {Marigo}, {Salasnich}, \& {Weiss}}]{Girardi2002}
{Girardi}, L., {Bertelli}, G., {Bressan}, A., {et~al.} 2002, \aap, 391, 195

\bibitem[{{Girardi} \& {Marigo}(2007)}]{Girardi2007}
{Girardi}, L., \& {Marigo}, P. 2007, \aap, 462, 237

\bibitem[{{Girardi} {et~al.}(2013){Girardi}, {Marigo}, {Bressan}, \&
  {Rosenfield}}]{Girardi2013}
{Girardi}, L., {Marigo}, P., {Bressan}, A., \& {Rosenfield}, P. 2013, \apj,
  777, 142

\bibitem[{{Girardi} {et~al.}(2008){Girardi}, {Dalcanton}, {Williams}, {de
  Jong}, {Gallart}, {Monelli}, {Groenewegen}, {Holtzman}, {Olsen}, {Seth},
  {Weisz}, \& {ANGST/ANGRRR Collaboration}}]{Girardi2008}
{Girardi}, L., {Dalcanton}, J., {Williams}, B., {et~al.} 2008, \pasp, 120, 583

\bibitem[{{Girardi} {et~al.}(2010){Girardi}, {Williams}, {Gilbert},
  {Rosenfield}, {Dalcanton}, {Marigo}, {Boyer}, {Dolphin}, {Weisz},
  {Melbourne}, {Olsen}, {Seth}, \& {Skillman}}]{Girardi2010}
{Girardi}, L., {Williams}, B.~F., {Gilbert}, K.~M., {et~al.} 2010, \apj, 724,
  1030

\bibitem[{{Glebbeek} {et~al.}(2009){Glebbeek}, {Gaburov}, {de Mink}, {Pols}, \&
  {Portegies Zwart}}]{Glebbeek2009}
{Glebbeek}, E., {Gaburov}, E., {de Mink}, S.~E., {Pols}, O.~R., \& {Portegies
  Zwart}, S.~F. 2009, \aap, 497, 255

\bibitem[{{Gordon} {et~al.}(2011){Gordon}, {Meixner}, {Meade}, {Whitney},
  {Engelbracht}, {Bot}, {Boyer}, {Lawton}, {Sewi{\l}o}, {Babler}, {Bernard},
  {Bracker}, {Block}, {Blum}, {Bolatto}, {Bonanos}, {Harris}, {Hora},
  {Indebetouw}, {Misselt}, {Reach}, {Shiao}, {Tielens}, {Carlson},
  {Churchwell}, {Clayton}, {Chen}, {Cohen}, {Fukui}, {Gorjian}, {Hony},
  {Israel}, {Kawamura}, {Kemper}, {Leroy}, {Li}, {Madden}, {Marble},
  {McDonald}, {Mizuno}, {Mizuno}, {Muller}, {Oliveira}, {Olsen}, {Onishi},
  {Paladini}, {Paradis}, {Points}, {Robitaille}, {Rubin}, {Sandstrom}, {Sato},
  {Shibai}, {Simon}, {Smith}, {Srinivasan}, {Vijh}, {Van Dyk}, {van Loon}, \&
  {Zaritsky}}]{Gordon2011}
{Gordon}, K.~D., {Meixner}, M., {Meade}, M.~R., {et~al.} 2011, \aj, 142, 102

\bibitem[{{Graboske} {et~al.}(1973){Graboske}, {Dewitt}, {Grossman}, \&
  {Cooper}}]{Graboske1973}
{Graboske}, H.~C., {Dewitt}, H.~E., {Grossman}, A.~S., \& {Cooper}, M.~S. 1973,
  \apj, 181, 457

\bibitem[{{Gr{\"a}fener} {et~al.}(2011){Gr{\"a}fener}, {Vink}, {de Koter}, \&
  {Langer}}]{Grafener2011}
{Gr{\"a}fener}, G., {Vink}, J.~S., {de Koter}, A., \& {Langer}, N. 2011, \aap,
  535, A56

\bibitem[{{Gratton} {et~al.}(2004){Gratton}, {Sneden}, \&
  {Carretta}}]{Gratton2004}
{Gratton}, R., {Sneden}, C., \& {Carretta}, E. 2004, \araa, 42, 385

\bibitem[{{Gratton} {et~al.}(2012){Gratton}, {Carretta}, \&
  {Bragaglia}}]{Gratton2012}
{Gratton}, R.~G., {Carretta}, E., \& {Bragaglia}, A. 2012, \aapr, 20, 50

\bibitem[{{Grevesse} \& {Sauval}(1998)}]{Grevesse1998}
{Grevesse}, N., \& {Sauval}, A.~J. 1998, \ssr, 85, 161

\bibitem[{{Groenewegen} \& {de Jong}(1993)}]{Groenewegen1993}
{Groenewegen}, M.~A.~T., \& {de Jong}, T. 1993, \aap, 267, 410

\bibitem[{{Grossman} \& {Taam}(1996)}]{Grossman1996}
{Grossman}, S.~A., \& {Taam}, R.~E. 1996, \mnras, 283, 1165

\bibitem[{{Harris} \& {Zaritsky}(2004)}]{Harris2004}
{Harris}, J., \& {Zaritsky}, D. 2004, \aj, 127, 1531

\bibitem[{{Harris} \& {Zaritsky}(2009)}]{Harris2009}
---. 2009, \aj, 138, 1243

\bibitem[{{Hartwick}(1970)}]{Hartwick1970}
{Hartwick}, F.~D.~A. 1970, \aplett, 7, 151

\bibitem[{{Hauschildt} {et~al.}(1999){Hauschildt}, {Allard}, \&
  {Baron}}]{Hauschildt1999a}
{Hauschildt}, P.~H., {Allard}, F., \& {Baron}, E. 1999, \apj, 512, 377

\bibitem[{{Heger} {et~al.}(2000){Heger}, {Langer}, \& {Woosley}}]{Heger2000}
{Heger}, A., {Langer}, N., \& {Woosley}, S.~E. 2000, \apj, 528, 368

\bibitem[{{Heger} \& {Woosley}(2002)}]{Heger2002}
{Heger}, A., \& {Woosley}, S.~E. 2002, \apj, 567, 532

\bibitem[{{Heger} {et~al.}(2005){Heger}, {Woosley}, \& {Spruit}}]{Heger2005}
{Heger}, A., {Woosley}, S.~E., \& {Spruit}, H.~C. 2005, \apj, 626, 350

\bibitem[{{Henriques} {et~al.}(2011){Henriques}, {Maraston}, {Monaco},
  {Fontanot}, {Menci}, {De Lucia}, \& {Tonini}}]{Henriques2011}
{Henriques}, B., {Maraston}, C., {Monaco}, P., {et~al.} 2011, \mnras, 415, 3571

\bibitem[{{Henyey} {et~al.}(1965){Henyey}, {Vardya}, \&
  {Bodenheimer}}]{Henyey1965}
{Henyey}, L., {Vardya}, M.~S., \& {Bodenheimer}, P. 1965, \apj, 142, 841

\bibitem[{{Henyey} {et~al.}(1964){Henyey}, {Forbes}, \& {Gould}}]{Henyey1964}
{Henyey}, L.~G., {Forbes}, J.~E., \& {Gould}, N.~L. 1964, \apj, 139, 306

\bibitem[{{Herbig}(1962)}]{Herbig1962}
{Herbig}, G.~H. 1962, \apj, 135, 736

\bibitem[{{Herczeg} \& {Hillenbrand}(2015)}]{Herczeg2015}
{Herczeg}, G.~J., \& {Hillenbrand}, L.~A. 2015, \apj, 808, 23

\bibitem[{{Herwig}(2000)}]{Herwig2000}
{Herwig}, F. 2000, \aap, 360, 952

\bibitem[{{Herwig}(2004)}]{Herwig2004a}
---. 2004, \apjs, 155, 651

\bibitem[{{Herwig}(2005)}]{Herwig2005}
---. 2005, \araa, 43, 435

\bibitem[{{Herwig} \& {Austin}(2004)}]{Herwig2004b}
{Herwig}, F., \& {Austin}, S.~M. 2004, \apjl, 613, L73

\bibitem[{{Hewett} {et~al.}(2006){Hewett}, {Warren}, {Leggett}, \&
  {Hodgkin}}]{Hewett2006}
{Hewett}, P.~C., {Warren}, S.~J., {Leggett}, S.~K., \& {Hodgkin}, S.~T. 2006,
  \mnras, 367, 454

\bibitem[{{Hirschi} {et~al.}(2004){Hirschi}, {Meynet}, \&
  {Maeder}}]{Hirschi2004}
{Hirschi}, R., {Meynet}, G., \& {Maeder}, A. 2004, \aap, 425, 649

\bibitem[{{Holtzman} {et~al.}(1995){Holtzman}, {Burrows}, {Casertano},
  {Hester}, {Trauger}, {Watson}, \& {Worthey}}]{Holtzman1995}
{Holtzman}, J.~A., {Burrows}, C.~J., {Casertano}, S., {et~al.} 1995, \pasp,
  107, 1065

\bibitem[{{Hosokawa} {et~al.}(2011){Hosokawa}, {Offner}, \&
  {Krumholz}}]{Hosokawa2011}
{Hosokawa}, T., {Offner}, S.~S.~R., \& {Krumholz}, M.~R. 2011, \apj, 738, 140

\bibitem[{{Howell} {et~al.}(2014){Howell}, {Sobeck}, {Haas}, {Still},
  {Barclay}, {Mullally}, {Troeltzsch}, {Aigrain}, {Bryson}, {Caldwell},
  {Chaplin}, {Cochran}, {Huber}, {Marcy}, {Miglio}, {Najita}, {Smith},
  {Twicken}, \& {Fortney}}]{Howell2014}
{Howell}, S.~B., {Sobeck}, C., {Haas}, M., {et~al.} 2014, \pasp, 126, 398

\bibitem[{{Hu} {et~al.}(2011){Hu}, {Tout}, {Glebbeek}, \& {Dupret}}]{Hu2011}
{Hu}, H., {Tout}, C.~A., {Glebbeek}, E., \& {Dupret}, M.-A. 2011, \mnras, 418,
  195

\bibitem[{{Huang} {et~al.}(2010){Huang}, {Gies}, \& {McSwain}}]{Huang2010}
{Huang}, W., {Gies}, D.~R., \& {McSwain}, M.~V. 2010, \apj, 722, 605

\bibitem[{{Humphreys}(1979)}]{Humphreys1979}
{Humphreys}, R.~M. 1979, \apjs, 39, 389

\bibitem[{{Hunter} {et~al.}(2007){Hunter}, {Dufton}, {Smartt}, {Ryans},
  {Evans}, {Lennon}, {Trundle}, {Hubeny}, \& {Lanz}}]{Hunter2007}
{Hunter}, I., {Dufton}, P.~L., {Smartt}, S.~J., {et~al.} 2007, \aap, 466, 277

\bibitem[{{Hunter} {et~al.}(2008){Hunter}, {Brott}, {Lennon}, {Langer},
  {Dufton}, {Trundle}, {Smartt}, {de Koter}, {Evans}, \& {Ryans}}]{Hunter2008}
{Hunter}, I., {Brott}, I., {Lennon}, D.~J., {et~al.} 2008, \apjl, 676, L29

\bibitem[{{Hunter} {et~al.}(2009){Hunter}, {Brott}, {Langer}, {Lennon},
  {Dufton}, {Howarth}, {Ryans}, {Trundle}, {Evans}, {de Koter}, \&
  {Smartt}}]{Hunter2009}
{Hunter}, I., {Brott}, I., {Langer}, N., {et~al.} 2009, \aap, 496, 841

\bibitem[{{Hurley} {et~al.}(2002){Hurley}, {Tout}, \& {Pols}}]{Hurley2002}
{Hurley}, J.~R., {Tout}, C.~A., \& {Pols}, O.~R. 2002, \mnras, 329, 897

\bibitem[{{Iben}(1982)}]{Iben1982}
{Iben}, Jr., I. 1982, \apj, 260, 821

\bibitem[{{Iben}(1983)}]{Iben1983b}
---. 1983, \apjl, 275, L65

\bibitem[{{Iben} {et~al.}(1983){Iben}, {Kaler}, {Truran}, \&
  {Renzini}}]{Iben1983a}
{Iben}, Jr., I., {Kaler}, J.~B., {Truran}, J.~W., \& {Renzini}, A. 1983, \apj,
  264, 605

\bibitem[{{Iglesias} \& {Rogers}(1991)}]{Iglesias1991}
{Iglesias}, C.~A., \& {Rogers}, F.~J. 1991, \apj, 371, 408

\bibitem[{{Iglesias} \& {Rogers}(1993)}]{Iglesias1993}
---. 1993, \apj, 412, 752

\bibitem[{{Iglesias} \& {Rogers}(1996)}]{Iglesias1996}
---. 1996, \apj, 464, 943

\bibitem[{{Irwin} {et~al.}(2011){Irwin}, {Quinn}, {Berta}, {Latham}, {Torres},
  {Burke}, {Charbonneau}, {Dittmann}, {Esquerdo}, {Stefanik}, {Oksanen},
  {Buchhave}, {Nutzman}, {Berlind}, {Calkins}, \& {Falco}}]{Irwin2011}
{Irwin}, J.~M., {Quinn}, S.~N., {Berta}, Z.~K., {et~al.} 2011, \apj, 742, 123

\bibitem[{{Itoh} {et~al.}(1979){Itoh}, {Totsuji}, {Ichimaru}, \&
  {Dewitt}}]{Itoh1979}
{Itoh}, N., {Totsuji}, H., {Ichimaru}, S., \& {Dewitt}, H.~E. 1979, \apj, 234,
  1079

\bibitem[{{Ivezic} {et~al.}(2008){Ivezic}, {Tyson}, {Abel}, {Acosta},
  {Allsman}, {AlSayyad}, {Anderson}, {Andrew}, {Angel}, {Angeli}, {Ansari},
  {Antilogus}, {Arndt}, {Astier}, {Aubourg}, {Axelrod}, {Bard}, {Barr},
  {Barrau}, {Bartlett}, {Bauman}, {Beaumont}, {Becker}, {Becla}, {Beldica},
  {Bellavia}, {Blanc}, {Blandford}, {Bloom}, {Bogart}, {Borne}, {Bosch},
  {Boutigny}, {Brandt}, {Brown}, {Bullock}, {Burchat}, {Burke}, {Cagnoli},
  {Calabrese}, {Chandrasekharan}, {Chesley}, {Cheu}, {Chiang}, {Claver},
  {Connolly}, {Cook}, {Cooray}, {Covey}, {Cribbs}, {Cui}, {Cutri}, {Daubard},
  {Daues}, {Delgado}, {Digel}, {Doherty}, {Dubois}, {Dubois-Felsmann},
  {Durech}, {Eracleous}, {Ferguson}, {Frank}, {Freemon}, {Gangler}, {Gawiser},
  {Geary}, {Gee}, {Geha}, {Gibson}, {Gilmore}, {Glanzman}, {Goodenow},
  {Gressler}, {Gris}, {Guyonnet}, {Hascall}, {Haupt}, {Hernandez}, {Hogan},
  {Huang}, {Huffer}, {Innes}, {Jacoby}, {Jain}, {Jee}, {Jernigan},
  {Jevremovic}, {Johns}, {Jones}, {Juramy-Gilles}, {Juric}, {Kahn}, {Kalirai},
  {Kallivayalil}, {Kalmbach}, {Kantor}, {Kasliwal}, {Kessler}, {Kirkby},
  {Knox}, {Kotov}, {Krabbendam}, {Krughoff}, {Kubanek}, {Kuczewski},
  {Kulkarni}, {Lambert}, {Le Guillou}, {Levine}, {Liang}, {Lim}, {Lintott},
  {Lupton}, {Mahabal}, {Marshall}, {Marshall}, {May}, {McKercher}, {Migliore},
  {Miller}, {Mills}, {Monet}, {Moniez}, {Neill}, {Nief}, {Nomerotski},
  {Nordby}, {O'Connor}, {Oliver}, {Olivier}, {Olsen}, {Ortiz}, {Owen}, {Pain},
  {Peterson}, {Petry}, {Pierfederici}, {Pietrowicz}, {Pike}, {Pinto}, {Plante},
  {Plate}, {Price}, {Prouza}, {Radeka}, {Rajagopal}, {Rasmussen}, {Regnault},
  {Ridgway}, {Ritz}, {Rosing}, {Roucelle}, {Rumore}, {Russo}, {Saha},
  {Sassolas}, {Schalk}, {Schindler}, {Schneider}, {Schumacher}, {Sebag},
  {Sembroski}, {Seppala}, {Shipsey}, {Silvestri}, {Smith}, {Smith}, {Strauss},
  {Stubbs}, {Sweeney}, {Szalay}, {Takacs}, {Thaler}, {Van Berg}, {Vanden Berk},
  {Vetter}, {Virieux}, {Xin}, {Walkowicz}, {Walter}, {Wang}, {Warner},
  {Willman}, {Wittman}, {Wolff}, {Wood-Vasey}, {Yoachim}, {Zhan}, \& {for the
  LSST Collaboration}}]{Ivezic2008}
{Ivezic}, Z., {Tyson}, J.~A., {Abel}, B., {et~al.} 2008, ArXiv e-prints,
  arXiv:0805.2366

\bibitem[{{Izzard} {et~al.}(2006){Izzard}, {Dray}, {Karakas}, {Lugaro}, \&
  {Tout}}]{Izzard2006}
{Izzard}, R.~G., {Dray}, L.~M., {Karakas}, A.~I., {Lugaro}, M., \& {Tout},
  C.~A. 2006, \aap, 460, 565

\bibitem[{{Jackson} \& {Jeffries}(2014)}]{Jackson2014}
{Jackson}, R.~J., \& {Jeffries}, R.~D. 2014, \mnras, 441, 2111

\bibitem[{{Jeffries} \& {Oliveira}(2005)}]{Jeffries2005}
{Jeffries}, R.~D., \& {Oliveira}, J.~M. 2005, \mnras, 358, 13

\bibitem[{{Jiang} {et~al.}(2015){Jiang}, {Cantiello}, {Bildsten}, {Quataert},
  \& {Blaes}}]{Jiang2015}
{Jiang}, Y.-F., {Cantiello}, M., {Bildsten}, L., {Quataert}, E., \& {Blaes}, O.
  2015, \apj, 813, 74

\bibitem[{{Juarez} {et~al.}(2014){Juarez}, {Cargile}, {James}, \&
  {Stassun}}]{Juarez2014}
{Juarez}, A.~J., {Cargile}, P.~A., {James}, D.~J., \& {Stassun}, K.~G. 2014,
  \apj, 795, 143

\bibitem[{{Kaiser} {et~al.}(2010){Kaiser}, {Burgett}, {Chambers}, {Denneau},
  {Heasley}, {Jedicke}, {Magnier}, {Morgan}, {Onaka}, \& {Tonry}}]{Kaiser2010}
{Kaiser}, N., {Burgett}, W., {Chambers}, K., {et~al.} 2010, in Society of
  Photo-Optical Instrumentation Engineers (SPIE) Conference Series, Vol. 7733,
  Society of Photo-Optical Instrumentation Engineers (SPIE) Conference Series,
  0

\bibitem[{{Kalirai} {et~al.}(2007){Kalirai}, {Bergeron}, {Hansen}, {Kelson},
  {Reitzel}, {Rich}, \& {Richer}}]{Kalirai2007}
{Kalirai}, J.~S., {Bergeron}, P., {Hansen}, B.~M.~S., {et~al.} 2007, \apj, 671,
  748

\bibitem[{{Kalirai} {et~al.}(2008){Kalirai}, {Hansen}, {Kelson}, {Reitzel},
  {Rich}, \& {Richer}}]{Kalirai2008}
{Kalirai}, J.~S., {Hansen}, B.~M.~S., {Kelson}, D.~D., {et~al.} 2008, \apj,
  676, 594

\bibitem[{{Kalirai} {et~al.}(2014){Kalirai}, {Marigo}, \&
  {Tremblay}}]{Kalirai2014}
{Kalirai}, J.~S., {Marigo}, P., \& {Tremblay}, P.-E. 2014, \apj, 782, 17

\bibitem[{{Kalirai} {et~al.}(2009){Kalirai}, {Saul Davis}, {Richer},
  {Bergeron}, {Catelan}, {Hansen}, \& {Rich}}]{Kalirai2009}
{Kalirai}, J.~S., {Saul Davis}, D., {Richer}, H.~B., {et~al.} 2009, \apj, 705,
  408

\bibitem[{{Kaluzny} {et~al.}(1995){Kaluzny}, {Krzeminski}, \&
  {Mazur}}]{Kaluzny1995}
{Kaluzny}, J., {Krzeminski}, W., \& {Mazur}, B. 1995, \aj, 110, 2206

\bibitem[{{Kamai} {et~al.}(2014){Kamai}, {Vrba}, {Stauffer}, \&
  {Stassun}}]{Kamai2014}
{Kamai}, B.~L., {Vrba}, F.~J., {Stauffer}, J.~R., \& {Stassun}, K.~G. 2014,
  \aj, 148, 30

\bibitem[{{Kawaler} {et~al.}(1996){Kawaler}, {Novikov}, {Srinivasan}, {Meynet},
  \& {Schaerer}}]{Kawaler1996}
{Kawaler}, S.~D., {Novikov}, I., {Srinivasan}, G., {Meynet}, G., \& {Schaerer},
  D. 1996, {Stellar Remnants}

\bibitem[{{Kepler} {et~al.}(2015){Kepler}, {Pelisoli}, {Koester}, {Ourique},
  {Kleinman}, {Romero}, {Nitta}, {Eisenstein}, {Costa}, {K{\"u}lebi}, {Jordan},
  {Dufour}, {Giommi}, \& {Rebassa-Mansergas}}]{Kepler2015}
{Kepler}, S.~O., {Pelisoli}, I., {Koester}, D., {et~al.} 2015, \mnras, 446,
  4078

\bibitem[{{Kilian}(1992)}]{Kilian1992}
{Kilian}, J. 1992, \aap, 262, 171

\bibitem[{{Kim} {et~al.}(2002){Kim}, {Demarque}, {Yi}, \&
  {Alexander}}]{Kim2002}
{Kim}, Y.-C., {Demarque}, P., {Yi}, S.~K., \& {Alexander}, D.~R. 2002, \apjs,
  143, 499

\bibitem[{{Kippenhahn} {et~al.}(1980){Kippenhahn}, {Ruschenplatt}, \&
  {Thomas}}]{Kippenhahn1980}
{Kippenhahn}, R., {Ruschenplatt}, G., \& {Thomas}, H.-C. 1980, \aap, 91, 175

\bibitem[{{Kippenhahn} \& {Thomas}(1970)}]{Kippenhahn1970}
{Kippenhahn}, R., \& {Thomas}, H.-C. 1970, in IAU Colloq. 4: Stellar Rotation,
  ed. A.~{Slettebak}, 20

\bibitem[{{Kjeldsen} \& {Bedding}(1995)}]{Kjeldsen1995}
{Kjeldsen}, H., \& {Bedding}, T.~R. 1995, \aap, 293, 87

\bibitem[{{Kleinman} {et~al.}(2013){Kleinman}, {Kepler}, {Koester}, {Pelisoli},
  {Pe{\c c}anha}, {Nitta}, {Costa}, {Krzesinski}, {Dufour}, {Lachapelle},
  {Bergeron}, {Yip}, {Harris}, {Eisenstein}, {Althaus}, \&
  {C{\'o}rsico}}]{Kleinman2013}
{Kleinman}, S.~J., {Kepler}, S.~O., {Koester}, D., {et~al.} 2013, \apjs, 204, 5

\bibitem[{{Koester}(2010)}]{Koester2010}
{Koester}, D. 2010, \memsai, 81, 921

\bibitem[{{K{\"o}hler} {et~al.}(2015){K{\"o}hler}, {Langer}, {de Koter}, {de
  Mink}, {Crowther}, {Evans}, {Gr{\"a}fener}, {Sana}, {Sanyal}, {Schneider}, \&
  {Vink}}]{Koehler2015}
{K{\"o}hler}, K., {Langer}, N., {de Koter}, A., {et~al.} 2015, \aap, 573, A71

\bibitem[{{Kraft}(1967)}]{Kraft1967}
{Kraft}, R.~P. 1967, \apj, 150, 551

\bibitem[{{Kraft}(1994)}]{Kraft1994}
---. 1994, \pasp, 106, 553

\bibitem[{{Kraus} {et~al.}(2015){Kraus}, {Cody}, {Covey}, {Rizzuto}, {Mann}, \&
  {Ireland}}]{Kraus2015}
{Kraus}, A.~L., {Cody}, A.~M., {Covey}, K.~R., {et~al.} 2015, \apj, 807, 3

\bibitem[{{Kraus} \& {Hillenbrand}(2007)}]{Kraus2007}
{Kraus}, A.~L., \& {Hillenbrand}, L.~A. 2007, \aj, 134, 2340

\bibitem[{{Kraus} {et~al.}(2011){Kraus}, {Tucker}, {Thompson}, {Craine}, \&
  {Hillenbrand}}]{Kraus2011}
{Kraus}, A.~L., {Tucker}, R.~A., {Thompson}, M.~I., {Craine}, E.~R., \&
  {Hillenbrand}, L.~A. 2011, \apj, 728, 48

\bibitem[{{Kroupa}(2001)}]{Kroupa2001}
{Kroupa}, P. 2001, \mnras, 322, 231

\bibitem[{{Kurucz}(1970)}]{Kurucz1970}
{Kurucz}, R.~L. 1970, SAO Special Report, 309

\bibitem[{{Kurucz}(1993)}]{Kurucz1993}
---. 1993, {SYNTHE spectrum synthesis programs and line data}

\bibitem[{{Lafreni{\`e}re} {et~al.}(2008){Lafreni{\`e}re}, {Jayawardhana}, \&
  {van Kerkwijk}}]{Lafreniere2008}
{Lafreni{\`e}re}, D., {Jayawardhana}, R., \& {van Kerkwijk}, M.~H. 2008, \apjl,
  689, L153

\bibitem[{{Langer}(1991)}]{Langer1991}
{Langer}, N. 1991, \aap, 252, 669

\bibitem[{{Langer}(1998)}]{Langer1998}
---. 1998, \aap, 329, 551

\bibitem[{{Langer}(2012)}]{Langer2012}
---. 2012, \araa, 50, 107

\bibitem[{{Langer} {et~al.}(1989){Langer}, {El Eid}, \& {Baraffe}}]{Langer1989}
{Langer}, N., {El Eid}, M.~F., \& {Baraffe}, I. 1989, \aap, 224, L17

\bibitem[{{Langer} {et~al.}(1985){Langer}, {El Eid}, \& {Fricke}}]{Langer1985}
{Langer}, N., {El Eid}, M.~F., \& {Fricke}, K.~J. 1985, \aap, 145, 179

\bibitem[{{Langer} {et~al.}(1983){Langer}, {Fricke}, \&
  {Sugimoto}}]{Langer1983}
{Langer}, N., {Fricke}, K.~J., \& {Sugimoto}, D. 1983, \aap, 126, 207

\bibitem[{{Langer} \& {Kudritzki}(2014)}]{Langer2014}
{Langer}, N., \& {Kudritzki}, R.~P. 2014, \aap, 564, A52

\bibitem[{{Langer} \& {Maeder}(1995)}]{Langer1995}
{Langer}, N., \& {Maeder}, A. 1995, \aap, 295, 685

\bibitem[{{Lattanzio}(2007)}]{Lattanzio2007}
{Lattanzio}, J.~C. 2007, in Astronomical Society of the Pacific Conference
  Series, Vol. 378, Why Galaxies Care About AGB Stars: Their Importance as
  Actors and Probes, ed. F.~{Kerschbaum}, C.~{Charbonnel}, \& R.~F. {Wing}, 3

\bibitem[{{Lau} {et~al.}(2009){Lau}, {Stancliffe}, \& {Tout}}]{Lau2009}
{Lau}, H.~H.~B., {Stancliffe}, R.~J., \& {Tout}, C.~A. 2009, \mnras, 396, 1046

\bibitem[{{Law} {et~al.}(2009){Law}, {Kulkarni}, {Dekany}, {Ofek}, {Quimby},
  {Nugent}, {Surace}, {Grillmair}, {Bloom}, {Kasliwal}, {Bildsten}, {Brown},
  {Cenko}, {Ciardi}, {Croner}, {Djorgovski}, {van Eyken}, {Filippenko}, {Fox},
  {Gal-Yam}, {Hale}, {Hamam}, {Helou}, {Henning}, {Howell}, {Jacobsen},
  {Laher}, {Mattingly}, {McKenna}, {Pickles}, {Poznanski}, {Rahmer}, {Rau},
  {Rosing}, {Shara}, {Smith}, {Starr}, {Sullivan}, {Velur}, {Walters}, \&
  {Zolkower}}]{Law2009}
{Law}, N.~M., {Kulkarni}, S.~R., {Dekany}, R.~G., {et~al.} 2009, \pasp, 121,
  1395

\bibitem[{{Levesque} {et~al.}(2005){Levesque}, {Massey}, {Olsen}, {Plez},
  {Josselin}, {Maeder}, \& {Meynet}}]{Levesque2005}
{Levesque}, E.~M., {Massey}, P., {Olsen}, K.~A.~G., {et~al.} 2005, \apj, 628,
  973

\bibitem[{{Levesque} {et~al.}(2006){Levesque}, {Massey}, {Olsen}, {Plez},
  {Meynet}, \& {Maeder}}]{Levesque2006}
---. 2006, \apj, 645, 1102

\bibitem[{{Liebert} {et~al.}(2005){Liebert}, {Bergeron}, \&
  {Holberg}}]{Liebert2005}
{Liebert}, J., {Bergeron}, P., \& {Holberg}, J.~B. 2005, \apjs, 156, 47

\bibitem[{{Lodders} {et~al.}(2009){Lodders}, {Palme}, \& {Gail}}]{Lodders2009}
{Lodders}, K., {Palme}, H., \& {Gail}, H.-P. 2009, Landolt B{\"o}rnstein, 44

\bibitem[{{Lodieu}(2013)}]{Lodieu2013}
{Lodieu}, N. 2013, \mnras, 431, 3222

\bibitem[{{Lodieu} {et~al.}(2008){Lodieu}, {Hambly}, {Jameson}, \&
  {Hodgkin}}]{Lodieu2008}
{Lodieu}, N., {Hambly}, N.~C., {Jameson}, R.~F., \& {Hodgkin}, S.~T. 2008,
  \mnras, 383, 1385

\bibitem[{{Lucy} \& {Solomon}(1970)}]{Lucy1970}
{Lucy}, L.~B., \& {Solomon}, P.~M. 1970, \apj, 159, 879

\bibitem[{{MacDonald} \& {Mullan}(2012)}]{MacDonald2012}
{MacDonald}, J., \& {Mullan}, D.~J. 2012, \mnras, 421, 3084

\bibitem[{{MacDonald} \& {Mullan}(2014)}]{MacDonald2014}
---. 2014, \apj, 787, 70

\bibitem[{{Maeder}(1975)}]{Maeder1975}
{Maeder}, A. 1975, \aap, 40, 303

\bibitem[{{Maeder} \& {Conti}(1994)}]{Maeder1994}
{Maeder}, A., \& {Conti}, P.~S. 1994, \araa, 32, 227

\bibitem[{{Maeder} \& {Meynet}(1994)}]{Maeder1994b}
{Maeder}, A., \& {Meynet}, G. 1994, \aap, 287, 803

\bibitem[{{Maeder} \& {Meynet}(2000)}]{Maeder2000}
---. 2000, \araa, 38, 143

\bibitem[{{Maeder} \& {Meynet}(2001)}]{Maeder2001}
---. 2001, \aap, 373, 555

\bibitem[{{Maeder} \& {Meynet}(2012)}]{Maeder2012}
---. 2012, Reviews of Modern Physics, 84, 25

\bibitem[{{Magic} {et~al.}(2010){Magic}, {Serenelli}, {Weiss}, \&
  {Chaboyer}}]{Magic2010}
{Magic}, Z., {Serenelli}, A., {Weiss}, A., \& {Chaboyer}, B. 2010, \apj, 718,
  1378

\bibitem[{{Magic} {et~al.}(2015){Magic}, {Weiss}, \& {Asplund}}]{Magic2015}
{Magic}, Z., {Weiss}, A., \& {Asplund}, M. 2015, \aap, 573, A89

\bibitem[{{Maraston}(2005)}]{Maraston2005}
{Maraston}, C. 2005, \mnras, 362, 799

\bibitem[{{Marigo}(2002)}]{Marigo2002}
{Marigo}, P. 2002, \aap, 387, 507

\bibitem[{{Marigo} {et~al.}(2013){Marigo}, {Bressan}, {Nanni}, {Girardi}, \&
  {Pumo}}]{Marigo2013}
{Marigo}, P., {Bressan}, A., {Nanni}, A., {Girardi}, L., \& {Pumo}, M.~L. 2013,
  \mnras, 434, 488

\bibitem[{{Marigo} \& {Girardi}(2001)}]{Marigo2001}
{Marigo}, P., \& {Girardi}, L. 2001, \aap, 377, 132

\bibitem[{{Marigo} \& {Girardi}(2007)}]{Marigo2007}
---. 2007, \aap, 469, 239

\bibitem[{{Marigo} {et~al.}(1999){Marigo}, {Girardi}, \&
  {Bressan}}]{Marigo1999}
{Marigo}, P., {Girardi}, L., \& {Bressan}, A. 1999, \aap, 344, 123

\bibitem[{{Marigo} {et~al.}(2008){Marigo}, {Girardi}, {Bressan}, {Groenewegen},
  {Silva}, \& {Granato}}]{Marigo2008}
{Marigo}, P., {Girardi}, L., {Bressan}, A., {et~al.} 2008, \aap, 482, 883

\bibitem[{{Marigo} {et~al.}(2003){Marigo}, {Girardi}, \& {Chiosi}}]{Marigo2003}
{Marigo}, P., {Girardi}, L., \& {Chiosi}, C. 2003, \aap, 403, 225

\bibitem[{{Massey}(2002)}]{Massey2002}
{Massey}, P. 2002, \apjs, 141, 81

\bibitem[{{Massey} \& {Olsen}(2003)}]{Massey2003}
{Massey}, P., \& {Olsen}, K.~A.~G. 2003, \aj, 126, 2867

\bibitem[{{Massey} {et~al.}(2009){Massey}, {Zangari}, {Morrell}, {Puls},
  {DeGioia-Eastwood}, {Bresolin}, \& {Kudritzki}}]{Massey2009}
{Massey}, P., {Zangari}, A.~M., {Morrell}, N.~I., {et~al.} 2009, \apj, 692, 618

\bibitem[{{Mauron} \& {Josselin}(2011)}]{Mauron2011}
{Mauron}, N., \& {Josselin}, E. 2011, \aap, 526, A156

\bibitem[{{McQuinn} {et~al.}(2011){McQuinn}, {Skillman}, {Dalcanton},
  {Dolphin}, {Holtzman}, {Weisz}, \& {Williams}}]{McQuinn2011}
{McQuinn}, K.~B.~W., {Skillman}, E.~D., {Dalcanton}, J.~J., {et~al.} 2011,
  \apj, 740, 48

\bibitem[{{Meixner} {et~al.}(2006){Meixner}, {Gordon}, {Indebetouw}, {Hora},
  {Whitney}, {Blum}, {Reach}, {Bernard}, {Meade}, {Babler}, {Engelbracht},
  {For}, {Misselt}, {Vijh}, {Leitherer}, {Cohen}, {Churchwell}, {Boulanger},
  {Frogel}, {Fukui}, {Gallagher}, {Gorjian}, {Harris}, {Kelly}, {Kawamura},
  {Kim}, {Latter}, {Madden}, {Markwick-Kemper}, {Mizuno}, {Mizuno}, {Mould},
  {Nota}, {Oey}, {Olsen}, {Onishi}, {Paladini}, {Panagia}, {Perez-Gonzalez},
  {Shibai}, {Sato}, {Smith}, {Staveley-Smith}, {Tielens}, {Ueta}, {van Dyk},
  {Volk}, {Werner}, \& {Zaritsky}}]{Meixner2006}
{Meixner}, M., {Gordon}, K.~D., {Indebetouw}, R., {et~al.} 2006, \aj, 132, 2268

\bibitem[{{Melbourne} {et~al.}(2012){Melbourne}, {Williams}, {Dalcanton},
  {Rosenfield}, {Girardi}, {Marigo}, {Weisz}, {Dolphin}, {Boyer}, {Olsen},
  {Skillman}, \& {Seth}}]{Melbourne2012}
{Melbourne}, J., {Williams}, B.~F., {Dalcanton}, J.~J., {et~al.} 2012, \apj,
  748, 47

\bibitem[{{Melis} {et~al.}(2014){Melis}, {Reid}, {Mioduszewski}, {Stauffer}, \&
  {Bower}}]{Melis2014}
{Melis}, C., {Reid}, M.~J., {Mioduszewski}, A.~J., {Stauffer}, J.~R., \&
  {Bower}, G.~C. 2014, Science, 345, 1029

\bibitem[{{Mestel}(1968)}]{Mestel1968}
{Mestel}, L. 1968, \mnras, 138, 359

\bibitem[{{Mestel} \& {Weiss}(1987)}]{Mestel1987}
{Mestel}, L., \& {Weiss}, N.~O. 1987, \mnras, 226, 123

\bibitem[{{Meylan} \& {Maeder}(1982)}]{Meylan1982}
{Meylan}, G., \& {Maeder}, A. 1982, \aap, 108, 148

\bibitem[{{Meynet} {et~al.}(2011){Meynet}, {Eggenberger}, \&
  {Maeder}}]{Meynet2011}
{Meynet}, G., {Eggenberger}, P., \& {Maeder}, A. 2011, \aap, 525, L11

\bibitem[{{Meynet} \& {Maeder}(1997)}]{Meynet1997}
{Meynet}, G., \& {Maeder}, A. 1997, \aap, 321, 465

\bibitem[{{Meynet} \& {Maeder}(2000)}]{Meynet2000}
---. 2000, \aap, 361, 101

\bibitem[{{Meynet} \& {Maeder}(2005)}]{Meynet2005}
---. 2005, \aap, 429, 581

\bibitem[{{Michaud} {et~al.}(1984){Michaud}, {Fontaine}, \&
  {Beaudet}}]{Michaud1984}
{Michaud}, G., {Fontaine}, G., \& {Beaudet}, G. 1984, \apj, 282, 206

\bibitem[{{Michaud} {et~al.}(2004){Michaud}, {Richard}, {Richer}, \&
  {VandenBerg}}]{Michaud2004}
{Michaud}, G., {Richard}, O., {Richer}, J., \& {VandenBerg}, D.~A. 2004, \apj,
  606, 452

\bibitem[{{Miglio} {et~al.}(2012){Miglio}, {Brogaard}, {Stello}, {Chaplin},
  {D'Antona}, {Montalb{\'a}n}, {Basu}, {Bressan}, {Grundahl}, {Pinsonneault},
  {Serenelli}, {Elsworth}, {Hekker}, {Kallinger}, {Mosser}, {Ventura},
  {Bonanno}, {Noels}, {Silva Aguirre}, {Szabo}, {Li}, {McCauliff}, {Middour},
  \& {Kjeldsen}}]{Miglio2012}
{Miglio}, A., {Brogaard}, K., {Stello}, D., {et~al.} 2012, \mnras, 419, 2077

\bibitem[{{Miller Bertolami}(2016)}]{MillerBertolami2016}
{Miller Bertolami}, M.~M. 2016, \aap, 588, A25

\bibitem[{{Morales} {et~al.}(2010){Morales}, {Gallardo}, {Ribas}, {Jordi},
  {Baraffe}, \& {Chabrier}}]{Morales2010}
{Morales}, J.~C., {Gallardo}, J., {Ribas}, I., {et~al.} 2010, \apj, 718, 502

\bibitem[{{Morales} {et~al.}(2008){Morales}, {Ribas}, \& {Jordi}}]{Morales2008}
{Morales}, J.~C., {Ribas}, I., \& {Jordi}, C. 2008, \aap, 478, 507

\bibitem[{{Morel} \& {Baglin}(1999)}]{Morel1999}
{Morel}, P., \& {Baglin}, A. 1999, \aap, 345, 156

\bibitem[{{Morel} \& {Th{\'e}venin}(2002)}]{Morel2002}
{Morel}, P., \& {Th{\'e}venin}, F. 2002, \aap, 390, 611

\bibitem[{{Morel} {et~al.}(2008){Morel}, {Hubrig}, \& {Briquet}}]{Morel2008}
{Morel}, T., {Hubrig}, S., \& {Briquet}, M. 2008, \aap, 481, 453

\bibitem[{{Neugent} \& {Massey}(2011)}]{Neugent2011}
{Neugent}, K.~F., \& {Massey}, P. 2011, \apj, 733, 123

\bibitem[{{Neugent} {et~al.}(2012){Neugent}, {Massey}, \&
  {Georgy}}]{Neugent2012_wr}
{Neugent}, K.~F., {Massey}, P., \& {Georgy}, C. 2012, \apj, 759, 11

\bibitem[{{Nieuwenhuijzen} \& {de Jager}(1988)}]{Nieuwenhuijzen1988}
{Nieuwenhuijzen}, H., \& {de Jager}, C. 1988, \aap, 203, 355

\bibitem[{{Nieuwenhuijzen} \& {de Jager}(1990)}]{Nieuwenhuijzen1990}
---. 1990, \aap, 231, 134

\bibitem[{{No{\"e}l} {et~al.}(2013){No{\"e}l}, {Greggio}, {Renzini}, {Carollo},
  \& {Maraston}}]{Noel2013}
{No{\"e}l}, N.~E.~D., {Greggio}, L., {Renzini}, A., {Carollo}, C.~M., \&
  {Maraston}, C. 2013, \apj, 772, 58

\bibitem[{{Nugis} \& {Lamers}(2000)}]{Nugis2000}
{Nugis}, T., \& {Lamers}, H.~J.~G.~L.~M. 2000, \aap, 360, 227

\bibitem[{{Origlia} {et~al.}(2006){Origlia}, {Valenti}, {Rich}, \&
  {Ferraro}}]{Origlia2006}
{Origlia}, L., {Valenti}, E., {Rich}, R.~M., \& {Ferraro}, F.~R. 2006, \apj,
  646, 499

\bibitem[{{Paczy{\'n}ski}(1970)}]{Paczynski1970}
{Paczy{\'n}ski}, B. 1970, \actaa, 20, 47

\bibitem[{{Palacios} {et~al.}(2003){Palacios}, {Talon}, {Charbonnel}, \&
  {Forestini}}]{Palacios2003}
{Palacios}, A., {Talon}, S., {Charbonnel}, C., \& {Forestini}, M. 2003, \aap,
  399, 603

\bibitem[{{Paquette} {et~al.}(1986){Paquette}, {Pelletier}, {Fontaine}, \&
  {Michaud}}]{Paquette1986}
{Paquette}, C., {Pelletier}, C., {Fontaine}, G., \& {Michaud}, G. 1986, \apjs,
  61, 177

\bibitem[{{Paxton} {et~al.}(2011){Paxton}, {Bildsten}, {Dotter}, {Herwig},
  {Lesaffre}, \& {Timmes}}]{Paxton2011}
{Paxton}, B., {Bildsten}, L., {Dotter}, A., {et~al.} 2011, \apjs, 192, 3

\bibitem[{{Paxton} {et~al.}(2013){Paxton}, {Cantiello}, {Arras}, {Bildsten},
  {Brown}, {Dotter}, {Mankovich}, {Montgomery}, {Stello}, {Timmes}, \&
  {Townsend}}]{Paxton2013}
{Paxton}, B., {Cantiello}, M., {Arras}, P., {et~al.} 2013, \apjs, 208, 4

\bibitem[{{Paxton} {et~al.}(2015){Paxton}, {Marchant}, {Schwab}, {Bauer},
  {Bildsten}, {Cantiello}, {Dessart}, {Farmer}, {Hu}, {Langer}, {Townsend},
  {Townsley}, \& {Timmes}}]{Paxton2015}
{Paxton}, B., {Marchant}, P., {Schwab}, J., {et~al.} 2015, \apjs, 220, 15

\bibitem[{{Pecaut} {et~al.}(2012){Pecaut}, {Mamajek}, \& {Bubar}}]{Pecaut2012}
{Pecaut}, M.~J., {Mamajek}, E.~E., \& {Bubar}, E.~J. 2012, \apj, 746, 154

\bibitem[{{Petrovic} {et~al.}(2005){Petrovic}, {Langer}, {Yoon}, \&
  {Heger}}]{Petrovic2005}
{Petrovic}, J., {Langer}, N., {Yoon}, S.-C., \& {Heger}, A. 2005, \aap, 435,
  247

\bibitem[{{Pietrinferni} {et~al.}(2004){Pietrinferni}, {Cassisi}, {Salaris}, \&
  {Castelli}}]{Pietrinferni2004}
{Pietrinferni}, A., {Cassisi}, S., {Salaris}, M., \& {Castelli}, F. 2004, \apj,
  612, 168

\bibitem[{{Pinsonneault}(1997)}]{Pinsonneault1997}
{Pinsonneault}, M. 1997, \araa, 35, 557

\bibitem[{{Pinsonneault} {et~al.}(1989){Pinsonneault}, {Kawaler}, {Sofia}, \&
  {Demarque}}]{Pinsonneault1989}
{Pinsonneault}, M.~H., {Kawaler}, S.~D., {Sofia}, S., \& {Demarque}, P. 1989,
  \apj, 338, 424

\bibitem[{{Piotto} {et~al.}(2012){Piotto}, {Milone}, {Anderson}, {Bedin},
  {Bellini}, {Cassisi}, {Marino}, {Aparicio}, \& {Nascimbeni}}]{Piotto2012}
{Piotto}, G., {Milone}, A.~P., {Anderson}, J., {et~al.} 2012, \apj, 760, 39

\bibitem[{{Planck Collaboration} {et~al.}(2015){Planck Collaboration}, {Ade},
  {Aghanim}, {Arnaud}, {Ashdown}, {Aumont}, {Baccigalupi}, {Banday},
  {Barreiro}, {Bartlett}, \& et~al.}]{Planck2015}
{Planck Collaboration}, {Ade}, P.~A.~R., {Aghanim}, N., {et~al.} 2015, ArXiv
  e-prints, arXiv:1502.01589

\bibitem[{{Pols} {et~al.}(1995){Pols}, {Tout}, {Eggleton}, \& {Han}}]{Pols1995}
{Pols}, O.~R., {Tout}, C.~A., {Eggleton}, P.~P., \& {Han}, Z. 1995, \mnras,
  274, 964

\bibitem[{{Potekhin} \& {Chabrier}(2010)}]{Potekhin2010}
{Potekhin}, A.~Y., \& {Chabrier}, G. 2010, Contributions to Plasma Physics, 50,
  82

\bibitem[{{Potter} {et~al.}(2012){Potter}, {Tout}, \& {Eldridge}}]{Potter2012a}
{Potter}, A.~T., {Tout}, C.~A., \& {Eldridge}, J.~J. 2012, \mnras, 419, 748

\bibitem[{{Prada Moroni} \& {Straniero}(2002)}]{Prada2002}
{Prada Moroni}, P.~G., \& {Straniero}, O. 2002, \apj, 581, 585

\bibitem[{{Preibisch} {et~al.}(2002){Preibisch}, {Brown}, {Bridges},
  {Guenther}, \& {Zinnecker}}]{Preibisch2002}
{Preibisch}, T., {Brown}, A.~G.~A., {Bridges}, T., {Guenther}, E., \&
  {Zinnecker}, H. 2002, \aj, 124, 404

\bibitem[{{Preibisch} \& {Zinnecker}(1999)}]{Preibisch1999}
{Preibisch}, T., \& {Zinnecker}, H. 1999, \aj, 117, 2381

\bibitem[{{Przybilla} {et~al.}(2010){Przybilla}, {Firnstein}, {Nieva},
  {Meynet}, \& {Maeder}}]{Przybilla2010}
{Przybilla}, N., {Firnstein}, M., {Nieva}, M.~F., {Meynet}, G., \& {Maeder}, A.
  2010, \aap, 517, A38

\bibitem[{{Rauscher} {et~al.}(2002){Rauscher}, {Heger}, {Hoffman}, \&
  {Woosley}}]{Rauscher2002}
{Rauscher}, T., {Heger}, A., {Hoffman}, R.~D., \& {Woosley}, S.~E. 2002, \apj,
  576, 323

\bibitem[{{Reimers}(1975)}]{Reimers1975}
{Reimers}, D. 1975, Memoires of the Societe Royale des Sciences de Liege, 8,
  369

\bibitem[{{Renzini} \& {Fusi Pecci}(1988)}]{Renzini1988}
{Renzini}, A., \& {Fusi Pecci}, F. 1988, \araa, 26, 199

\bibitem[{{Rhodes} {et~al.}(1997){Rhodes}, {Kosovichev}, {Schou}, {Scherrer},
  \& {Reiter}}]{Rhodes1997}
{Rhodes}, Jr., E.~J., {Kosovichev}, A.~G., {Schou}, J., {Scherrer}, P.~H., \&
  {Reiter}, J. 1997, \solphys, 175, 287

\bibitem[{{Richard} {et~al.}(1996){Richard}, {Vauclair}, {Charbonnel}, \&
  {Dziembowski}}]{Richard1996}
{Richard}, O., {Vauclair}, S., {Charbonnel}, C., \& {Dziembowski}, W.~A. 1996,
  \aap, 312, 1000

\bibitem[{{Rivero Gonz{\'a}lez} {et~al.}(2012){Rivero Gonz{\'a}lez}, {Puls},
  {Najarro}, \& {Brott}}]{RiveroGonzalez2012}
{Rivero Gonz{\'a}lez}, J.~G., {Puls}, J., {Najarro}, F., \& {Brott}, I. 2012,
  \aap, 537, A79

\bibitem[{{Rogers} \& {Nayfonov}(2002)}]{Rogers2002}
{Rogers}, F.~J., \& {Nayfonov}, A. 2002, \apj, 576, 1064

\bibitem[{{Romero} {et~al.}(2015){Romero}, {Campos}, \& {Kepler}}]{Romero2015}
{Romero}, A.~D., {Campos}, F., \& {Kepler}, S.~O. 2015, \mnras, 450, 3708

\bibitem[{{Rosen} {et~al.}(2012){Rosen}, {Krumholz}, \&
  {Ramirez-Ruiz}}]{Rosen2012}
{Rosen}, A.~L., {Krumholz}, M.~R., \& {Ramirez-Ruiz}, E. 2012, \apj, 748, 97

\bibitem[{{Rosenfield} {et~al.}(2014){Rosenfield}, {Marigo}, {Girardi},
  {Dalcanton}, {Bressan}, {Gullieuszik}, {Weisz}, {Williams}, {Dolphin}, \&
  {Aringer}}]{Rosenfield2014}
{Rosenfield}, P., {Marigo}, P., {Girardi}, L., {et~al.} 2014, \apj, 790, 22

\bibitem[{{Roxburgh}(1978)}]{Roxburgh1978}
{Roxburgh}, I.~W. 1978, \aap, 65, 281

\bibitem[{{R{\"u}diger} {et~al.}(2015){R{\"u}diger}, {Gellert}, {Spada}, \&
  {Tereshin}}]{Rudiger2015}
{R{\"u}diger}, G., {Gellert}, M., {Spada}, F., \& {Tereshin}, I. 2015, \aap,
  573, A80

\bibitem[{{Salaris} \& {Cassisi}(2015)}]{Salaris2015}
{Salaris}, M., \& {Cassisi}, S. 2015, \aap, 577, A60

\bibitem[{{Salaris} {et~al.}(2000){Salaris}, {Groenewegen}, \&
  {Weiss}}]{Salaris2000}
{Salaris}, M., {Groenewegen}, M.~A.~T., \& {Weiss}, A. 2000, \aap, 355, 299

\bibitem[{{Sana} {et~al.}(2012){Sana}, {de Mink}, {de Koter}, {Langer},
  {Evans}, {Gieles}, {Gosset}, {Izzard}, {Le Bouquin}, \&
  {Schneider}}]{Sana2012}
{Sana}, H., {de Mink}, S.~E., {de Koter}, A., {et~al.} 2012, Science, 337, 444

\bibitem[{{Sana} {et~al.}(2013){Sana}, {de Koter}, {de Mink}, {Dunstall},
  {Evans}, {H{\'e}nault-Brunet}, {Ma{\'{\i}}z Apell{\'a}niz},
  {Ram{\'{\i}}rez-Agudelo}, {Taylor}, {Walborn}, {Clark}, {Crowther},
  {Herrero}, {Gieles}, {Langer}, {Lennon}, \& {Vink}}]{Sana2013}
{Sana}, H., {de Koter}, A., {de Mink}, S.~E., {et~al.} 2013, \aap, 550, A107

\bibitem[{{Sanders} {et~al.}(2012){Sanders}, {Caldwell}, {McDowell}, \&
  {Harding}}]{Sanders2012}
{Sanders}, N.~E., {Caldwell}, N., {McDowell}, J., \& {Harding}, P. 2012, \apj,
  758, 133

\bibitem[{{Sandquist}(2004)}]{Sandquist2004}
{Sandquist}, E.~L. 2004, \mnras, 347, 101

\bibitem[{{Sarajedini} {et~al.}(2009){Sarajedini}, {Dotter}, \&
  {Kirkpatrick}}]{Sarajedini2009}
{Sarajedini}, A., {Dotter}, A., \& {Kirkpatrick}, A. 2009, \apj, 698, 1872

\bibitem[{{Saumon} {et~al.}(1995){Saumon}, {Chabrier}, \& {van
  Horn}}]{Saumon1995}
{Saumon}, D., {Chabrier}, G., \& {van Horn}, H.~M. 1995, \apjs, 99, 713

\bibitem[{{Schaller} {et~al.}(1992){Schaller}, {Schaerer}, {Meynet}, \&
  {Maeder}}]{Schaller1992}
{Schaller}, G., {Schaerer}, D., {Meynet}, G., \& {Maeder}, A. 1992, \aaps, 96,
  269

\bibitem[{{Schoenberner}(1979)}]{Schoenberner1979}
{Schoenberner}, D. 1979, \aap, 79, 108

\bibitem[{{Schwarzschild}(1958)}]{Schwarzschild1958}
{Schwarzschild}, M. 1958, {Structure and evolution of the stars.}

\bibitem[{{Schwarzschild} \& {H{\"a}rm}(1962)}]{Schwarzschild1962}
{Schwarzschild}, M., \& {H{\"a}rm}, R. 1962, \apj, 136, 158

\bibitem[{{Schwarzschild} \& {H{\"a}rm}(1965)}]{Schwarzschild1965}
---. 1965, \apj, 142, 855

\bibitem[{{Searle} {et~al.}(2008){Searle}, {Prinja}, {Massa}, \&
  {Ryans}}]{Searle2008}
{Searle}, S.~C., {Prinja}, R.~K., {Massa}, D., \& {Ryans}, R. 2008, \aap, 481,
  777

\bibitem[{{Seaton}(2005)}]{Seaton2005}
{Seaton}, M.~J. 2005, \mnras, 362, L1

\bibitem[{{Serenelli} {et~al.}(2009){Serenelli}, {Basu}, {Ferguson}, \&
  {Asplund}}]{Serenelli2009}
{Serenelli}, A.~M., {Basu}, S., {Ferguson}, J.~W., \& {Asplund}, M. 2009,
  \apjl, 705, L123

\bibitem[{{Sestito} \& {Randich}(2005)}]{Sestito2005}
{Sestito}, P., \& {Randich}, S. 2005, \aap, 442, 615

\bibitem[{{Shaviv} \& {Salpeter}(1973)}]{Shaviv1973}
{Shaviv}, G., \& {Salpeter}, E.~E. 1973, \apj, 184, 191

\bibitem[{{Siess}(2009)}]{Siess2009}
{Siess}, L. 2009, \aap, 497, 463

\bibitem[{{Siess} {et~al.}(2000){Siess}, {Dufour}, \& {Forestini}}]{Siess2000}
{Siess}, L., {Dufour}, E., \& {Forestini}, M. 2000, \aap, 358, 593

\bibitem[{{Silva Aguirre} {et~al.}(2011){Silva Aguirre}, {Ballot}, {Serenelli},
  \& {Weiss}}]{Silva2011}
{Silva Aguirre}, V., {Ballot}, J., {Serenelli}, A.~M., \& {Weiss}, A. 2011,
  \aap, 529, A63

\bibitem[{{Slesnick} {et~al.}(2008){Slesnick}, {Hillenbrand}, \&
  {Carpenter}}]{Slesnick2008}
{Slesnick}, C.~L., {Hillenbrand}, L.~A., \& {Carpenter}, J.~M. 2008, \apj, 688,
  377

\bibitem[{{Smartt}(2009)}]{Smartt2009}
{Smartt}, S.~J. 2009, \araa, 47, 63

\bibitem[{{Smith}(2014)}]{Smith2014}
{Smith}, N. 2014, \araa, 52, 487

\bibitem[{{Smith} \& {Owocki}(2006)}]{Smith2006}
{Smith}, N., \& {Owocki}, S.~P. 2006, \apjl, 645, L45

\bibitem[{{Soderblom} {et~al.}(2009){Soderblom}, {Laskar}, {Valenti},
  {Stauffer}, \& {Rebull}}]{Soderblom2009}
{Soderblom}, D.~R., {Laskar}, T., {Valenti}, J.~A., {Stauffer}, J.~R., \&
  {Rebull}, L.~M. 2009, \aj, 138, 1292

\bibitem[{{Soderblom} {et~al.}(2005){Soderblom}, {Nelan}, {Benedict},
  {McArthur}, {Ramirez}, {Spiesman}, \& {Jones}}]{Soderblom2005}
{Soderblom}, D.~R., {Nelan}, E., {Benedict}, G.~F., {et~al.} 2005, \aj, 129,
  1616

\bibitem[{S{\"o}derlind \& Wang(2006)}]{Soederlind2006}
S{\"o}derlind, G., \& Wang, L. 2006, Journal of Computational and Applied
  Mathematics, 185, 244 , special Issue: International Workshop on the
  Technological Aspects of Mathematics

\bibitem[{{Somers} \& {Pinsonneault}(2014)}]{Somers2014}
{Somers}, G., \& {Pinsonneault}, M.~H. 2014, \apj, 790, 72

\bibitem[{{Song} {et~al.}(2012){Song}, {Zuckerman}, \& {Bessell}}]{Song2012}
{Song}, I., {Zuckerman}, B., \& {Bessell}, M.~S. 2012, \aj, 144, 8

\bibitem[{{Southworth}(2015)}]{Southworth2015}
{Southworth}, J. 2015, in Astronomical Society of the Pacific Conference
  Series, Vol. 496, Living Together: Planets, Host Stars and Binaries, ed.
  S.~M. {Rucinski}, G.~{Torres}, \& M.~{Zejda}, 164

\bibitem[{{Spada} {et~al.}(2013){Spada}, {Demarque}, {Kim}, \&
  {Sills}}]{Spada2013}
{Spada}, F., {Demarque}, P., {Kim}, Y.-C., \& {Sills}, A. 2013, \apj, 776, 87

\bibitem[{{Spruit}(2002)}]{Spruit2002}
{Spruit}, H.~C. 2002, \aap, 381, 923

\bibitem[{{Spruit}(2013)}]{Spruit2013}
---. 2013, \aap, 552, A76

\bibitem[{{Spruit} \& {Weiss}(1986)}]{Spruit1986}
{Spruit}, H.~C., \& {Weiss}, A. 1986, \aap, 166, 167

\bibitem[{{Stancliffe}(2010)}]{Stancliffe2010}
{Stancliffe}, R.~J. 2010, \mnras, 403, 505

\bibitem[{{Stancliffe} {et~al.}(2011){Stancliffe}, {Dearborn}, {Lattanzio},
  {Heap}, \& {Campbell}}]{Stancliffe2011}
{Stancliffe}, R.~J., {Dearborn}, D.~S.~P., {Lattanzio}, J.~C., {Heap}, S.~A.,
  \& {Campbell}, S.~W. 2011, \apj, 742, 121

\bibitem[{{Stancliffe} {et~al.}(2007){Stancliffe}, {Glebbeek}, {Izzard}, \&
  {Pols}}]{Stancliffe2007}
{Stancliffe}, R.~J., {Glebbeek}, E., {Izzard}, R.~G., \& {Pols}, O.~R. 2007,
  \aap, 464, L57

\bibitem[{{Stauffer} {et~al.}(2003){Stauffer}, {Jones}, {Backman}, {Hartmann},
  {Barrado y Navascu{\'e}s}, {Pinsonneault}, {Terndrup}, \&
  {Muench}}]{Stauffer2003}
{Stauffer}, J.~R., {Jones}, B.~F., {Backman}, D., {et~al.} 2003, \aj, 126, 833

\bibitem[{{Stauffer} {et~al.}(2007){Stauffer}, {Hartmann}, {Fazio}, {Allen},
  {Patten}, {Lowrance}, {Hurt}, {Rebull}, {Cutri}, {Ramirez}, {Young}, {Rieke},
  {Gorlova}, {Muzerolle}, {Slesnick}, \& {Skrutskie}}]{Stauffer2007}
{Stauffer}, J.~R., {Hartmann}, L.~W., {Fazio}, G.~G., {et~al.} 2007, \apjs,
  172, 663

\bibitem[{{Stetson} {et~al.}(2003){Stetson}, {Bruntt}, \&
  {Grundahl}}]{Stetson2003}
{Stetson}, P.~B., {Bruntt}, H., \& {Grundahl}, F. 2003, \pasp, 115, 413

\bibitem[{{Stothers} \& {Chin}(1975)}]{Stothers1975}
{Stothers}, R., \& {Chin}, C.-W. 1975, \apj, 198, 407

\bibitem[{{Stothers} \& {Simon}(1969)}]{Stothers1969}
{Stothers}, R., \& {Simon}, N.~R. 1969, \apj, 157, 673

\bibitem[{{Stothers} \& {Chin}(1991)}]{Stothers1991}
{Stothers}, R.~B., \& {Chin}, C.-W. 1991, \apjl, 381, L67

\bibitem[{{Strittmatter}(1969)}]{Strittmatter1969}
{Strittmatter}, P.~A. 1969, \araa, 7, 665

\bibitem[{{Suijs} {et~al.}(2008){Suijs}, {Langer}, {Poelarends}, {Yoon},
  {Heger}, \& {Herwig}}]{Suijs2008}
{Suijs}, M.~P.~L., {Langer}, N., {Poelarends}, A.-J., {et~al.} 2008, \aap, 481,
  L87

\bibitem[{{Sukhbold} \& {Woosley}(2014)}]{Sukhbold2014}
{Sukhbold}, T., \& {Woosley}, S.~E. 2014, \apj, 783, 10

\bibitem[{{Takeda} \& {Takada-Hidai}(2000)}]{Takeda2000}
{Takeda}, Y., \& {Takada-Hidai}, M. 2000, \pasj, 52, 113

\bibitem[{{Talon} \& {Charbonnel}(2005)}]{Talon2005}
{Talon}, S., \& {Charbonnel}, C. 2005, \aap, 440, 981

\bibitem[{{Tang} {et~al.}(2014){Tang}, {Bressan}, {Rosenfield}, {Slemer},
  {Marigo}, {Girardi}, \& {Bianchi}}]{Tang2014}
{Tang}, J., {Bressan}, A., {Rosenfield}, P., {et~al.} 2014, \mnras, 445, 4287

\bibitem[{{Tang} {et~al.}(2016){Tang}, {Bressan}, {Slemer}, {Marigo},
  {Girardi}, {Bianchi}, {Rosenfield}, \& {Momany}}]{Tang2016}
{Tang}, J., {Bressan}, A., {Slemer}, A., {et~al.} 2016, \mnras, 455, 3393

\bibitem[{{Tassoul}(1978)}]{Tassoul1978}
{Tassoul}, J.-L. 1978, {Theory of rotating stars}

\bibitem[{{Taylor}(2006)}]{Taylor2006}
{Taylor}, B.~J. 2006, \aj, 132, 2453

\bibitem[{{Taylor}(2007)}]{Taylor2007}
---. 2007, \aj, 133, 370

\bibitem[{{Thomas}(1967)}]{Thomas1967}
{Thomas}, H.-C. 1967, \zap, 67, 420

\bibitem[{{Thoul} {et~al.}(1994){Thoul}, {Bahcall}, \& {Loeb}}]{Thoul1994}
{Thoul}, A.~A., {Bahcall}, J.~N., \& {Loeb}, A. 1994, \apj, 421, 828

\bibitem[{{Timmes} \& {Swesty}(2000)}]{Timmes2000}
{Timmes}, F.~X., \& {Swesty}, F.~D. 2000, \apjs, 126, 501

\bibitem[{{Tognelli} {et~al.}(2011){Tognelli}, {Prada Moroni}, \&
  {Degl'Innocenti}}]{Tognelli2011}
{Tognelli}, E., {Prada Moroni}, P.~G., \& {Degl'Innocenti}, S. 2011, \aap, 533,
  A109

\bibitem[{{Tonry} {et~al.}(2012){Tonry}, {Stubbs}, {Lykke}, {Doherty},
  {Shivvers}, {Burgett}, {Chambers}, {Hodapp}, {Kaiser}, {Kudritzki},
  {Magnier}, {Morgan}, {Price}, \& {Wainscoat}}]{Tonry2012}
{Tonry}, J.~L., {Stubbs}, C.~W., {Lykke}, K.~R., {et~al.} 2012, \apj, 750, 99

\bibitem[{{Torres}(2013)}]{Torres2013}
{Torres}, G. 2013, Astronomische Nachrichten, 334, 4

\bibitem[{{Torres} {et~al.}(2010){Torres}, {Andersen}, \&
  {Gim{\'e}nez}}]{Torres2010}
{Torres}, G., {Andersen}, J., \& {Gim{\'e}nez}, A. 2010, \aapr, 18, 67

\bibitem[{{Torres} {et~al.}(2013){Torres}, {Ru{\'{\i}}z-Rodr{\'{\i}}guez},
  {Badenas}, {Prato}, {Schaefer}, {Wasserman}, {Mathieu}, \&
  {Latham}}]{Torres2013b}
{Torres}, G., {Ru{\'{\i}}z-Rodr{\'{\i}}guez}, D., {Badenas}, M., {et~al.} 2013,
  \apj, 773, 40

\bibitem[{{Townsend} \& {Teitler}(2013)}]{Townsend2013}
{Townsend}, R.~H.~D., \& {Teitler}, S.~A. 2013, \mnras, 435, 3406

\bibitem[{{Trampedach}(2007)}]{Trampedach2007}
{Trampedach}, R. 2007, in American Institute of Physics Conference Series, Vol.
  948, Unsolved Problems in Stellar Physics: A Conference in Honor of Douglas
  Gough, ed. R.~J. {Stancliffe}, G.~{Houdek}, R.~G. {Martin}, \& C.~A. {Tout},
  141--148

\bibitem[{{Trampedach} {et~al.}(2014){Trampedach}, {Stein},
  {Christensen-Dalsgaard}, {Nordlund}, \& {Asplund}}]{Trampedach2014}
{Trampedach}, R., {Stein}, R.~F., {Christensen-Dalsgaard}, J., {Nordlund},
  {\AA}., \& {Asplund}, M. 2014, \mnras, 445, 4366

\bibitem[{{Traxler} {et~al.}(2011){Traxler}, {Garaud}, \&
  {Stellmach}}]{Traxler2011}
{Traxler}, A., {Garaud}, P., \& {Stellmach}, S. 2011, \apjl, 728, L29

\bibitem[{{Trundle} {et~al.}(2002){Trundle}, {Dufton}, {Lennon}, {Smartt}, \&
  {Urbaneja}}]{Trundle2002}
{Trundle}, C., {Dufton}, P.~L., {Lennon}, D.~J., {Smartt}, S.~J., \&
  {Urbaneja}, M.~A. 2002, \aap, 395, 519

\bibitem[{{Turcotte} {et~al.}(1998){Turcotte}, {Richer}, \&
  {Michaud}}]{Turcotte1998}
{Turcotte}, S., {Richer}, J., \& {Michaud}, G. 1998, \apj, 504, 559

\bibitem[{{ud-Doula} \& {Owocki}(2002)}]{udDoula2002}
{ud-Doula}, A., \& {Owocki}, S.~P. 2002, \apj, 576, 413

\bibitem[{{Ulrich}(1972)}]{Ulrich1972}
{Ulrich}, R.~K. 1972, \apj, 172, 165

\bibitem[{{Ulrich}(1986)}]{Ulrich1986}
---. 1986, \apjl, 306, L37

\bibitem[{{van den Bergh}(1968)}]{vandenBergh1968}
{van den Bergh}, S. 1968, \jrasc, 62, 145

\bibitem[{{van der Hucht}(2001)}]{vanderHucht2001}
{van der Hucht}, K.~A. 2001, NewAR, 45, 135

\bibitem[{{van Saders} \& {Pinsonneault}(2012)}]{vanSaders2012}
{van Saders}, J.~L., \& {Pinsonneault}, M.~H. 2012, \apj, 751, 98

\bibitem[{{VandenBerg} {et~al.}(2006){VandenBerg}, {Bergbusch}, \&
  {Dowler}}]{VandenBerg2006}
{VandenBerg}, D.~A., {Bergbusch}, P.~A., \& {Dowler}, P.~D. 2006, \apjs, 162,
  375

\bibitem[{{Vardya}(1985)}]{Vardya1985}
{Vardya}, M.~S. 1985, \apj, 299, 255

\bibitem[{{Varenne} \& {Monier}(1999)}]{Varenne1999}
{Varenne}, O., \& {Monier}, R. 1999, \aap, 351, 247

\bibitem[{{Vassiliadis} \& {Wood}(1993)}]{Vassiliadis1993}
{Vassiliadis}, E., \& {Wood}, P.~R. 1993, \apj, 413, 641

\bibitem[{{Vassiliadis} \& {Wood}(1994)}]{Vassiliadis1994}
---. 1994, \apjs, 92, 125

\bibitem[{{Vauclair}(1983)}]{Vauclair1983}
{Vauclair}, S. 1983, in Saas-Fee Advanced Course 13: Astrophysical Processes in
  Upper Main Sequence Stars, ed. A.~N. {Cox}, S.~{Vauclair}, \& J.~P. {Zahn},
  167

\bibitem[{{Venn} {et~al.}(2000){Venn}, {McCarthy}, {Lennon}, {Przybilla},
  {Kudritzki}, \& {Lemke}}]{Venn2000}
{Venn}, K.~A., {McCarthy}, J.~K., {Lennon}, D.~J., {et~al.} 2000, \apj, 541,
  610

\bibitem[{{Villamariz} \& {Herrero}(2005)}]{Villamariz2005}
{Villamariz}, M.~R., \& {Herrero}, A. 2005, \aap, 442, 263

\bibitem[{{Villamariz} {et~al.}(2002){Villamariz}, {Herrero}, {Becker}, \&
  {Butler}}]{Villamariz2002}
{Villamariz}, M.~R., {Herrero}, A., {Becker}, S.~R., \& {Butler}, K. 2002,
  \aap, 388, 940

\bibitem[{{Villanova} {et~al.}(2013){Villanova}, {Geisler}, {Carraro}, {Moni
  Bidin}, \& {Mu{\~n}oz}}]{Villanova2013}
{Villanova}, S., {Geisler}, D., {Carraro}, G., {Moni Bidin}, C., \&
  {Mu{\~n}oz}, C. 2013, \apj, 778, 186

\bibitem[{{Villaume} {et~al.}(2015){Villaume}, {Conroy}, \&
  {Johnson}}]{Villaume2015}
{Villaume}, A., {Conroy}, C., \& {Johnson}, B.~D. 2015, \apj, 806, 82

\bibitem[{{Vink} \& {de Koter}(2005)}]{Vink2005}
{Vink}, J.~S., \& {de Koter}, A. 2005, \aap, 442, 587

\bibitem[{{Vink} {et~al.}(2000){Vink}, {de Koter}, \& {Lamers}}]{Vink2000}
{Vink}, J.~S., {de Koter}, A., \& {Lamers}, H.~J.~G.~L.~M. 2000, \aap, 362, 295

\bibitem[{{Vink} {et~al.}(2001){Vink}, {de Koter}, \& {Lamers}}]{Vink2001}
---. 2001, \aap, 369, 574

\bibitem[{{Vink} {et~al.}(2011){Vink}, {Muijres}, {Anthonisse}, {de Koter},
  {Gr{\"a}fener}, \& {Langer}}]{Vink2011}
{Vink}, J.~S., {Muijres}, L.~E., {Anthonisse}, B., {et~al.} 2011, \aap, 531,
  A132

\bibitem[{{Wachlin} {et~al.}(2011){Wachlin}, {Miller Bertolami}, \&
  {Althaus}}]{Wachlin2011}
{Wachlin}, F.~C., {Miller Bertolami}, M.~M., \& {Althaus}, L.~G. 2011, \aap,
  533, A139

\bibitem[{{Wachlin} {et~al.}(2014){Wachlin}, {Vauclair}, \&
  {Althaus}}]{Wachlin2014}
{Wachlin}, F.~C., {Vauclair}, S., \& {Althaus}, L.~G. 2014, \aap, 570, A58

\bibitem[{{Wagenhuber} \& {Groenewegen}(1998)}]{Wagenhuber1998}
{Wagenhuber}, J., \& {Groenewegen}, M.~A.~T. 1998, \aap, 340, 183

\bibitem[{{Walker}(1964)}]{Walker1964}
{Walker}, M.~F. 1964, \aj, 69, 744

\bibitem[{{Wang} {et~al.}(2014){Wang}, {Chen}, {Lin}, {Pandey}, {Huang},
  {Panwar}, {Lee}, {Tsai}, {Tang}, {Goldman}, {Burgett}, {Chambers}, {Draper},
  {Flewelling}, {Grav}, {Heasley}, {Hodapp}, {Huber}, {Jedicke}, {Kaiser},
  {Kudritzki}, {Luppino}, {Lupton}, {Magnier}, {Metcalfe}, {Monet}, {Morgan},
  {Onaka}, {Price}, {Stubbs}, {Sweeney}, {Tonry}, {Wainscoat}, \&
  {Waters}}]{Wang2014}
{Wang}, P.~F., {Chen}, W.~P., {Lin}, C.~C., {et~al.} 2014, \apj, 784, 57

\bibitem[{{Weber} \& {Davis}(1967)}]{Weber1967}
{Weber}, E.~J., \& {Davis}, Jr., L. 1967, \apj, 148, 217

\bibitem[{{Weidemann}(1977)}]{Weidemann1977}
{Weidemann}, V. 1977, \aap, 59, 411

\bibitem[{{Weidemann}(2000)}]{Weidemann2000}
---. 2000, \aap, 363, 647

\bibitem[{{Weiss} \& {Ferguson}(2009)}]{Weiss2009}
{Weiss}, A., \& {Ferguson}, J.~W. 2009, \aap, 508, 1343

\bibitem[{{Weiss} \& {Schlattl}(2008)}]{Weiss2008}
{Weiss}, A., \& {Schlattl}, H. 2008, \apss, 316, 99

\bibitem[{{Wheeler} {et~al.}(2015){Wheeler}, {Kagan}, \&
  {Chatzopoulos}}]{Wheeler2015}
{Wheeler}, J.~C., {Kagan}, D., \& {Chatzopoulos}, E. 2015, \apj, 799, 85

\bibitem[{{Williams} \& {Bolte}(2007)}]{Williams2007}
{Williams}, K.~A., \& {Bolte}, M. 2007, \aj, 133, 1490

\bibitem[{{Williams} {et~al.}(2009){Williams}, {Bolte}, \&
  {Koester}}]{Williams2009}
{Williams}, K.~A., {Bolte}, M., \& {Koester}, D. 2009, \apj, 693, 355

\bibitem[{{Willson}(2000)}]{Willson2000}
{Willson}, L.~A. 2000, \araa, 38, 573

\bibitem[{{Wolff} \& {Simon}(1997)}]{Wolff1997}
{Wolff}, S., \& {Simon}, T. 1997, \pasp, 109, 759

\bibitem[{{Wood} {et~al.}(2013){Wood}, {Garaud}, \& {Stellmach}}]{Wood2013}
{Wood}, T.~S., {Garaud}, P., \& {Stellmach}, S. 2013, \apj, 768, 157

\bibitem[{{Woodward} {et~al.}(2015){Woodward}, {Herwig}, \&
  {Lin}}]{Woodward2015}
{Woodward}, P.~R., {Herwig}, F., \& {Lin}, P.-H. 2015, \apj, 798, 49

\bibitem[{{Woosley} \& {Heger}(2006)}]{Woosley2006}
{Woosley}, S.~E., \& {Heger}, A. 2006, \apj, 637, 914

\bibitem[{{Wright} {et~al.}(2010){Wright}, {Eisenhardt}, {Mainzer}, {Ressler},
  {Cutri}, {Jarrett}, {Kirkpatrick}, {Padgett}, {McMillan}, {Skrutskie},
  {Stanford}, {Cohen}, {Walker}, {Mather}, {Leisawitz}, {Gautier}, {McLean},
  {Benford}, {Lonsdale}, {Blain}, {Mendez}, {Irace}, {Duval}, {Liu}, {Royer},
  {Heinrichsen}, {Howard}, {Shannon}, {Kendall}, {Walsh}, {Larsen}, {Cardon},
  {Schick}, {Schwalm}, {Abid}, {Fabinsky}, {Naes}, \& {Tsai}}]{Wright2010}
{Wright}, E.~L., {Eisenhardt}, P.~R.~M., {Mainzer}, A.~K., {et~al.} 2010, \aj,
  140, 1868

\bibitem[{{Yi} {et~al.}(2001){Yi}, {Demarque}, {Kim}, {Lee}, {Ree}, {Lejeune},
  \& {Barnes}}]{Yi2001}
{Yi}, S., {Demarque}, P., {Kim}, Y.-C., {et~al.} 2001, \apjs, 136, 417

\bibitem[{{Yi} {et~al.}(2003){Yi}, {Kim}, \& {Demarque}}]{Yi2003}
{Yi}, S.~K., {Kim}, Y.-C., \& {Demarque}, P. 2003, \apjs, 144, 259

\bibitem[{{Yoon} \& {Cantiello}(2010)}]{Yoon2010}
{Yoon}, S.-C., \& {Cantiello}, M. 2010, \apjl, 717, L62

\bibitem[{{Yoon} \& {Langer}(2005)}]{Yoon2005}
{Yoon}, S.-C., \& {Langer}, N. 2005, \aap, 443, 643

\bibitem[{{Yoon} {et~al.}(2006){Yoon}, {Langer}, \& {Norman}}]{Yoon2006}
{Yoon}, S.-C., {Langer}, N., \& {Norman}, C. 2006, \aap, 460, 199

\bibitem[{{Young} \& {Arnett}(2005)}]{Young2005}
{Young}, P.~A., \& {Arnett}, D. 2005, \apj, 618, 908

\bibitem[{{Yurchenko} {et~al.}(2011){Yurchenko}, {Barber}, \&
  {Tennyson}}]{Yurchenko2011}
{Yurchenko}, S.~N., {Barber}, R.~J., \& {Tennyson}, J. 2011, \mnras, 413, 1828

\bibitem[{{Zahn}(1983)}]{Zahn1983}
{Zahn}, J.-P. 1983, in Saas-Fee Advanced Course 13: Astrophysical Processes in
  Upper Main Sequence Stars, ed. A.~N. {Cox}, S.~{Vauclair}, \& J.~P. {Zahn},
  253

\bibitem[{{Zahn}(1992)}]{Zahn1992}
{Zahn}, J.-P. 1992, \aap, 265, 115

\bibitem[{{Zahn} {et~al.}(2007){Zahn}, {Brun}, \& {Mathis}}]{Zahn2007}
{Zahn}, J.-P., {Brun}, A.~S., \& {Mathis}, S. 2007, \aap, 474, 145

\bibitem[{{Zaritsky} {et~al.}(2002){Zaritsky}, {Harris}, {Thompson}, {Grebel},
  \& {Massey}}]{Zaritsky2002}
{Zaritsky}, D., {Harris}, J., {Thompson}, I.~B., {Grebel}, E.~K., \& {Massey},
  P. 2002, \aj, 123, 855

\bibitem[{{Zasowski} {et~al.}(2013){Zasowski}, {Johnson}, {Frinchaboy},
  {Majewski}, {Nidever}, {Rocha Pinto}, {Girardi}, {Andrews}, {Chojnowski},
  {Cudworth}, {Jackson}, {Munn}, {Skrutskie}, {Beaton}, {Blake}, {Covey},
  {Deshpande}, {Epstein}, {Fabbian}, {Fleming}, {Garcia Hernandez}, {Herrero},
  {Mahadevan}, {M{\'e}sz{\'a}ros}, {Schultheis}, {Sellgren}, {Terrien}, {van
  Saders}, {Allende Prieto}, {Bizyaev}, {Burton}, {Cunha}, {da Costa},
  {Hasselquist}, {Hearty}, {Holtzman}, {Garc{\'{\i}}a P{\'e}rez}, {Maia},
  {O'Connell}, {O'Donnell}, {Pinsonneault}, {Santiago}, {Schiavon}, {Shetrone},
  {Smith}, \& {Wilson}}]{Zasowski2013}
{Zasowski}, G., {Johnson}, J.~A., {Frinchaboy}, P.~M., {et~al.} 2013, \aj, 146,
  81

\bibitem[{{Zhang} {et~al.}(2013){Zhang}, {Li}, {Han}, {Zhuang}, \&
  {Kang}}]{Zhang2013}
{Zhang}, F., {Li}, L., {Han}, Z., {Zhuang}, Y., \& {Kang}, X. 2013, \mnras,
  428, 3390

\bibitem[{{Zhao} {et~al.}(2012){Zhao}, {Oswalt}, {Willson}, {Wang}, \&
  {Zhao}}]{Zhao2012}
{Zhao}, J.~K., {Oswalt}, T.~D., {Willson}, L.~A., {Wang}, Q., \& {Zhao}, G.
  2012, \apj, 746, 144

\bibitem[{{Zurita} \& {Bresolin}(2012)}]{Zurita2012}
{Zurita}, A., \& {Bresolin}, F. 2012, \mnras, 427, 1463

\end{thebibliography}

\end{document}